\pdfoutput=1
\documentclass[11pt,twoside,a4paper,cmspaper,final,collab]{cms-tdr}
\def\svnVersion{722ba7f}\def\svnDate{2025/07/17}\def\cmsCernNoTag{CERN-EP-2025-043}\def\cmsCernDate{\today}\def\cmsMessage{Published in the Journal of High Energy Physics as \href{http://dx.doi.org/10.1007/JHEP07(2025)118}{\doi{10.1007/JHEP07(2025)118}.}}
\begin{document}\cmsNoteHeader{HIN-23-010}

\cmsNoteHeader{HIN-23-010}
\title{Search for jet quenching with dijets from high-multiplicity pPb collisions at \texorpdfstring{$\sqrtsNN = 8.16\TeV$}{sqrt(s[NN]) = 8.16 TeV}}

\date{\today}
\newcommand{\Ntroff}{\ensuremath{N_{\text{trk}}^{\text{offline}}}\xspace}
\newcommand{\etalab}{\ensuremath{\eta_{\text{lab}}}\xspace}
\newcommand{\etacm}{\ensuremath{\eta_{\text{CM}}}\xspace}
\newcommand{\xj}{\ensuremath{x_{j}}\xspace}
\newcommand{\jone}{\ensuremath{j_{1}}\xspace}
\newcommand{\jtwo}{\ensuremath{j_{2}}\xspace}
\newcommand{\x}{\ensuremath{x}\xspace}
\newcommand{\xp}{\ensuremath{x_{\Pp}}\xspace}
\newcommand{\xpb}{\ensuremath{x_{\mathrm{Pb}}}\xspace}
\newcommand{\xja}{\ensuremath{\langle x_{j} \rangle}\xspace}
\newcommand{\EPOS}{\textsc{epos lhc}\xspace}
\newcommand{\roots}{\ensuremath{\sqrt{s}}\xspace}
\newcommand{\ppb}{\ensuremath{\Pp\mathrm{Pb}}\xspace}
\newcommand{\pp}{\ensuremath{\Pp\Pp}\xspace}

\abstract{
The first measurement of the dijet transverse momentum balance \xj in proton-lead (\ppb) collisions at a nucleon-nucleon center-of-mass energy of $\sqrtsNN = 8.16\TeV$ is presented. The \xj observable, defined as the ratio of the subleading over leading jet transverse momentum in a dijet pair, is used to search for jet quenching effects. The data, corresponding to an integrated luminosity of 174.6\nbinv, were collected with the CMS detector in 2016. The \xj distributions and their average values are studied as functions of the charged-particle multiplicity of the events and for various dijet rapidity selections. The latter enables probing hard scattering of partons carrying distinct nucleon momentum fractions \x in the proton- and lead-going directions. The former, aided by the high-multiplicity triggers, allows probing for potential jet quenching effects in high-multiplicity events (with up to 400 charged particles), for which collective phenomena consistent with quark-gluon plasma (QGP) droplet formation were previously observed. The ratios of \xj distributions for high- to low-multiplicity events are used to quantify the possible medium effects. These ratios are consistent with simulations of the hard-scattering process that do not include QGP production. These measurements set an upper limit on medium-induced energy loss of the subleading jet of 1.26\% of its transverse momentum at the 90\% confidence level in high multiplicity \ppb events.}

\hypersetup{
pdfauthor={CMS Collaboration},
pdftitle={Search for jet quenching with dijets from high-multiplicity pPb collisions at sqrt(s[NN]) = 8.16 TeV},
pdfsubject={CMS},
pdfkeywords={CMS, dijets in pPb}}

\maketitle 

\section{Introduction}
\label{intro}

High-energy jets consist of particles produced by the fragmentation and hadronization of partons (quarks and gluons) that have undergone hard scattering in the early stages of a collision. Jets have been successfully employed to probe the properties of the quark-gluon plasma (QGP), a unique form of hot and dense matter formed in high-energy heavy ion collisions~\cite{bjorkenquench}. One of the most recognized signatures of the QGP is the jet quenching phenomenon, which refers to the suppression or modification of the energy and shape of jets caused by partonic interaction in the hot QGP medium. The presence of jet quenching has been demonstrated using a wide range of measurements in lead-lead (PbPb) and gold-gold (AuAu) collisions~\cite{Chatrchyan:2011sx, ATLAS:2010isq, STAR:2017hhs, STAR:2017ieb, CMS:2018dqf, CMS:2024krd}. A recent review of CMS jet quenching measurements is presented in Ref.~\cite{CMS:2024krd}.

More recently, possible indications of QGP production have been observed in small colliding systems. For example, particle collectivity attributed to hydrodynamic flow within a QGP has been measured in high-multiplicity proton-proton (\pp) and proton-lead (\ppb) collisions at the CERN LHC~\cite{CMS:2015fgy, Khachatryan:2016txc, CMS:2012qk}, as well as in pAu, deuteron-Au, and helium-Au collisions at the BNL RHIC~\cite{PHENIX:2018lia}. In addition, strangeness enhancement, another QGP signature, has been observed by ALICE in \pp and \ppb collisions~\cite{ALICE:2016fzo}. However, the observations of jet quenching, which is expected in a hot and dense medium, remain elusive in such small systems~\cite{CMS:2014qvs,Adare:2015gla,CMS:2016svx,Acharya:2017okq,ATLAS:2022iyq,STAR:2024nwm}.

Motivated by the measurements of dijet transverse momentum (\pt) balance in PbPb collisions~\cite{Chatrchyan:2011sx,CMS:2018dqf}, this analysis aims to study the same quantity in different multiplicity classes of the \ppb data sample. The study utilizes CMS high-multiplicity events to probe different pseudorapidity regions in the center-of-mass frame (\etacm) of the collision, allowing investigations of possible jet quenching effects for the hard scattering processes initiated by partons carrying different nucleon momentum fraction \x in the $\Pp$- and Pb-going directions. The data, collected in 2016 during LHC Run2 at a center-of-mass energy of 8.16\TeV, correspond to a total integrated luminosity of 174.6\nbinv~\cite{CMS:LUM-17-003, CMS:LUM-17-004, CMS:LUM-18-002}. During part of the data-taking period, the directions of the proton and lead beams were reversed. For consistency, positive \etalab is defined to always point in the direction of the proton beam. Because the two beams have different energy per nucleon, a massless particle emitted at $\etacm = 0$ in the nucleon-nucleon center-of-mass frame will be detected at $\etalab = +0.465$ in the laboratory frame.

This work presents searches for jet quenching signals using dijets, which are composed of the two highest energy, back-to-back jets in the event.
The observable under study, dijet \pt balance (\xj), is defined as a ratio of transverse momenta of subleading (less energetic) to leading (more energetic) jets in a dijet.
The analysis is performed by selecting leading and subleading jets with transverse momenta of at least 100\GeV and 50\GeV, respectively, and in different kinematic regions (detailed below) to isolate mid-, forward, and backward rapidities.
The measured \xj distributions for events with different charged-particle multiplicities are compared with Monte Carlo (MC) simulations of the hard scattering process without quenching effects.
Tabulated results are provided in the HEPData record for this analysis~\cite{hepdata}.

\section{The CMS detector and event reconstruction}

The central feature of the CMS apparatus is a superconducting solenoid with a 6\unit{m} internal diameter, providing a magnetic field of 3.8\unit{T}. Within the solenoid volume are the silicon pixel and strip trackers, a lead tungstate crystal electromagnetic calorimeter, and a brass and scintillator hadron calorimeter, each composed of barrel and two endcap sections. The CMS silicon tracker measures charged-particle tracks within $\abs{\eta} < 2.5$. It consists of 1,856 silicon pixel and 15,148 silicon strip detector modules. Two hadron forward (HF) steel and quartz-fiber calorimeters complement the barrel and endcap detectors, extending the calorimeter coverage from the $\abs{\eta} < 3.0$ range provided by the barrel and endcap out to $\abs{\eta} < 5.2$. The HF calorimeters are subdivided in azimuth ($\phi$) into $20^{\circ}$ modular wedges and further segmented to form $0.175\times0.175$ radians ($\Delta\eta\times\Delta\phi$) ``towers''. Muons are measured in the range $\abs{\eta} < 2.4$, with detection planes located outside the solenoid core. A detailed description of the CMS detector, along with a definition of the coordinate system used and the relevant kinematic variables, can be found in Ref.~\cite{Collaboration_2008}.

A particle-flow algorithm~\cite{Sirunyan_2017} aims to reconstruct and identify each individual particle in an event with an optimized combination of information from the various elements of the CMS detector. The energy of photons is obtained from the ECAL measurement. The energy of an electron is determined from a combination of the electron momentum at the primary interaction vertex as determined by the tracker, the energy of the corresponding ECAL cluster, and the energy sum of all bremsstrahlung photons spatially compatible with originating from the electron track. The energy of a muon is obtained from the curvature of the corresponding track. The energy of a charged hadron is determined from a combination of the momentum measured in the tracker and the matching ECAL and HCAL energy deposits, corrected for the response function of the calorimeters to hadronic showers. Finally, the energy of neutral hadrons is obtained from the corresponding corrected ECAL and HCAL energies.

Events are selected using a two-tiered trigger system. The first level (L1), composed of custom hardware processors, uses information from the calorimeters and muon detectors to select events at a rate of around 100\unit{kHz}~\cite{CMS:2020cmk}. The second level, known as the high-level trigger (HLT), consists of a farm of processors running a version of the full event reconstruction software optimized for fast processing, reducing the event rate to around 1\unit{kHz} in \pp collisions~\cite{CMS:2016ngn} and up to 31\unit{kHz} and \ppb collisions, before data storage.

\section{Event selection}

The triggers and event selections are similar to those described in Refs.~\cite{CMS:2017lrw,CMS:2017kcs}. Minimum bias (MB) events are triggered by requiring energy deposits in at least one of the two HF calorimeters above a threshold of approximately 1\GeV and the presence of at least one track with $\pt > 0.4\GeV$ in the pixel tracker. Because of the high collision rate, the integrated luminosity collected by this trigger in 2016 data-taking was approximately 5.8\nbinv.

To ensure the collection of a large sample of high-multiplicity (HM) \ppb collisions, a dedicated trigger was implemented using both the CMS L1 and HLT systems. At L1, the total number of towers in the electromagnetic and hadronic calorimeters above a transverse energy threshold of 0.5\GeV is required to be greater than 120 and 150 towers, respectively. Track reconstruction is performed online as part of the HLT trigger with the identical reconstruction algorithm used offline~\cite{Chatrchyan:2014fea}. For each event, the reconstructed vertex with the highest number of associated tracks is selected as the primary vertex. The number of tracks with $\abs{\etalab} < 2.4$, $\pt > 0.4~\GeV$, and a distance of closest approach less than 0.12\unit{cm} to the primary vertex is determined for each event and is required to exceed a certain threshold to enrich the sample with HM events. This analysis employs two distinct HM triggers designed to select events with charged-particle multiplicities exceeding 185 and 250 with approximate integrated luminosities of 96.2\nbinv and 174.6\nbinv, respectively.

Events are required to have at least one interaction vertex, reconstructed from two or more tracks, within 15\unit{cm} of the nominal interaction point along the beam axis and 0.2\unit{cm} in the transverse direction. A requirement of at least one calorimeter tower with energy above 3\GeV on each side of the interaction point in an HF calorimeter is imposed, effectively removing ultraperipheral events and electromagnetic interactions. Events with exceptionally high occupancy in the tracker without a common vertex are removed. Additionally, selections are applied to remove events with more than one interaction per bunch crossing (pileup)~\cite{CMS:2017lrw, CMS:2017kcs}. The average pileup probability, defining the chance to have more than one interaction recorded in a minimum bias event, in \ppb collisions at 8.16\TeV varied between 10--25\%, depending on the LHC operation conditions. Rejecting those events is essential to reduce possible contamination that can distort the jet energy, momentum, and shape, as well as introduce misidentified jets from random energy deposits.

For the charged-particle multiplicity (\Ntroff), high-purity tracks~\cite{Chatrchyan:2014fea} are utilized, where \Ntroff represents the number of offline-selected tracks~\cite{Khachatryan:2016txc,Chatrchyan:2014fea} with $\abs{\eta} < 2.4$ and $\pt > 0.4\GeV$. Additionally, a reconstructed track is considered a primary-track candidate if the impact parameter significance in the plane transverse to the beams ($d_{\mathrm{xy}}/\sigma(d_{{\mathrm{xy}}})$) and the significance of the separation in $z$ between the track and the best reconstructed primary vertex ($d_{\mathrm{z}}/\sigma(d_{\mathrm{z}})$) are both less than 3. The primary vertex is defined as the vertex associated with the largest number of tracks. Tracks with poor momentum resolution are removed by requiring that the relative \pt uncertainty ($\sigma(\pt)/\pt$) is less than 10\%. Primary tracks that fall within the kinematic range $\abs{\eta}<2.4$ and $\pt > 0.4\GeV$ are selected in the analysis to ensure reasonable tracking efficiency ($\sim$90\%) and a low misreconstruction rate ($<$2\%).

Simulated events are employed in this analysis for cross-checks and to determine jet reconstruction performance, event selection, and other potential detector effects. Embedded {{\PYTHIA}8}+\EPOS~\cite{Sjostrand:2014zea,PhysRevC.92.034906} event generators are used, where {\PYTHIA}8 (version 212) with the CUETP8M1 tune~\cite{CMS:2015wcf} is used to model the hard scattering, parton showering, and hadronization processes for the production of the dijets of interest, while \EPOS simulates the soft interactions and underlying event background. Additionally, \GEANTfour~\cite{Agostinelli:2002hh} is used for detailed simulation of the CMS detector response.

\section{Analysis procedure}

The hadronic jets are reconstructed using the anti-\kt sequential recombination algorithm~\cite{Cacciari:2008gp} with a distance parameter of $R=0.4$, provided in the \FASTJET framework~\cite{Cacciari:2011ma}, clustered from all particles reconstructed by the PF algorithm. An underlying event (UE)-subtraction method~\cite{Berta:2014eza}, called constituent subtraction, is applied to remove contributions from the UE soft fragments (constituents) within the jets. Jet energy corrections~\cite{Khachatryan_2017} are derived from simulation studies to ensure that the average energy of reconstructed jets matches that of particle-level jets, which are defined as jets clustered from all stable particles produced in the collision, excluding neutrinos. In situ measurements of the momentum balance in dijet, \GAMJET, \PZ{+}jet, and multijet events from \pp data are used to account for any residual differences in the jet energy scale (JES) between data and simulation. The jet energy resolution (JER) typically amounts to 15--20\% at 30\GeV, 10\% at 100\GeV, and less than 7\% at 400\GeV~\cite{Khachatryan_2017}.

The final-state observable is the dijet momentum balance, defined as the \pt ratio between the leading \jone and subleading jets \jtwo,
\begin{equation}
\xj= \frac{\pt^{\jone}}{\pt^{\jtwo}},
\end{equation}
where $\pt^{\jone}$ and $\pt^{\jtwo}$ are of the highest and second-highest \pt jets, respectively. This observable is always less than unity because $\pt^{\jone} > \pt^{\jtwo}$. The average dijet momentum balance (\xja) is obtained by taking the mean value of the data. These observables have been frequently employed to explore QGP production in nucleus-nucleus collisions~\cite{CMS:2012ulu,CMS:2018dqf,ATLAS:2022zbu}. In those systems, we observed that the \xj distributions are significantly imbalanced, compared to \pp, shifting toward lower values, indicating that the subleading jet loses more energy as it traverses the medium.

The analysis requires $\pt^{\jone} > 100\GeV$, $\pt^{\jtwo} > 50\GeV$ to guarantee a jet resolution less than 15\% and an absolute azimuthal separation ($\Delta\phi_{\text{dijets}} = \abs{\phi^{\jone} - \phi^{\jtwo}}) > 5\pi/6$ to ensure that the two jets of the pair are back-to-back. In addition, because of the detector acceptance, it is required that $\abs{\etalab} < 3.0$ for both jets. To remove jets potentially dominated by spurious contributions from various subdetector components or reconstruction failures, selection criteria similar to those used Ref.~\cite{CMS:2020ebo} are applied. In the case of dijet events, there is a non-zero possibility that the selection criteria might remove a leading or subleading jet, resulting in the pairing of other jets in the event. To avoid this issue, the entire event is excluded if the leading or subleading jets do not pass the jet selections criteria. Implementing this approach has resulted in the removal of at most 3\% of dijet events.

The analysis is performed by selecting leading and subleading jets in different \etacm configurations to study the forward and backward \etacm regions, probing jets in both forward (proton-going) and backward (Pb-going) directions, respectively. Three different measurement regions are defined: the forward proton direction $(1.2 < \etacm < 2.4)$, the backward lead direction $(-3.3 < \etacm < -1.2)$, and the midrapidity region $(\abs{\etacm}<1)$. The boundaries are chosen because of the detector acceptance and the small gap added is introduced to reduce possible correlations between jets and multiplicity. This analysis requires that at least one jet (leading or subleading) is located in the midrapidity region to ensure a sufficient number of events. As a result, five regions are used in the combinations listed below:
\begin{enumerate}
\item Both the leading and subleading jets are in the midrapidity region.
\item The leading jet is in the midrapidity region and the subleading jet is in the forward region.
\item The leading jet is in the midrapidity region and the subleading jet is in the backward region.
\item The leading jet is in the forward region and the subleading jet is in the midrapidity region.
\item The leading jet is in the backward region and the subleading jet is in the midrapidity region.
\end{enumerate}

Jets traveling in the backward direction must experience effects from the fragmented nuclei, while jets in the forward direction are exposed to the proton remnant interaction. These regions correspond to different $x$ values, with \xpb describing the nucleus-going side and \xp the proton-going side. Jets at backward rapidity probe high-\xpb partons, where initial-state energy loss and multiple scattering dominate, whereas forward jets access low-\xpb partons, where shadowing and gluon saturation effects become relevant. For each of the dijet pseudorapidity combinations, the measurement is performed as a function of the charged \Ntroff in the following ranges: [10, 60), [60, 120), [120, 185), [185, 250), and [250, 400]. The last two ranges require the use of specific HM triggers, while the data for the other ranges are collected with the MB trigger.

Given the limited detector resolution and additional experimental effects, such as reconstruction efficiency and misidentified contributions, the properties of reconstructed jets can differ from those of jets defined at the particle level. This discrepancy leads to a migration of dijet events within the ranges of the final-state observable, \xj. To enable a direct comparison between the measured data and theoretical particle-level calculations or between the measured data and results from other experiments, the effect of these migrations is accounted for as part of the unfolding procedure.

This analysis performs the first unfolding of \xj using data collected by the CMS experiment. The \xj observable is unfolded using a two-dimensional reconstructed-versus-generated response matrix from {{\PYTHIA}8}+\EPOS, built for each possible pseudorapidity combination and in the 10--60, 60--120, and $\geq120$ multiplicity ranges. The response matrix for multiplicity $\geq120$ is utilized for the data ranges 120--185, 185--250, and 250--400. The kinematic selections are applied to reconstructed quantities, while no requirements are applied to matched generator-level jets. A swapping effect, where the leading jet becomes subleading and vice versa, occurs in approximately 20\% of events. This swapping is accounted for in the response matrix by identifying the effect and filling the matrix with $1/\xj$. Generator-level \xj values that satisfy the analysis fiducial region but are not reconstructed because one or both jets do not satisfy the dijet kinematic selection, referred to as missing values, are incorporated into the response matrix using the \textsc{RooUnfold} software package~\cite{Adye:2011gm, Brenner:2019lmf}. The contributions from misidentified jets are negligible. The matrices are reweighted using two-dimensional probability distribution maps of the reconstructed $\pt^{\jone}$ and $\pt^{\jtwo}$ to take into account data and simulation differences. Once the final matrices are produced, the unfolding process employs the iterative method of D'Agostini~\cite{DAGOSTINI1995487,Richardson:72,Lucy:1974yx}, which is implemented in \textsc{RooUnfold} with five nominal iterations.

\section{Systematic uncertainties}

Six different sources of systematic uncertainty are studied in the present analysis for all of the observables and are outlined as follows:

\begin{itemize}
\item \textit{Pileup}: The impact of pileup is evaluated by adjusting the pileup event selection criteria, which consider the distance between reconstructed vertices and the number of associated tracks. The resulting systematic uncertainty is found to be less than 5\%.
\item \textit{JES}: To evaluate the uncertainty from the JES, the jet \pt in the data is shifted up and down according to the JES uncertainties. The sources of uncertainty considered are similar to those described in Ref.~\cite{collaboration_2011}. The resulting systematic errors vary from 0.05--10\% depending on the measurement, pseudorapidity configuration, and multiplicity range. Larger uncertainties are observed at low-\xj values ($<$0.5).
\item \textit{JER}: To evaluate the uncertainty from the JER, the jet \pt resolution is smeared by the estimated data-to-simulation difference. The amount of smearing depends on $\eta$ and is applied to account for differences between data and simulation. The difference from the nominal distribution is taken as the systematic uncertainty. The variation observed is between 0.5--10\% similarly to the JES uncertainty.
\item \textit{Underlying events}: As an alternative to the nominal method for correcting the underlying event contribution in jets, a jet-area-based background subtraction method, as described in Ref.~\cite{CMS:2021vui}, is used. The difference between the two methods is less than 1\% and is taken as a systematic uncertainty.
\item \textit{Trigger bias}: For the multiplicity dependency study, trigger bias is estimated using an approach that selects events in data from the multiplicity overlap region of the MB and HM triggers. Good agreement in \xj between MB and HM triggers in the same multiplicity range was observed, with deviations of 0.5\% for 185--250 and 2.0\% for 250--400 multiplicity ranges.
\item \textit{Unfolding}: Five iterations were used to perform the unfolding. The results were compared with those obtained using four and six iterations, and the difference of the level of 1\% was taken as the systematic uncertainty.
\end{itemize}

All systematic uncertainties were evaluated using a bin-by-bin approach for each observable, and the uncertainties were propagated to the final result. The dominant sources of systematic uncertainty are the JES and JER. All uncertainties are assumed to be uncorrelated and added in quadrature to calculate the total systematic uncertainty, resulting in values between 0.5--12\%.

\section{Results}

\begin{figure}[!hbpt]
  \centering
    \includegraphics[width=0.44\textwidth]{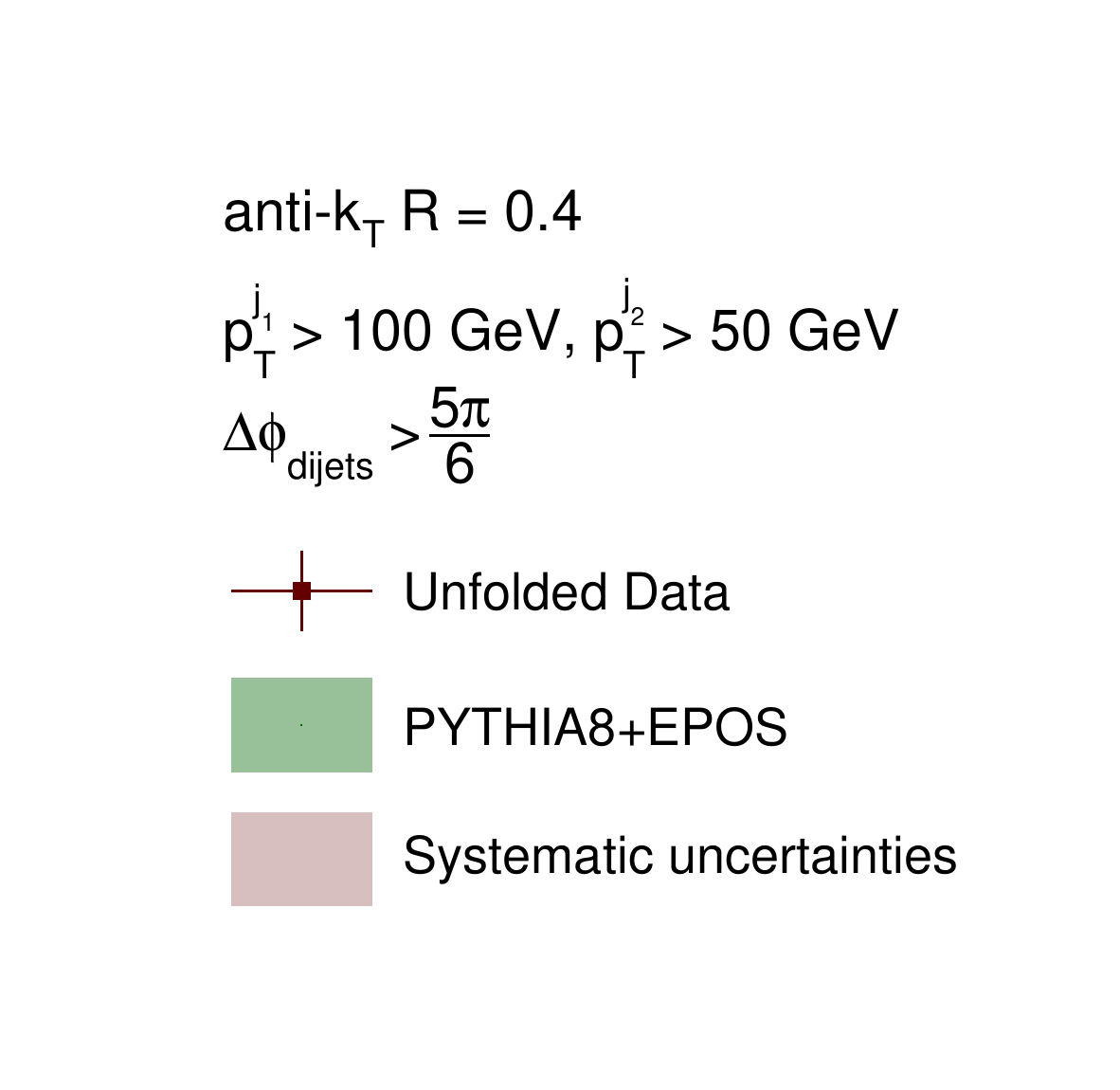} 
    \includegraphics[width=0.44\textwidth]{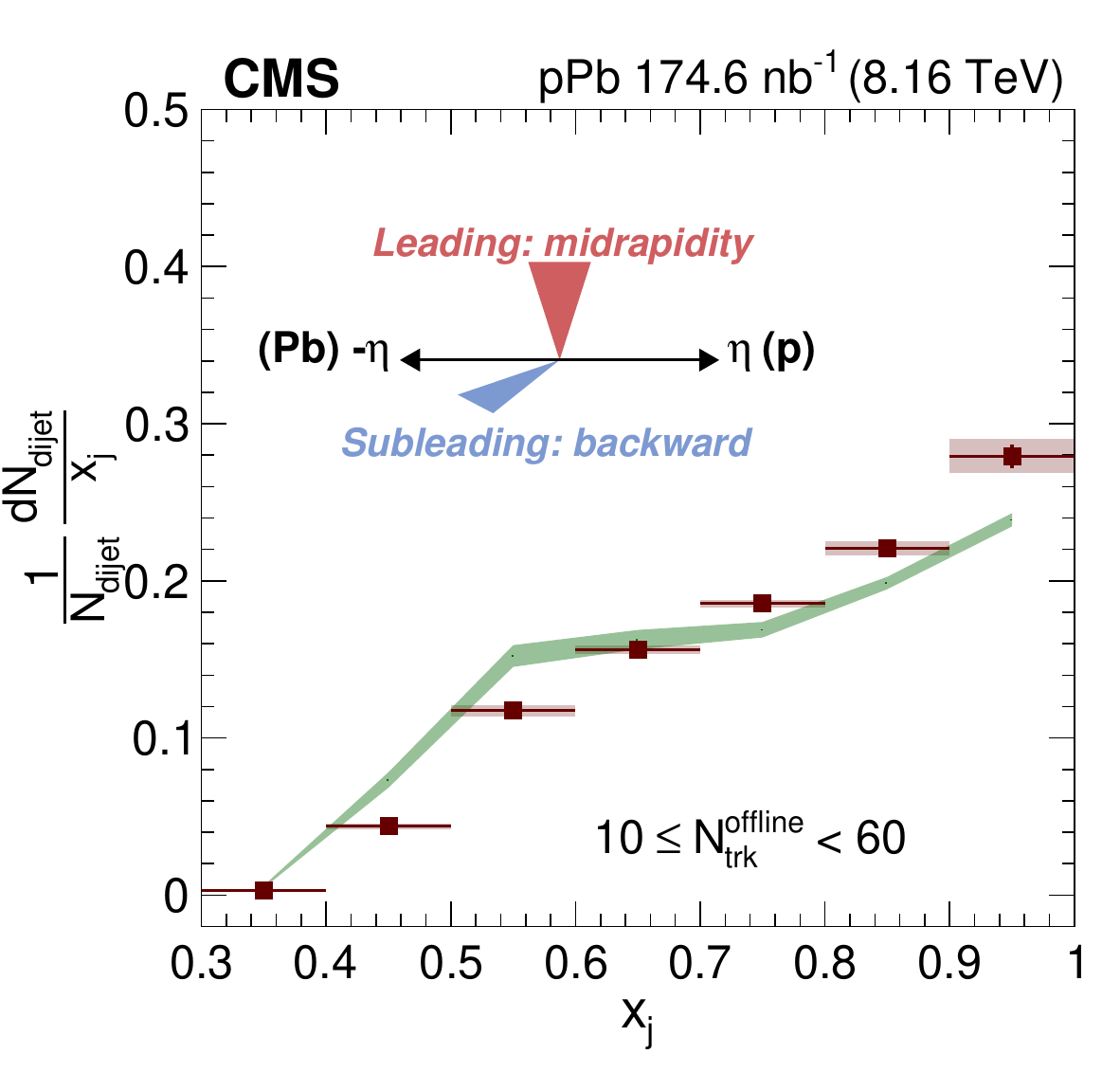} \\
    \includegraphics[width=0.44\textwidth]{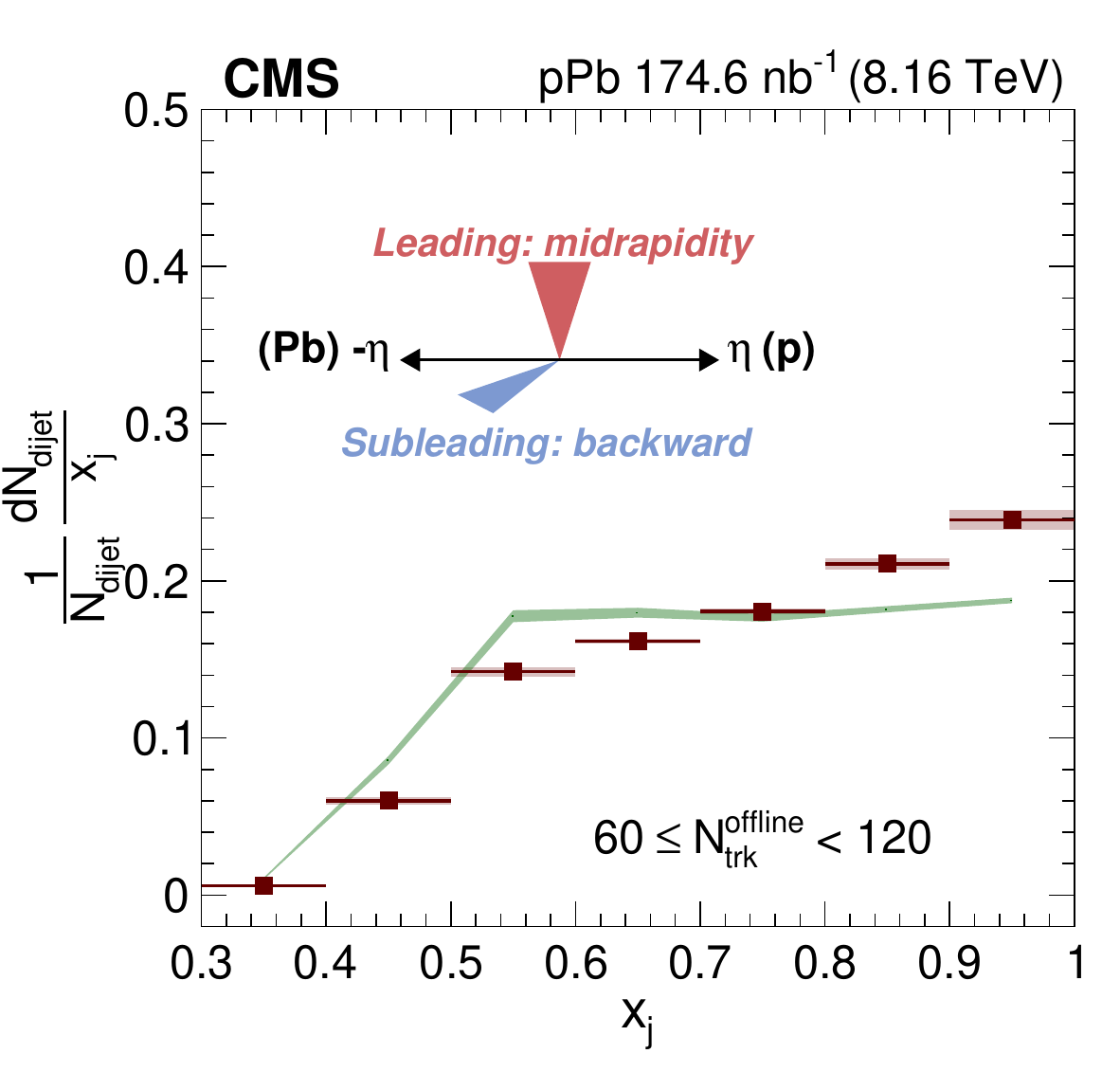} 
    \includegraphics[width=0.44\textwidth]{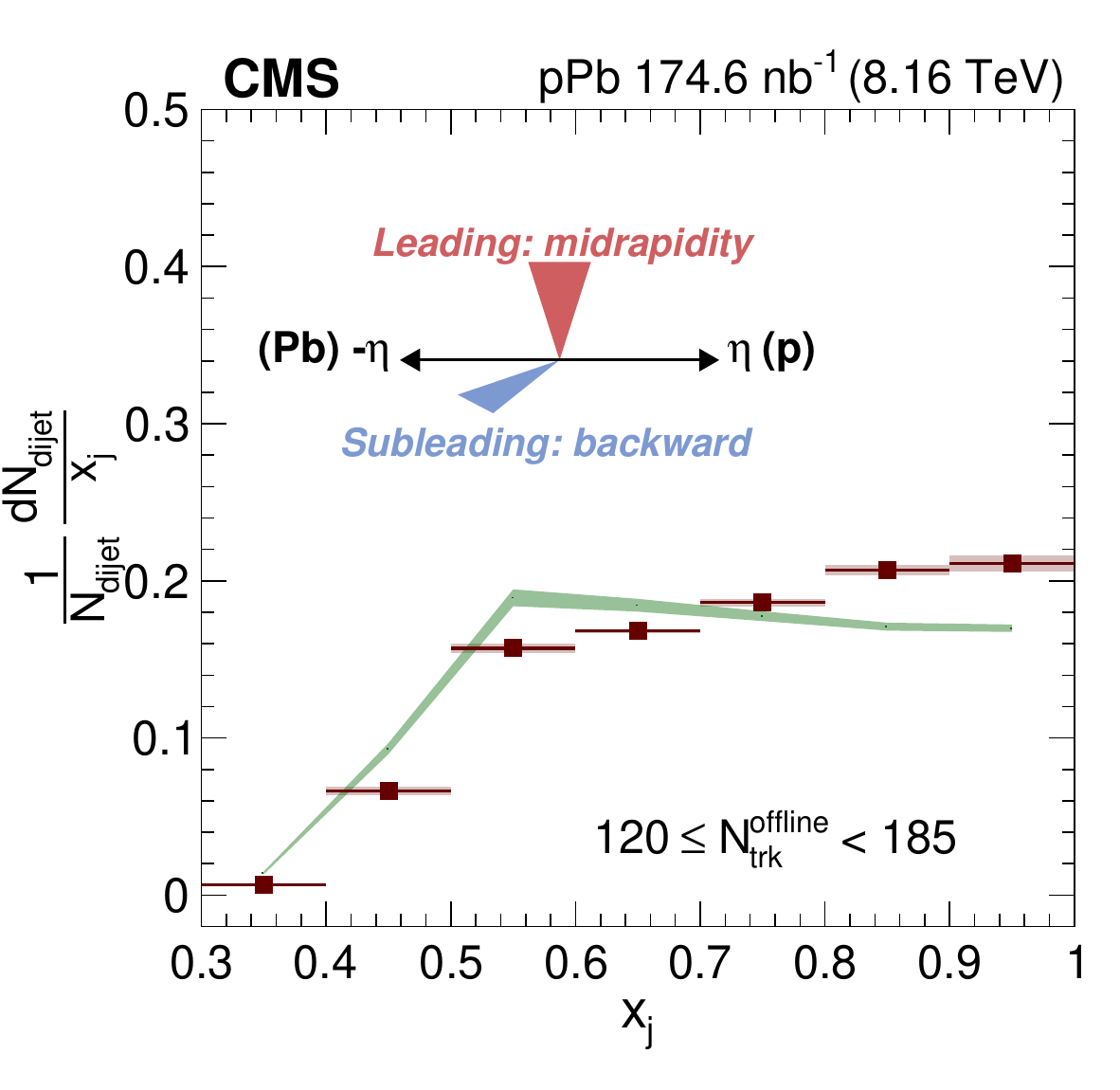} \\
    \includegraphics[width=0.44\textwidth]{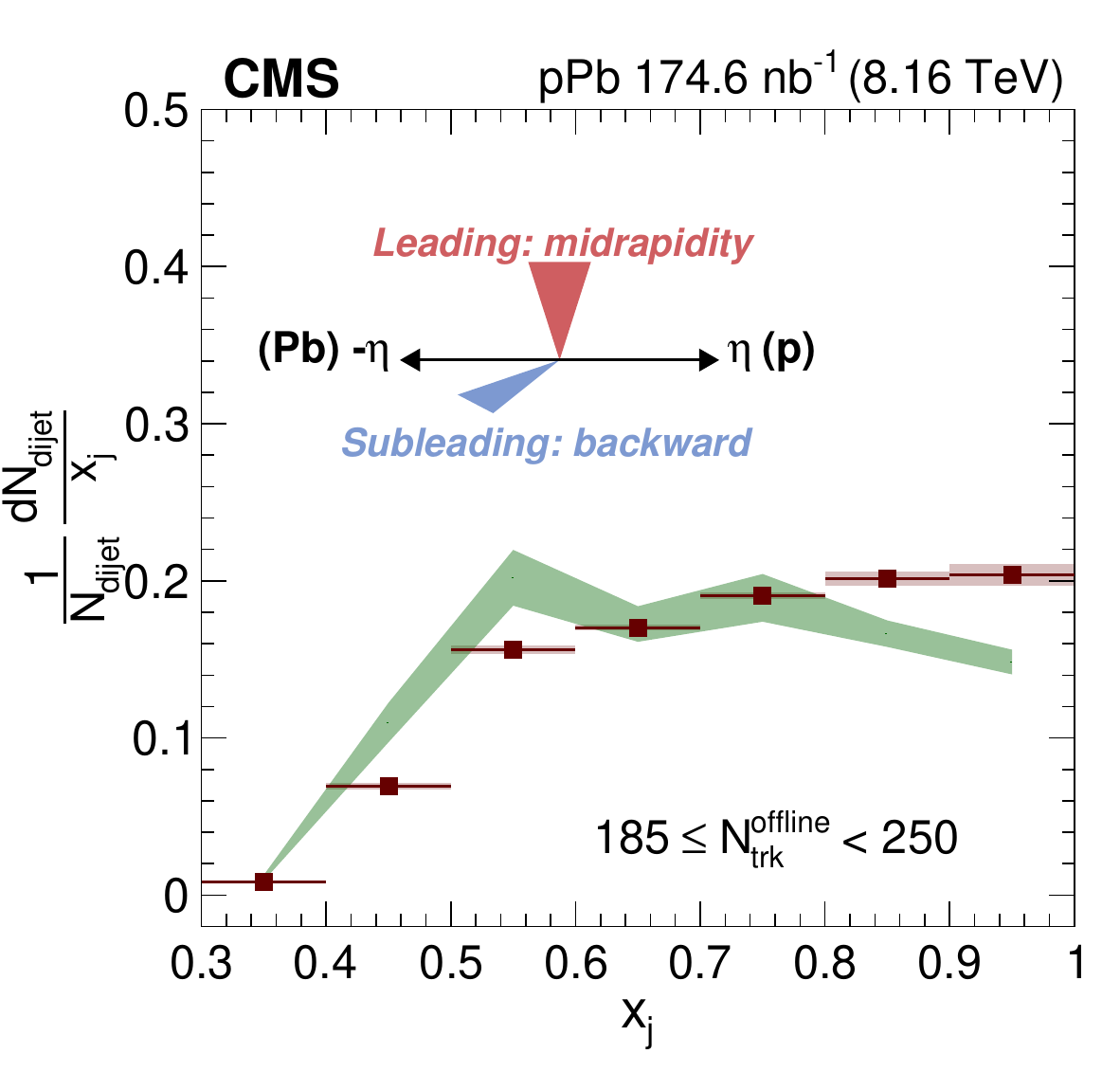} 
    \includegraphics[width=0.44\textwidth]{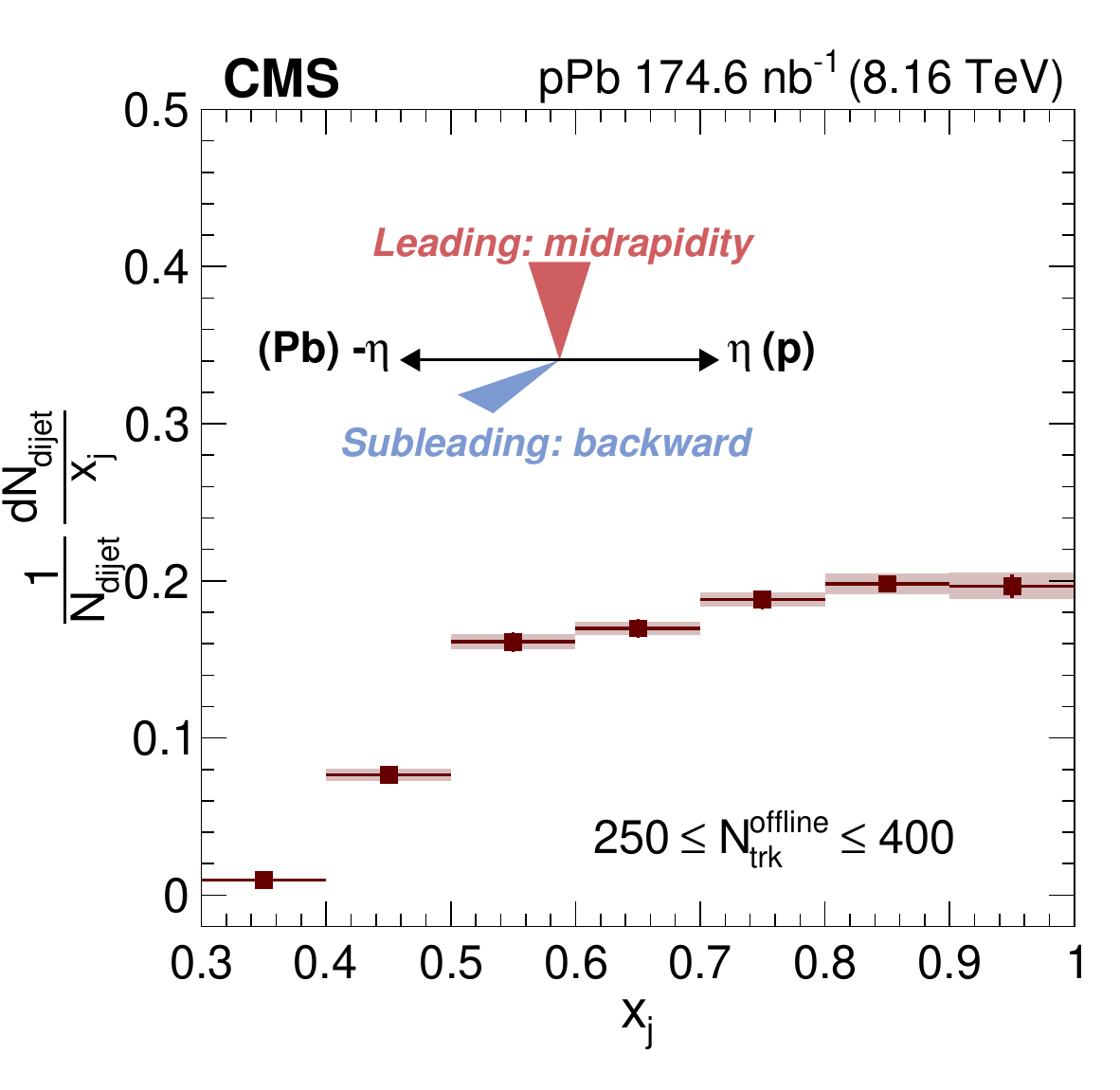} 
    \caption{The unfolded \xj distributions, from lower to higher multiplicities, and for leading and subleading jets in the midrapidity and backward regions, respectively, are shown. Comparisons with the generator-level {{\PYTHIA}8}+\EPOS simulation (green band) are also displayed. Monte Carlo event generation for multiplicities greater than 250 was deemed impractical.}
    \label{fig:xjmult_mid_bkw}
\end{figure}

\begin{figure}[!hbpt]
  \centering
    \includegraphics[width=0.44\textwidth]{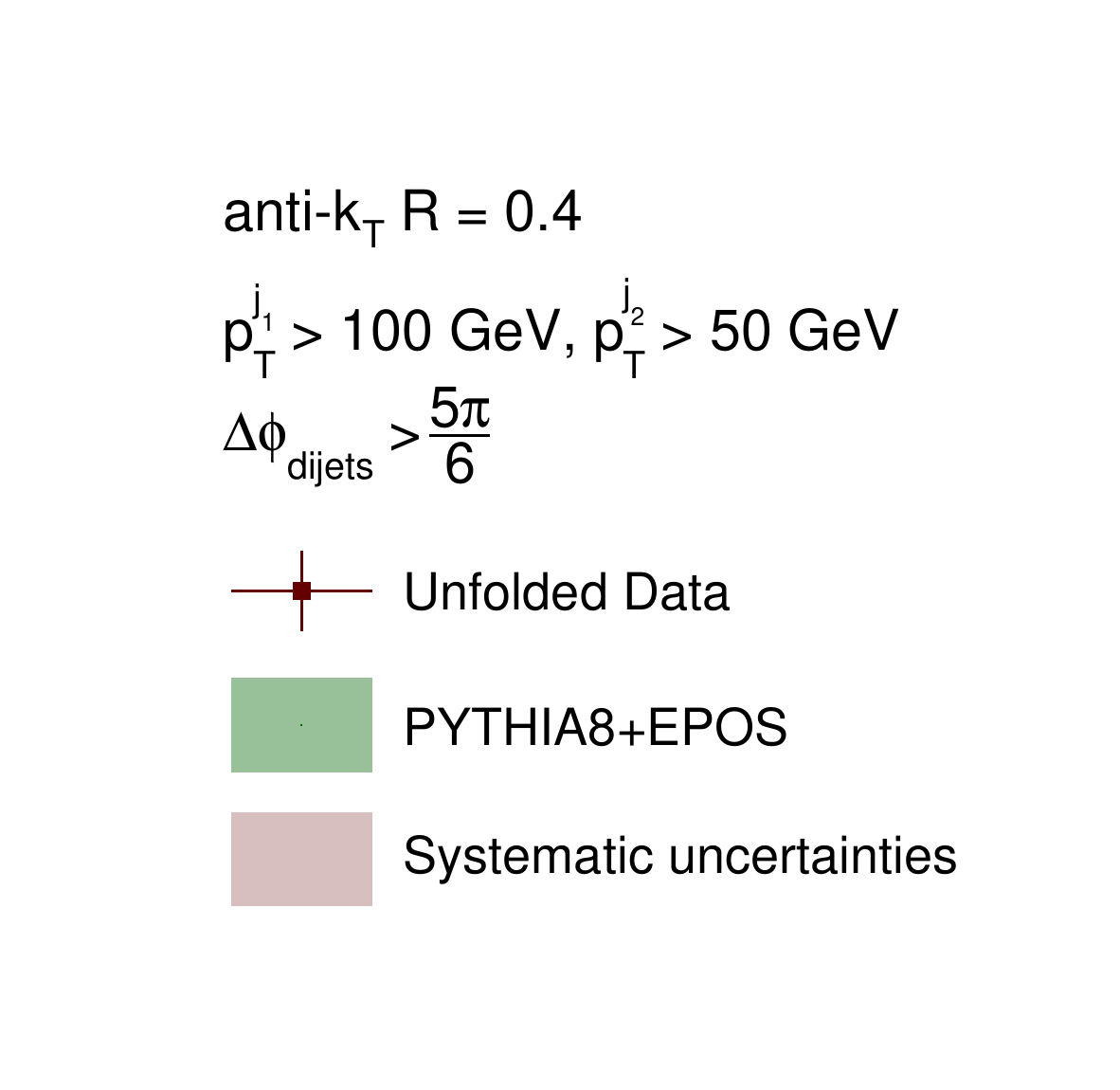} 
    \includegraphics[width=0.44\textwidth]{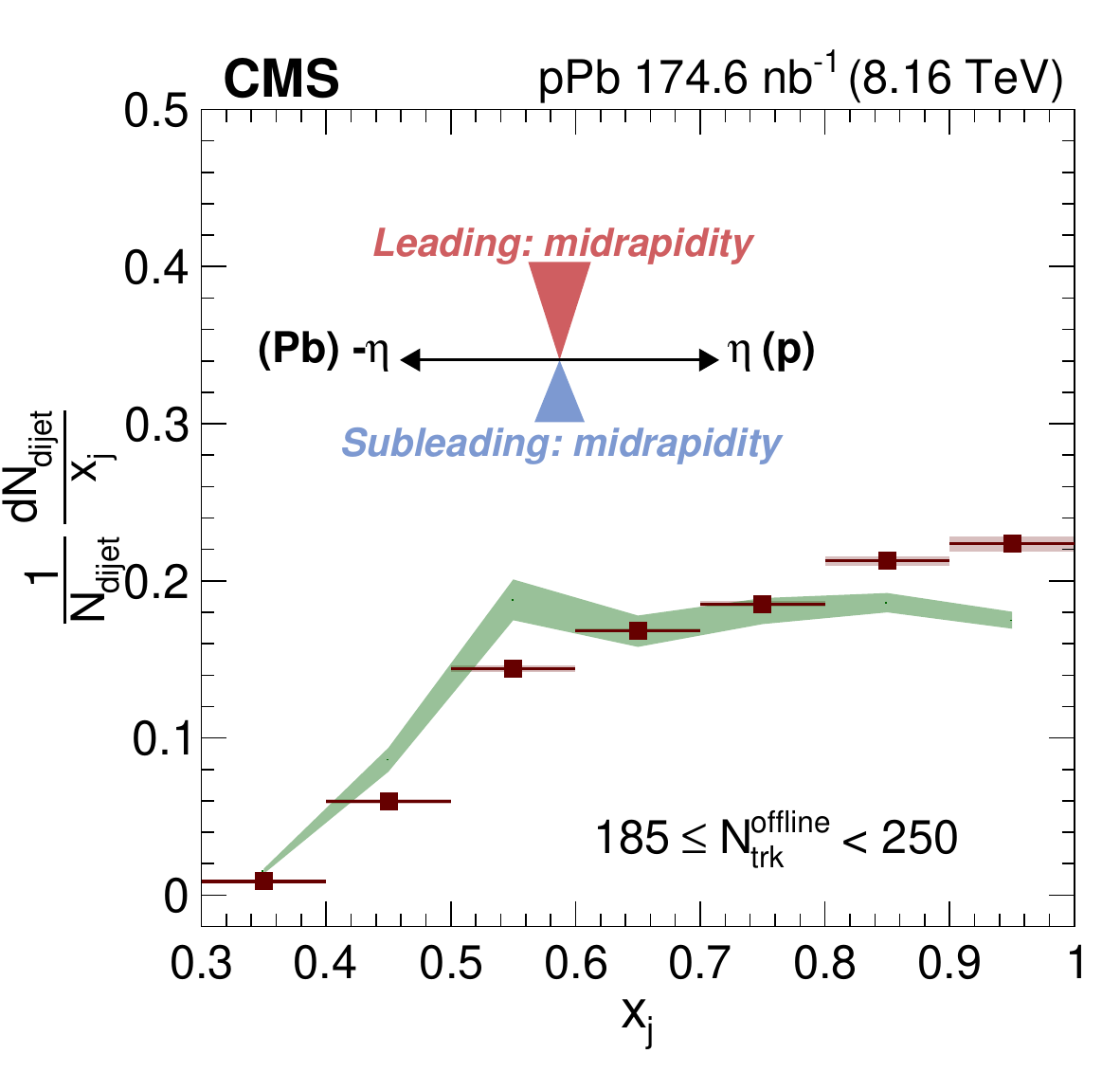} \\
    \includegraphics[width=0.44\textwidth]{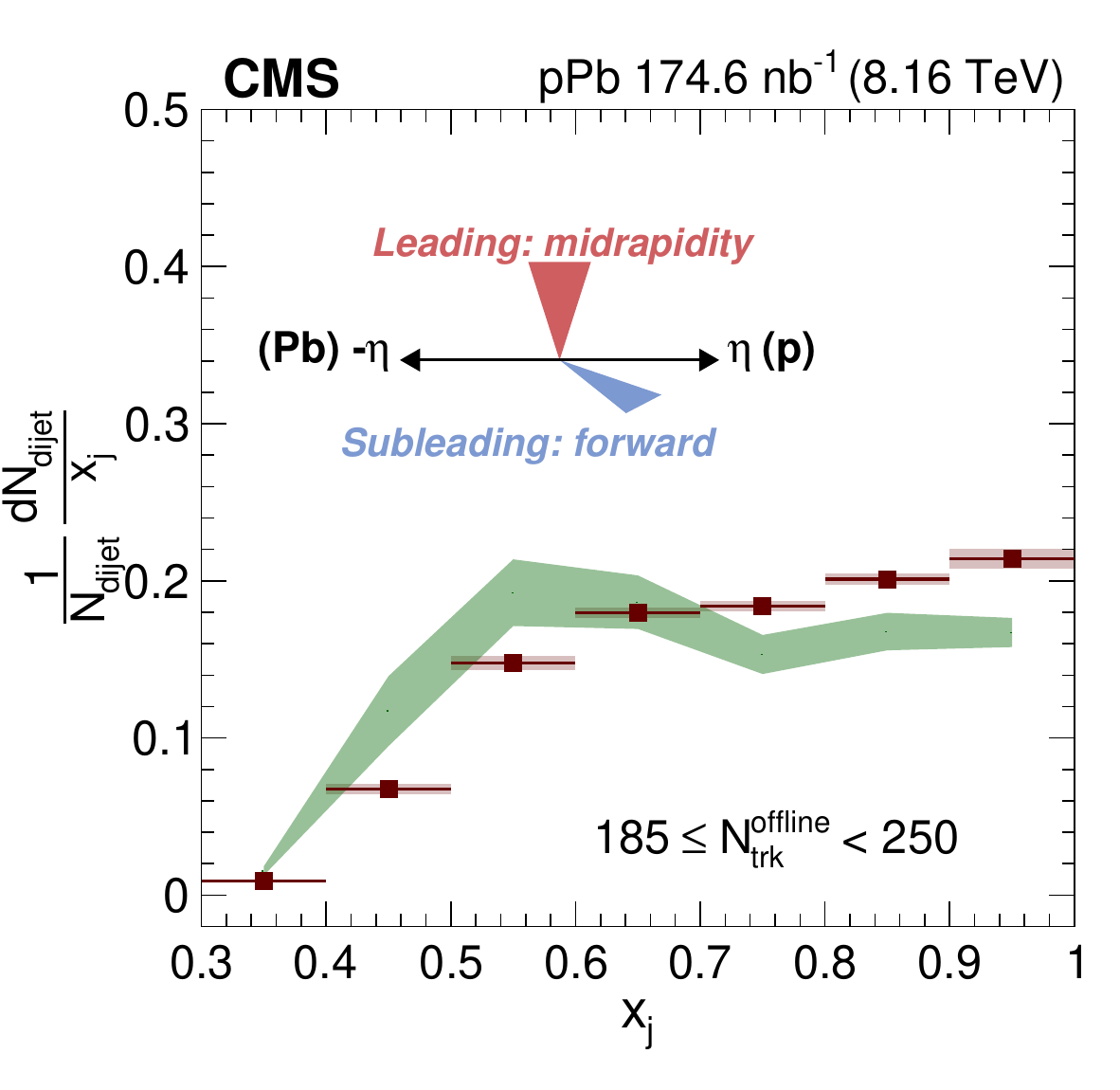} 
    \includegraphics[width=0.44\textwidth]{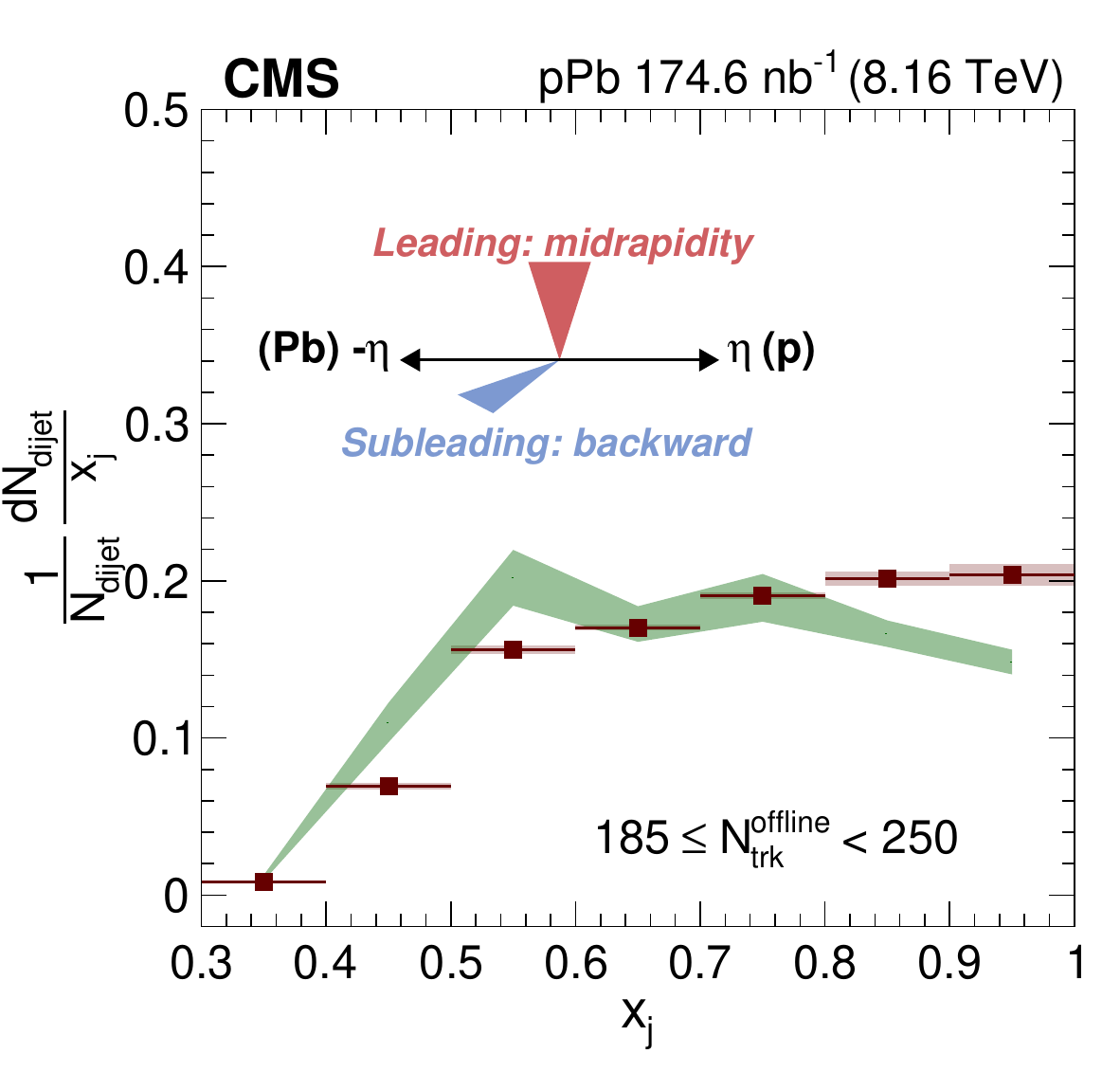} \\
    \includegraphics[width=0.44\textwidth]{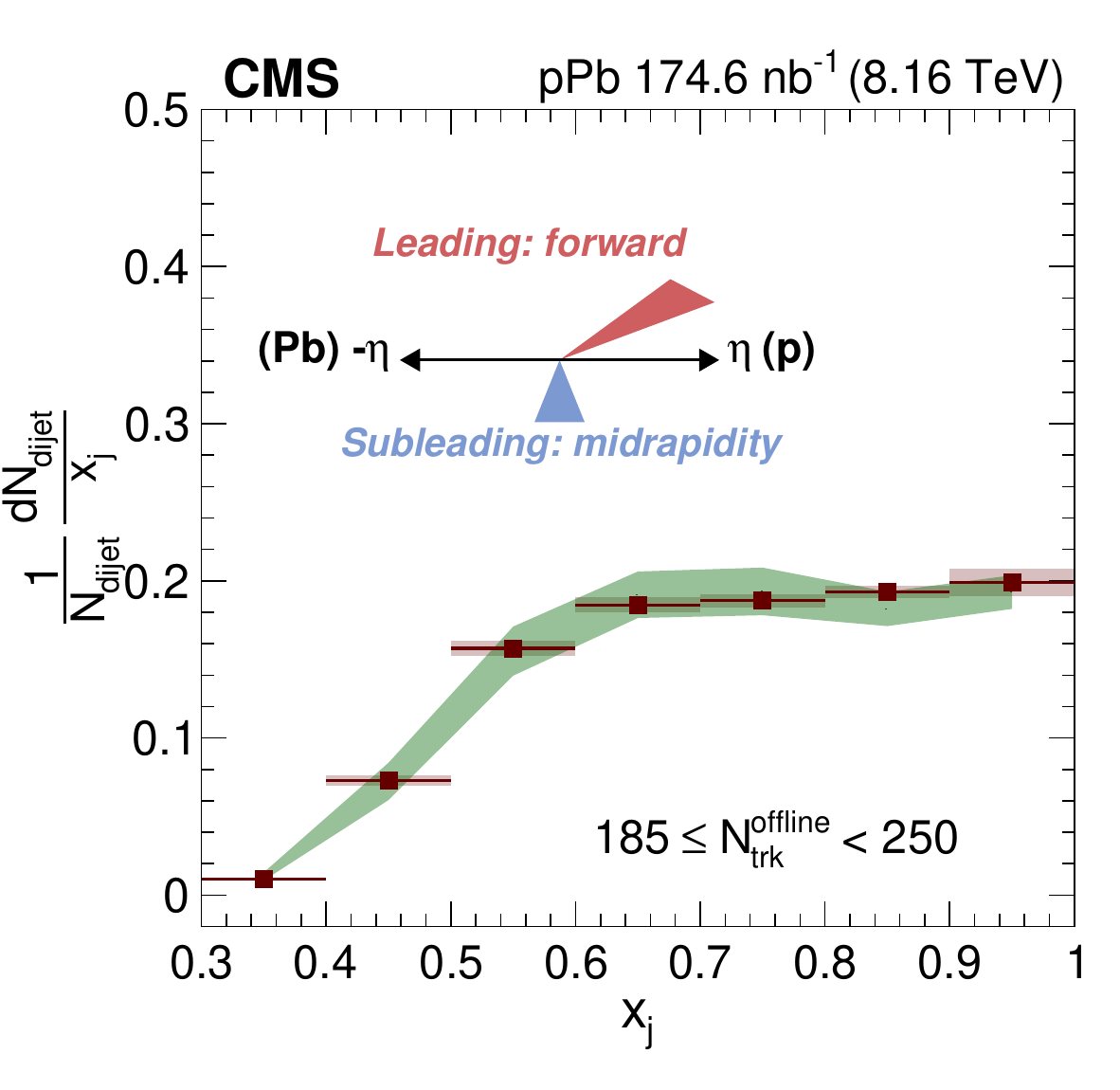} 
    \includegraphics[width=0.44\textwidth]{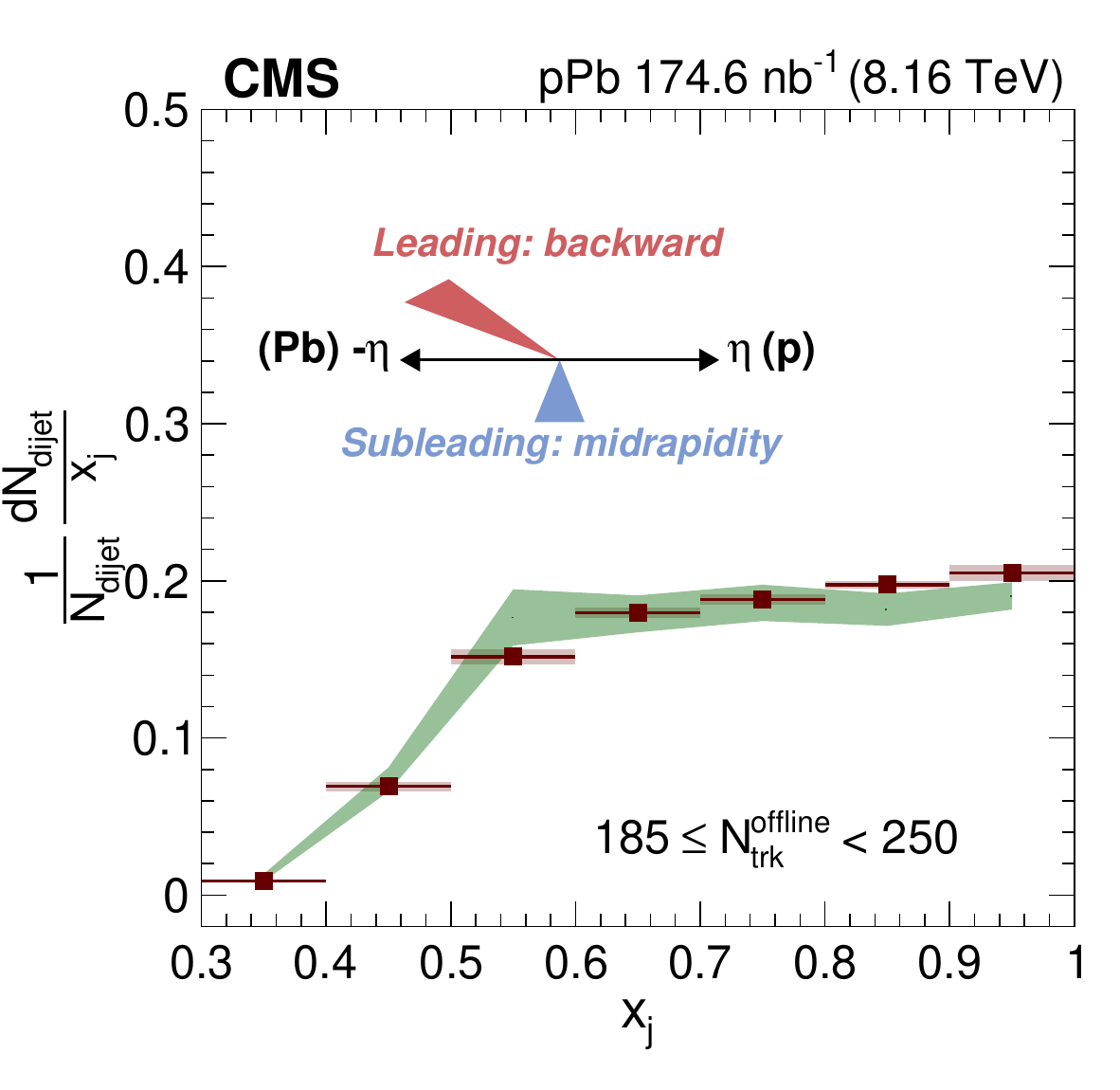} 
    \caption{The unfolded \xj distributions, in the multiplicity range $185 \leq \Ntroff < 250$ from different leading and subleading jet combinations for the midrapidity, forward, and backward regions, are shown. Comparisons with the generator-level {{\PYTHIA}8}+\EPOS simulation (green band) are also displayed.}
    \label{fig:xjeta}
\end{figure}

The unfolded \xj distributions across different multiplicity ranges for leading and subleading jets in midrapidity and backward regions, respectively, are shown in Fig.~\ref{fig:xjmult_mid_bkw}. In higher multiplicity events, the \xj distribution shows a suppression near unity, shifting towards lower \xj values, similar to trends seen in nucleus-nucleus collisions. This result suggests increased energy loss in these events, potentially due to the contributions from multijet production or energy conservation. Dijets in different pseudorapidity configurations, resulting in distinct \xpb and \xp regimes, for the multiplicity range $185 \leq \Ntroff < 250$ are shown in Fig.~\ref{fig:xjeta}. The various \etacm arrangements allow the exploration of \xpb and \xp within a range of $\sim$0.03 to $\sim$0.28. High-\xpb corresponds to cases where one of the jets is in the backward region; for example, \xpb is approximately 0.28 when the leading and subleading jets are in the backward and midrapidity regions, respectively. In contrast, low-\xpb is around 0.03 for jet pairs where the leading jet is at midrapidity, and the subleading jet is in the forward region. Overall we observe only a slight dependence of \xj across all dijet combinations. Comparisons with generator-level {{\PYTHIA}8}+\EPOS MC simulations are presented, showing similar behavior between the simulation and data. The observed deviations arise from nuclear effects that are not implemented in the {{\PYTHIA}8}+\EPOS MC. Monte Carlo event generation for multiplicities greater than 250 with EPOS was impractical. The remaining figures containing other multiplicity ranges and pseudorapidity combinations can be found in the appendix (Figs.~\ref{fig:xjmult_mid_mid} to~\ref{fig:xjmult_bkw_mid}).

\begin{figure}[!hbpt]
  \centering
    \includegraphics[width=0.44\textwidth]{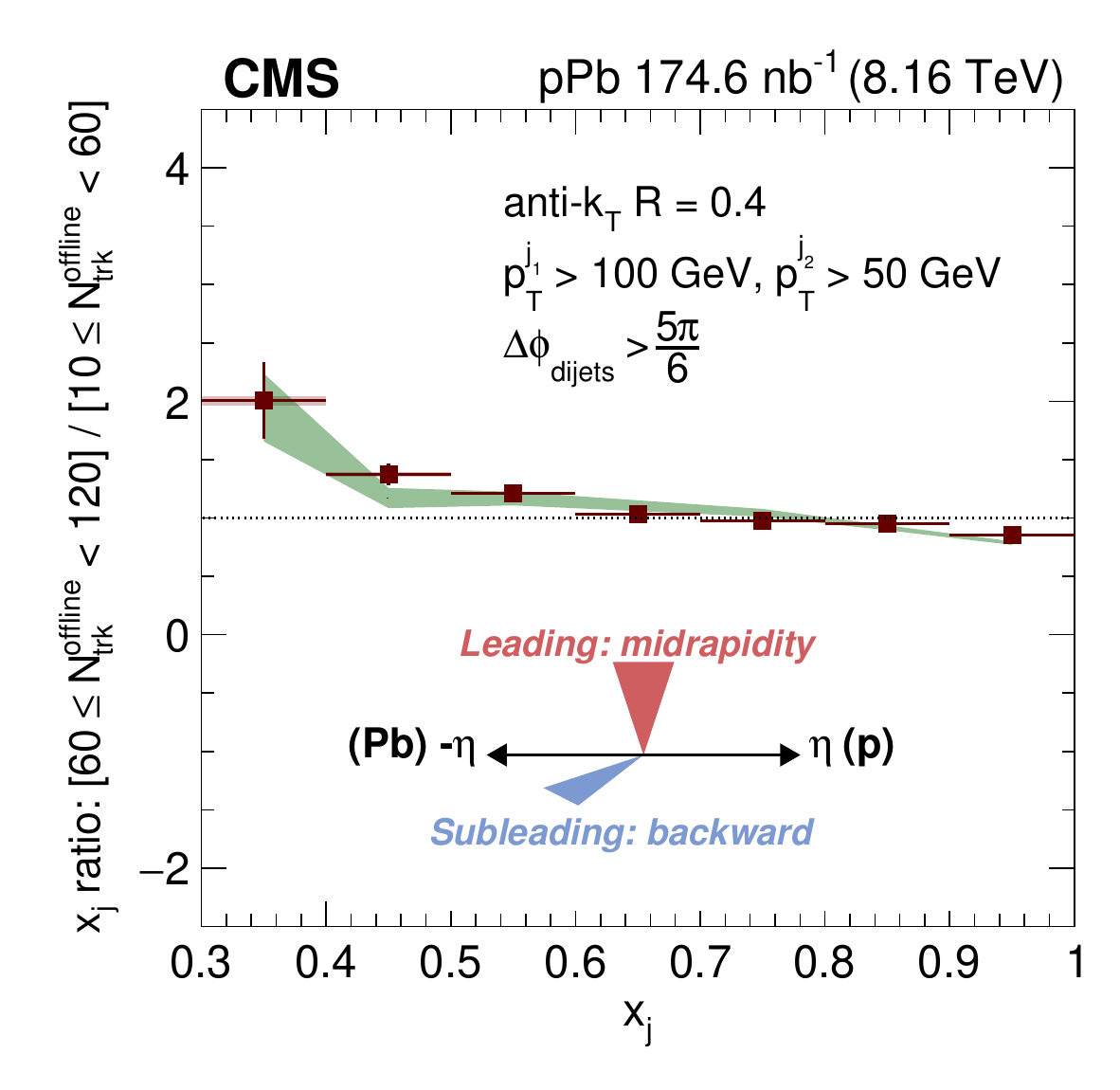} 
    \includegraphics[width=0.44\textwidth]{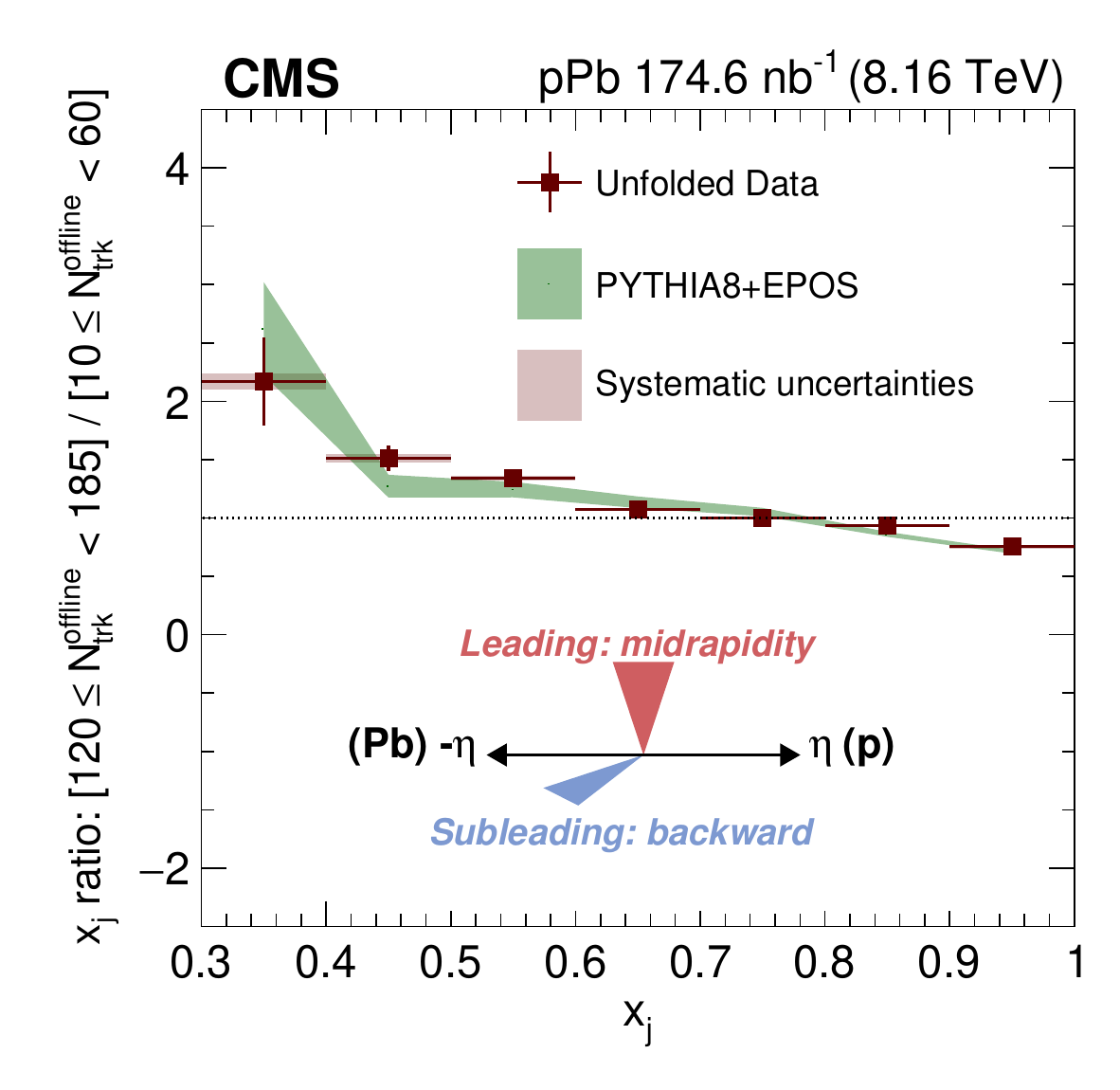} \\
    \includegraphics[width=0.44\textwidth]{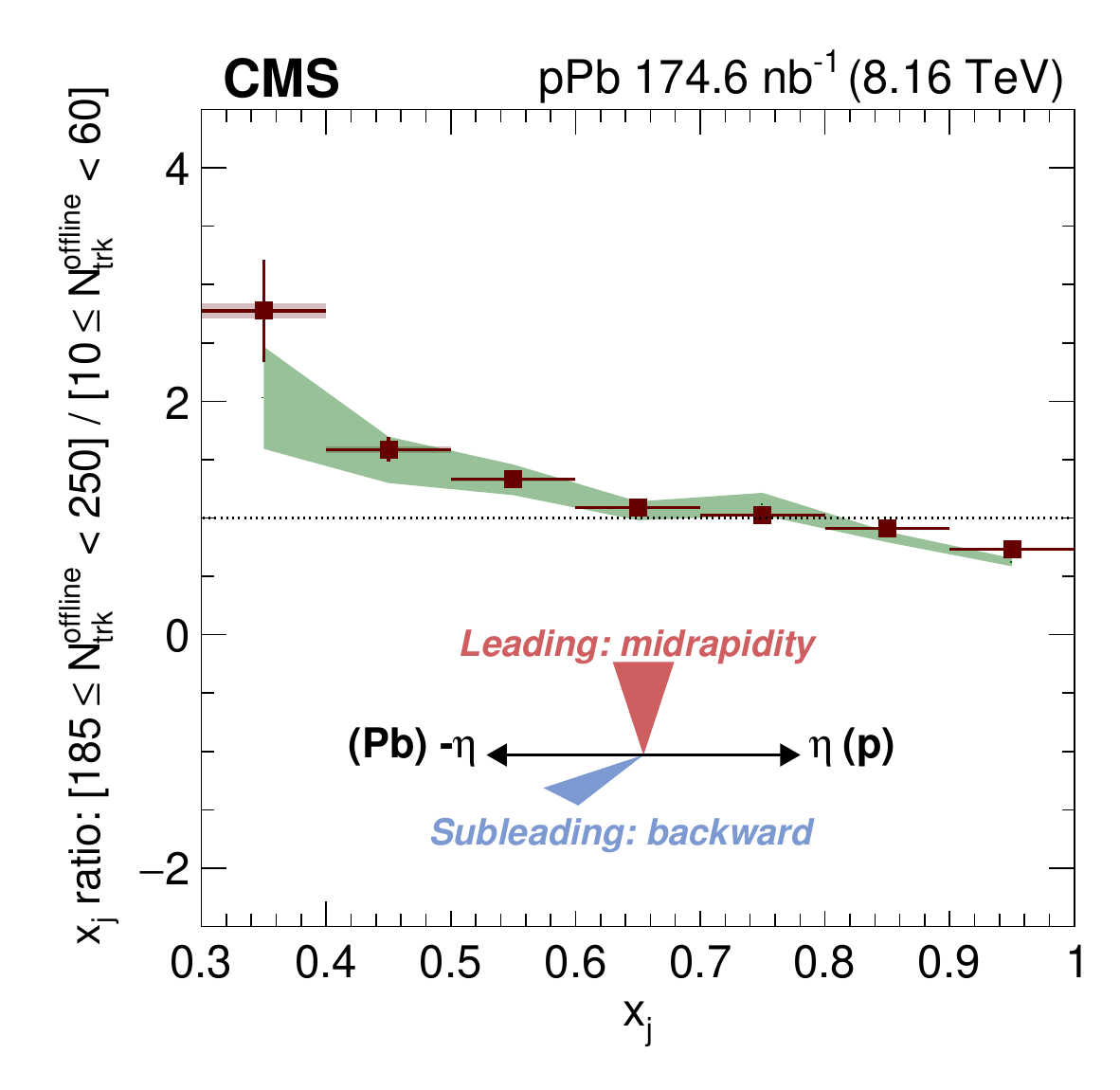} 
    \includegraphics[width=0.44\textwidth]{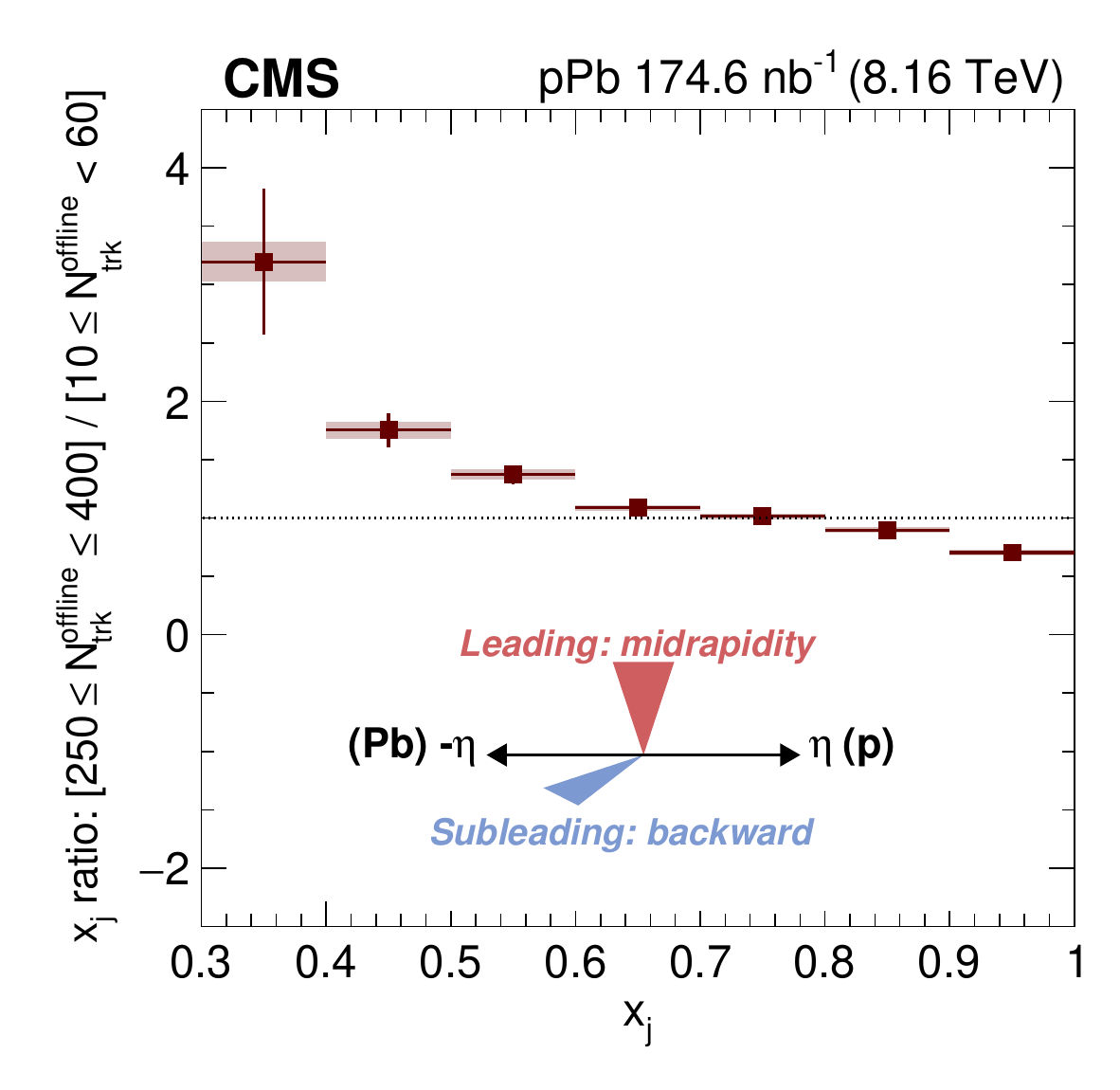} 
    \caption{The unfolded \xj ratio in the range $10 \leq \Ntroff < 60$, from lower to higher multiplicities for leading and subleading jets in the midrapidity and backward regions, respectively, are shown. Comparison with the generator-level {{\PYTHIA}8}+\EPOS simulation (green band) are also displayed. Monte Carlo event generation for multiplicities greater than 250 was deemed impractical.}
    \label{fig:xjratiomult_mid_bkw}
\end{figure}

\begin{figure}[!hbpt]
  \centering
    \includegraphics[width=0.44\textwidth]{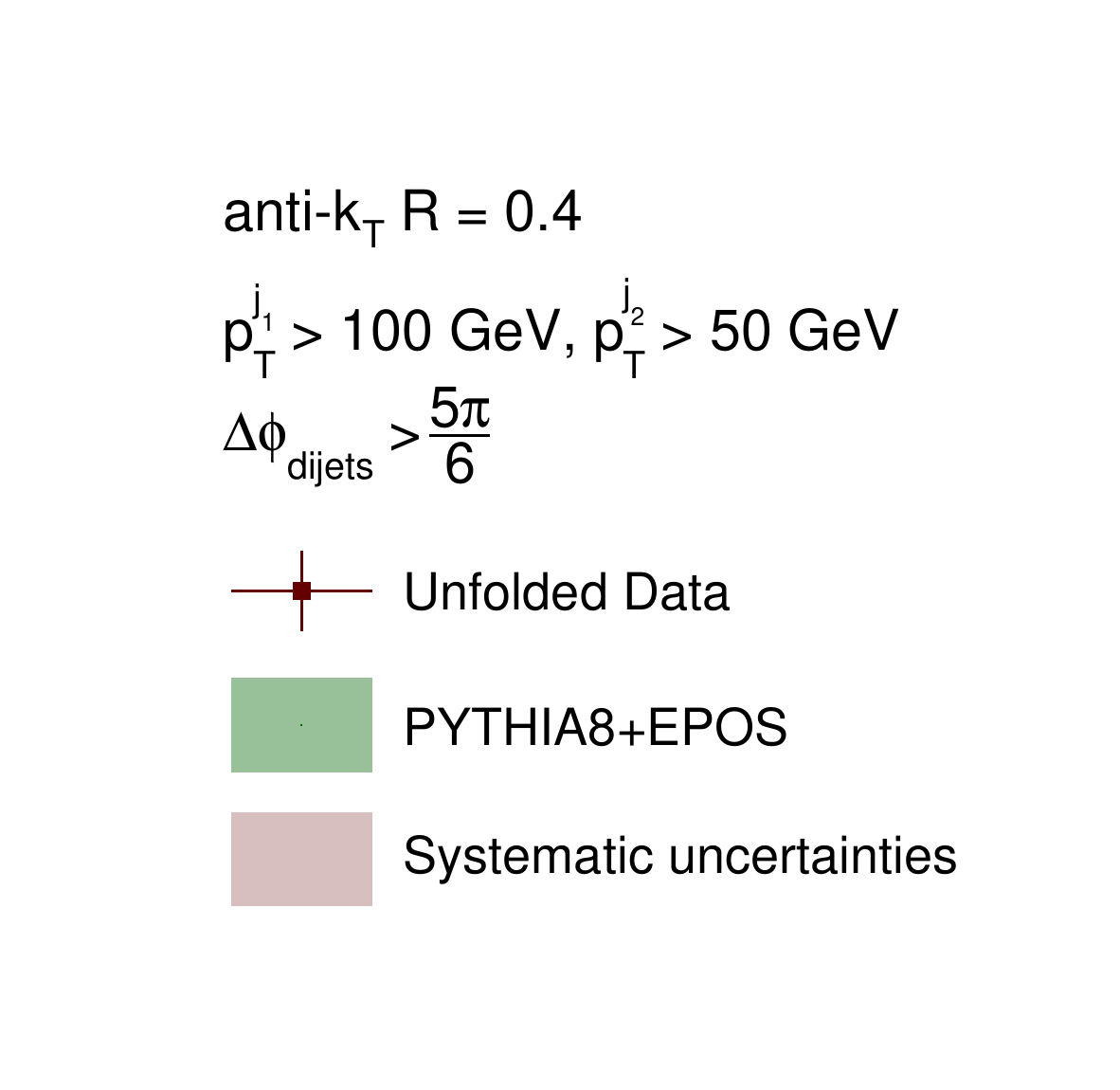} 
    \includegraphics[width=0.44\textwidth]{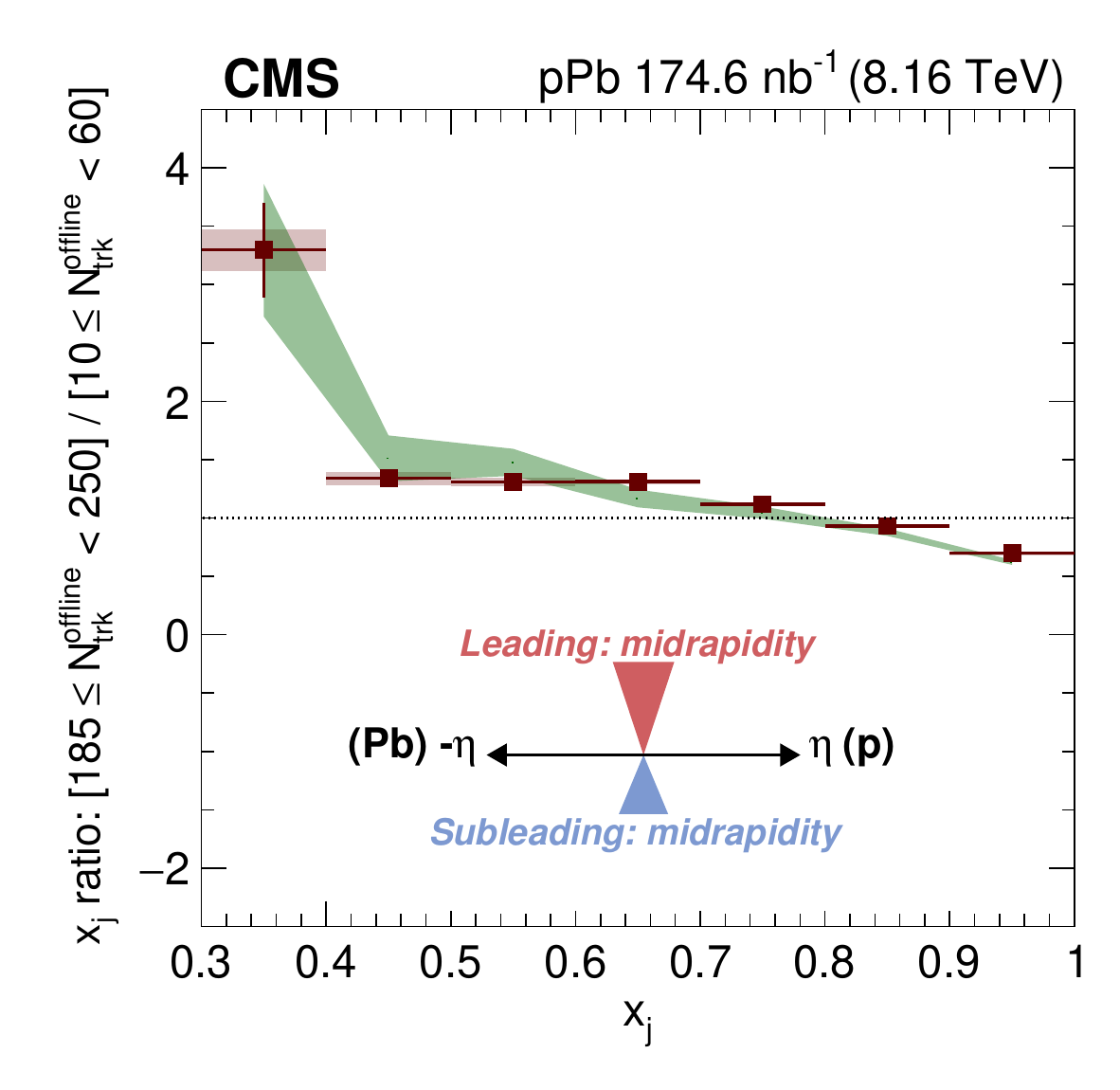} \\
    \includegraphics[width=0.44\textwidth]{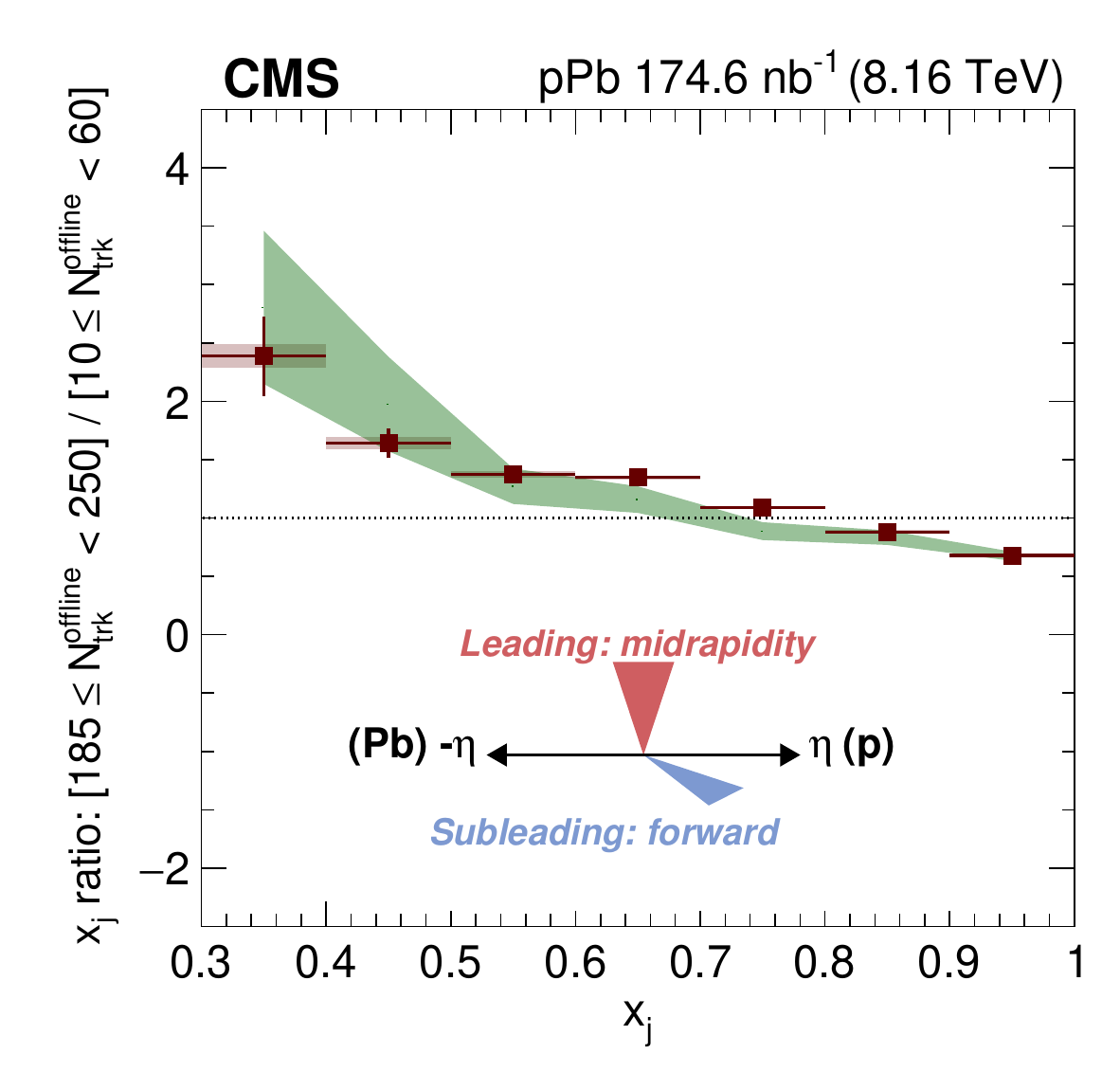} 
    \includegraphics[width=0.44\textwidth]{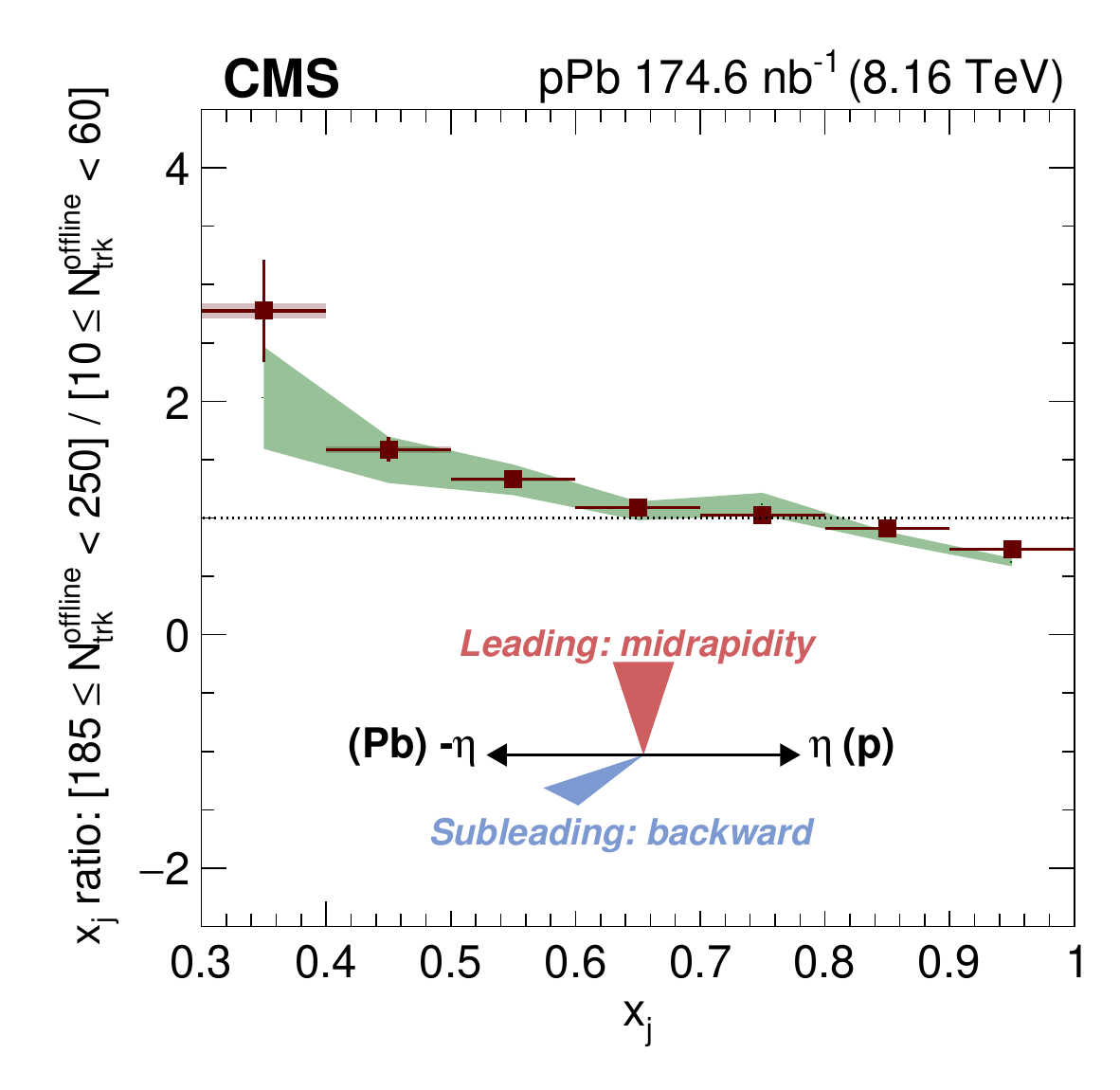} \\
    \includegraphics[width=0.44\textwidth]{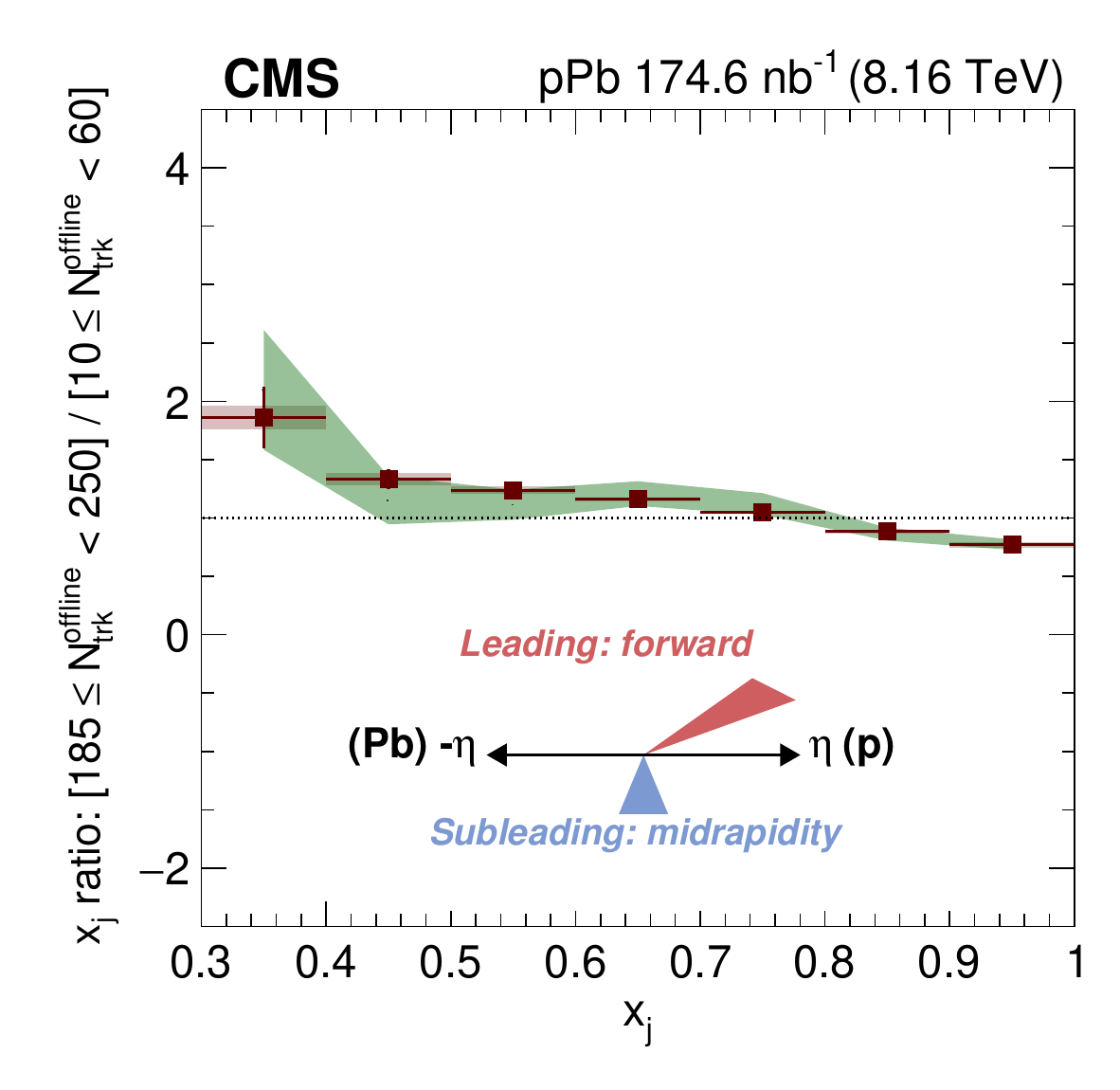} 
    \includegraphics[width=0.44\textwidth]{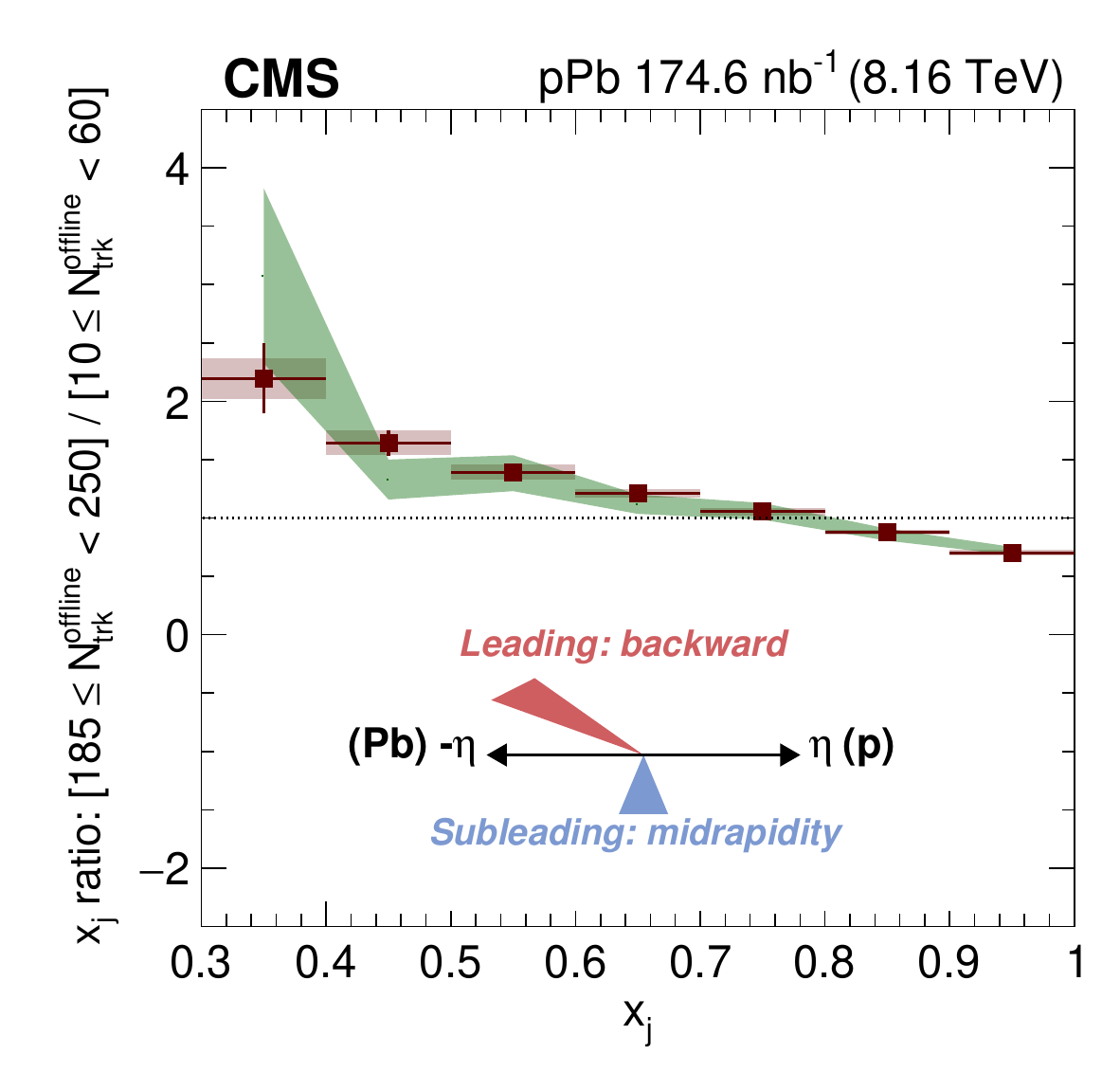} 
    \caption{The ratio of unfolded \xj distributions between dijet events from multiplicity range $185 \leq \Ntroff < 250$ to $10 \leq \Ntroff < 60 $ is shown for leading and subleading jet combinations at mid-, forward, and backward rapidity regions. Comparisons with the generator-level {{\PYTHIA}8}+\EPOS simulation (green band) are also displayed.}
    \label{fig:xjratioeta}
\end{figure}

The ratios between higher multiplicity intervals and the multiplicity range of $10 \leq \Ntroff < 60$ can be used to search for quenching effects since a QGP medium is not expected to be formed in such low multiplicity events. Figures~\ref{fig:xjratiomult_mid_bkw} and~\ref{fig:xjratioeta} show the high-to-low multiplicity ratios in different \Ntroff ranges and different \etacm configurations for the unfolded values, respectively. The ratio is close to unity, and the values decrease linearly from low to high \xj being less (greater) than unity for $\xj > 0.7$ ($<$0.6). The linear trend becomes more pronounced for higher multiplicities, showing the imbalance behavior discussed previously, and can be understood as a result of transverse momentum conservation~\cite{Kordell:2016njg, STAR:2024nwm} due to the presence of multijet events. Similar to the \xj distributions, a small overall multiplicity dependence, within the uncertainties, is observed across the different \etacm combinations, indicating a slight dependency of the ratio with \xpb and \xp. There is generally an agreement between simulations and data within the uncertainties. The additional figures corresponding to other multiplicity ranges and pseudorapidity configurations are provided in the appendix (Figs.~\ref{fig:xjratiomult_mid_mid} to~\ref{fig:xjratiomult_bkw_mid}). The measured (raw) distributions, which have not been unfolded, are also available in the appendix (Figs.~\ref{fig:xjmultrec_mid_mid} to~\ref{fig:xjmeanratioreco}) and support the same conclusions as discussed above for the unfolded distributions.

\begin{figure}[!hbpt]
  \centering
    \includegraphics[width=0.42\textwidth]{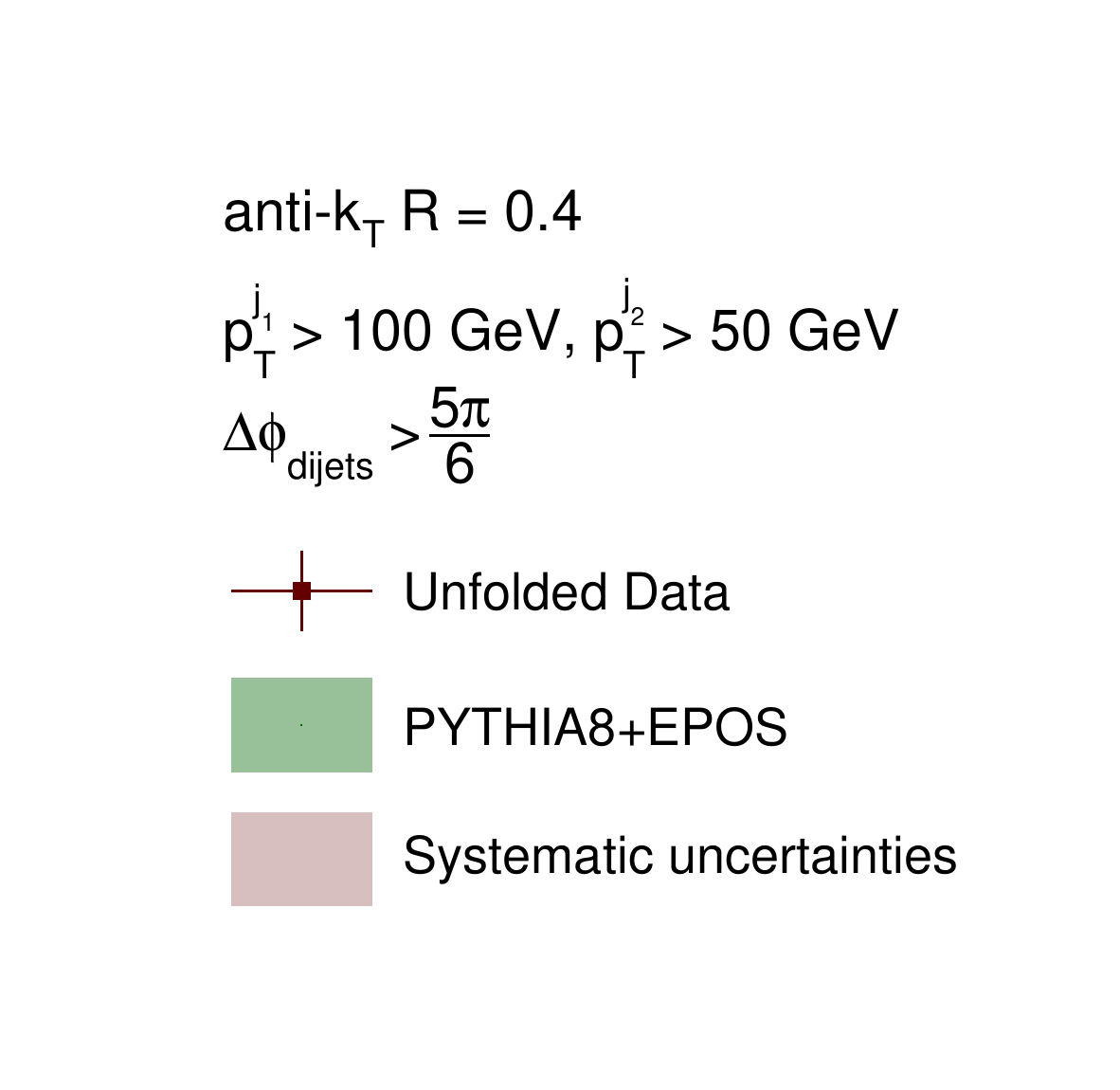} 
    \includegraphics[width=0.42\textwidth]{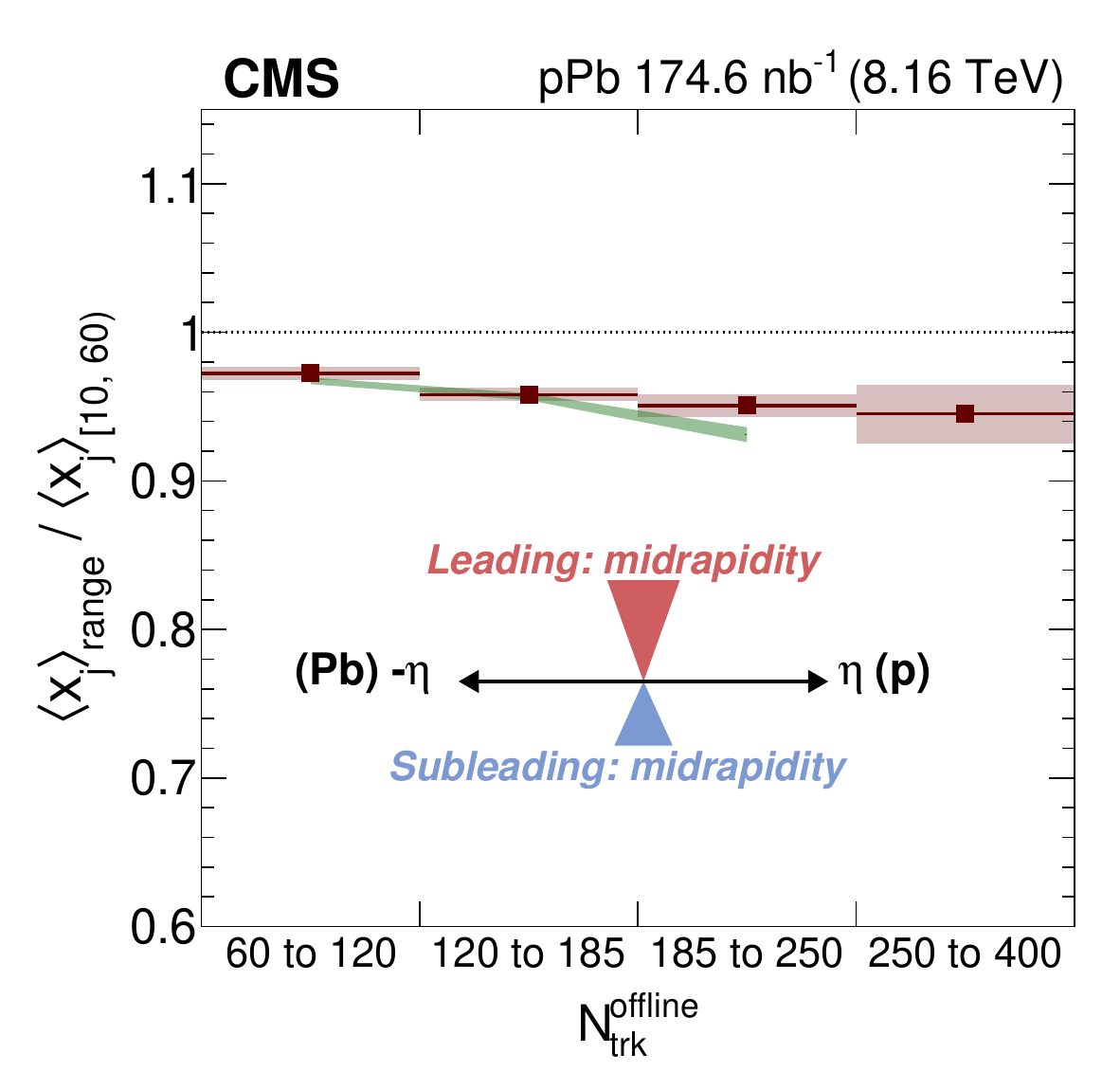} \\
    \includegraphics[width=0.42\textwidth]{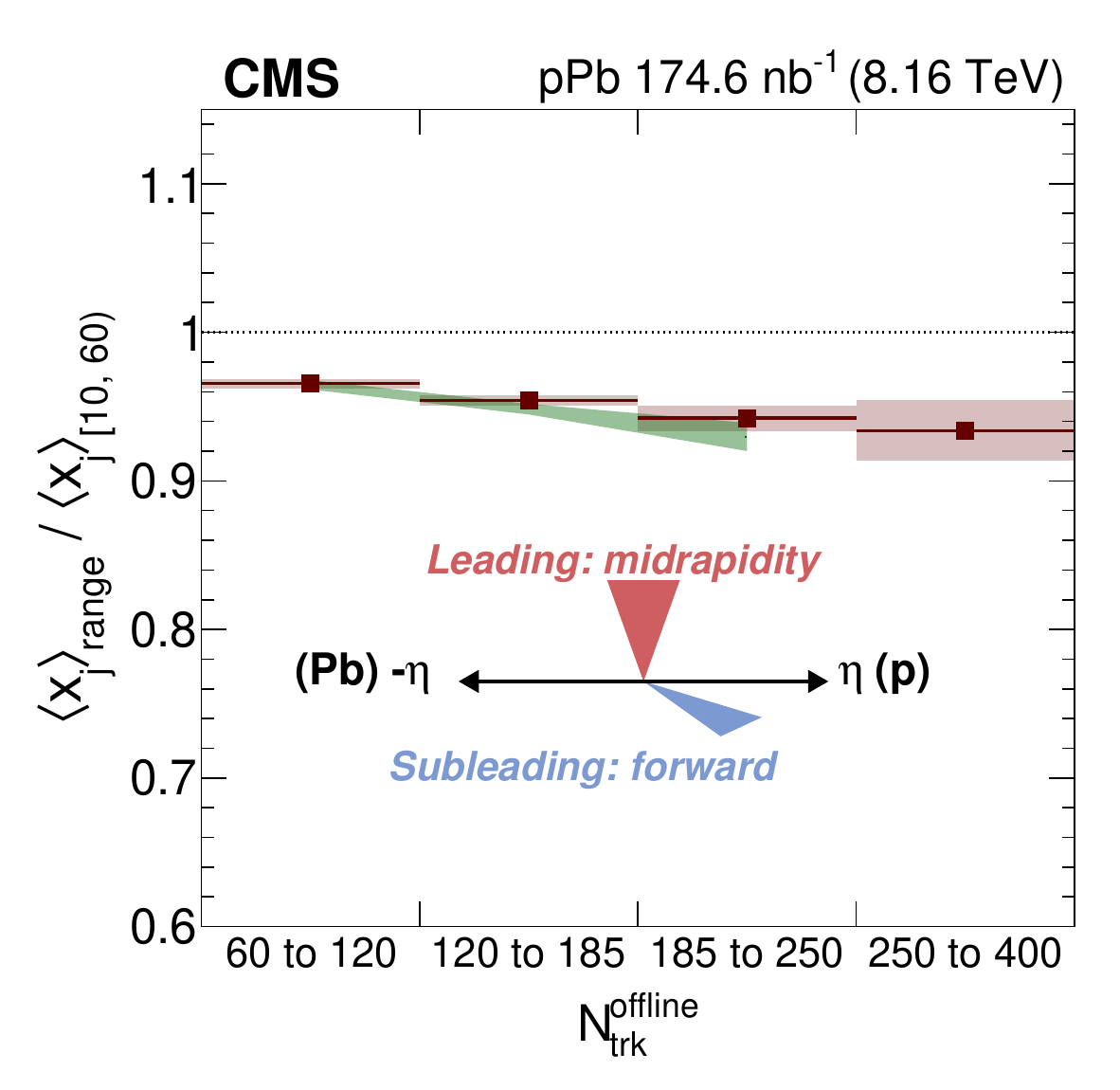} 
    \includegraphics[width=0.42\textwidth]{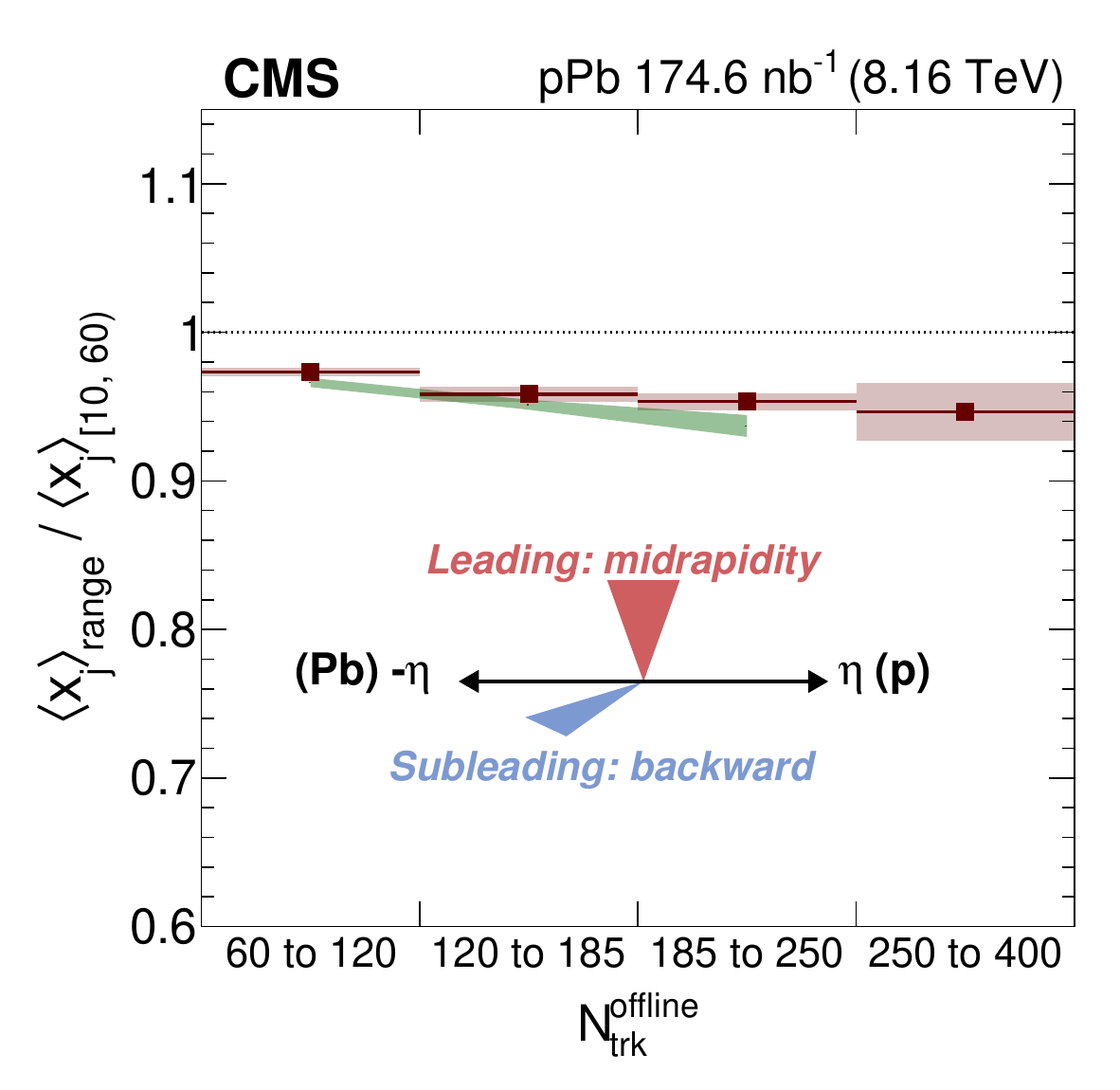} \\
    \includegraphics[width=0.42\textwidth]{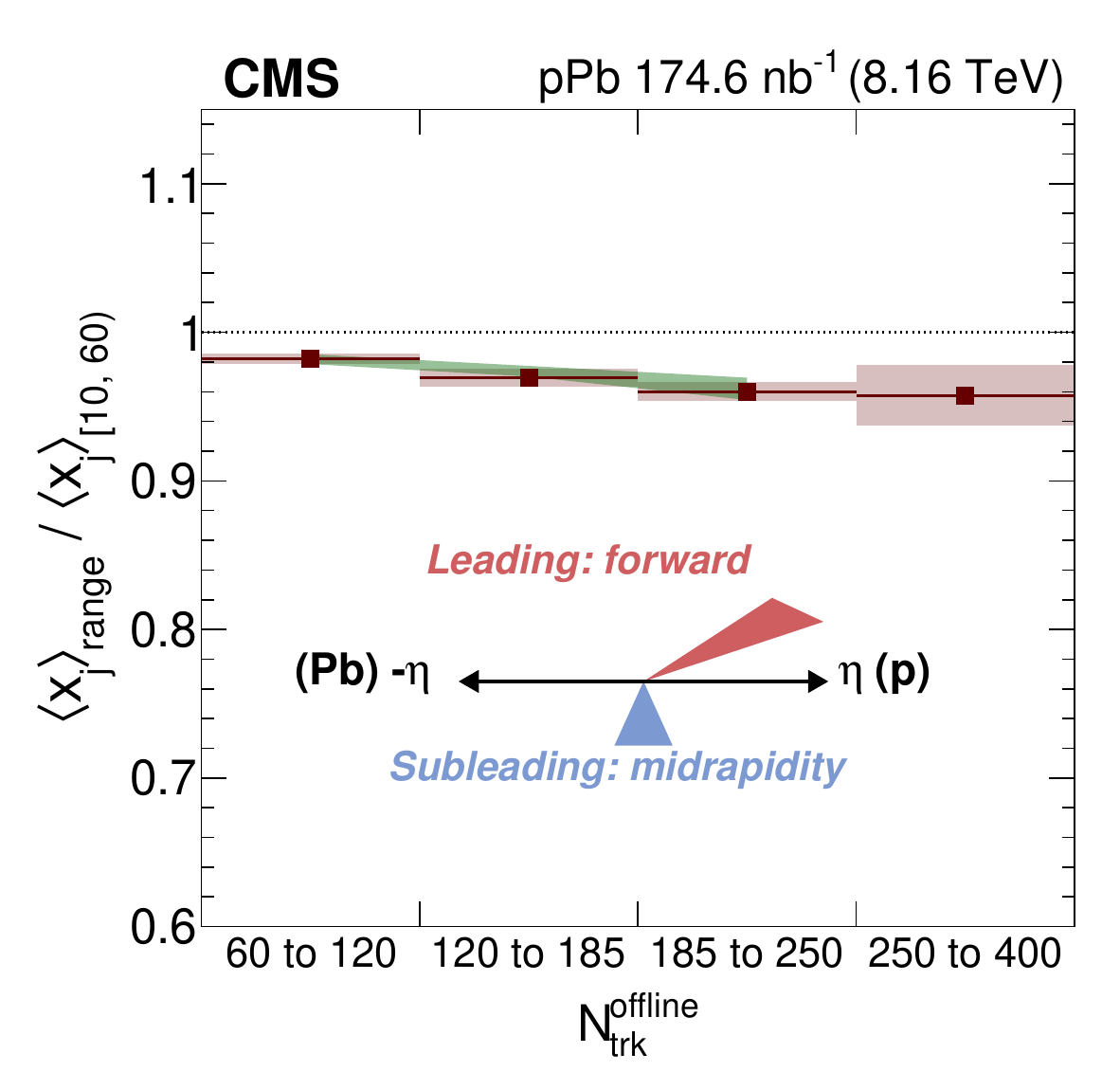} 
    \includegraphics[width=0.42\textwidth]{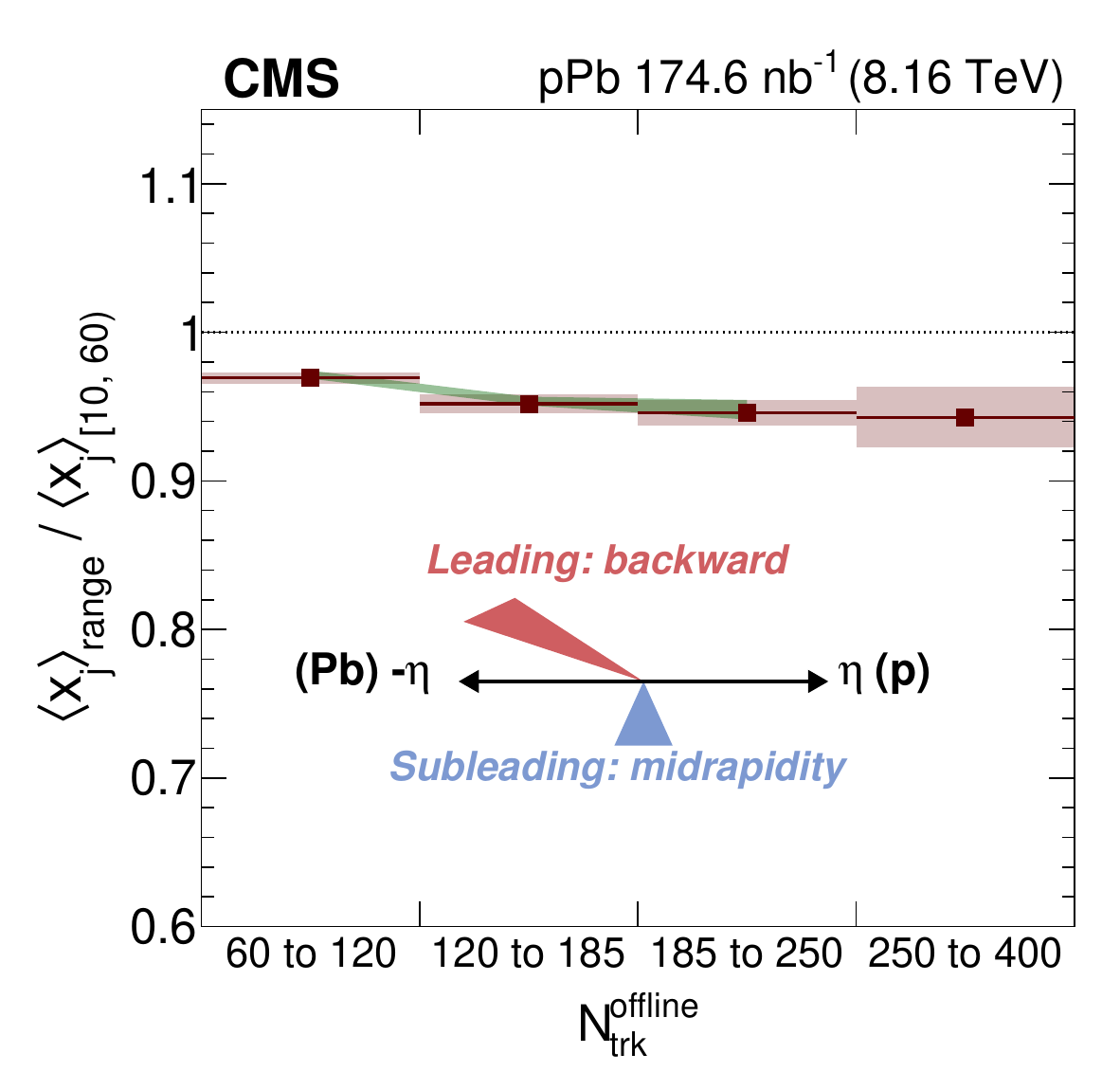} 
    \caption{The ratio of the mean \xj values ratio \xja for dijet events from different multiplicity classes to that of events with the lowest multiplicity selection in the range $10 \leq \Ntroff < 60$. The ratios of \xja values, extracted from unfolded \xj distributions, are presented for different leading and subleading jet selections from mid-, forward and backward rapidity regions. Comparison with the generator-level (green band) {{\PYTHIA}8}+\EPOS are also displayed. The multiplicity range shown on the x-axis represents the multiplicity selection for the numerator of the ratio.}
    \label{fig:xjmeanratio}
\end{figure}

The analysis also measures the \xja ratio from high-to-low \Ntroff, as shown in Fig.~\ref{fig:xjmeanratio}, in bins of \Ntroff for all jet \etacm configurations for unfolded distributions. The results are close to unity, and a slight decrease in trend is aligned with the {{\PYTHIA}8}+\EPOS simulation. This observation is at odds with PbPb collision data trends in~\cite{CMS:2018dqf}. Compared to \pp-based reference, which provides an estimate for resolution effects, the PbPb data showed a significant departure from that baseline with progressively lower average \xj values toward more central (higher multiplicity) events.
The study of dijet pseudorapidity combinations indicates a weak to no dependence of the dijet \pt balance trends on \xpb and \xp, with variations less than 2\%.

Comparing simulation to data, with simulation not including an energy loss mechanism, constrains the subleading jet energy loss, assuming that \xja remains unquenched in the multiplicity range of [10, 60). A chi-square test was performed comparing the measurements and MC simulations to extract the limit on the possible downward shift (expected from the energy loss scenario) in the mean \xj values. For high-multiplicity events [185, 250), where both leading and subleading jets are in the midrapidity region, the average subleading jet energy loss is constrained to be less than 1.26\% at the one-sided 90\% confidence level for an average \pt of 103.5\GeV. The result aligns with measurements from other LHC experiments~\cite{Acharya:2017okq, ATLAS:2022iyq} and represents the first constraint on jet energy loss in \ppb collisions using dijet observables. Interpreting the combination of this strong upper limit on jet quenching and the previously observed evidence for collective hydrodynamic flow in high multiplicity \ppb events will require further theoretical developments incorporating both physics phenomena.

\section{Summary}

The first search for jet quenching effects on dijet momentum balance in proton-lead collisions at a nucleon-nucleon center-of-mass energy of $\sqrtsNN = 8.16\TeV$ was presented.
The dijet transverse momentum balance (\xj), defined as the ratio of the subleading over the leading jet in a dijet pair, was studied as a function of charged-particle multiplicity for several dijet kinematic selections. The data collected by the CMS experiment with minimum bias and special high-multiplicity triggers during the 2016 data-taking period corresponds to an integrated luminosity of 174.6\nbinv. These special triggers aided the accumulation of a substantial sample of high multiplicity events (producing up to 400 charged-particles per event) and enabled jet-quenching signal search in high-multiplicity events.

An anti-\kt algorithm with a distance parameter of 0.4 was used to reconstruct jets using a combination of tracker and calorimeter information. The dijets were formed by the two highest transverse momentum (\pt) jets in the event with the minimum \pt requirements of 100\GeV and 50\GeV, respectively, that pass back-to-back selection (azimuthal separation greater than $5\pi/6$). The \xj distributions for such dijets were measured for events in several multiplicity ranges, and modifications of the measured distributions were quantified by taking ratios of the measurements from the higher multiplicity event selections to that of the lowest multiplicity class. These high-to-low multiplicity selection ratios were found to be close to unity and in agreement with simulations that do not include jet-quenching effects. These measurements set an upper limit on medium-induced energy loss of the subleading jet to be less than 1.26\% of its transverse momentum at the one-sided 90\% confidence level at an average \pt of 103.5\GeV for high-multiplicity \ppb events, where strong collective signals have been reported. Interpreting this upper limit on jet quenching and the previously observed evidence for collective behavior in small systems will require further theoretical developments to reproduce all observations.

\begin{acknowledgments}
We congratulate our colleagues in the CERN accelerator departments for the excellent performance of the LHC and thank the technical and administrative staffs at CERN and at other CMS institutes for their contributions to the success of the CMS effort. In addition, we gratefully acknowledge the computing centers and personnel of the Worldwide LHC Computing Grid and other centers for delivering so effectively the computing infrastructure essential to our analyses. Finally, we acknowledge the enduring support for the construction and operation of the LHC, the CMS detector, and the supporting computing infrastructure provided by the following funding agencies: SC (Armenia), BMBWF and FWF (Austria); FNRS and FWO (Belgium); CNPq, CAPES, FAPERJ, FAPERGS, and FAPESP (Brazil); MES and BNSF (Bulgaria); CERN; CAS, MoST, and NSFC (China); MINCIENCIAS (Colombia); MSES and CSF (Croatia); RIF (Cyprus); SENESCYT (Ecuador); ERC PRG, RVTT3 and MoER TK202 (Estonia); Academy of Finland, MEC, and HIP (Finland); CEA and CNRS/IN2P3 (France); SRNSF (Georgia); BMBF, DFG, and HGF (Germany); GSRI (Greece); NKFIH (Hungary); DAE and DST (India); IPM (Iran); SFI (Ireland); INFN (Italy); MSIP and NRF (Republic of Korea); MES (Latvia); LMTLT (Lithuania); MOE and UM (Malaysia); BUAP, CINVESTAV, CONACYT, LNS, SEP, and UASLP-FAI (Mexico); MOS (Montenegro); MBIE (New Zealand); PAEC (Pakistan); MES and NSC (Poland); FCT (Portugal); MESTD (Serbia); MICIU/AEI and PCTI (Spain); MOSTR (Sri Lanka); Swiss Funding Agencies (Switzerland); MST (Taipei); MHESI and NSTDA (Thailand); TUBITAK and TENMAK (Turkey); NASU (Ukraine); STFC (United Kingdom); DOE and NSF (USA).

\hyphenation{Rachada-pisek} Individuals have received support from the Marie-Curie program and the European Research Council and Horizon 2020 Grant, contract Nos.\ 675440, 724704, 752730, 758316, 765710, 824093, 101115353, 101002207, and COST Action CA16108 (European Union); the Leventis Foundation; the Alfred P.\ Sloan Foundation; the Alexander von Humboldt Foundation; the Science Committee, project no. 22rl-037 (Armenia); the Fonds pour la Formation \`a la Recherche dans l'Industrie et dans l'Agriculture (FRIA-Belgium); the Beijing Municipal Science \& Technology Commission, No. Z191100007219010 and Fundamental Research Funds for the Central Universities (China); the Ministry of Education, Youth and Sports (MEYS) of the Czech Republic; the Shota Rustaveli National Science Foundation, grant FR-22-985 (Georgia); the Deutsche Forschungsgemeinschaft (DFG), among others, under Germany's Excellence Strategy -- EXC 2121 ``Quantum Universe" -- 390833306, and under project number 400140256 - GRK2497; the Hellenic Foundation for Research and Innovation (HFRI), Project Number 2288 (Greece); the Hungarian Academy of Sciences, the New National Excellence Program - \'UNKP, the NKFIH research grants K 131991, K 133046, K 138136, K 143460, K 143477, K 146913, K 146914, K 147048, 2020-2.2.1-ED-2021-00181, TKP2021-NKTA-64, and 2021-4.1.2-NEMZ\_KI-2024-00036 (Hungary); the Council of Science and Industrial Research, India; ICSC -- National Research Center for High Performance Computing, Big Data and Quantum Computing and FAIR -- Future Artificial Intelligence Research, funded by the NextGenerationEU program (Italy); the Latvian Council of Science; the Ministry of Education and Science, project no. 2022/WK/14, and the National Science Center, contracts Opus 2021/41/B/ST2/01369 and 2021/43/B/ST2/01552 (Poland); the Funda\c{c}\~ao para a Ci\^encia e a Tecnologia, grant CEECIND/01334/2018 (Portugal); the National Priorities Research Program by Qatar National Research Fund; MICIU/AEI/10.13039/501100011033, ERDF/EU, "European Union NextGenerationEU/PRTR", and Programa Severo Ochoa del Principado de Asturias (Spain); the Chulalongkorn Academic into Its 2nd Century Project Advancement Project, and the National Science, Research and Innovation Fund via the Program Management Unit for Human Resources \& Institutional Development, Research and Innovation, grant B39G670016 (Thailand); the Kavli Foundation; the Nvidia Corporation; the SuperMicro Corporation; the Welch Foundation, contract C-1845; and the Weston Havens Foundation (USA).
\end{acknowledgments}

\bibliography{auto_generated}

\appendix
\numberwithin{figure}{section}
\section{The \texorpdfstring{\xj}{xj} distributions for various event multiplicities and jet rapidity selections}

\begin{figure}[!hbpt]
  \centering
    \includegraphics[width=0.43\textwidth]{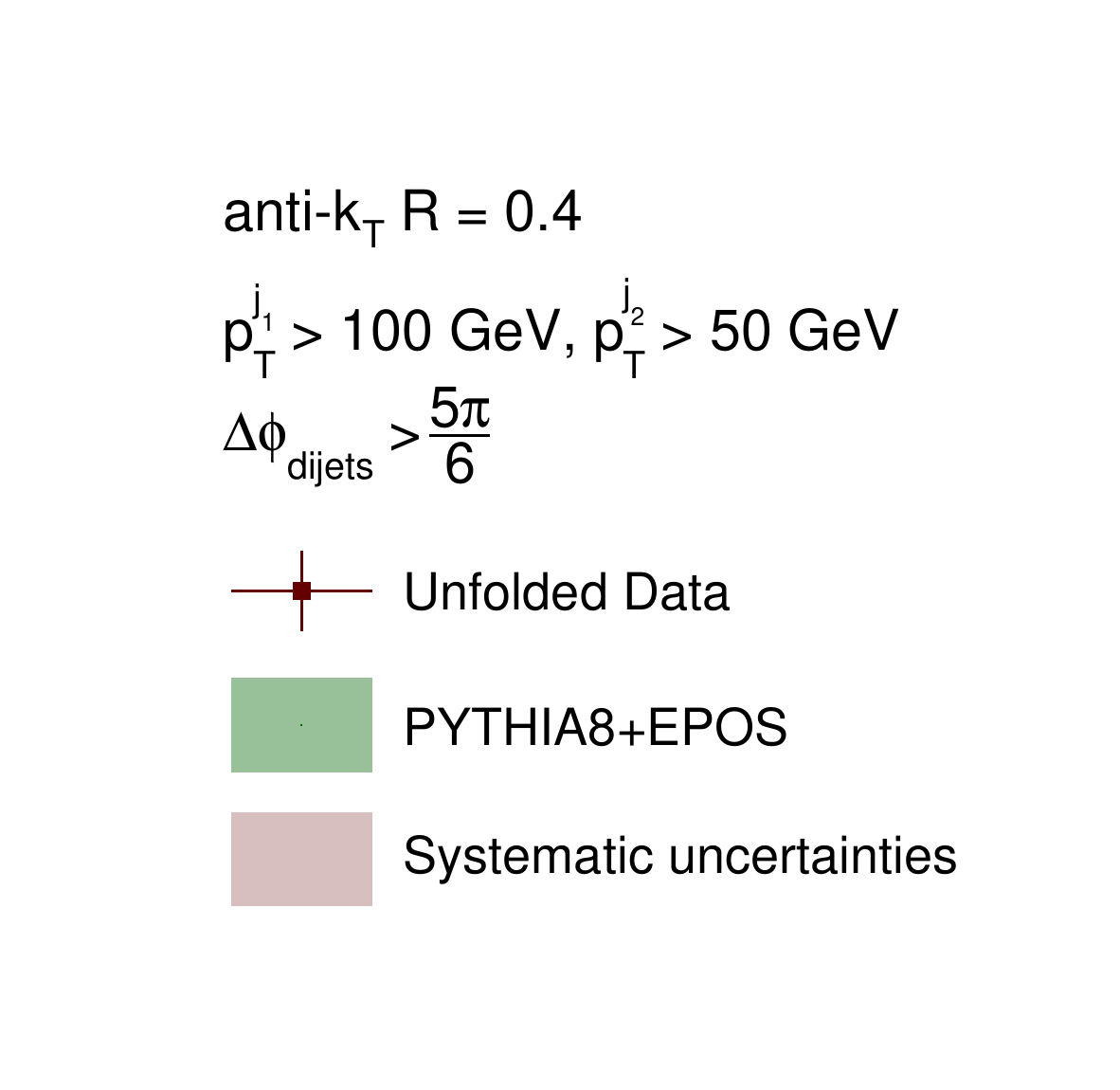} 
    \includegraphics[width=0.43\textwidth]{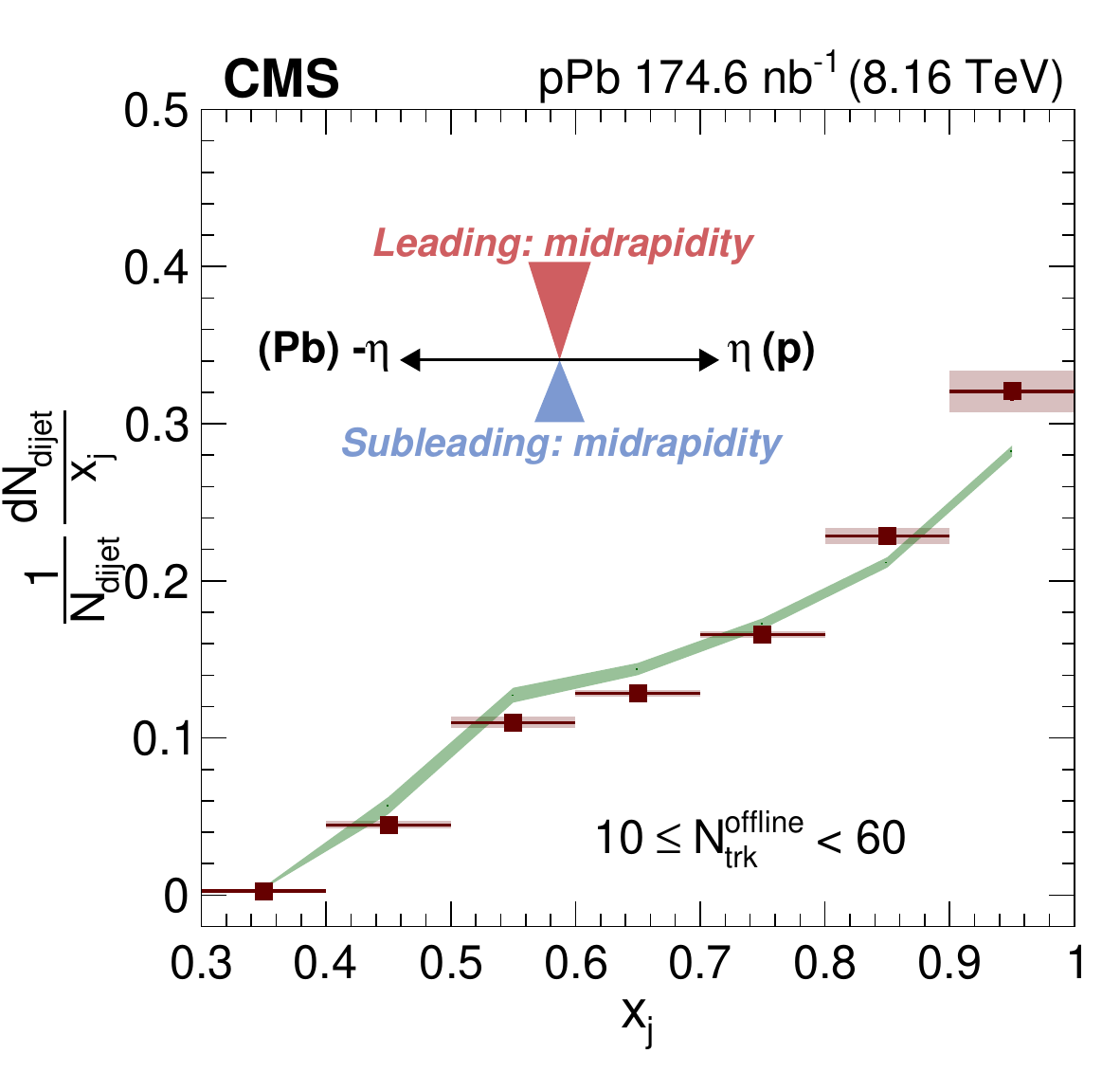} \\
    \includegraphics[width=0.43\textwidth]{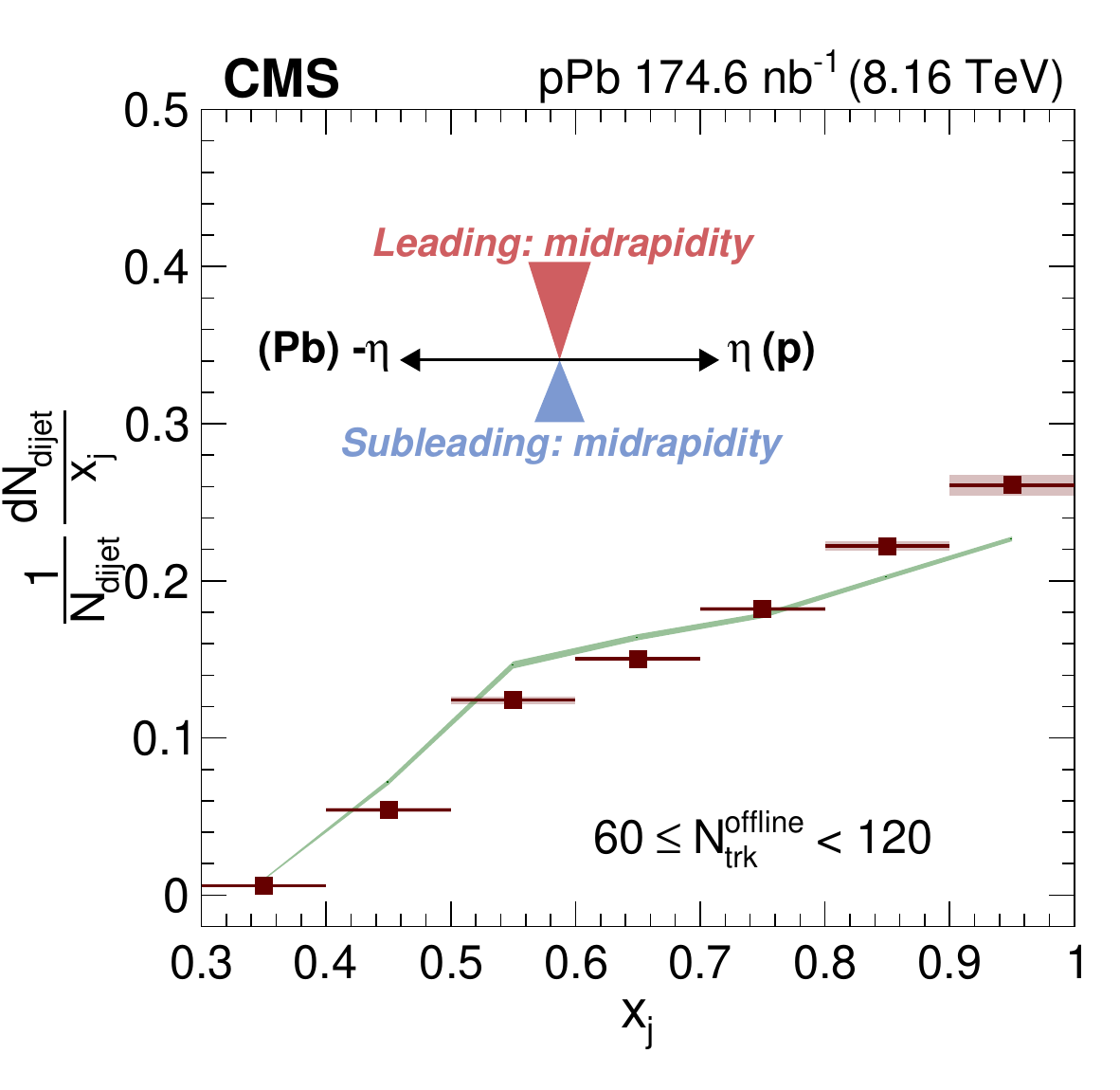} 
    \includegraphics[width=0.43\textwidth]{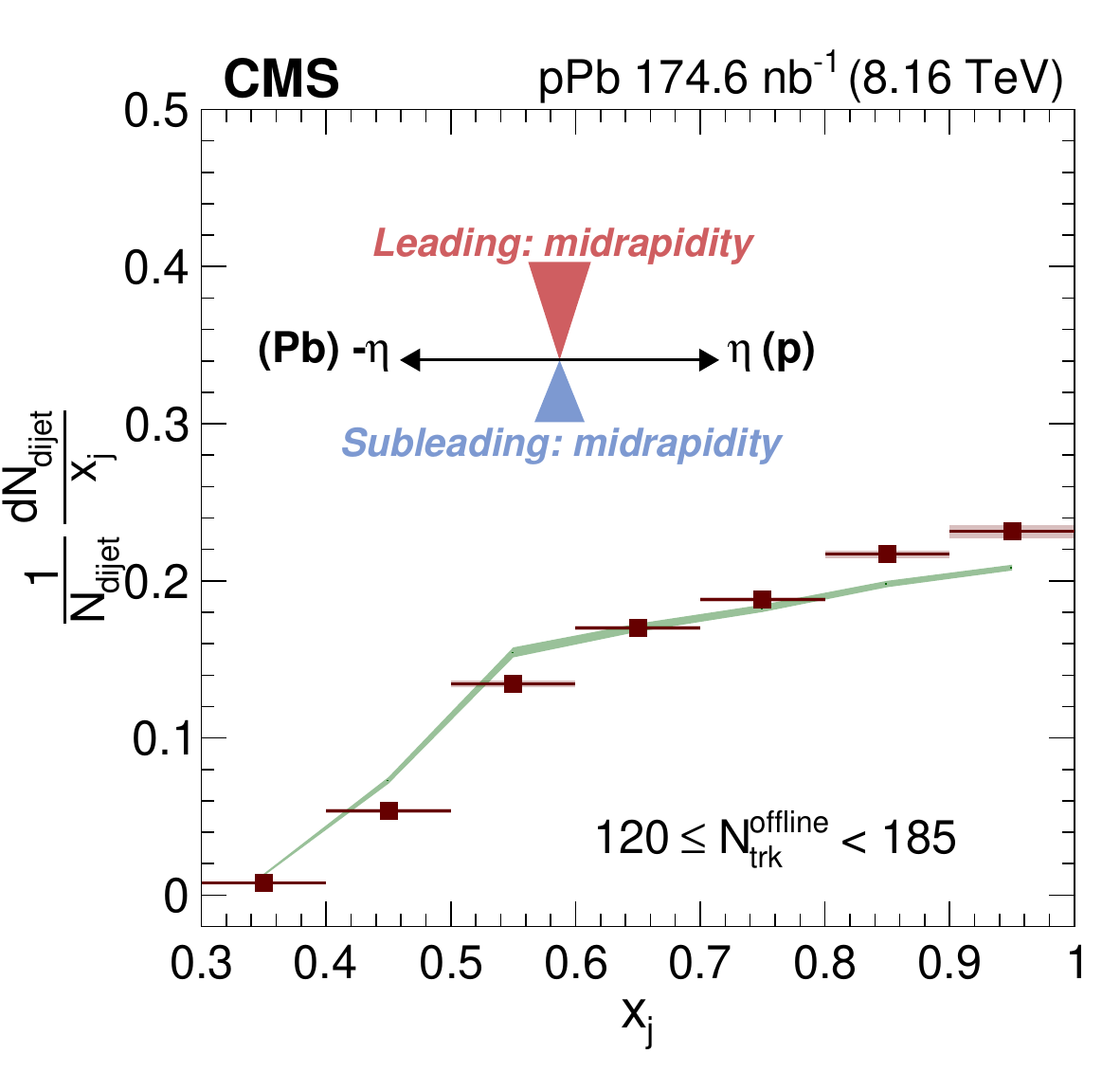} \\
    \includegraphics[width=0.43\textwidth]{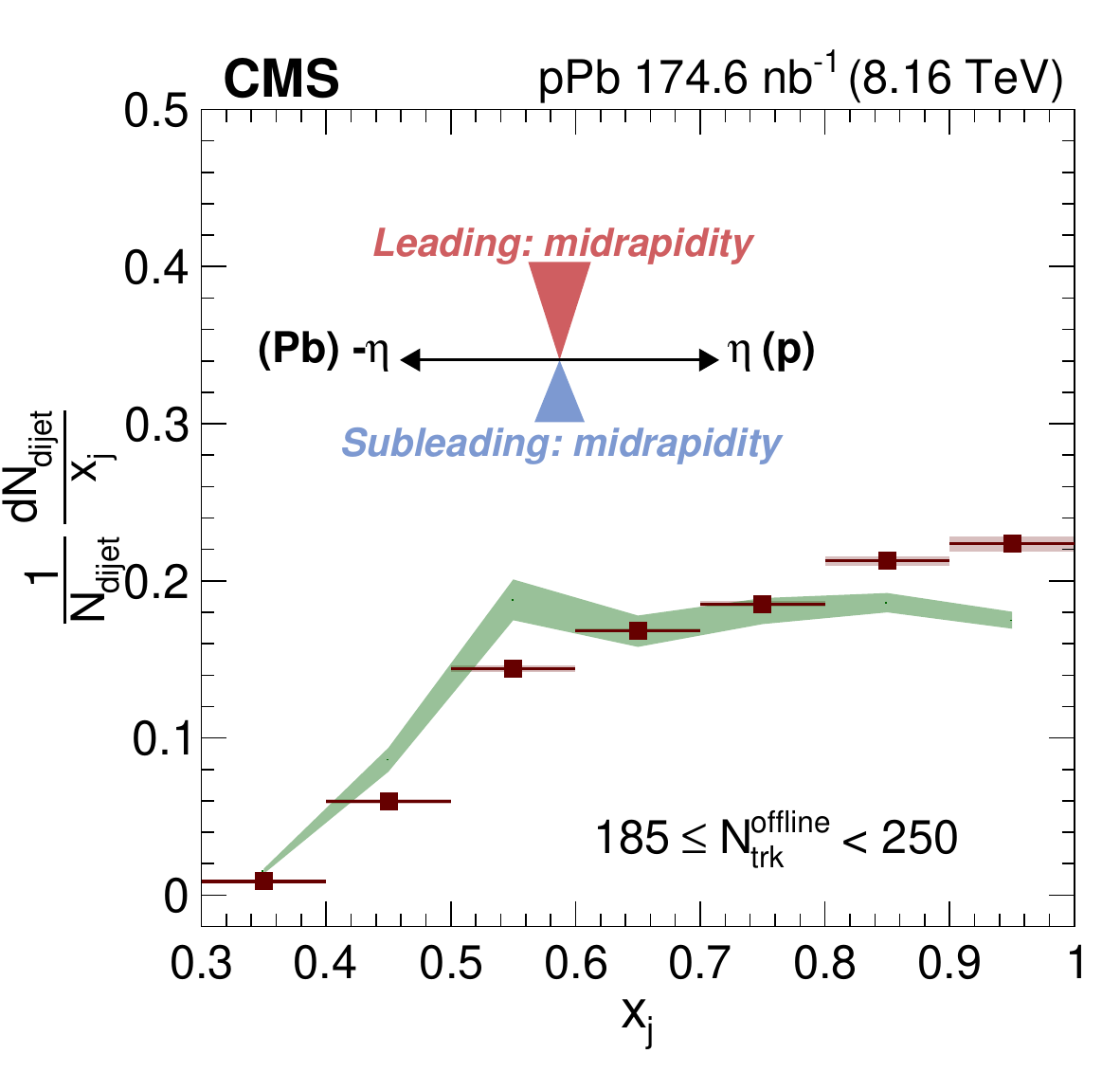} 
    \includegraphics[width=0.43\textwidth]{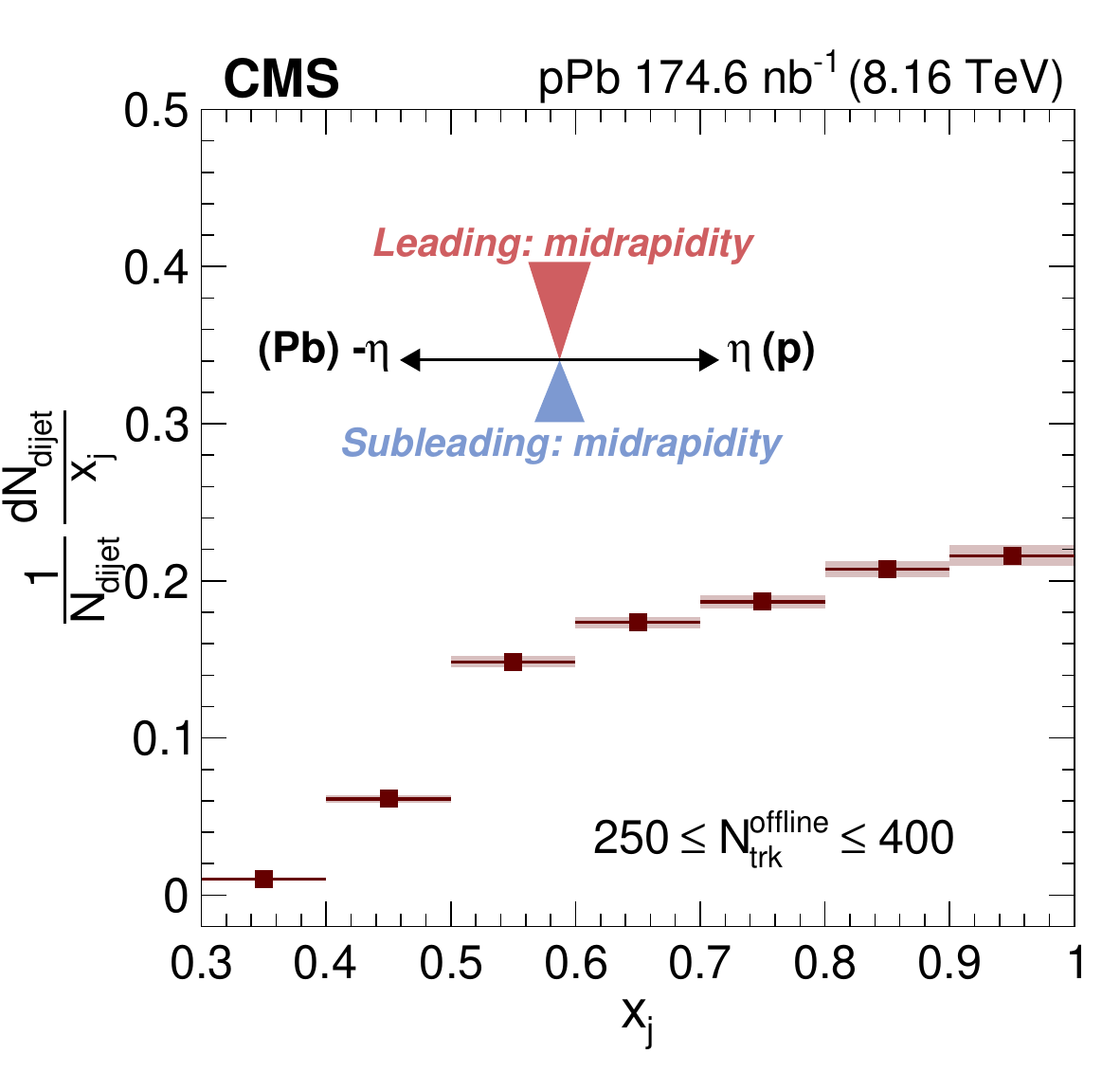} 
    \caption{The unfolded \xj distributions, from lower to higher multiplicities and for both leading and subleading jets in the midrapidity region, are shown. Comparisons with the generator-level {{\PYTHIA}8}+\EPOS simulation (green band) are also displayed. }
    \label{fig:xjmult_mid_mid}
\end{figure}

\begin{figure}[!hbpt]
  \centering
    \includegraphics[width=0.44\textwidth]{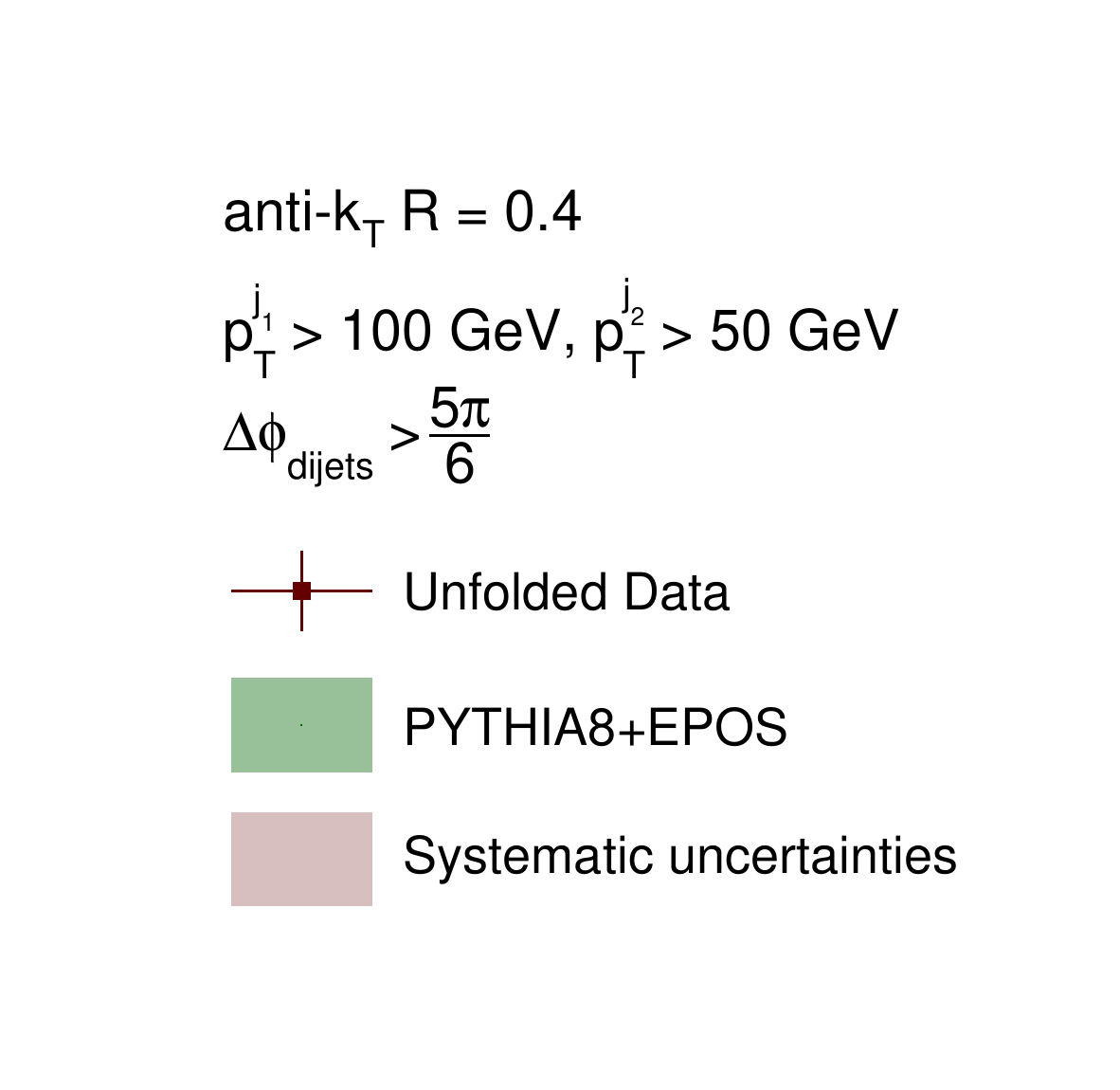} 
    \includegraphics[width=0.44\textwidth]{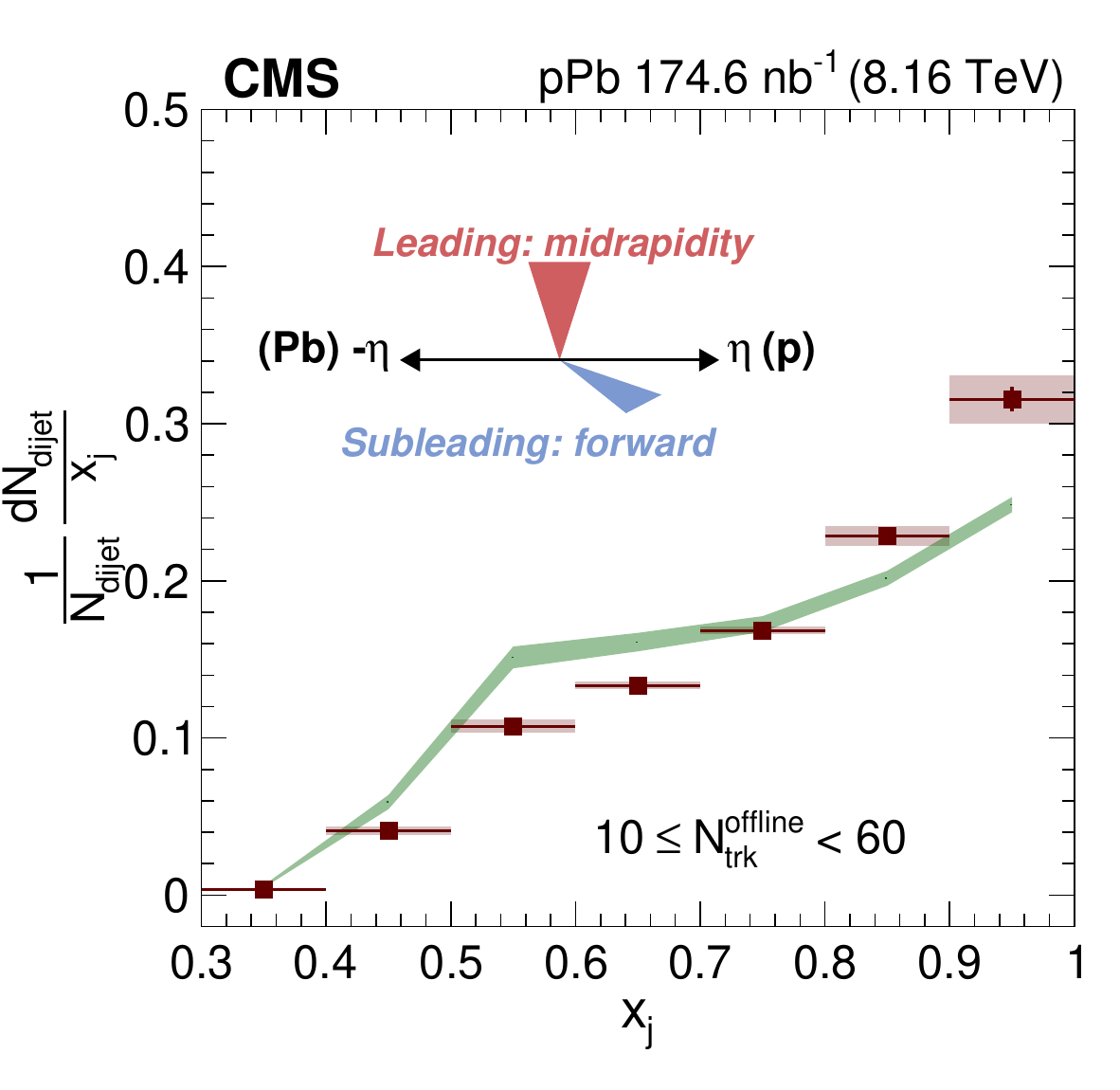} \\
    \includegraphics[width=0.44\textwidth]{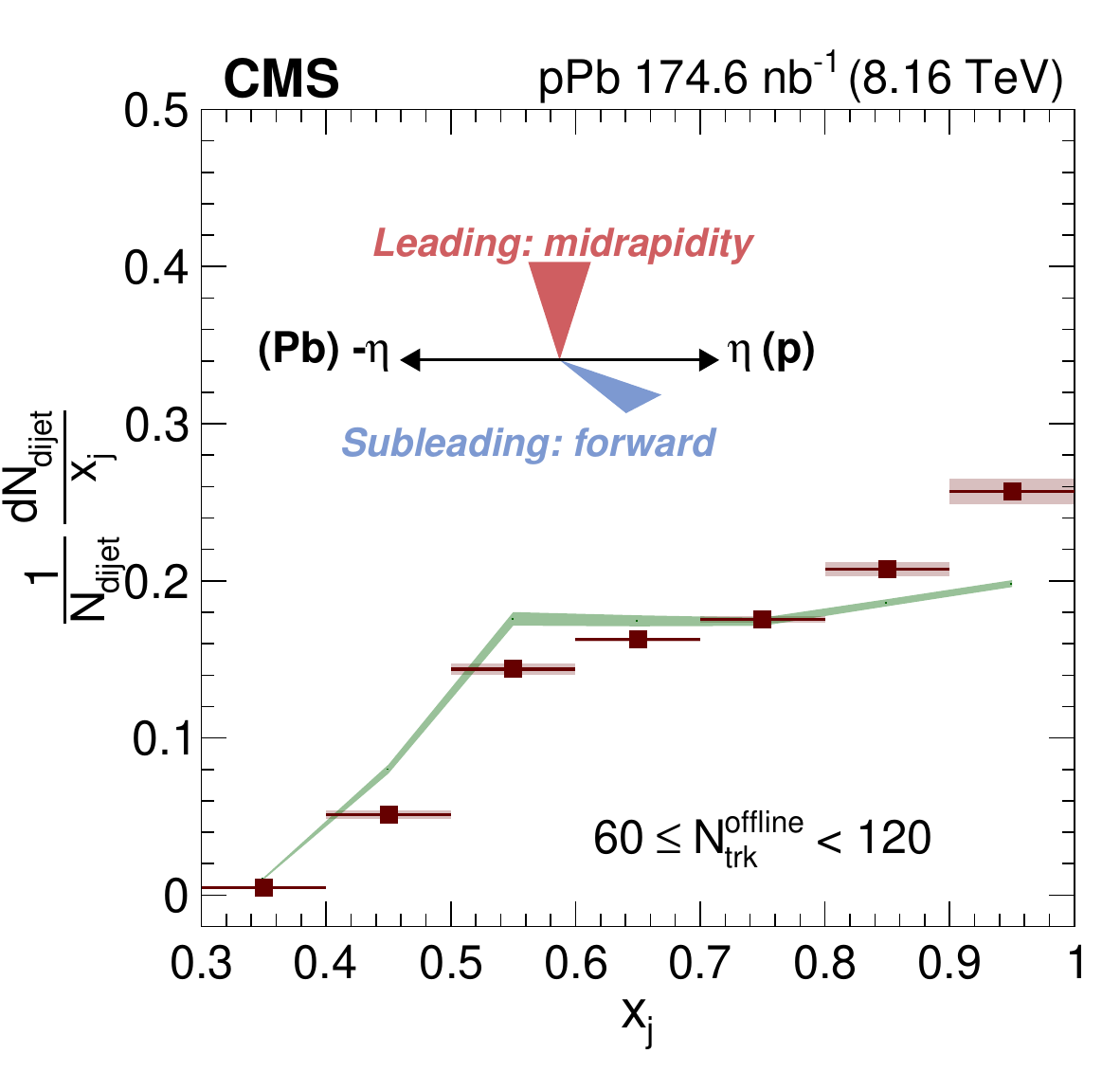} 
    \includegraphics[width=0.44\textwidth]{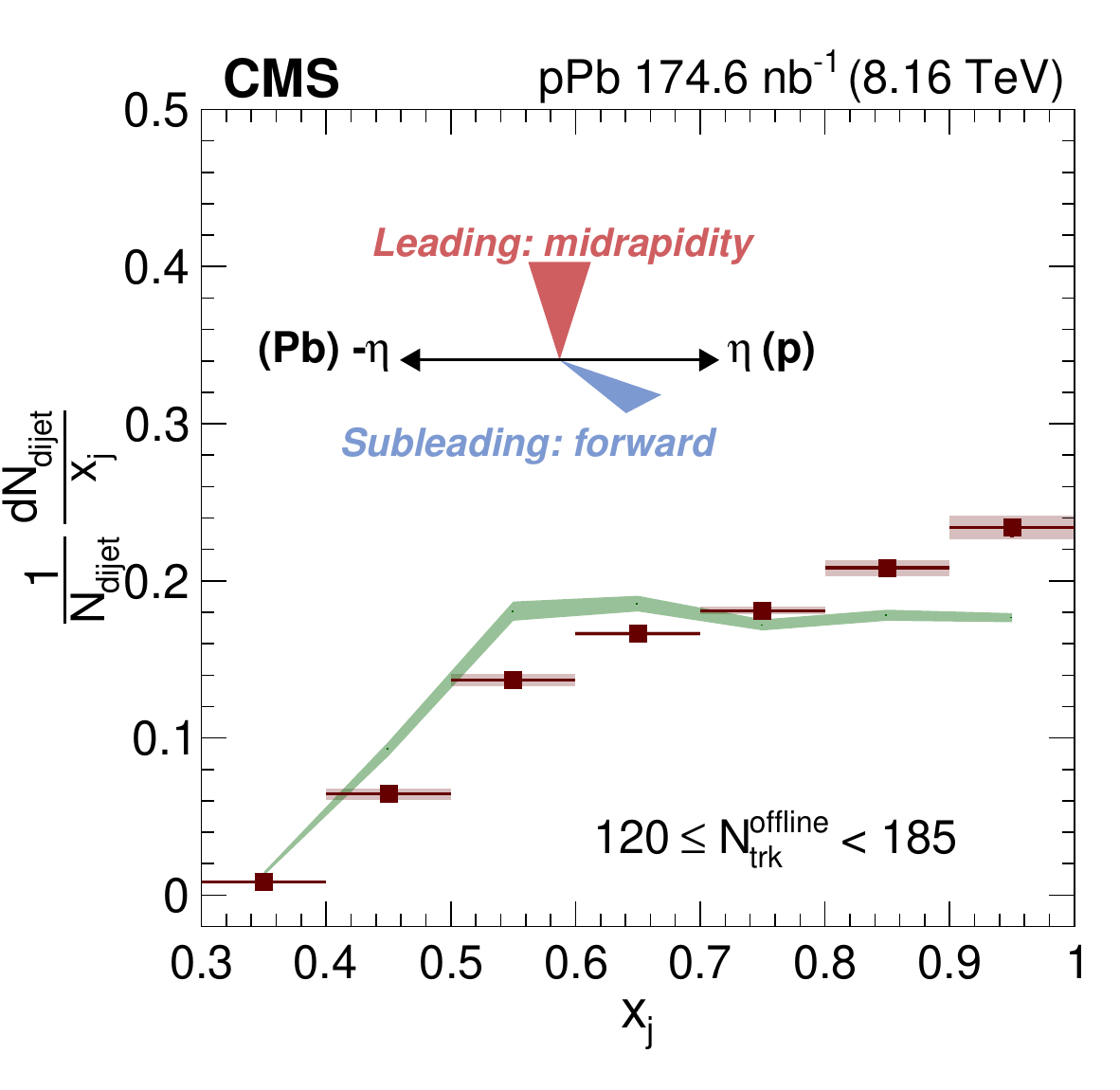} \\
    \includegraphics[width=0.44\textwidth]{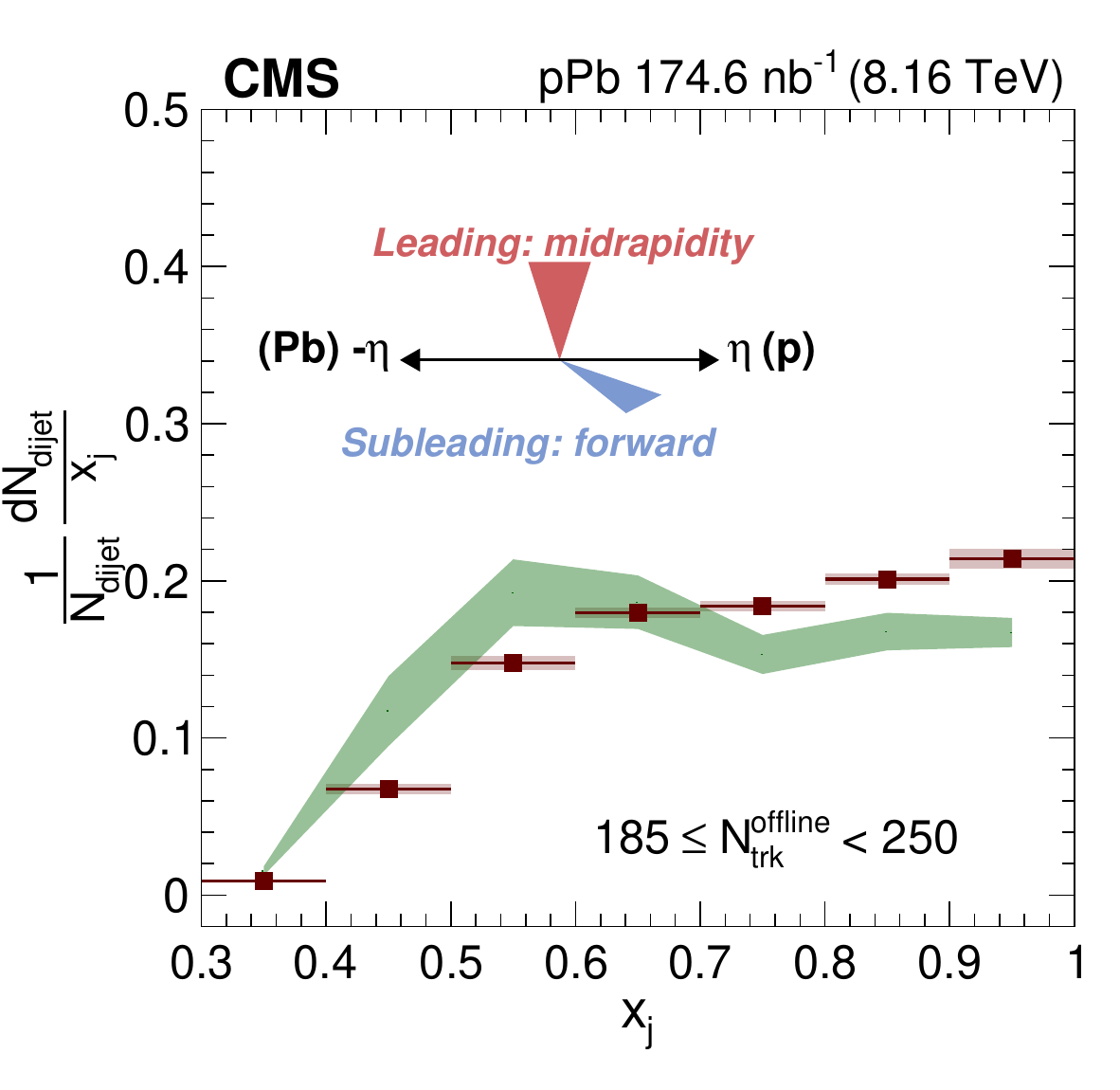} 
    \includegraphics[width=0.44\textwidth]{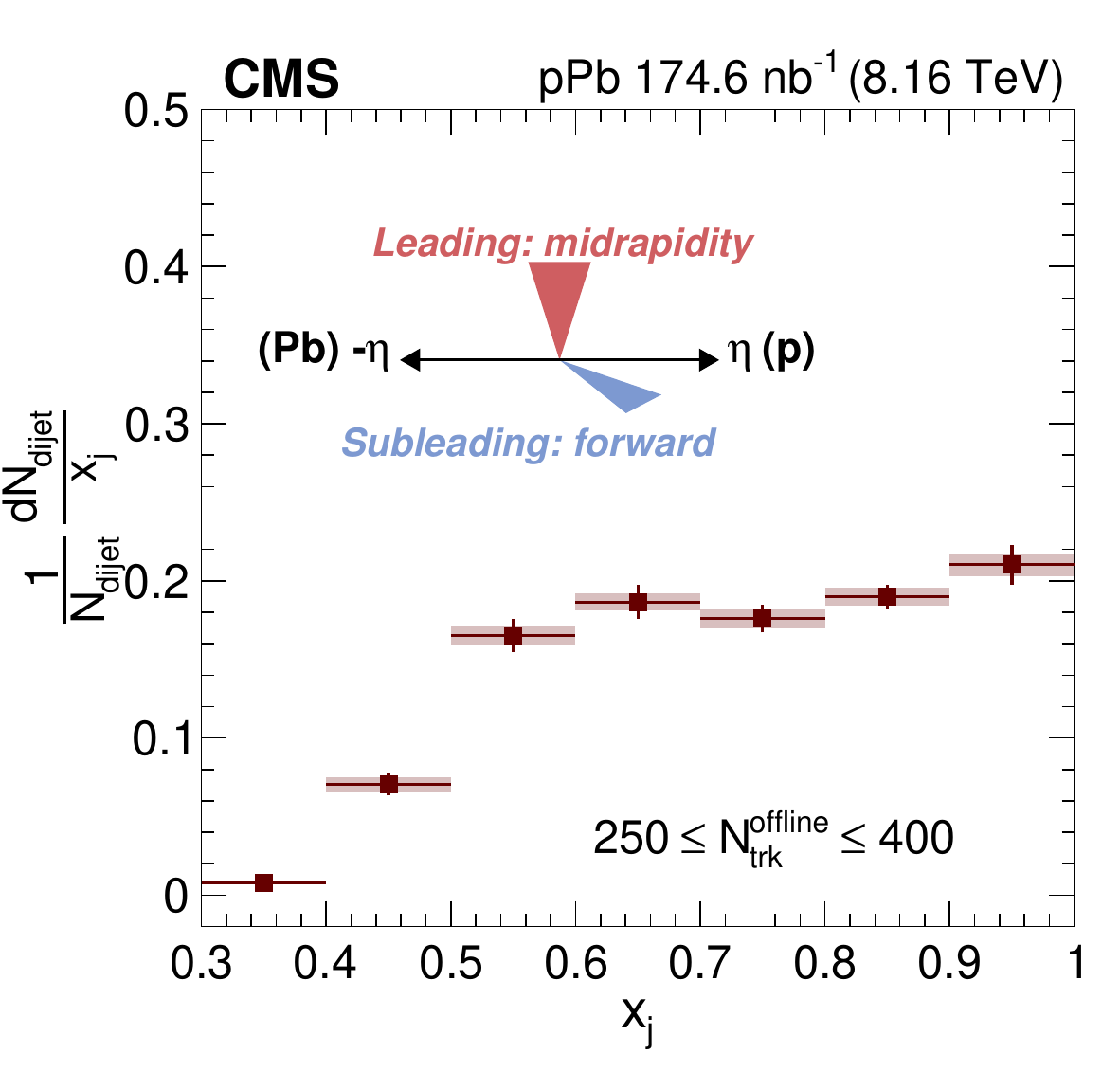} 
    \caption{The unfolded \xj distributions, from lower to higher multiplicities, and for leading and subleading jets in the midrapidity and forward regions, respectively, are shown. Comparisons with the generator-level {{\PYTHIA}8}+\EPOS simulation (green band) are also displayed. }
    \label{fig:xjmult_mid_fwd}
\end{figure}

\begin{figure}[!hbpt]
  \centering
    \includegraphics[width=0.44\textwidth]{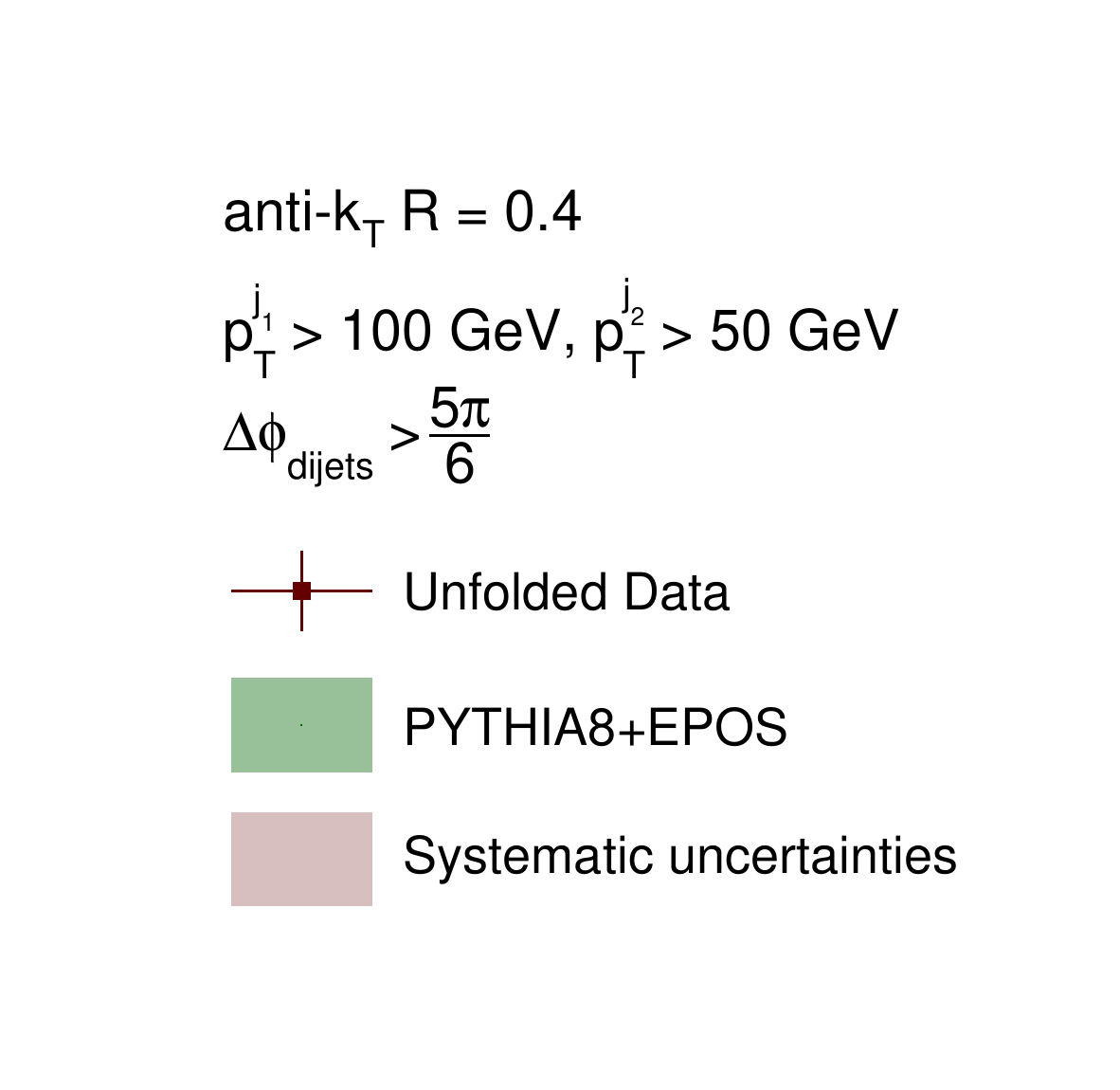} 
    \includegraphics[width=0.44\textwidth]{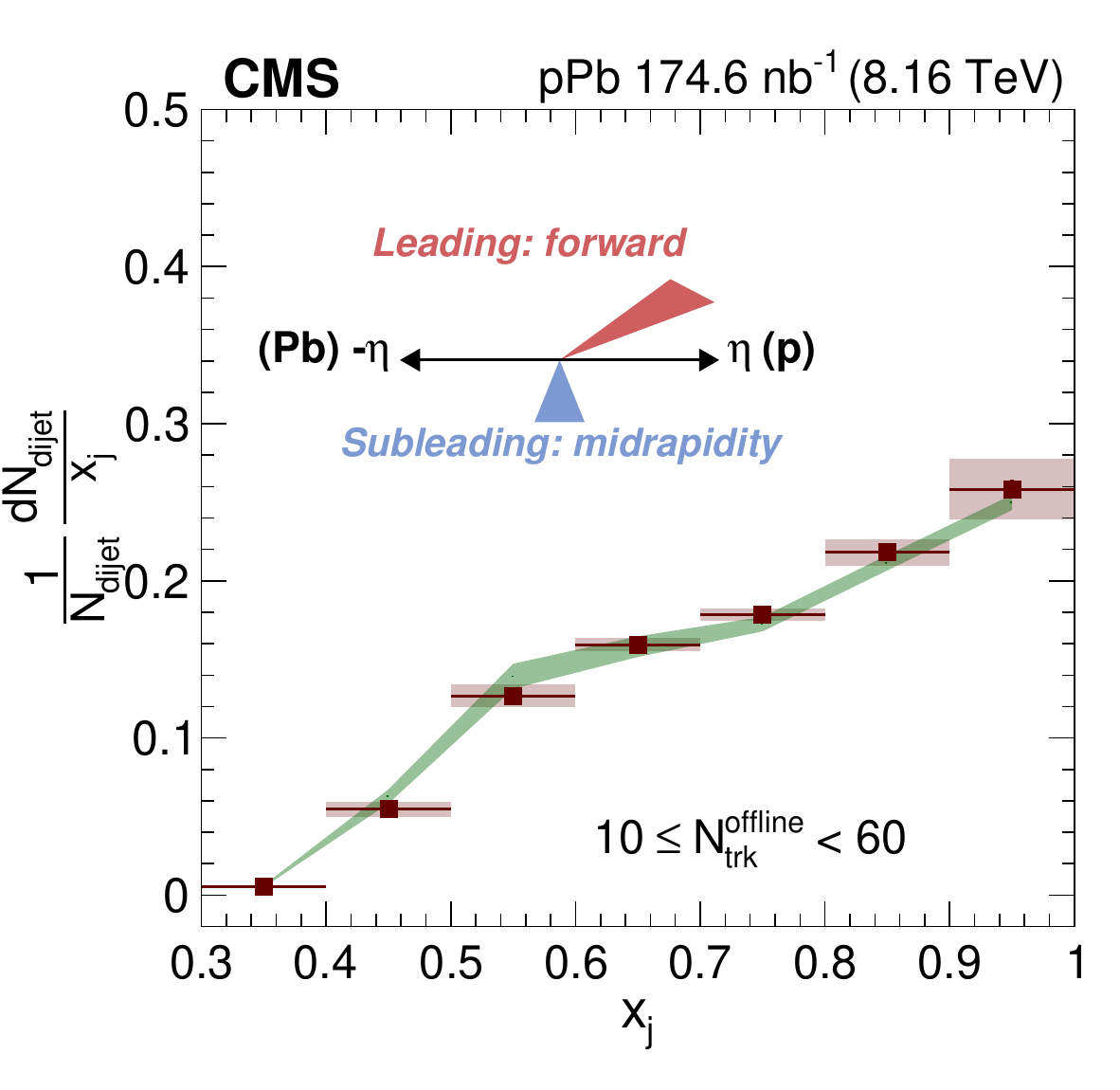} \\
    \includegraphics[width=0.44\textwidth]{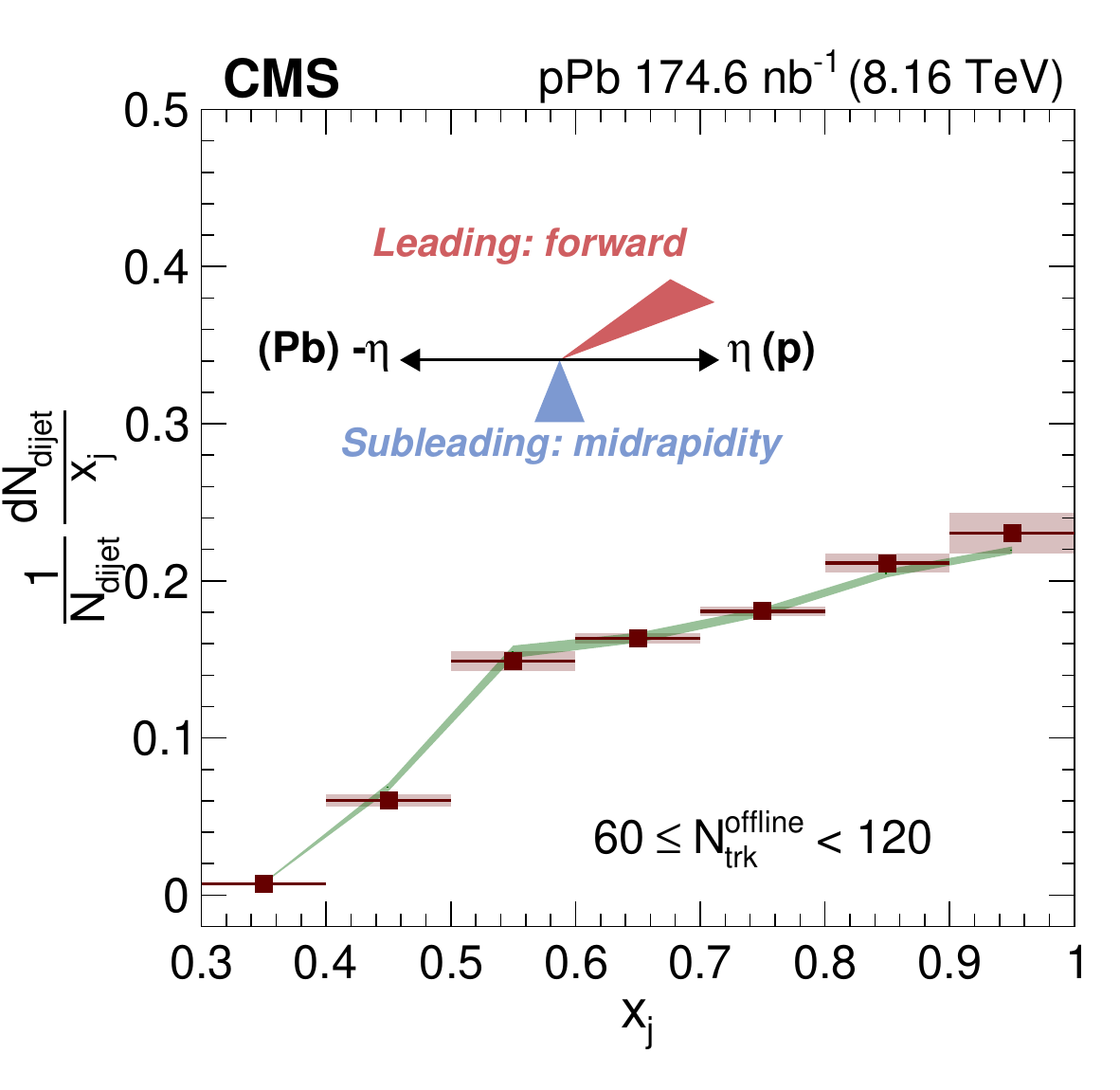} 
    \includegraphics[width=0.44\textwidth]{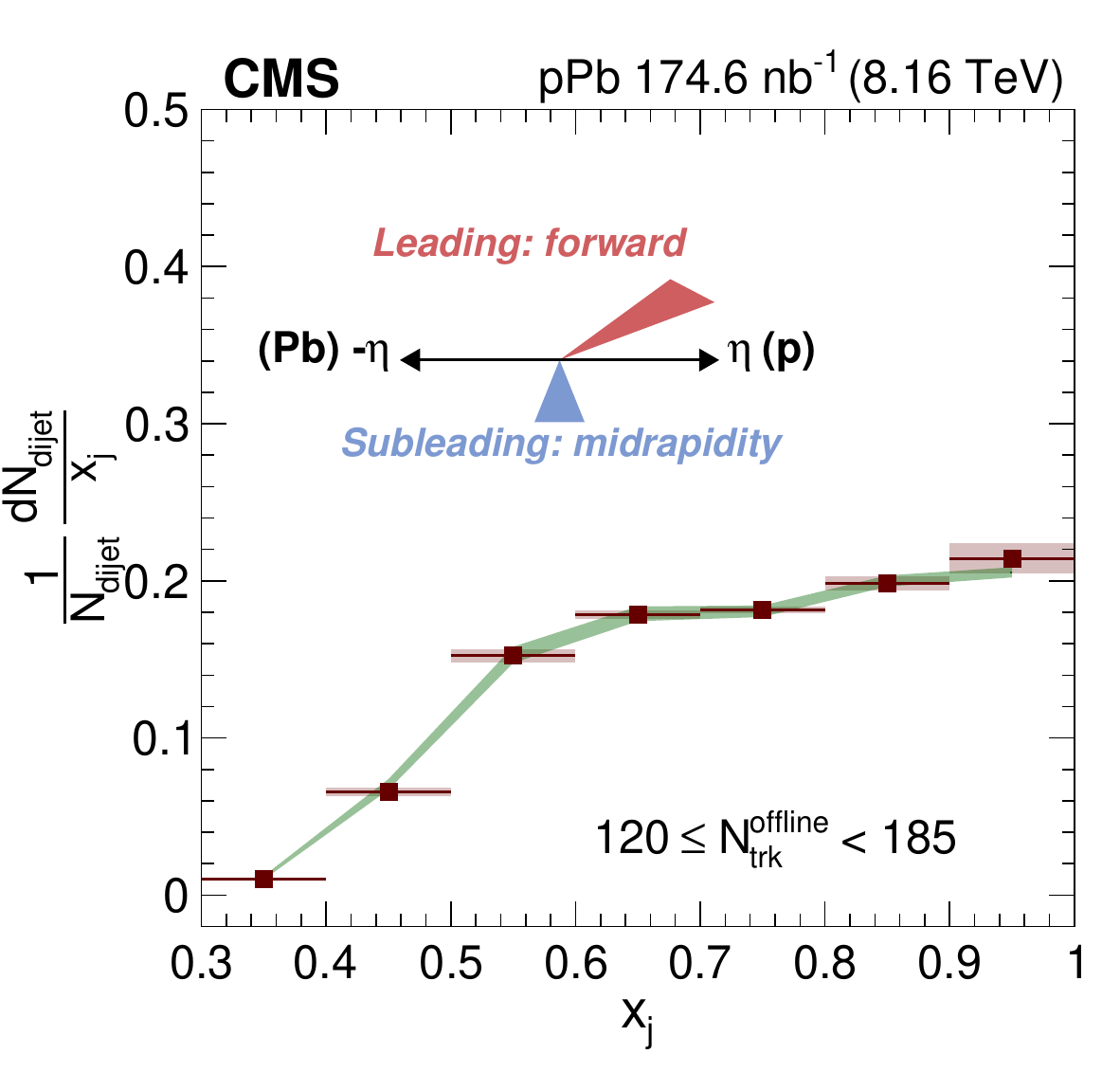} \\
    \includegraphics[width=0.44\textwidth]{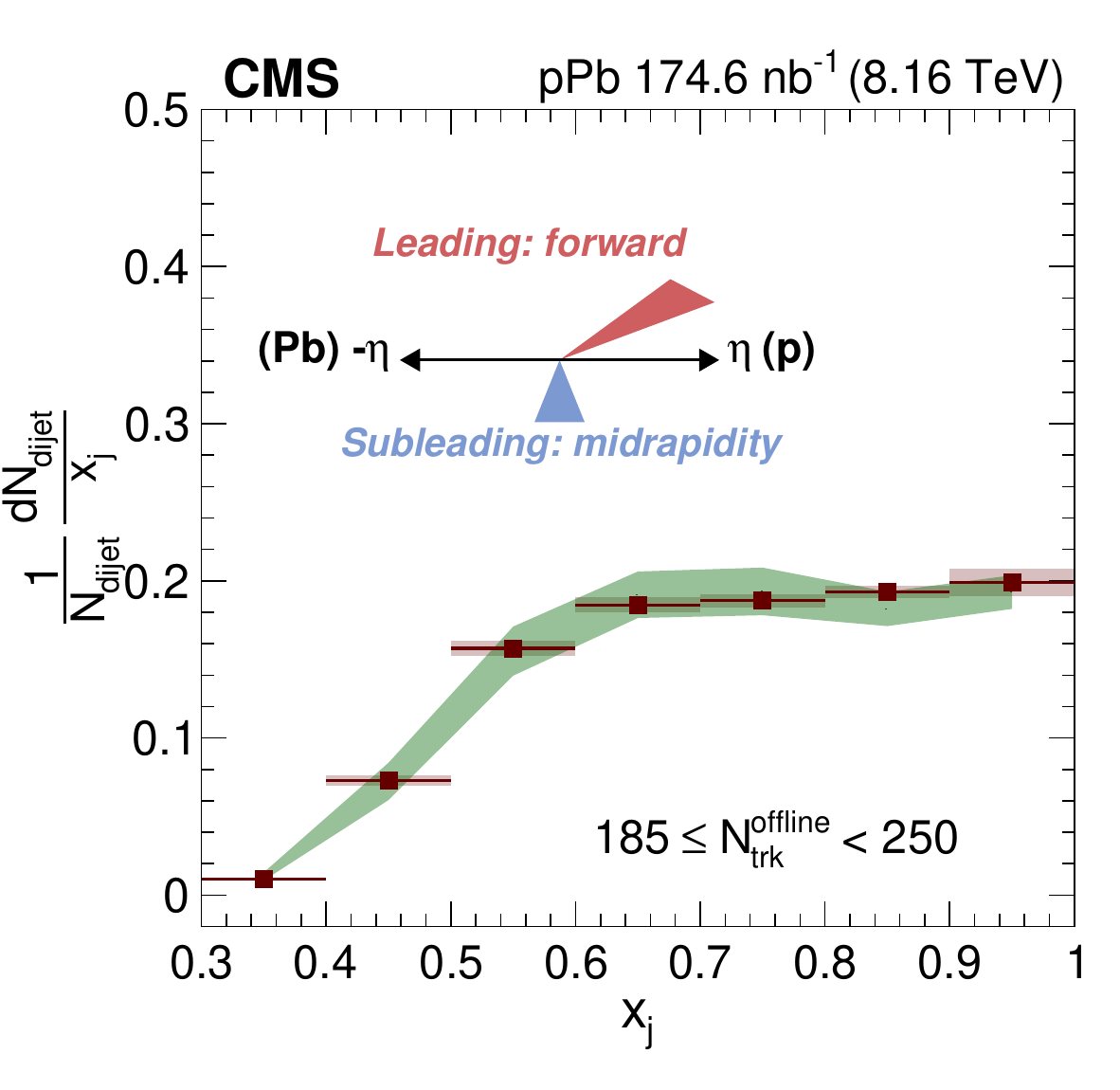} 
    \includegraphics[width=0.44\textwidth]{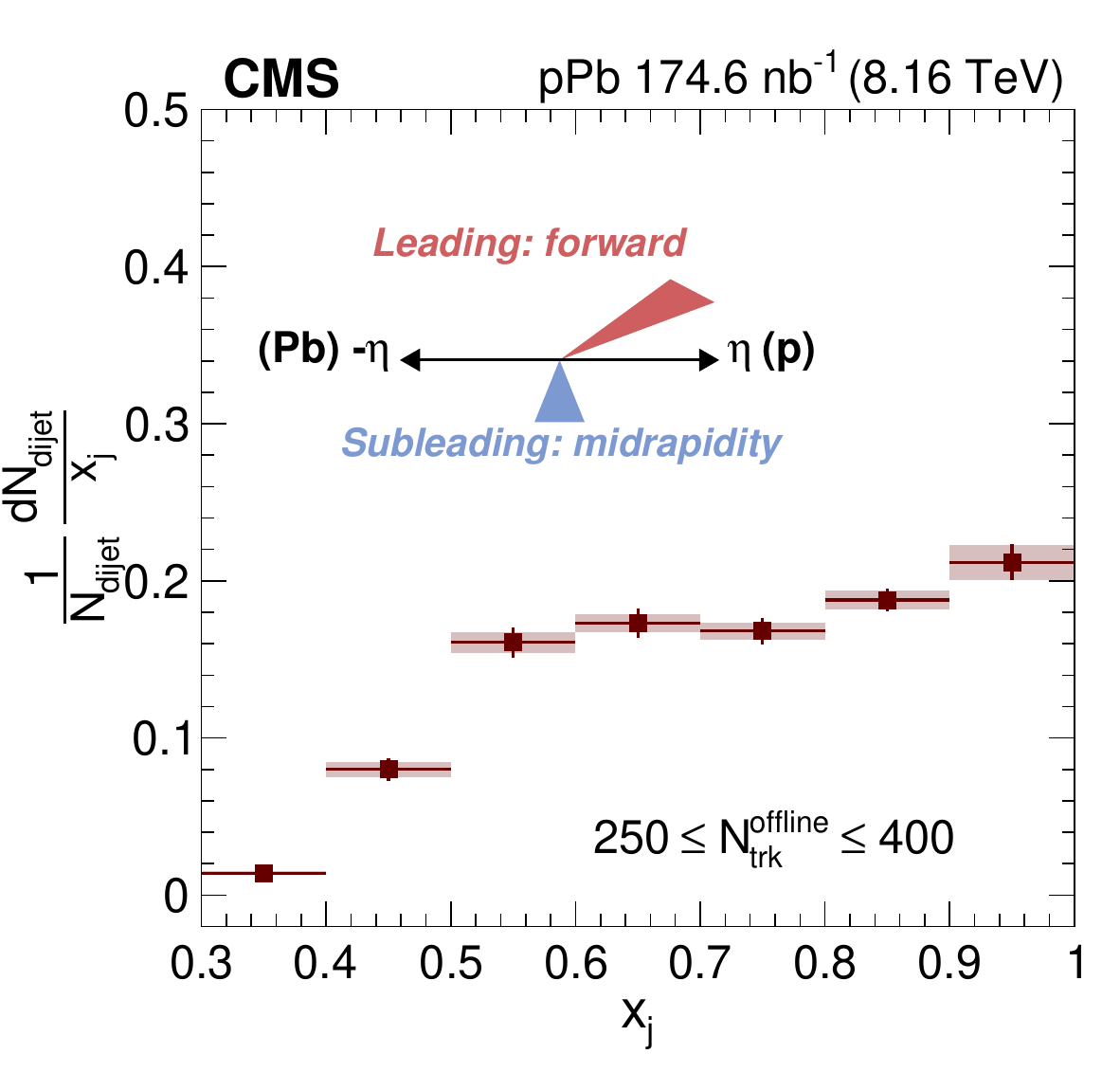} 
    \caption{The unfolded \xj distributions, from lower to higher multiplicities, and for leading and subleading jets in the forward and midrapidity regions, respectively, are shown. Comparisons with the generator-level {{\PYTHIA}8}+\EPOS simulation (green band) are also displayed.}
    \label{fig:xjmult_fwd_mid}
\end{figure}

\begin{figure}[!hbpt]
  \centering
    \includegraphics[width=0.44\textwidth]{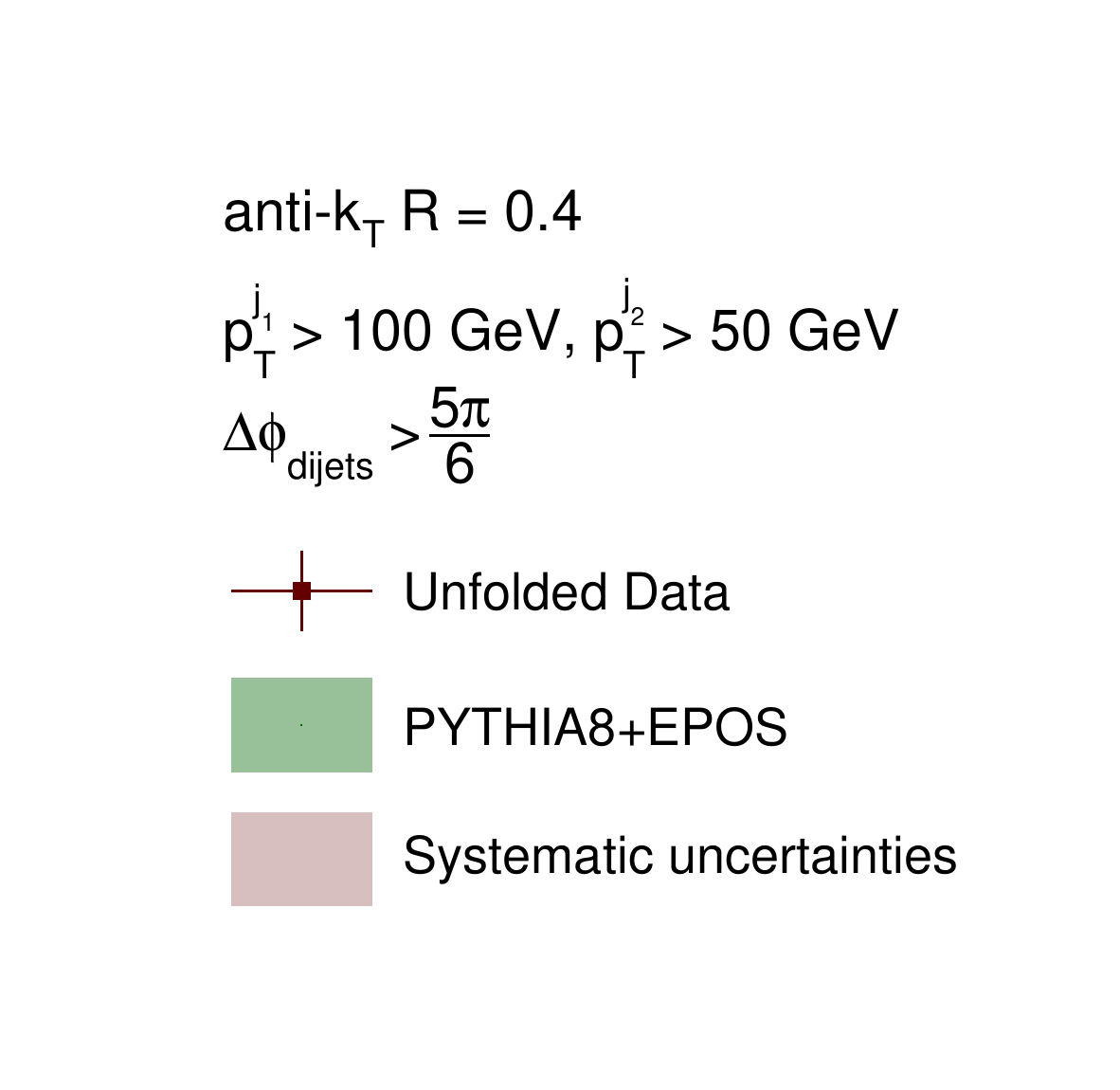} 
    \includegraphics[width=0.44\textwidth]{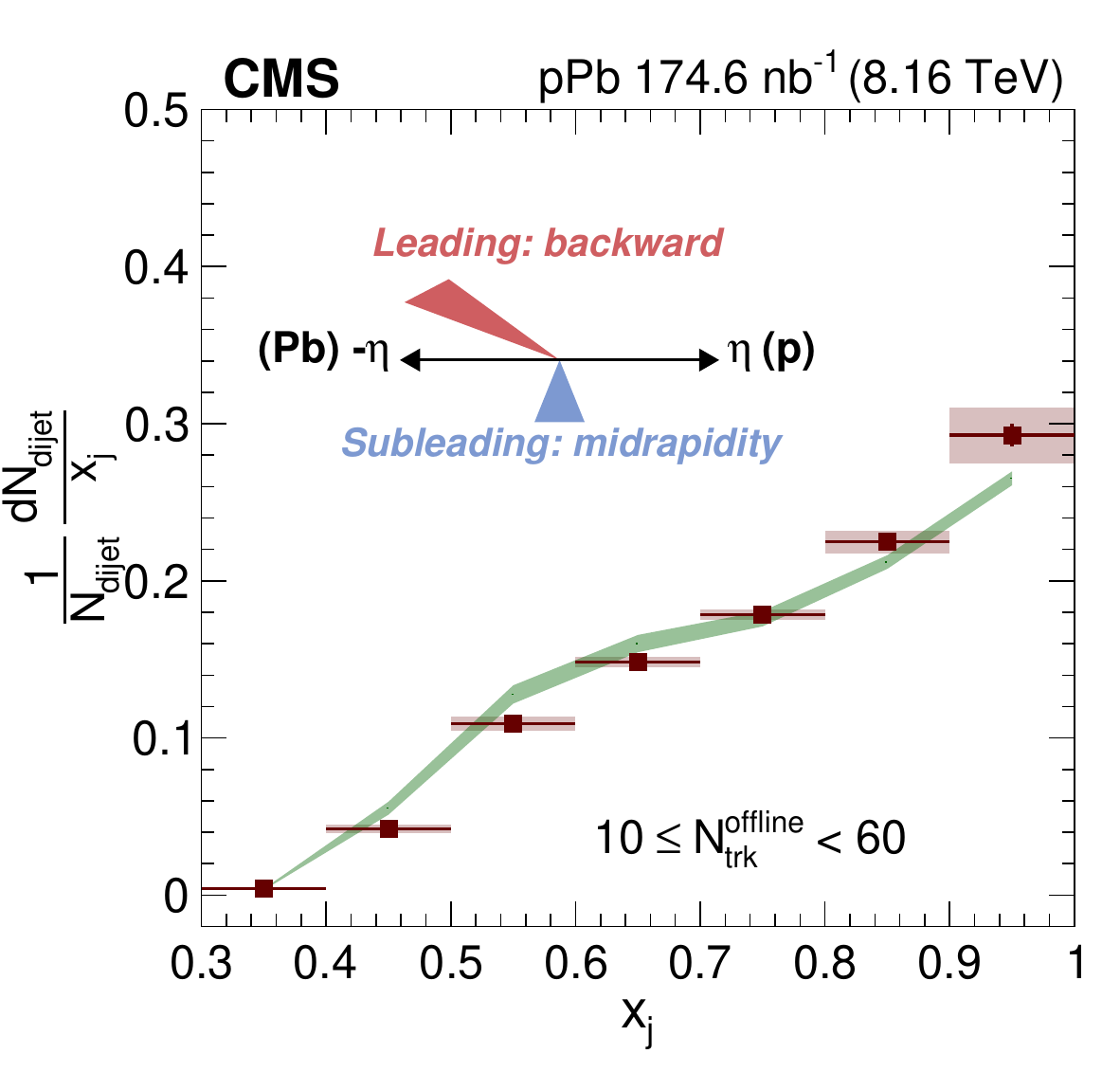} \\
    \includegraphics[width=0.44\textwidth]{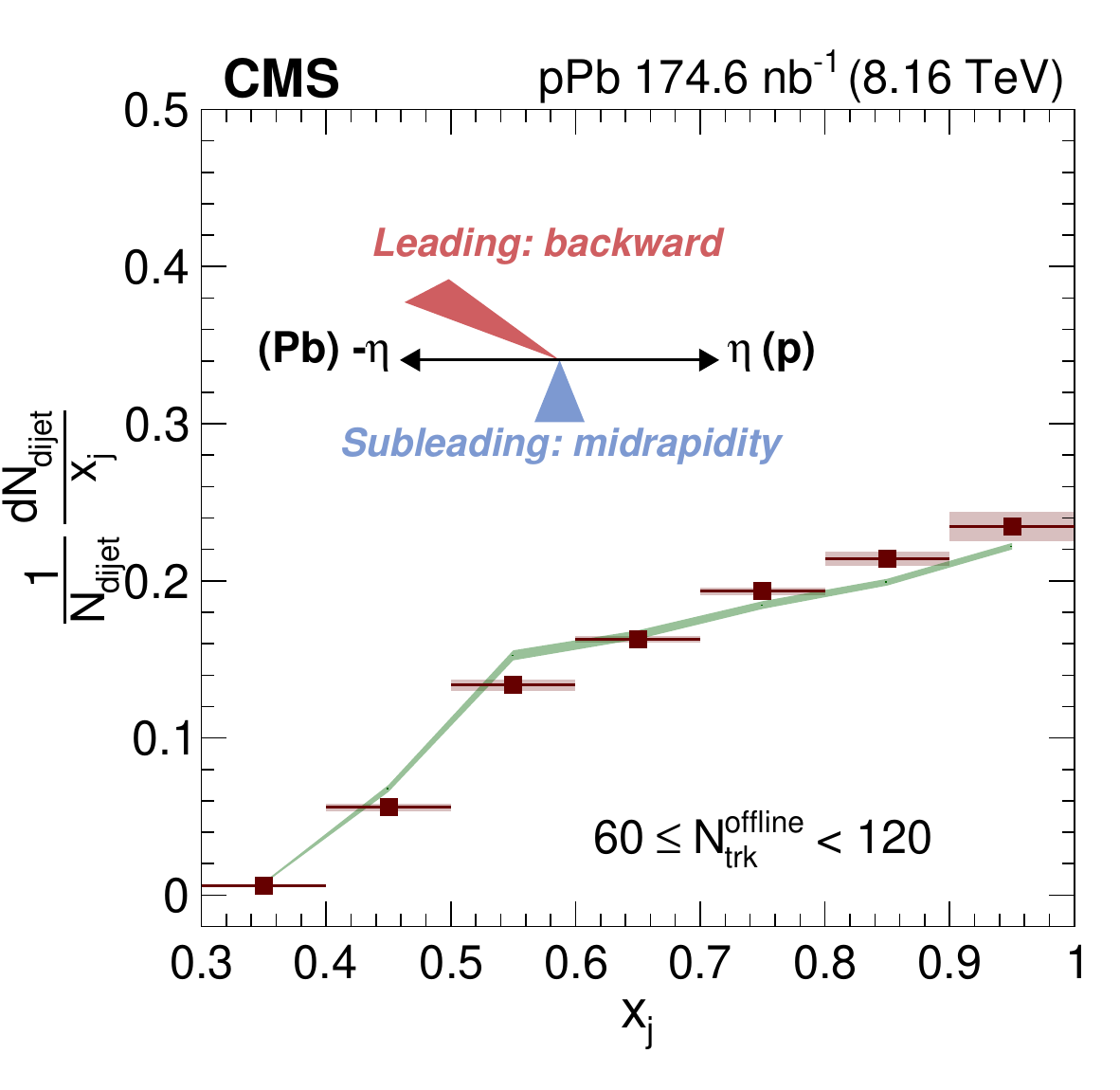} 
    \includegraphics[width=0.44\textwidth]{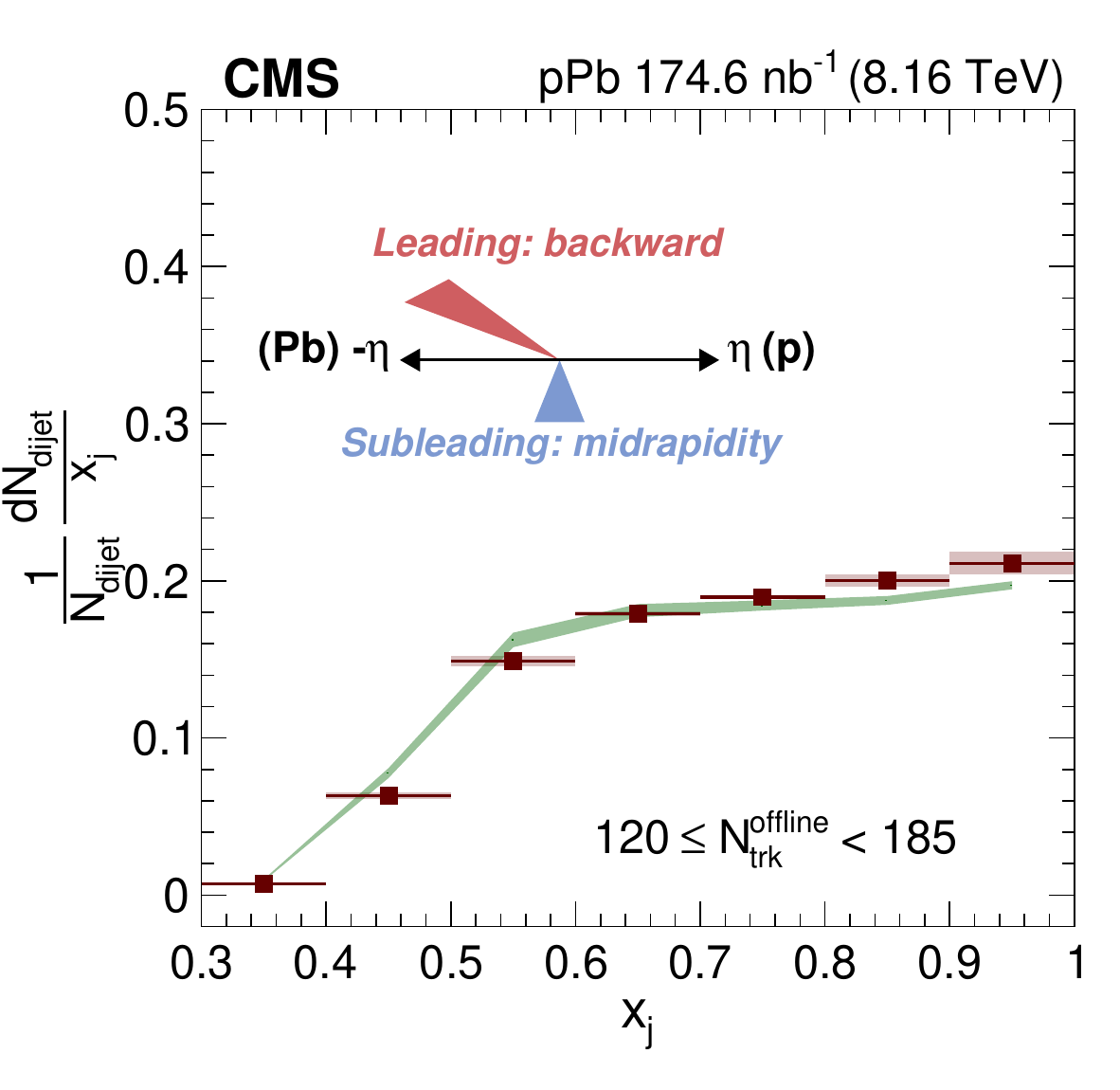} \\
    \includegraphics[width=0.44\textwidth]{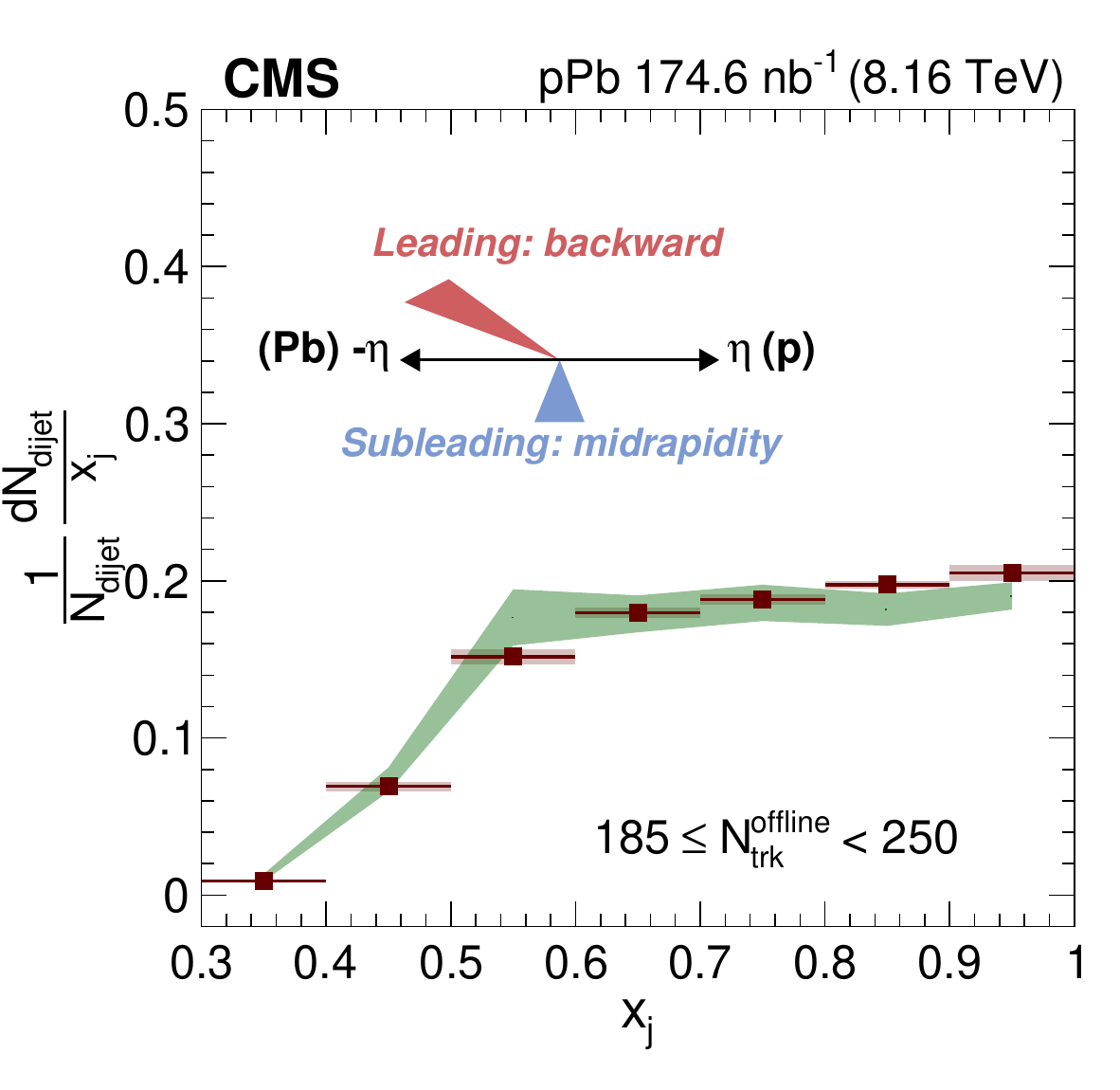} 
    \includegraphics[width=0.44\textwidth]{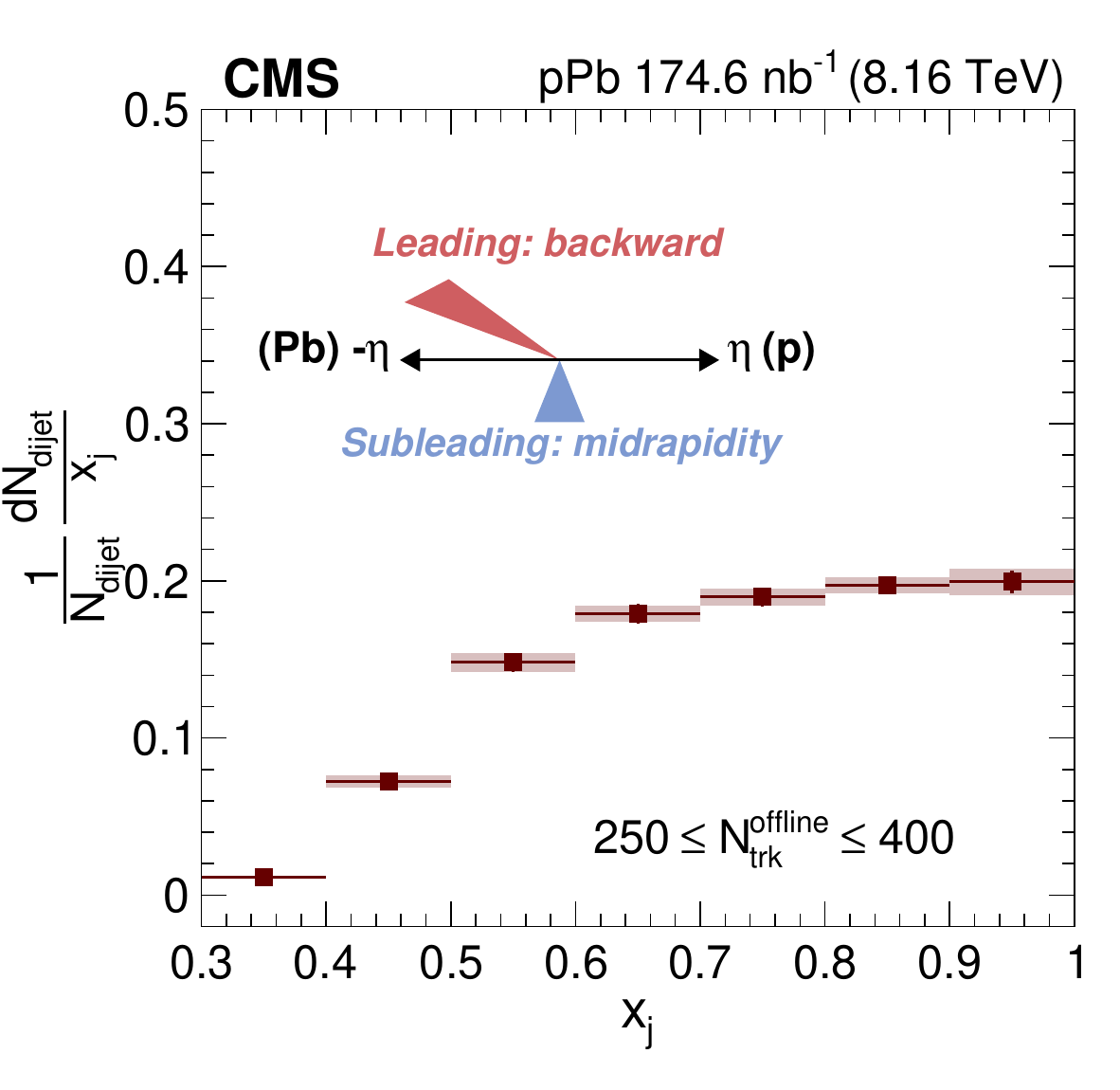} 
    \caption{The unfolded \xj distributions, from lower to higher multiplicities, and for leading and subleading jets in the backward and midrapidity regions, respectively, are shown. Comparisons with the generator-level {{\PYTHIA}8}+\EPOS simulation (green band) are also displayed. }
    \label{fig:xjmult_bkw_mid}
\end{figure}

\begin{figure}[!hbpt]
  \centering
    \includegraphics[width=0.44\textwidth]{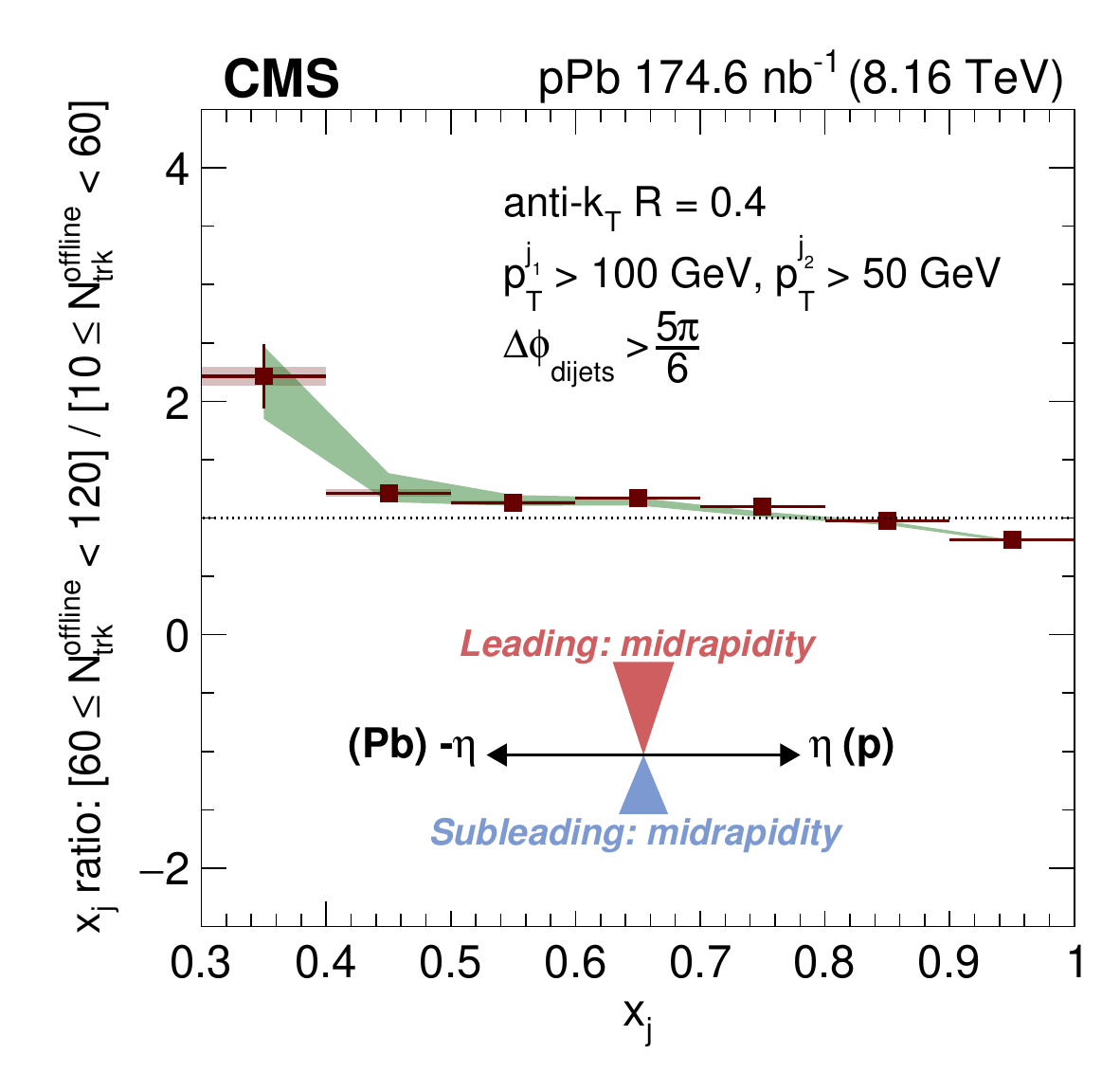} 
    \includegraphics[width=0.44\textwidth]{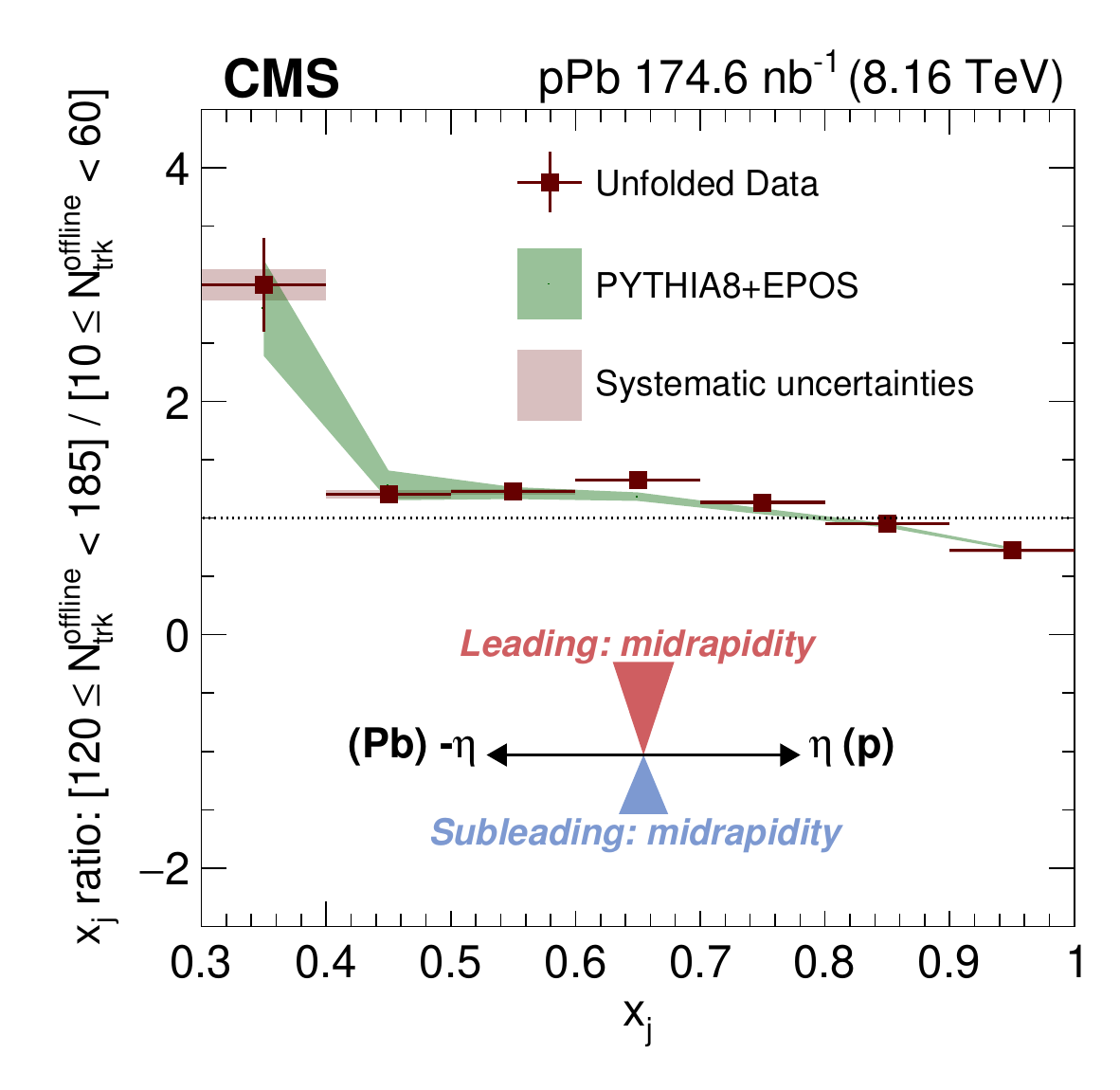} \\
    \includegraphics[width=0.44\textwidth]{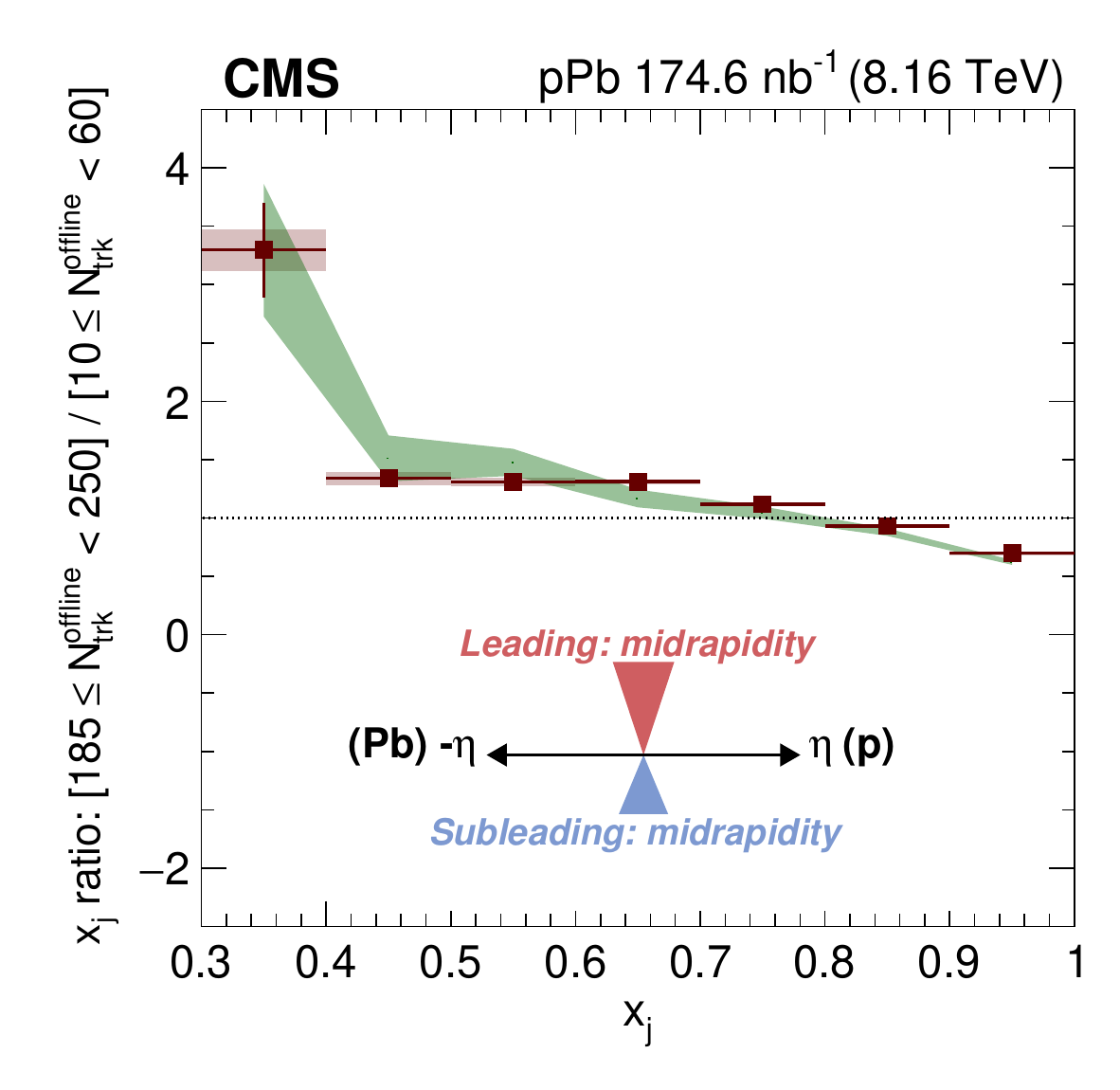} 
    \includegraphics[width=0.44\textwidth]{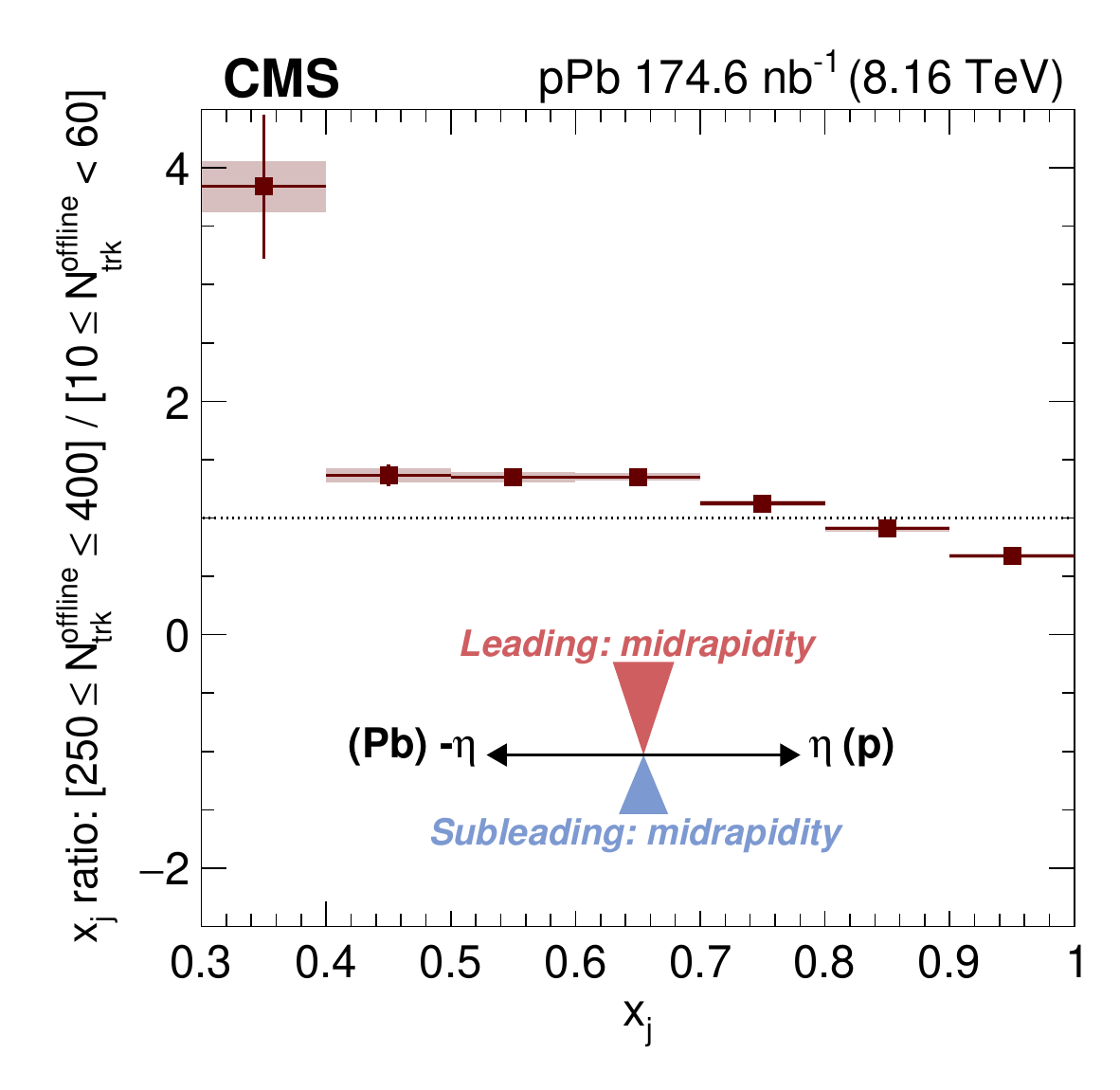} 
    \caption{The unfolded \xj ratio in the range $10 \leq \Ntroff < 60$, from lower to higher multiplicities for both leading and subleading jets in the midrapidity region, are shown. Comparison with the generator-level {{\PYTHIA}8}+\EPOS simulation (green band) are also displayed. }
    \label{fig:xjratiomult_mid_mid}
\end{figure}

\begin{figure}[!hbpt]
  \centering
    \includegraphics[width=0.44\textwidth]{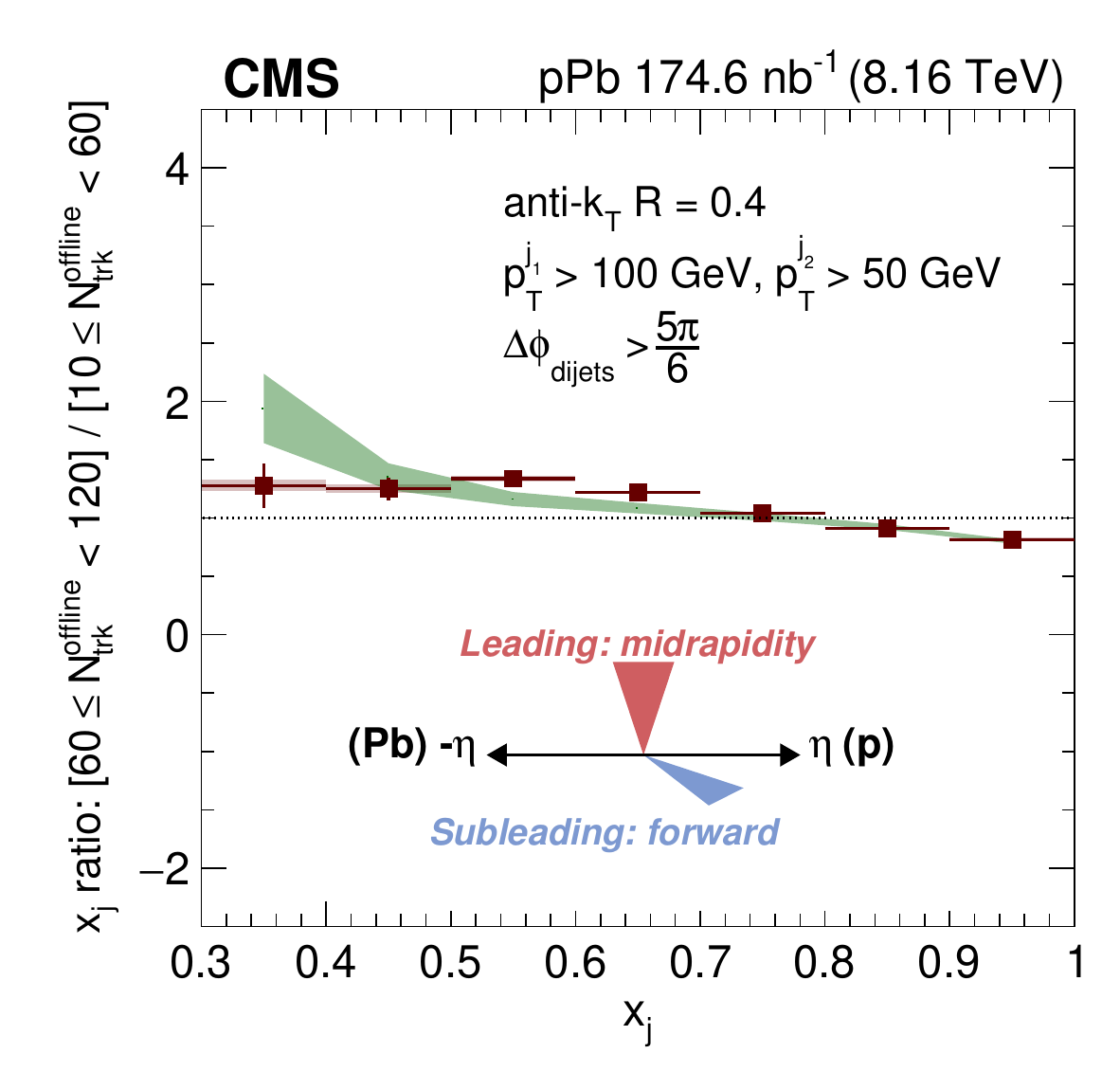} 
    \includegraphics[width=0.44\textwidth]{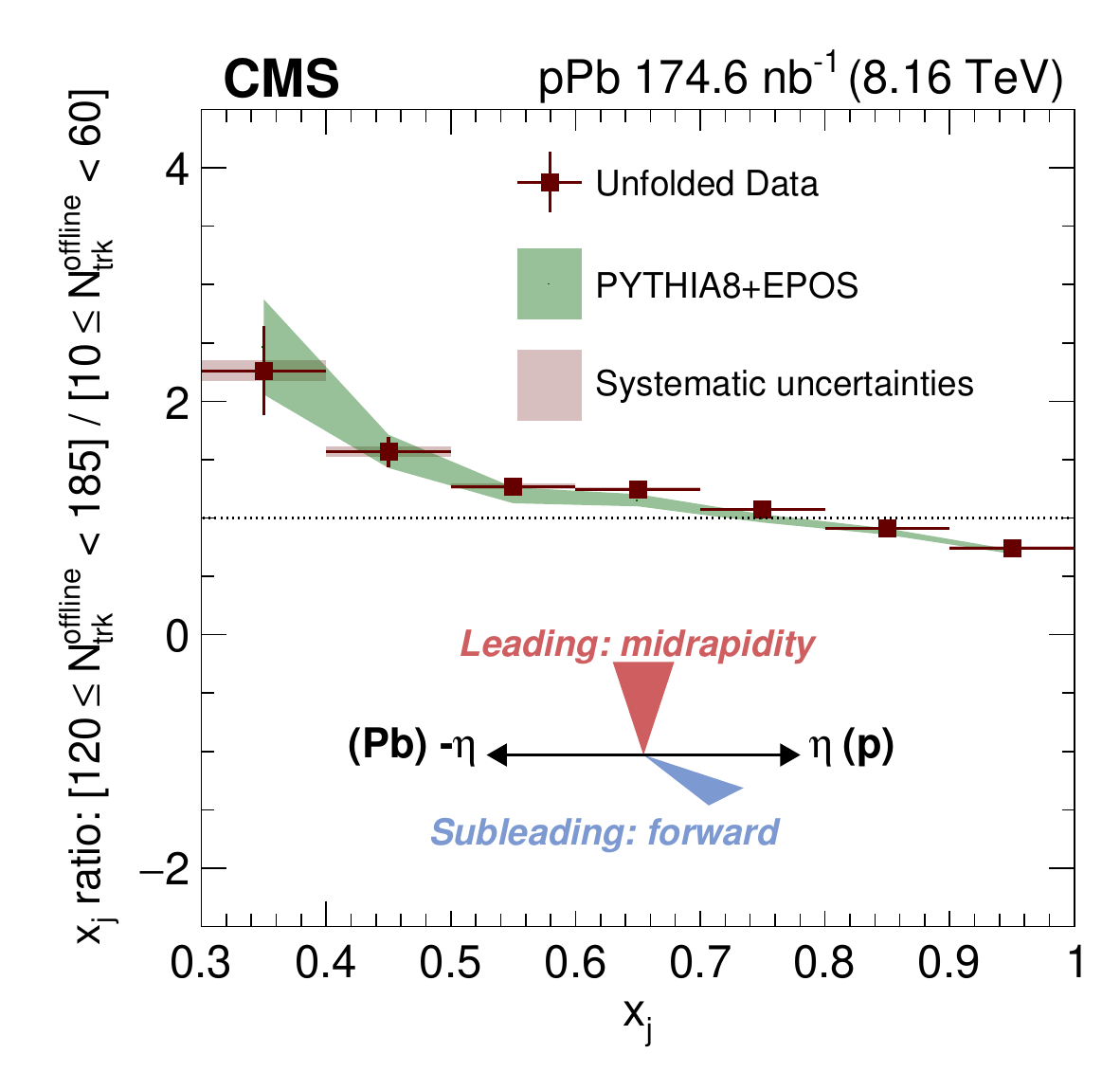} \\
    \includegraphics[width=0.44\textwidth]{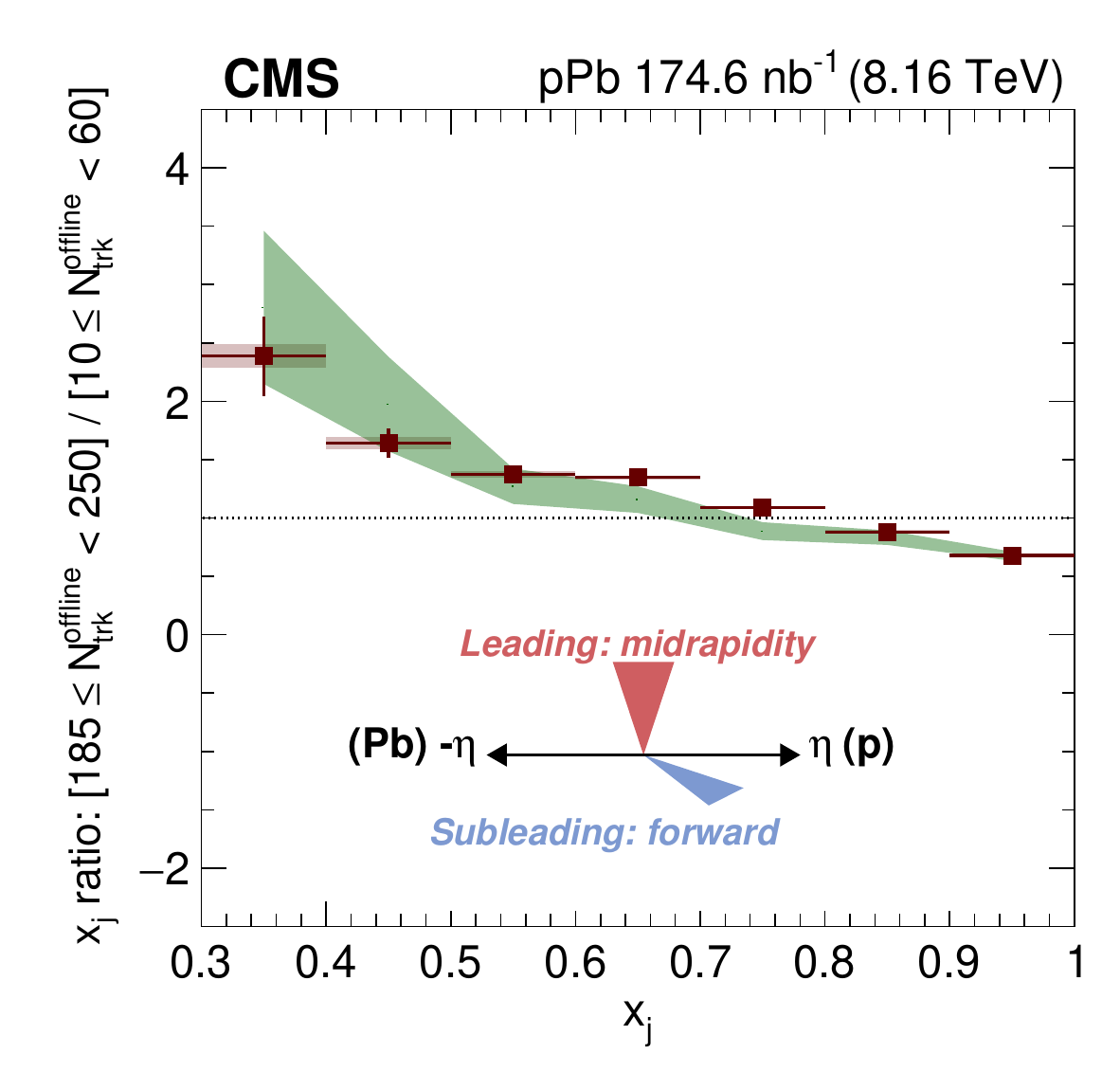} 
    \includegraphics[width=0.44\textwidth]{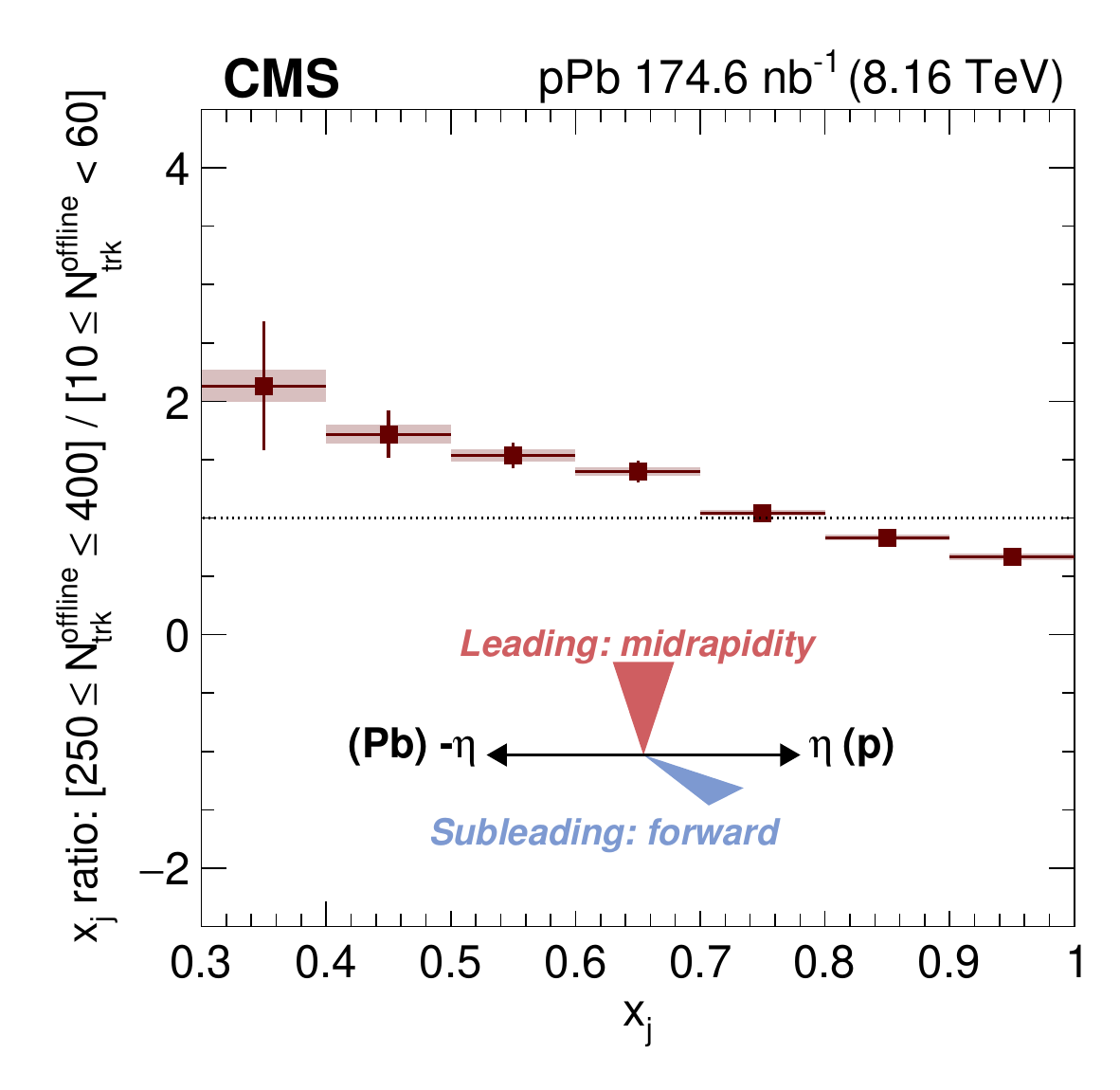} 
    \caption{The unfolded \xj ratio in the range $10 \leq \Ntroff < 60$, from lower to higher multiplicities for leading and subleading jets in the midrapidity and forward regions, respectively, are shown. Comparison with the generator-level {{\PYTHIA}8}+\EPOS simulation (green band) are also displayed. }
    \label{fig:xjratiomult_mid_fwd}
\end{figure}

\begin{figure}[!hbpt]
  \centering
    \includegraphics[width=0.44\textwidth]{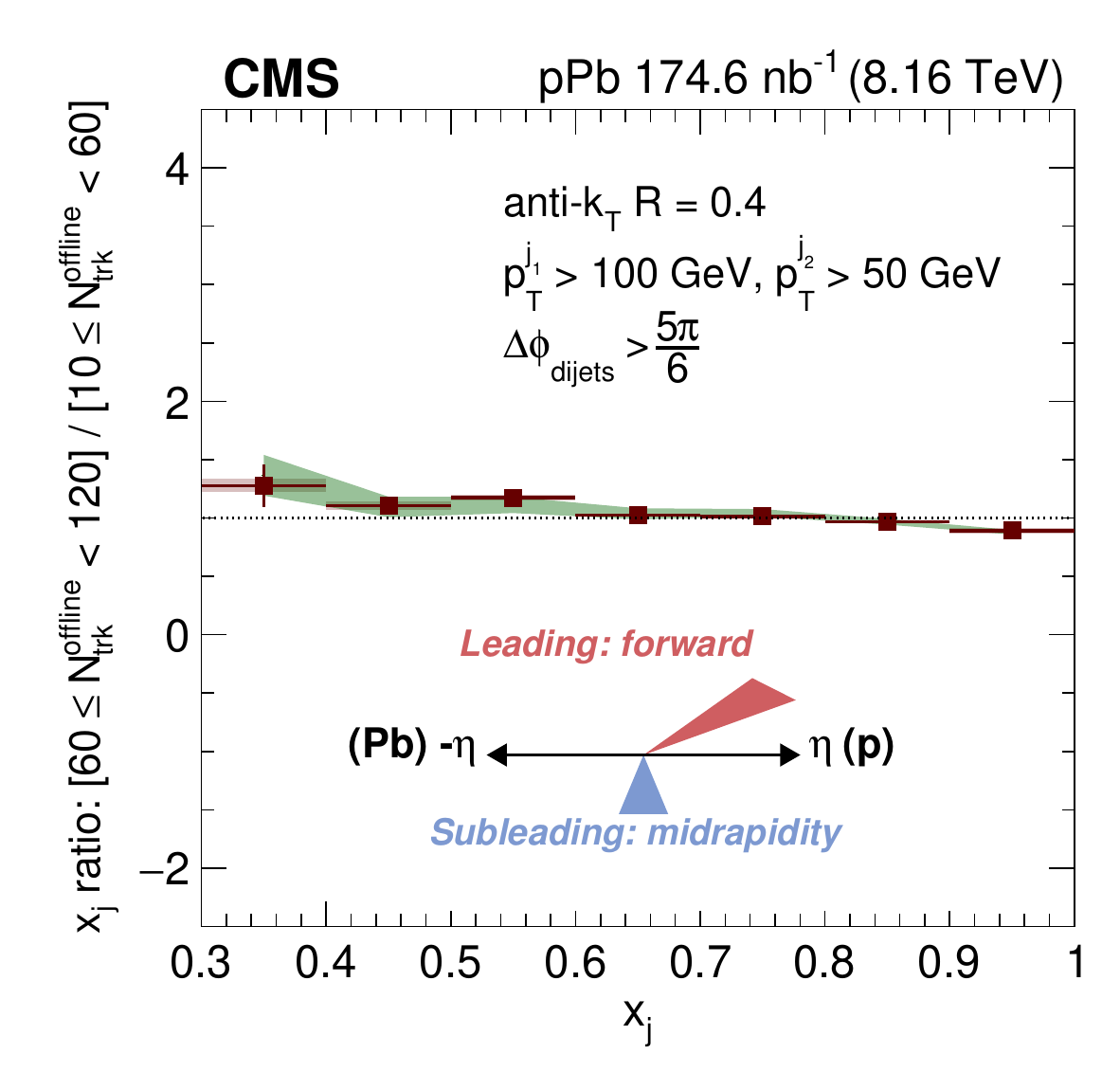} 
    \includegraphics[width=0.44\textwidth]{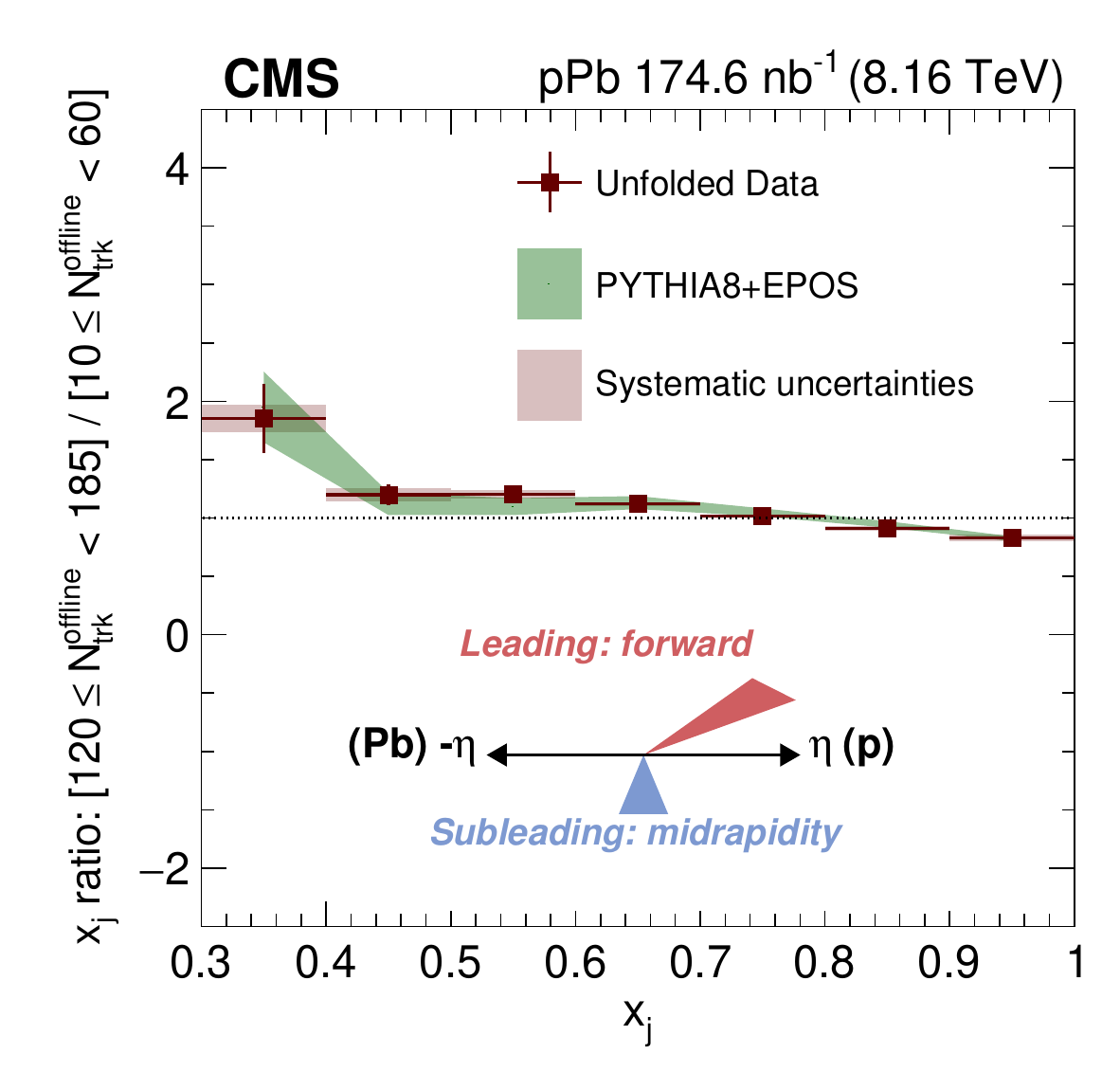} \\
    \includegraphics[width=0.44\textwidth]{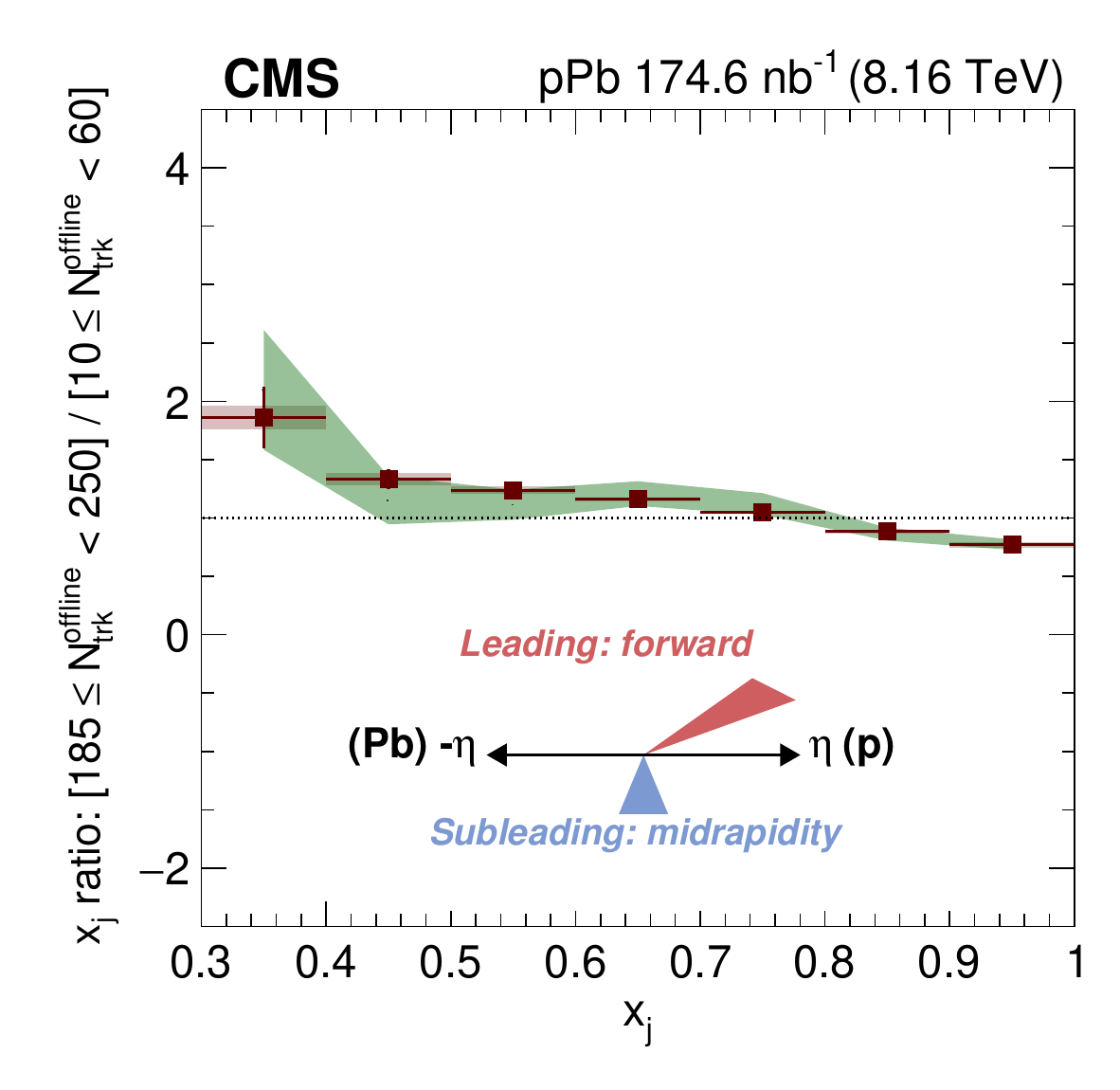} 
    \includegraphics[width=0.44\textwidth]{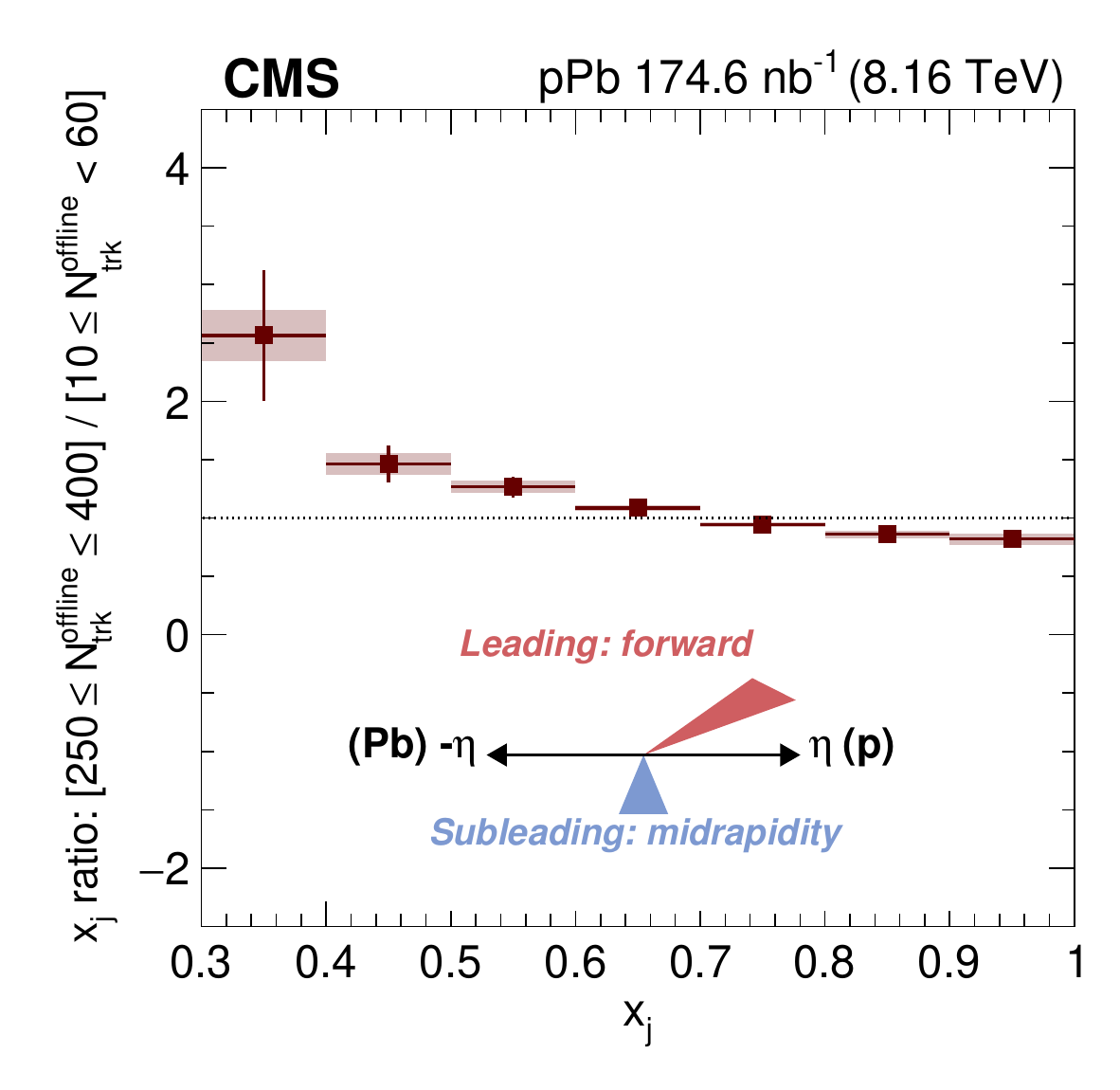} 
    \caption{The unfolded \xj ratio in the range $10 \leq \Ntroff < 60$, from lower to higher multiplicities for leading and subleading jets in the forward and midrapidity regions, respectively, are shown. Comparison with the generator-level {{\PYTHIA}8}+\EPOS simulation (green band) are also displayed. }
    \label{fig:xjratiomult_fwd_mid}
\end{figure}

\begin{figure}[!hbpt]
  \centering
    \includegraphics[width=0.44\textwidth]{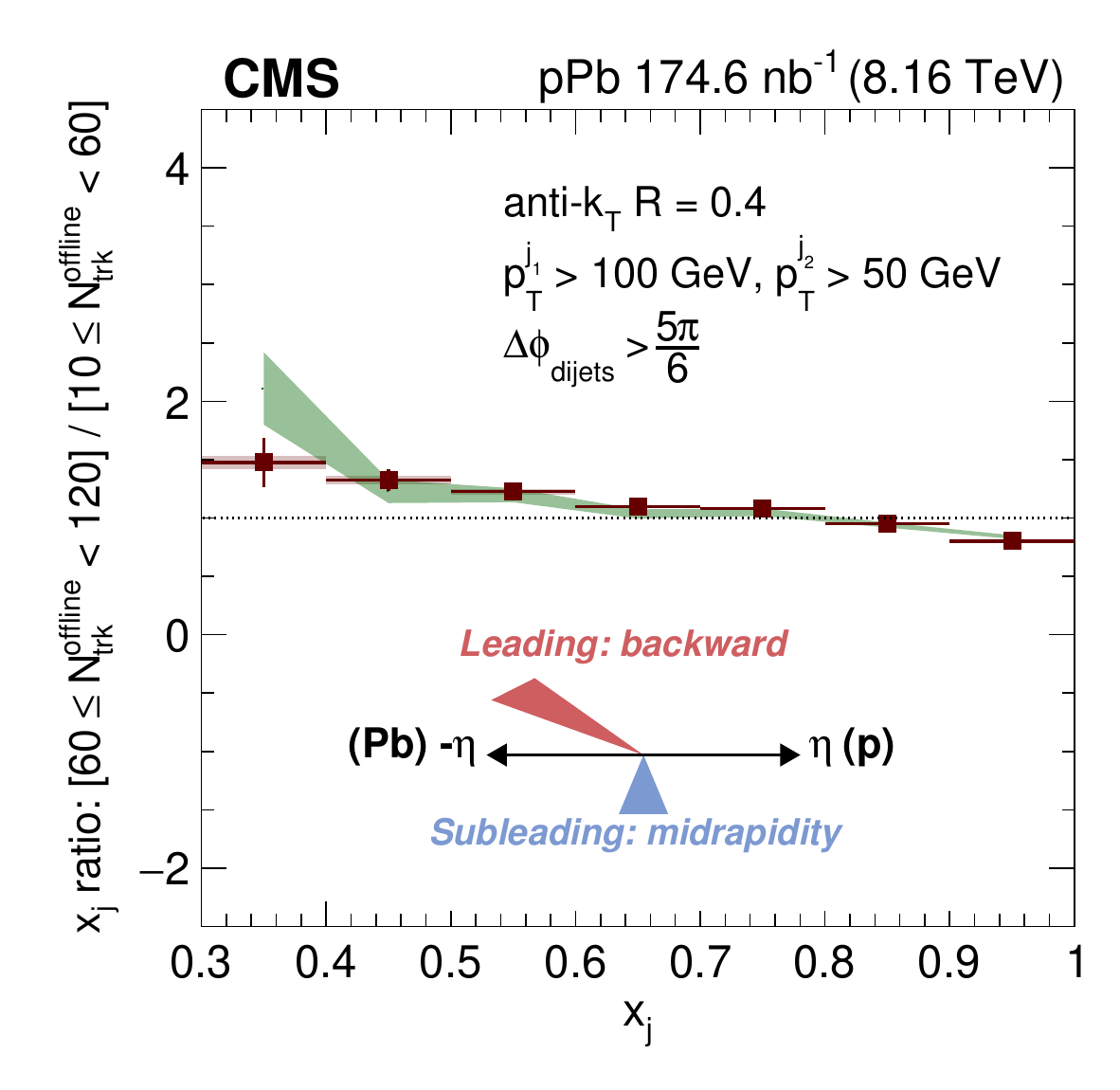} 
    \includegraphics[width=0.44\textwidth]{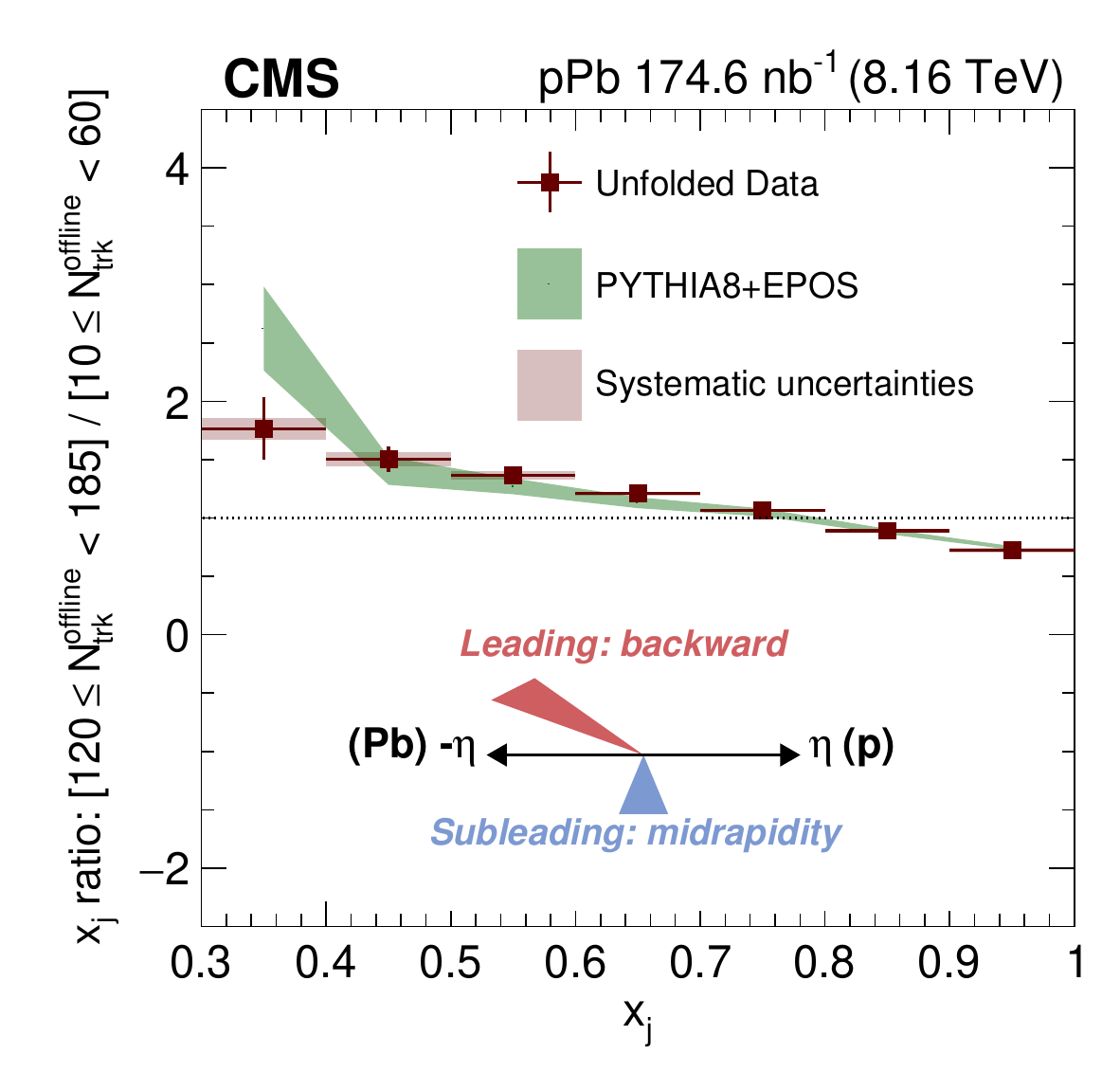} \\
    \includegraphics[width=0.44\textwidth]{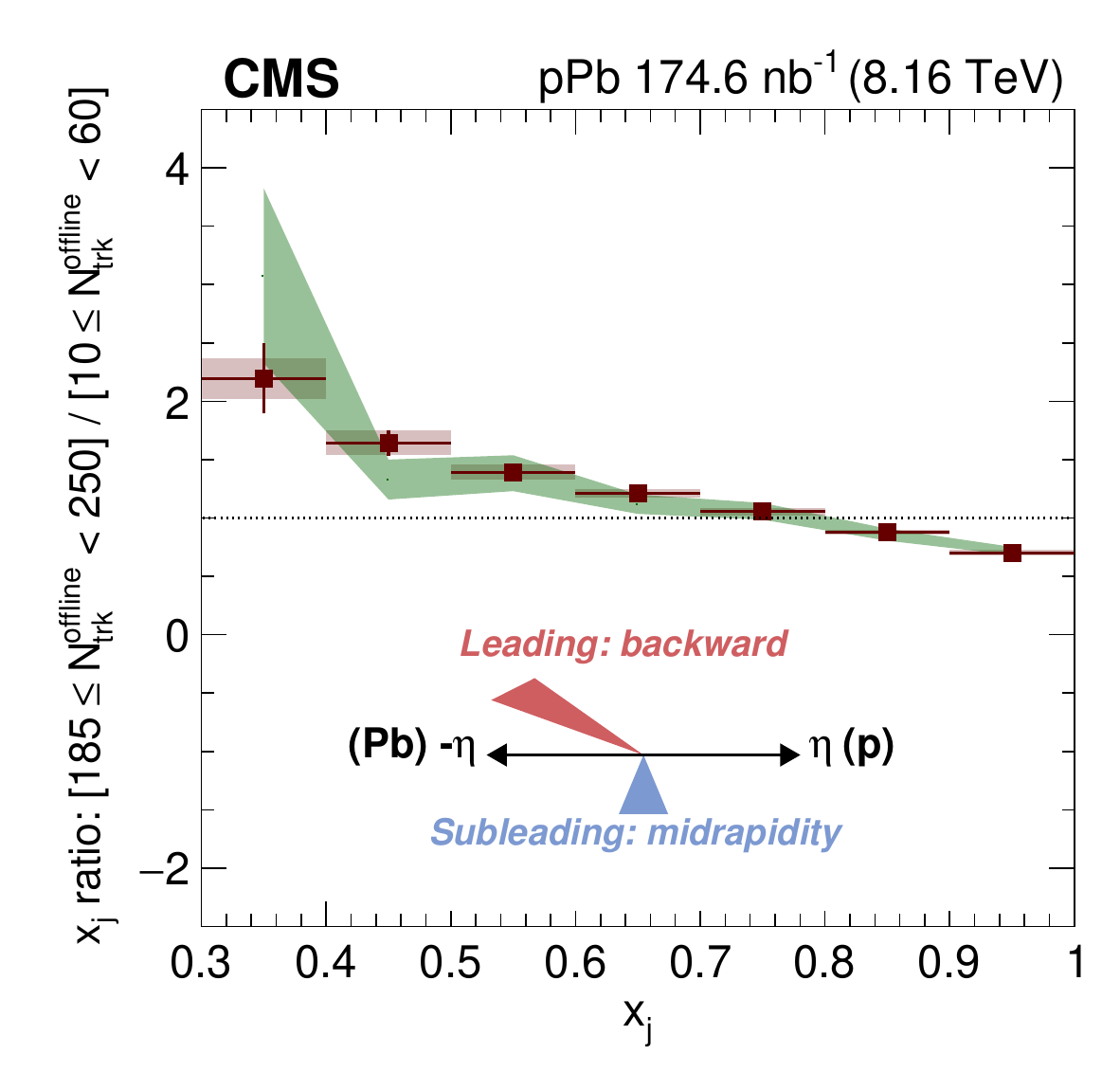} 
    \includegraphics[width=0.44\textwidth]{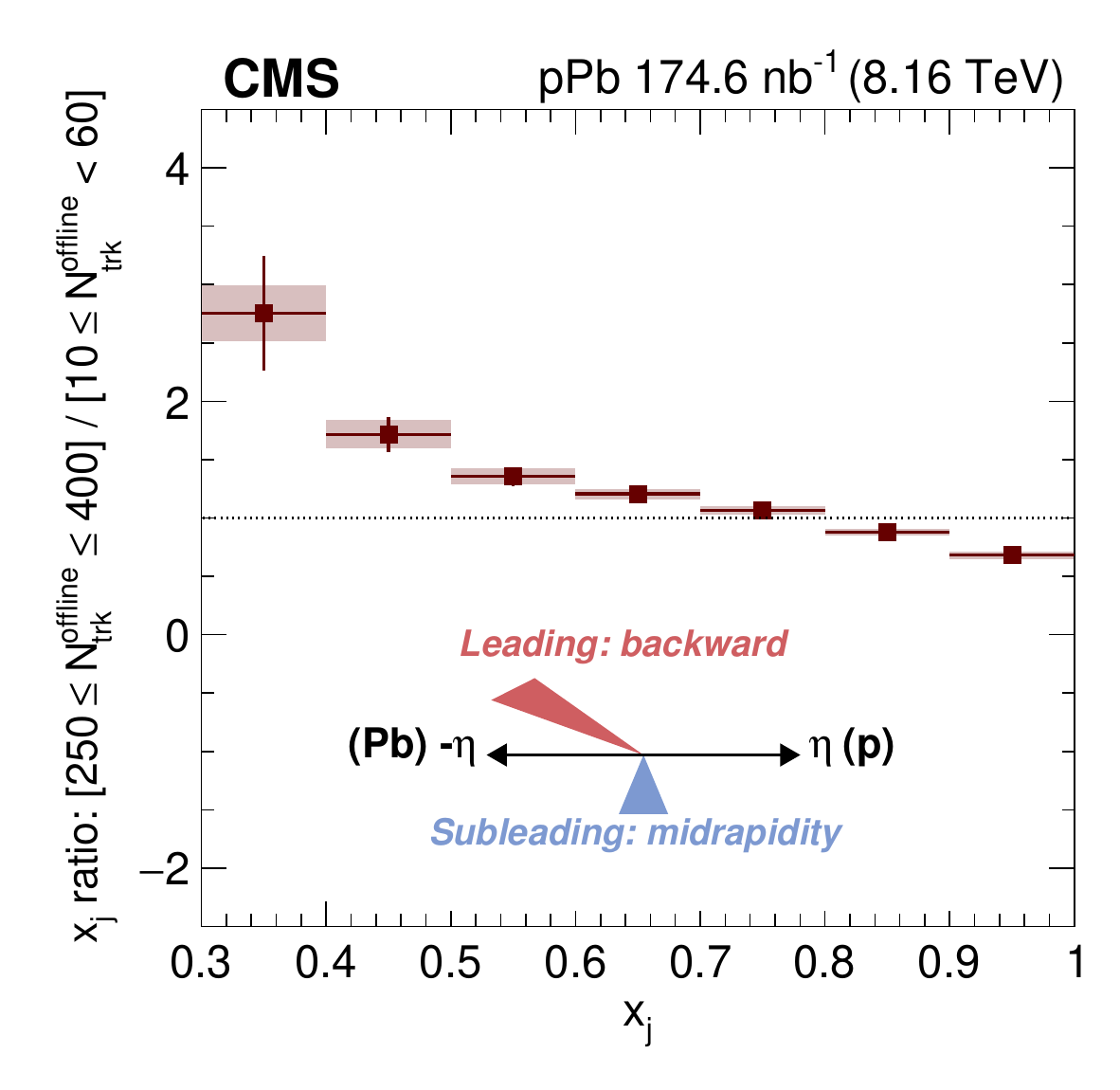} 
    \caption{The unfolded \xj ratio in the range $10 \leq \Ntroff < 60$, from lower to higher multiplicities for leading and subleading jets in the backward and midrapidity regions, respectively, are shown. Comparison with the generator-level {{\PYTHIA}8}+\EPOS simulation (green band) are also displayed. }
    \label{fig:xjratiomult_bkw_mid}
\end{figure}

\begin{figure}[!hbpt]
  \centering
    \includegraphics[width=0.44\textwidth]{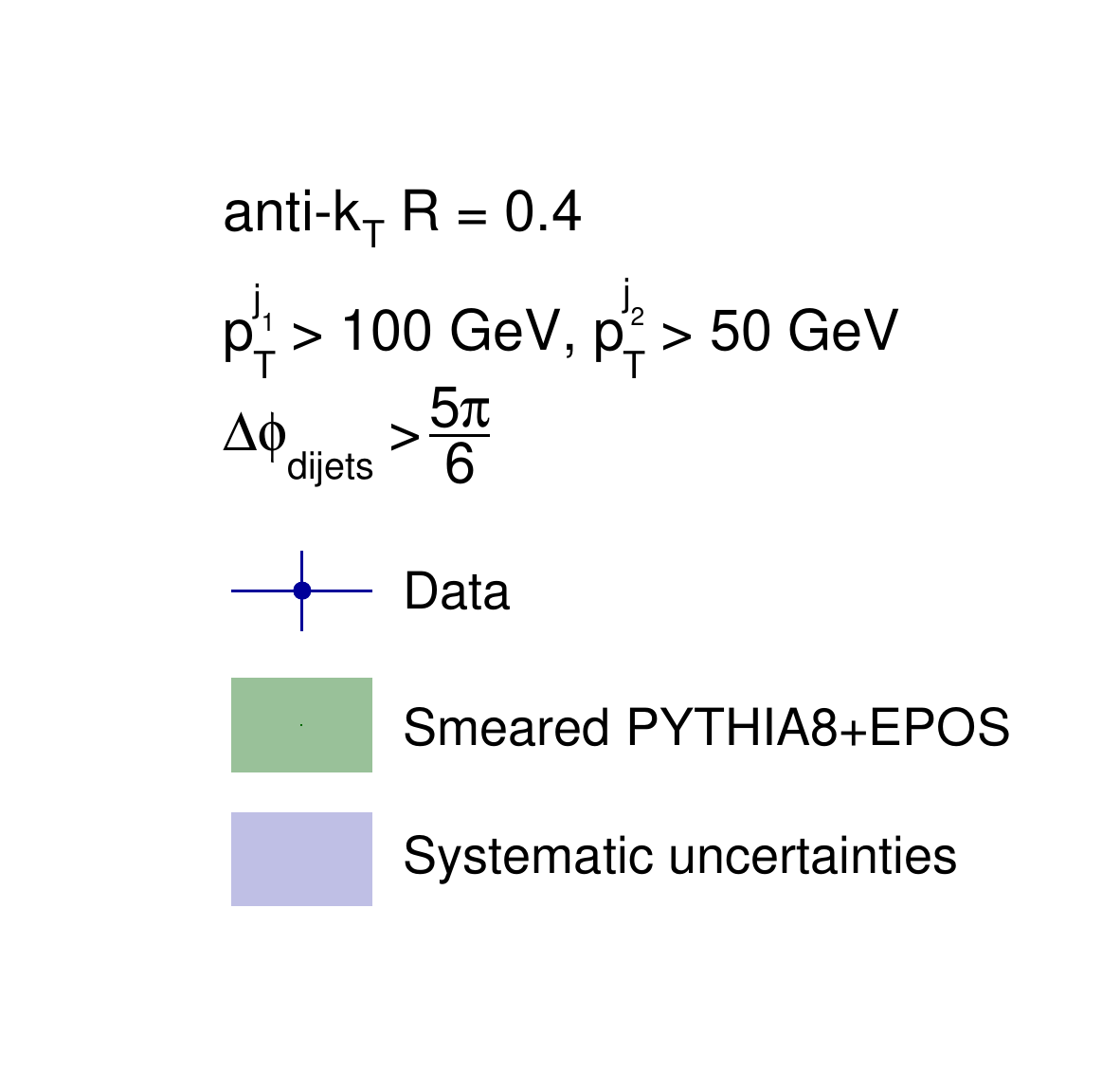} 
    \includegraphics[width=0.44\textwidth]{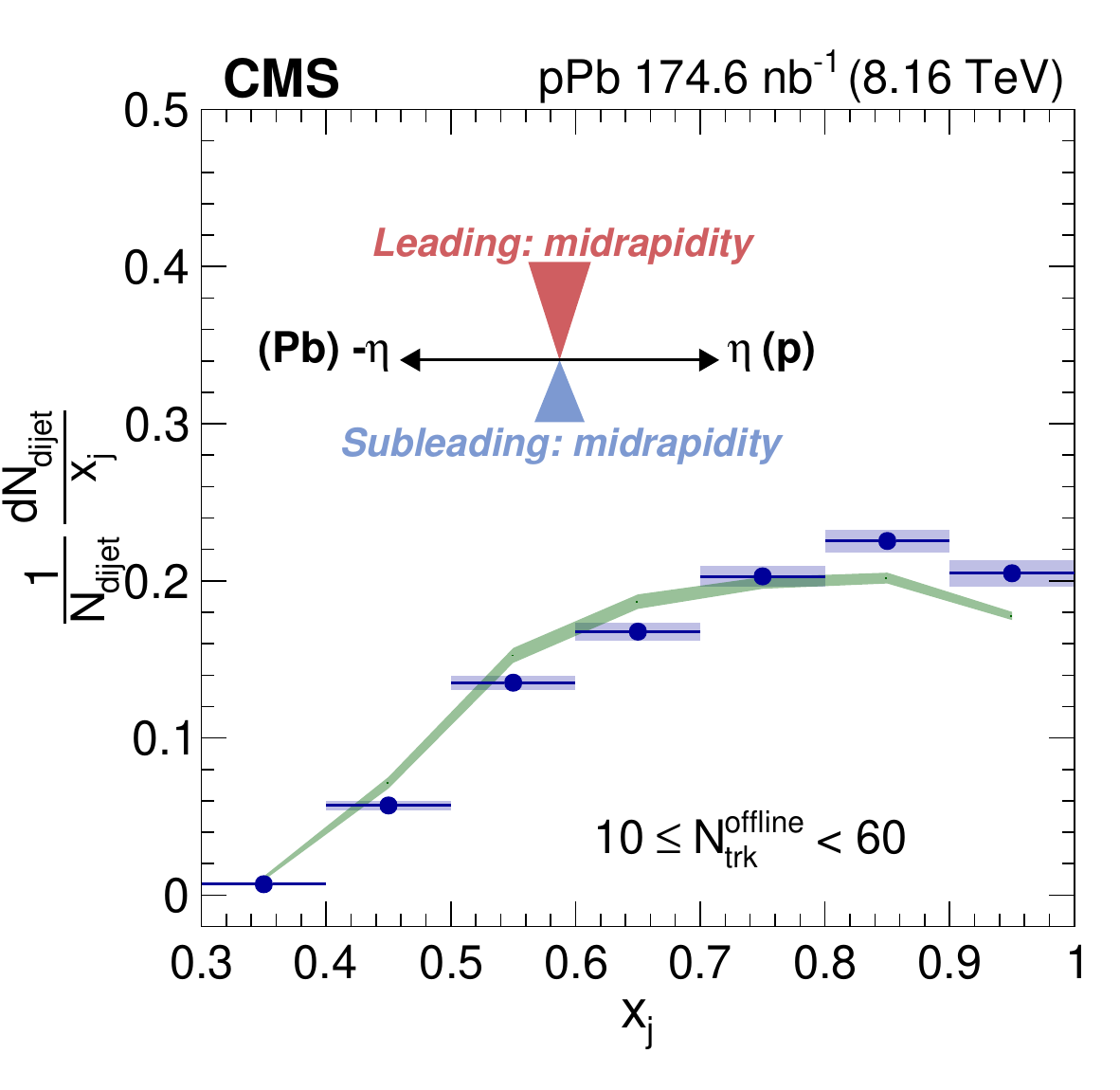} \\
    \includegraphics[width=0.44\textwidth]{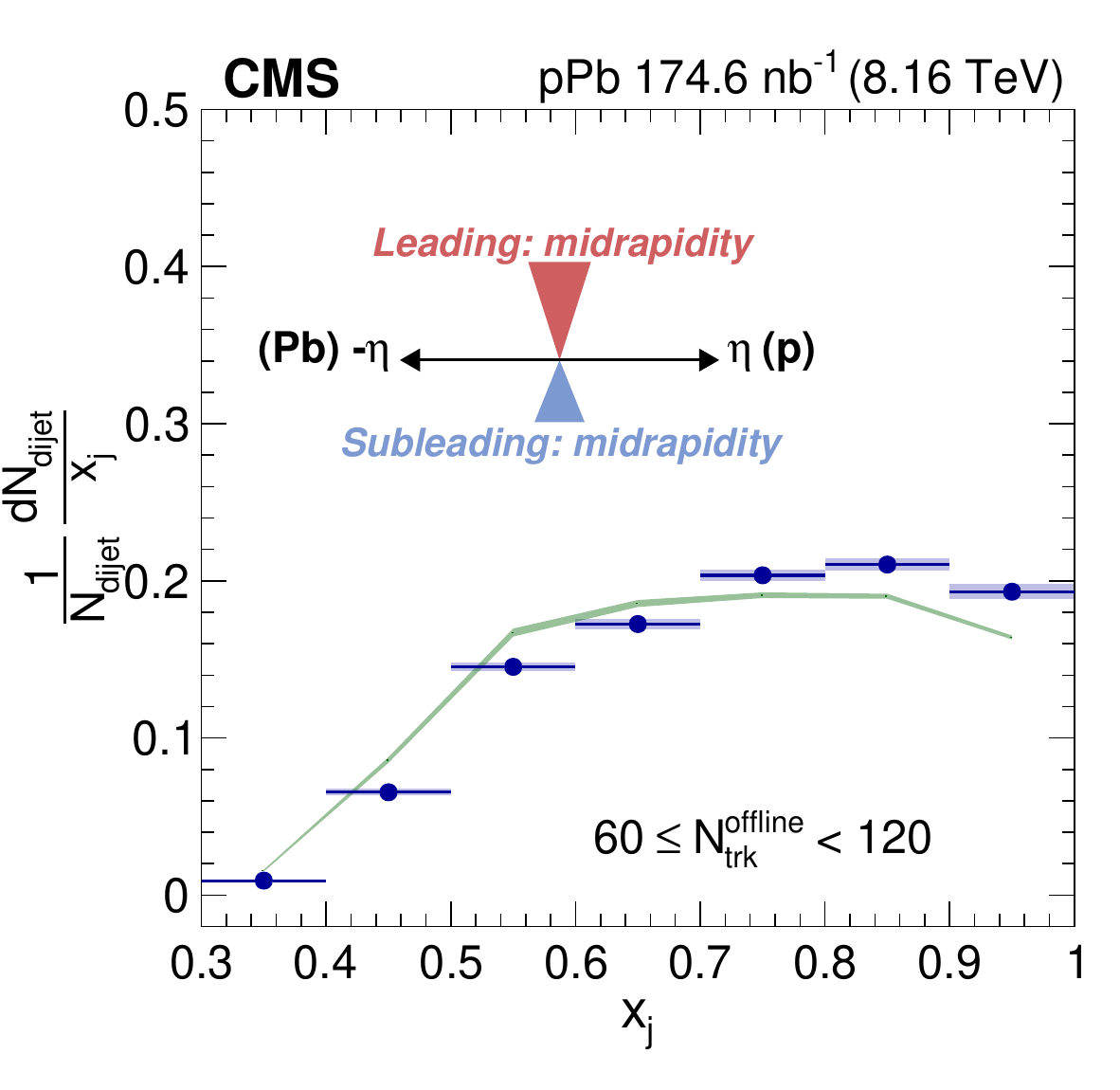} 
    \includegraphics[width=0.44\textwidth]{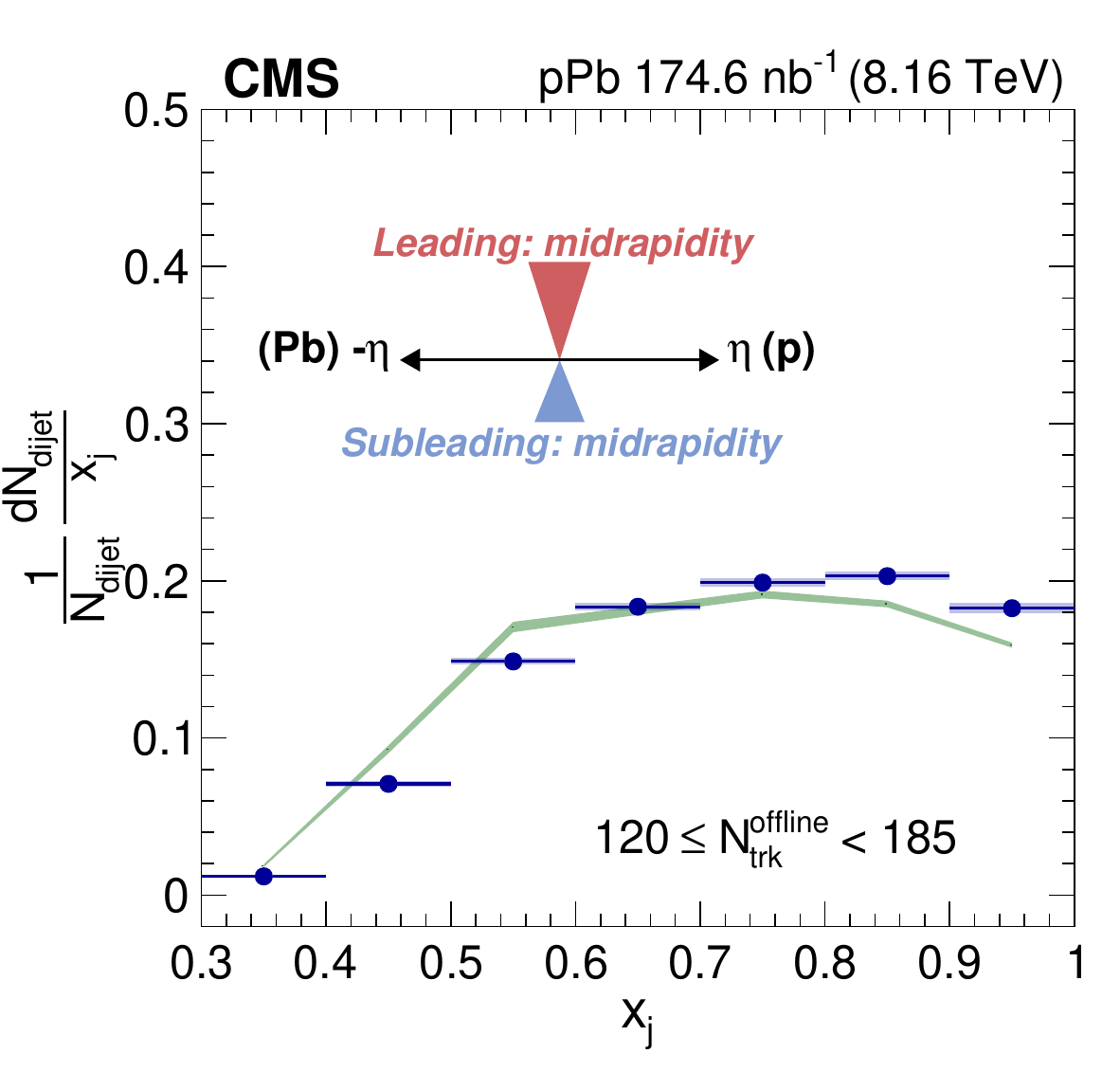} \\
    \includegraphics[width=0.44\textwidth]{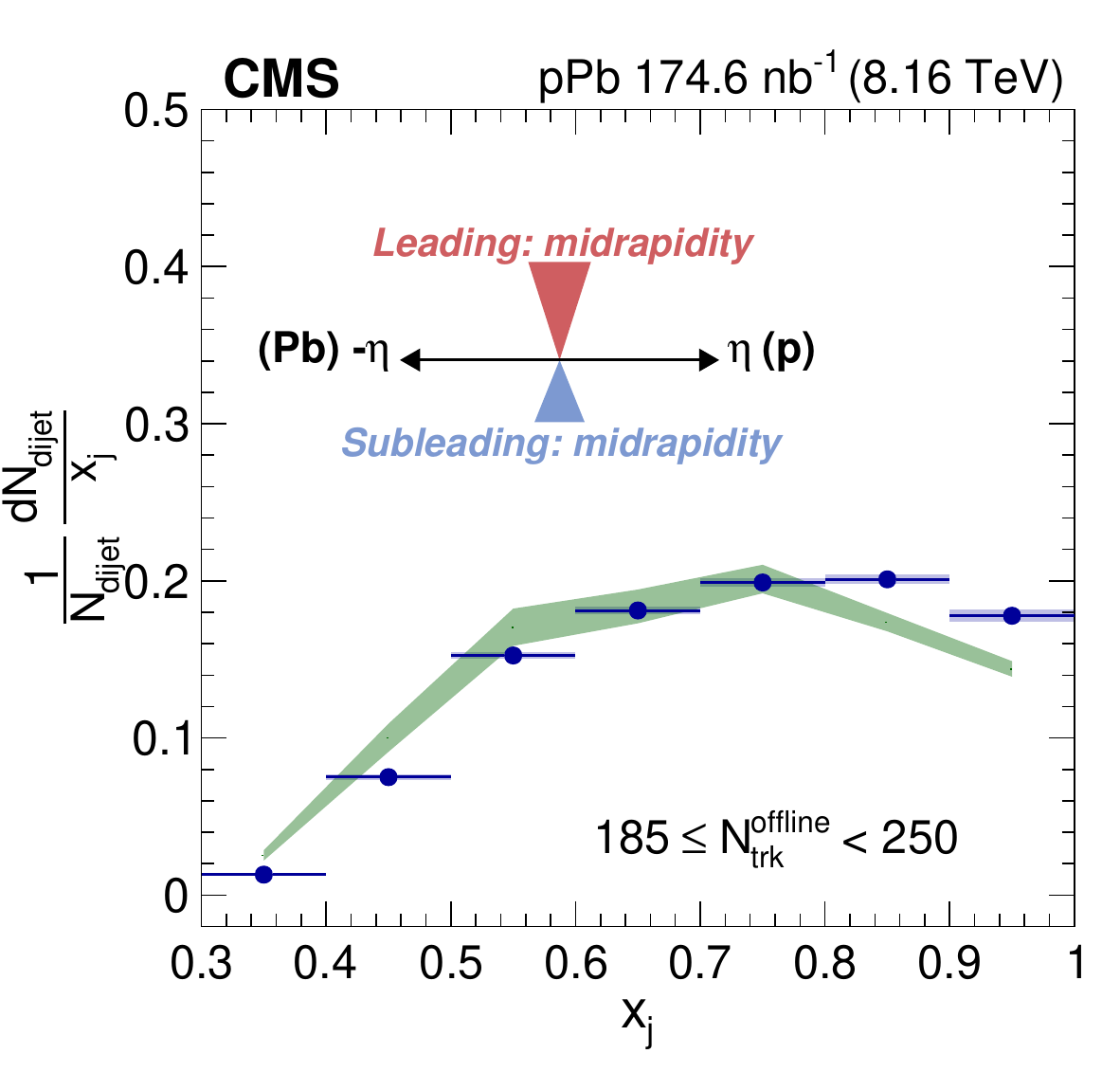} 
    \includegraphics[width=0.44\textwidth]{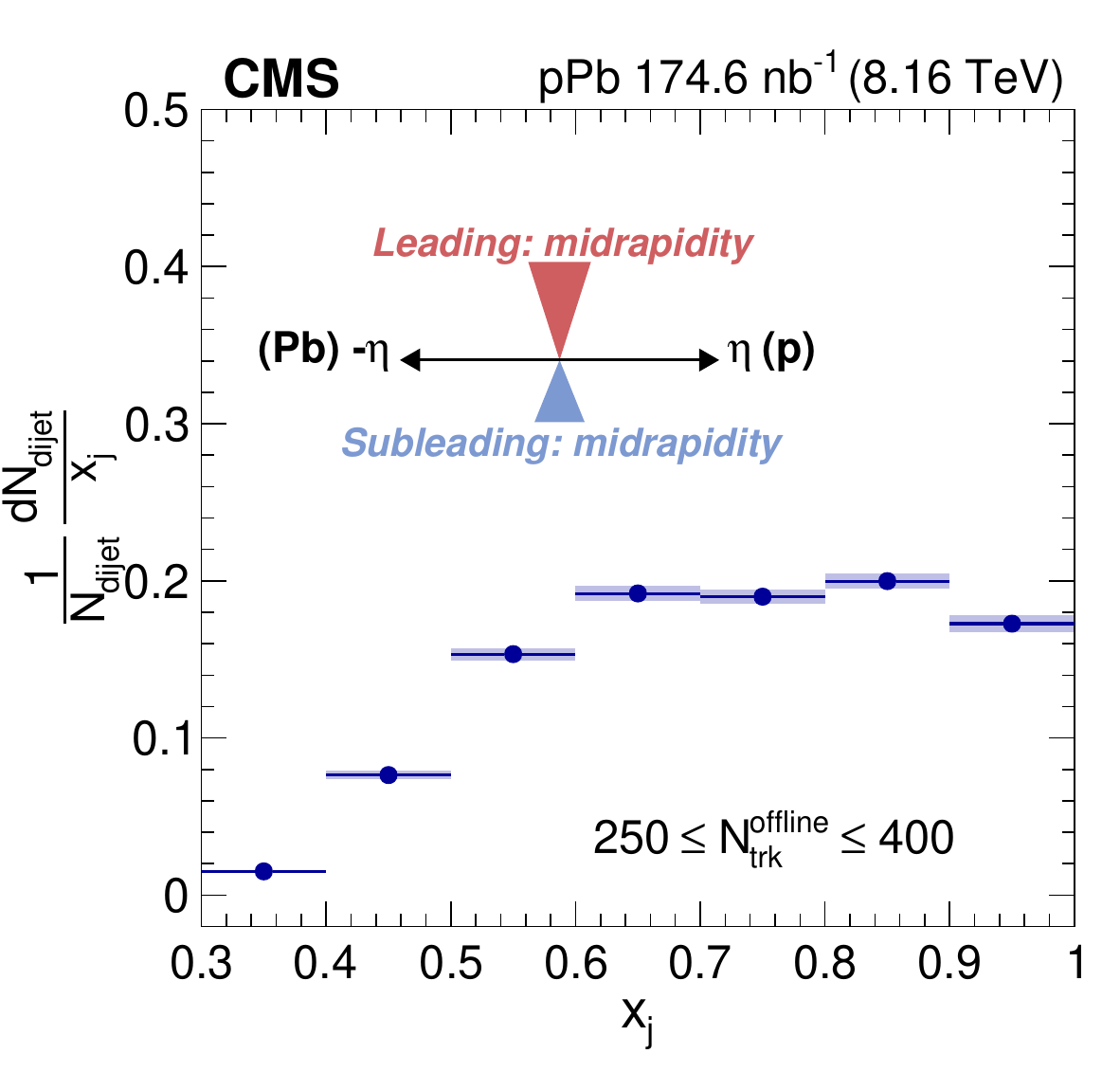} 
    \caption{The measured (raw) \xj distributions, from lower to higher multiplicities and for both leading and subleading jets in the midrapidity region, are shown. Comparisons with the reconstruction-level {{\PYTHIA}8}+\EPOS simulation (green band) are also displayed. }
    \label{fig:xjmultrec_mid_mid}
\end{figure}

\begin{figure}[!hbpt]
  \centering
    \includegraphics[width=0.44\textwidth]{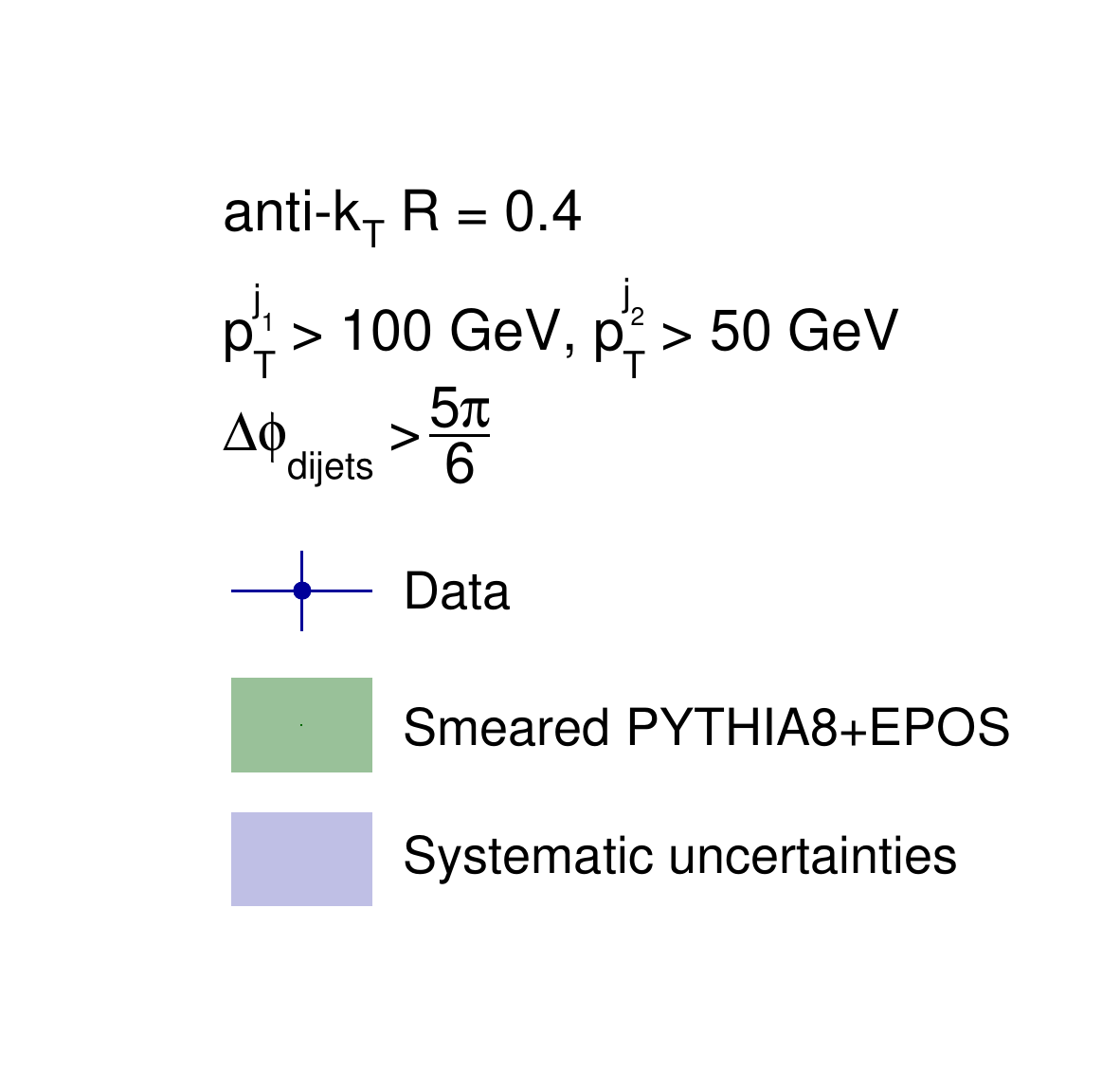} 
    \includegraphics[width=0.44\textwidth]{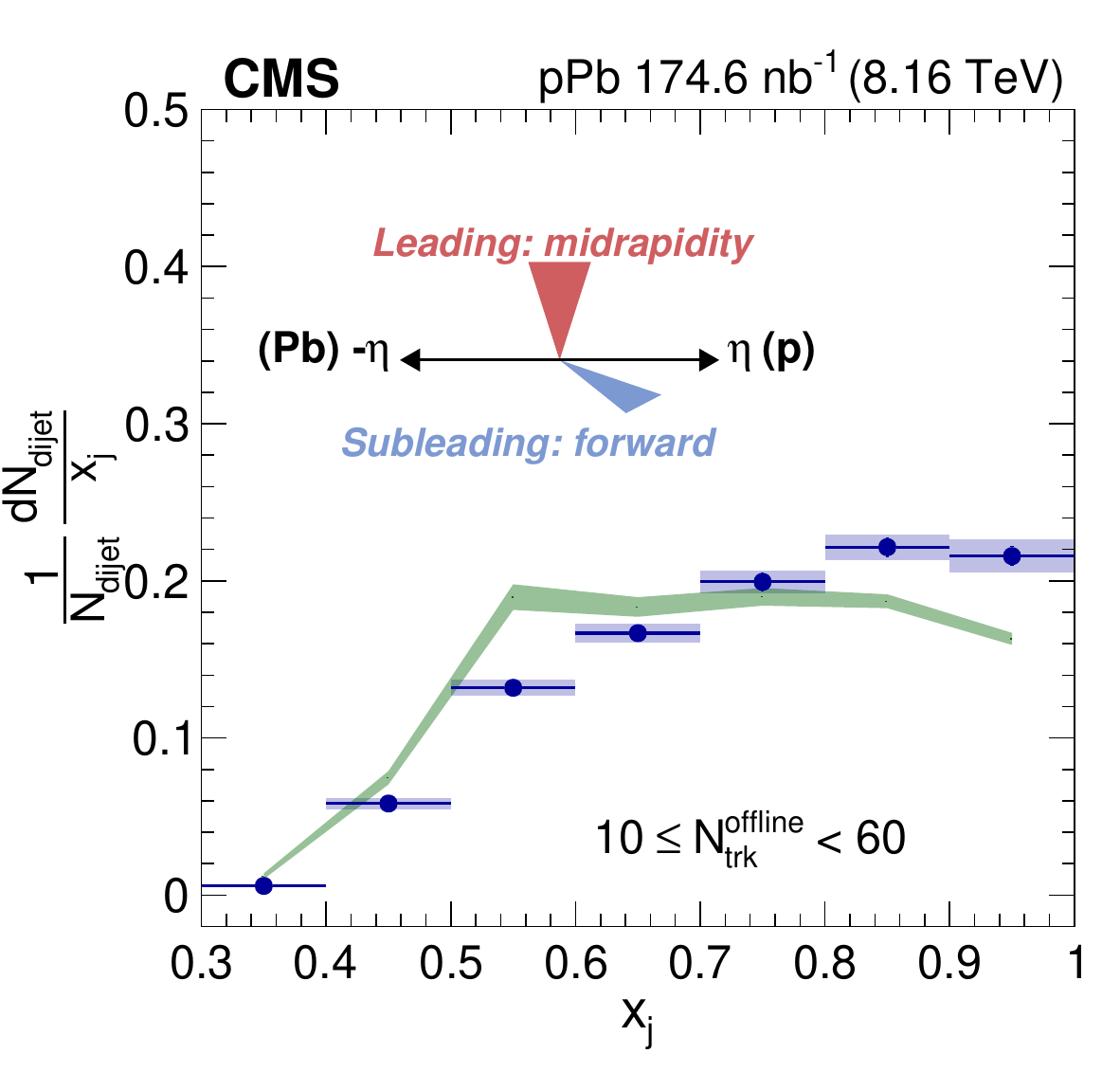} \\
    \includegraphics[width=0.44\textwidth]{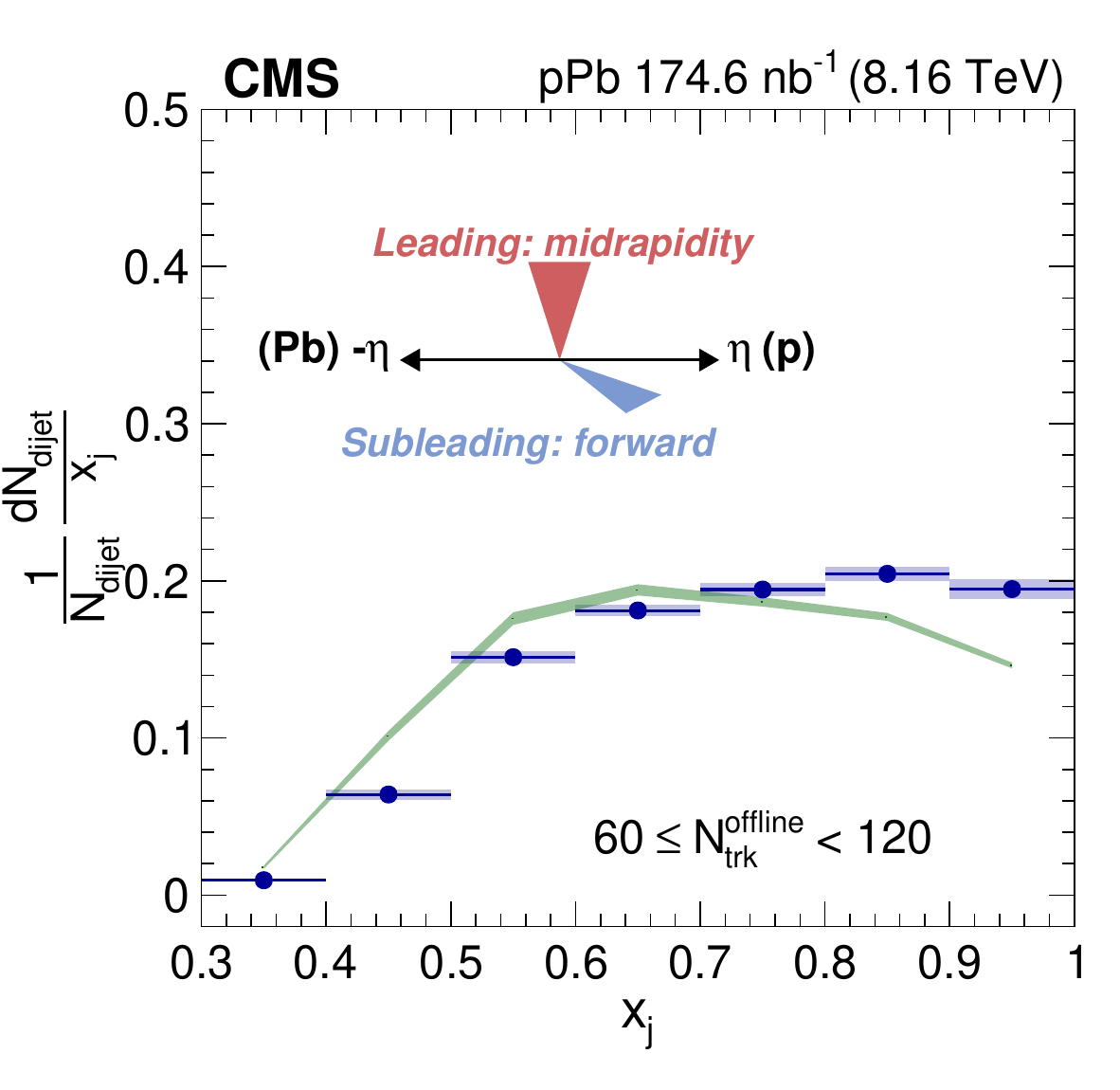} 
    \includegraphics[width=0.44\textwidth]{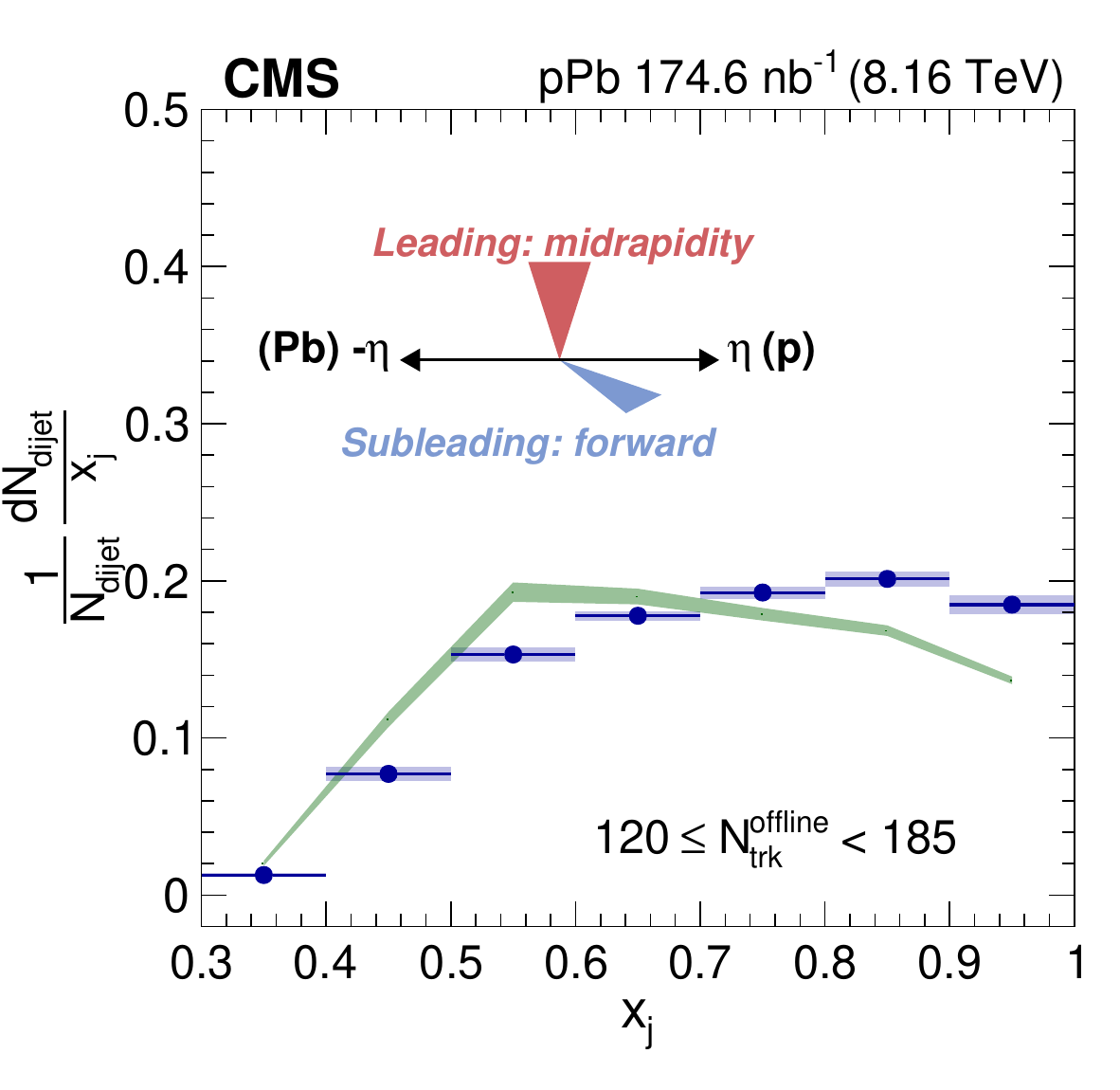} \\
    \includegraphics[width=0.44\textwidth]{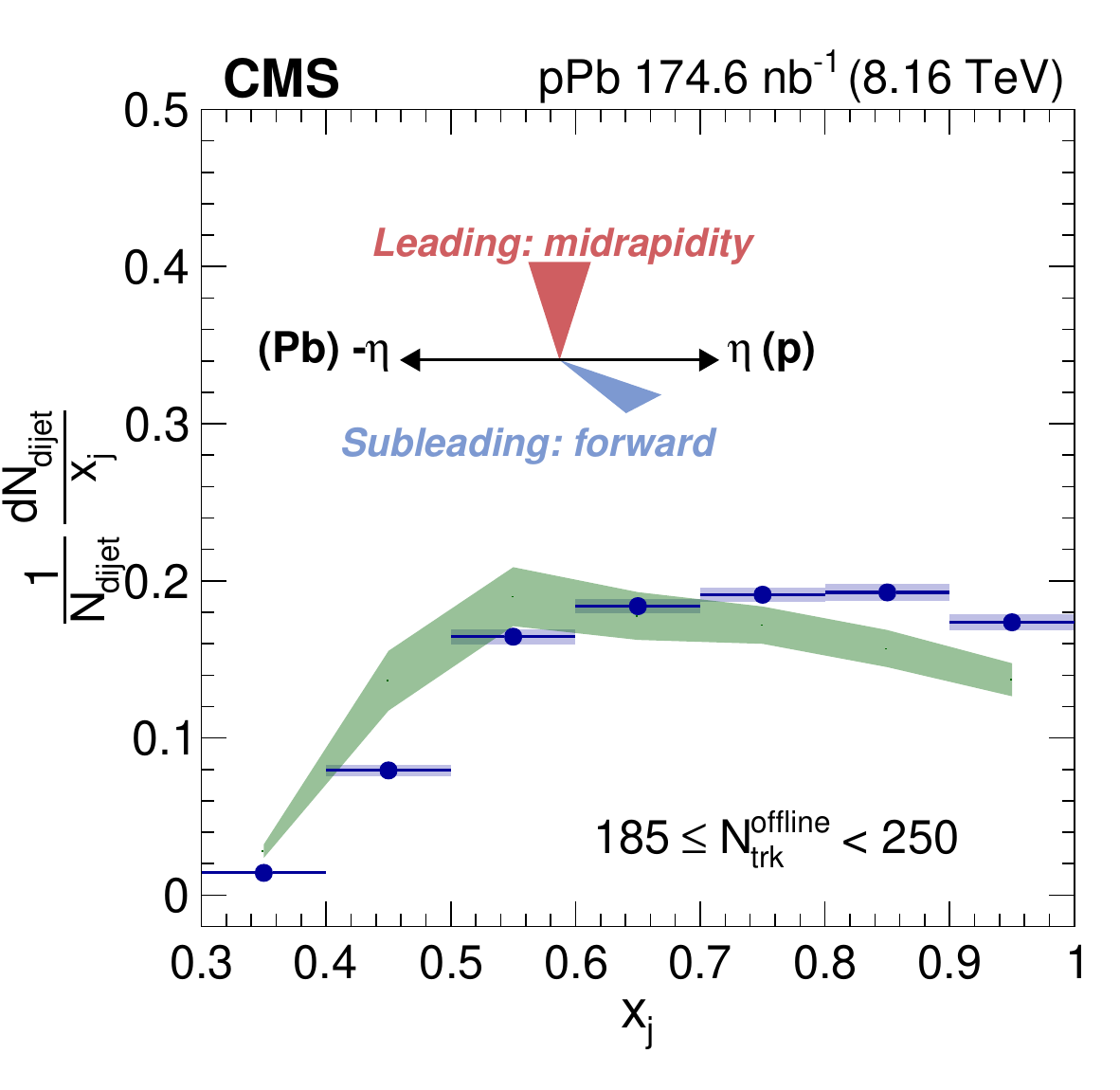} 
    \includegraphics[width=0.44\textwidth]{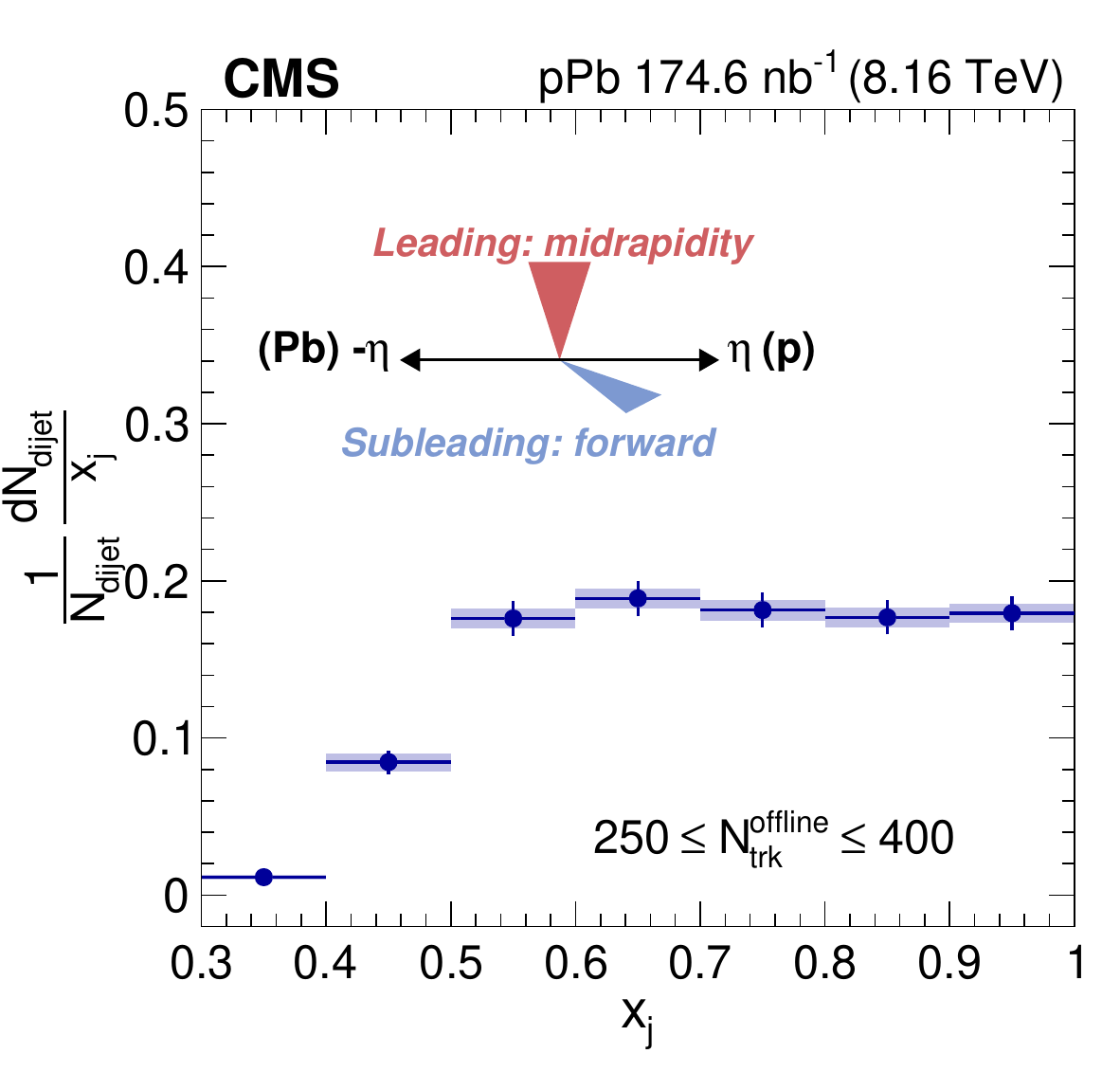} 
    \caption{The measured (raw) \xj distributions, from lower to higher multiplicities and for leading and subleading jets in the midrapidity and forward regions, respectively, are shown. Comparisons with the reconstruction-level {{\PYTHIA}8}+\EPOS simulation (green band) are also displayed. }
    \label{fig:xjmultrec_mid_fwd}
\end{figure}

\begin{figure}[!hbpt]
  \centering
    \includegraphics[width=0.44\textwidth]{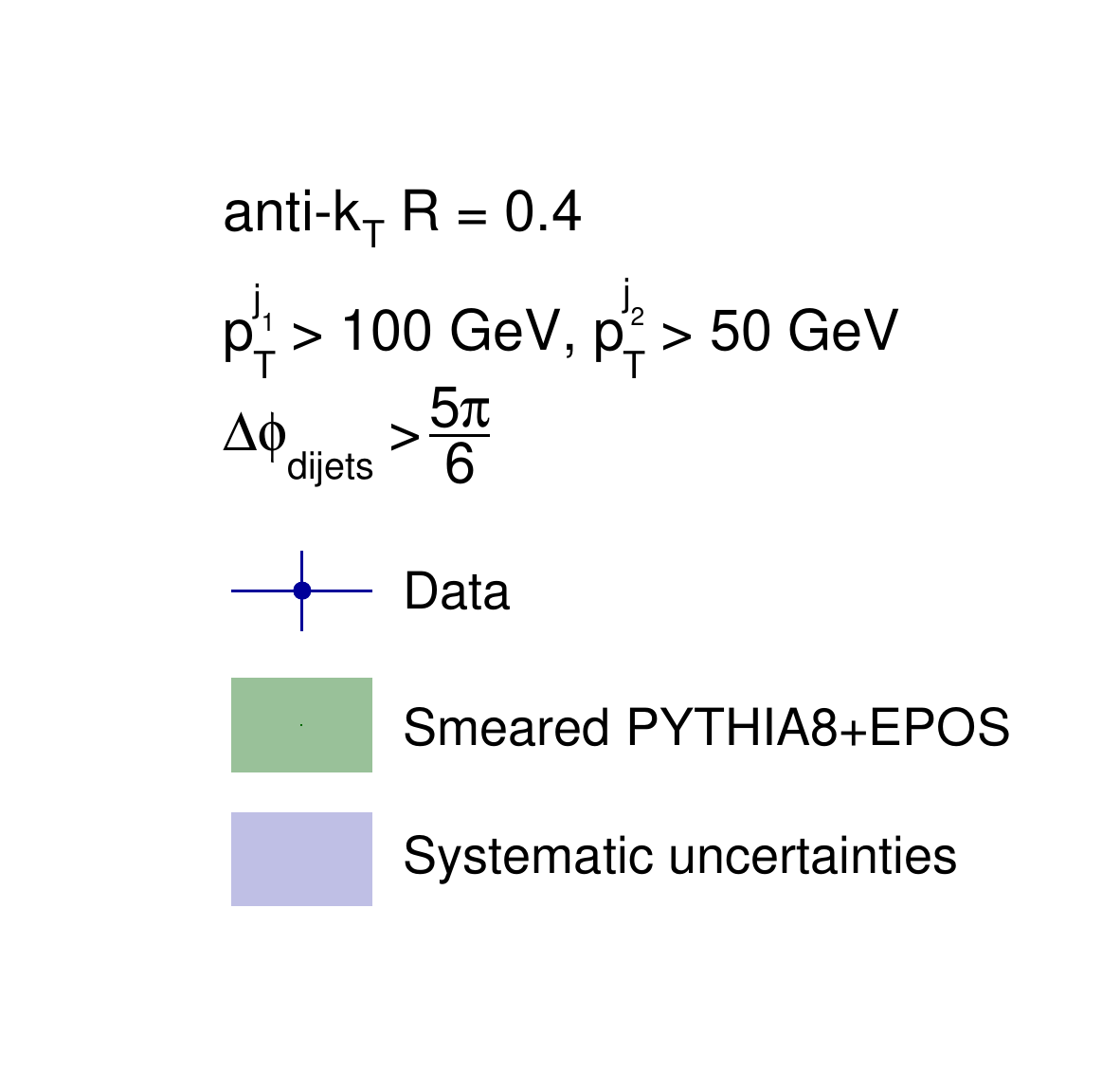} 
    \includegraphics[width=0.44\textwidth]{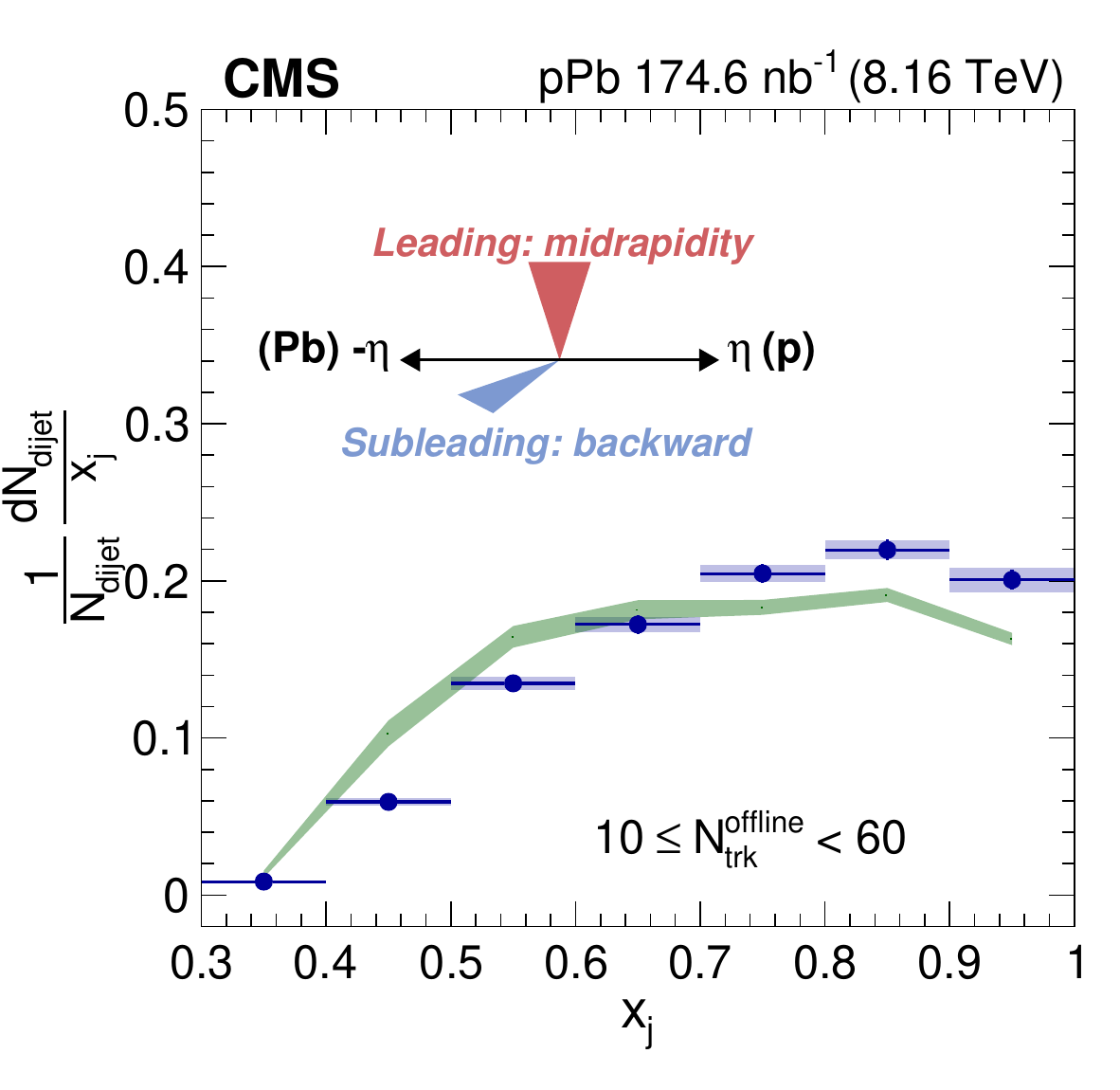} \\
    \includegraphics[width=0.44\textwidth]{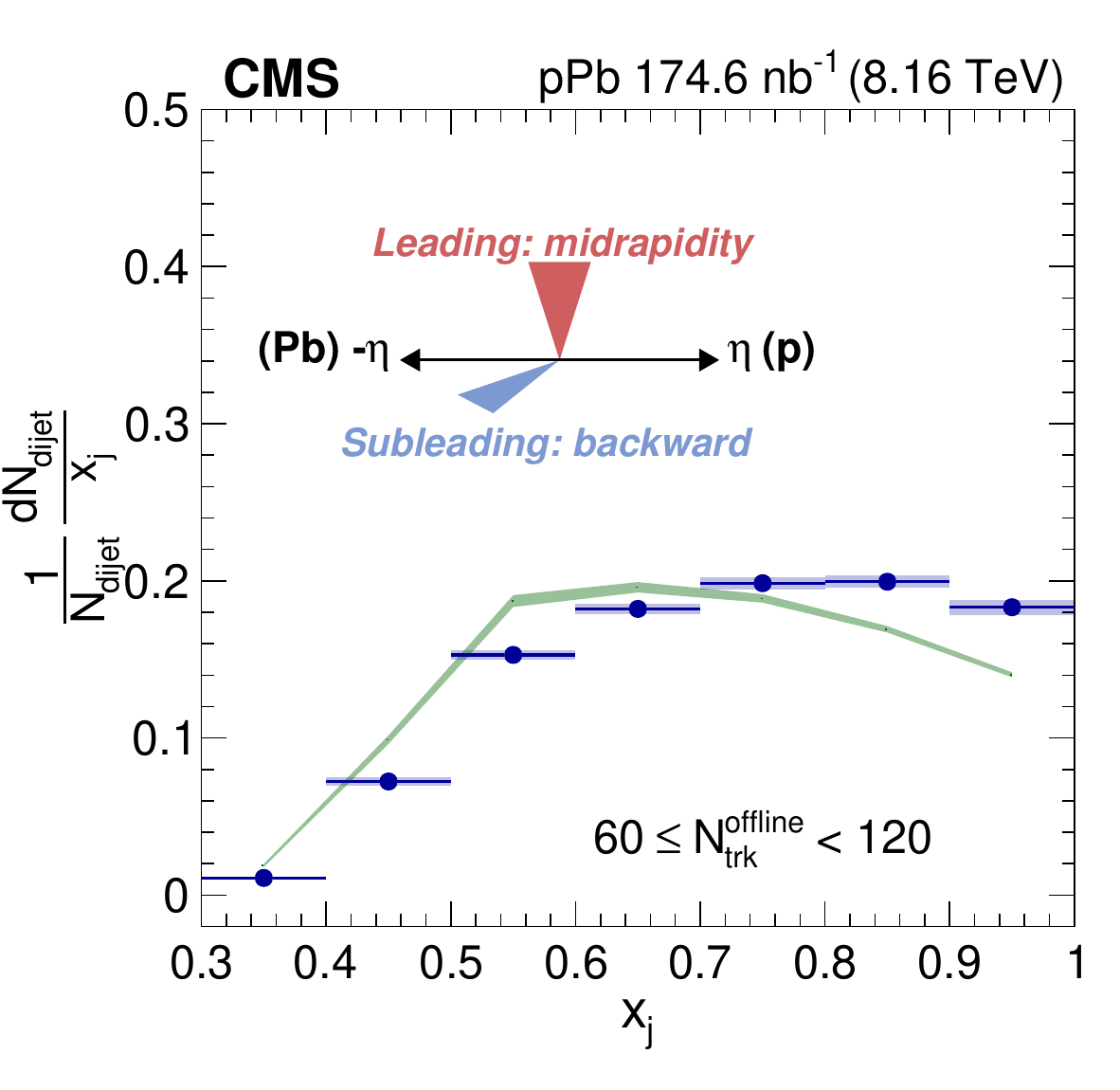} 
    \includegraphics[width=0.44\textwidth]{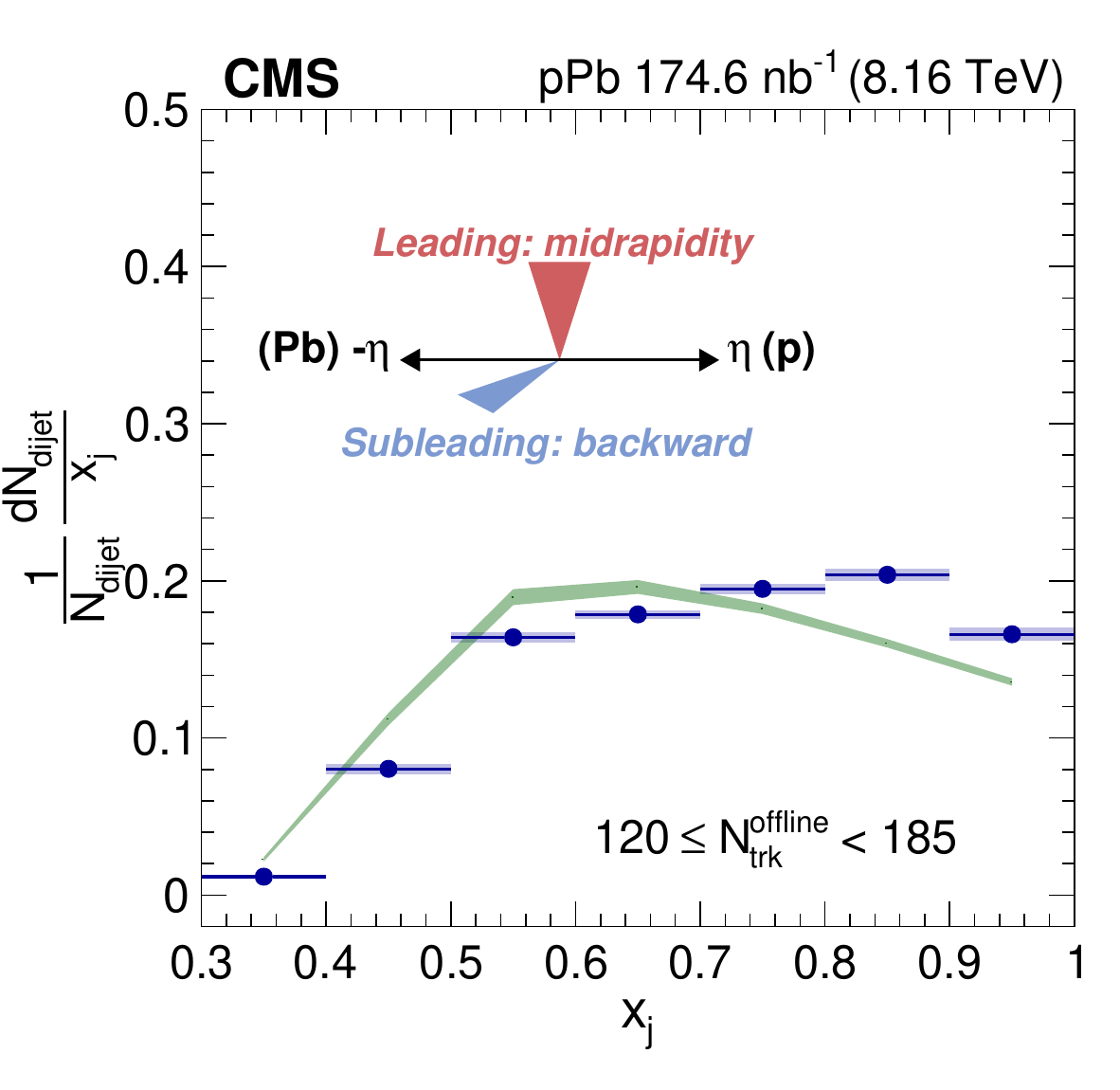} \\
    \includegraphics[width=0.44\textwidth]{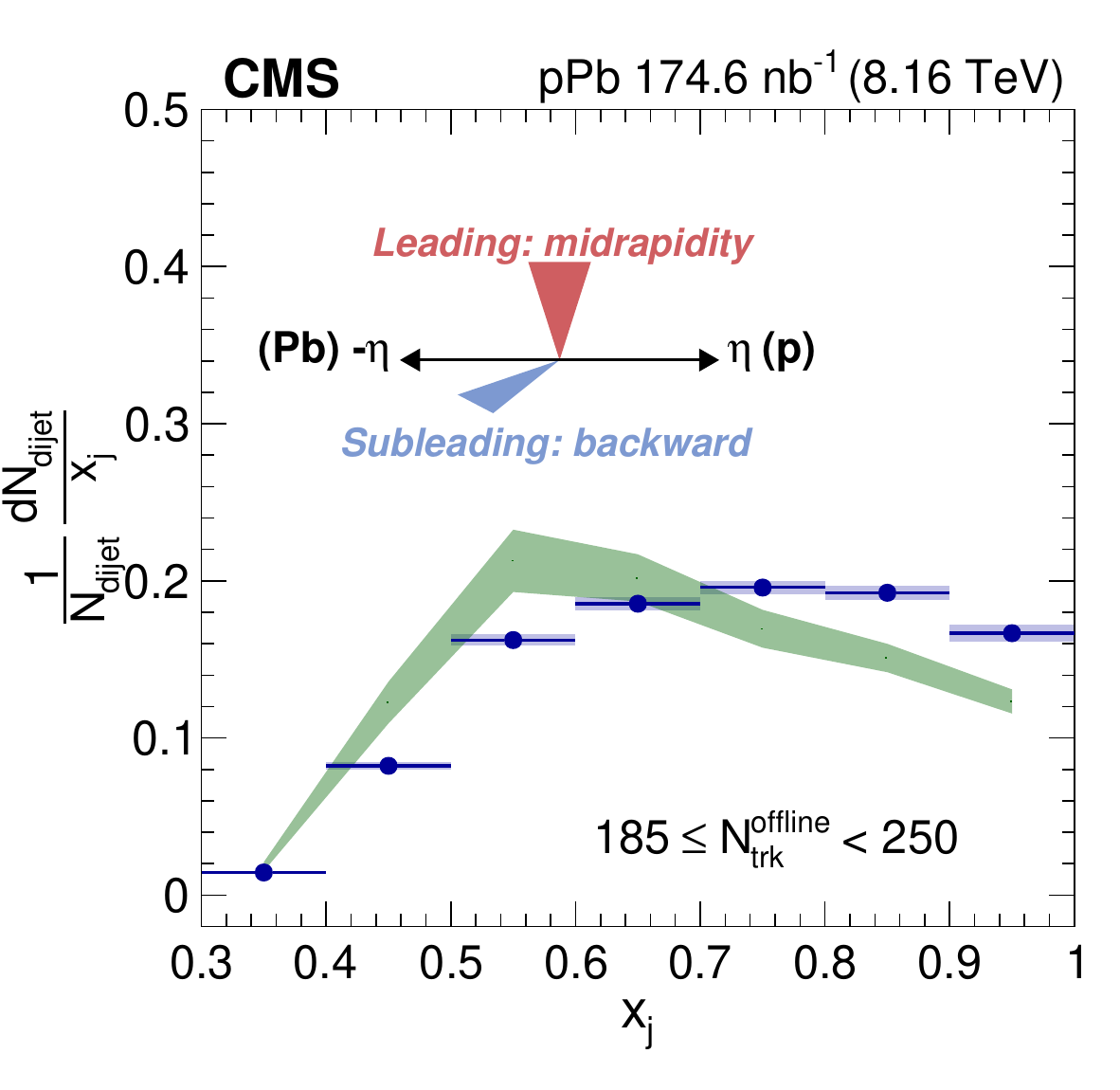} 
    \includegraphics[width=0.44\textwidth]{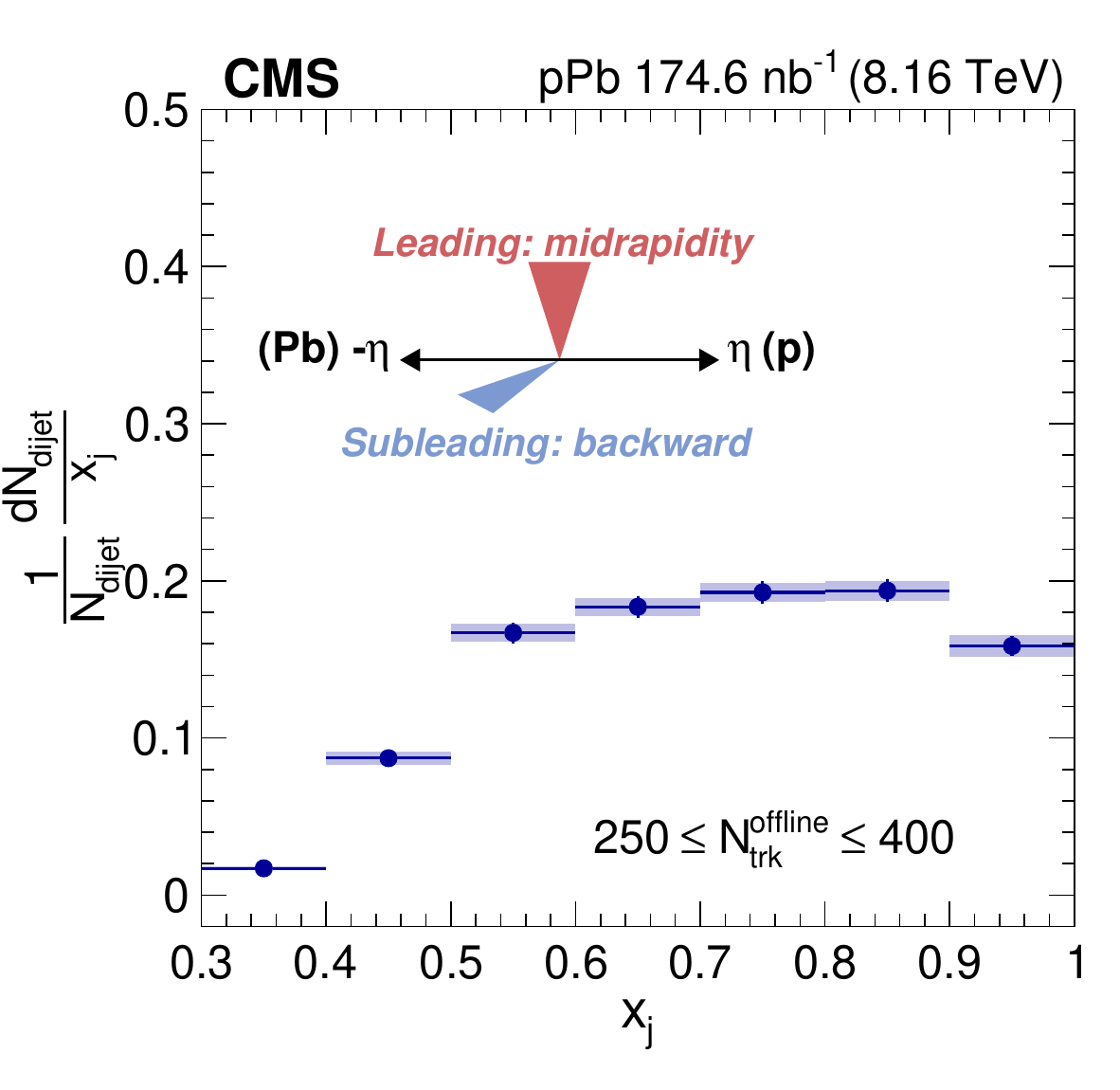} 
    \caption{The measured (raw) \xj distributions, from lower to higher multiplicities and for leading and subleading jets in the midrapidity and backward regions, respectively, are shown. Comparisons with the reconstruction-level {{\PYTHIA}8}+\EPOS simulation (green band) are also displayed.}
    \label{fig:xjmultrec_mid_bkw}
\end{figure}

\begin{figure}[!hbpt]
  \centering
    \includegraphics[width=0.44\textwidth]{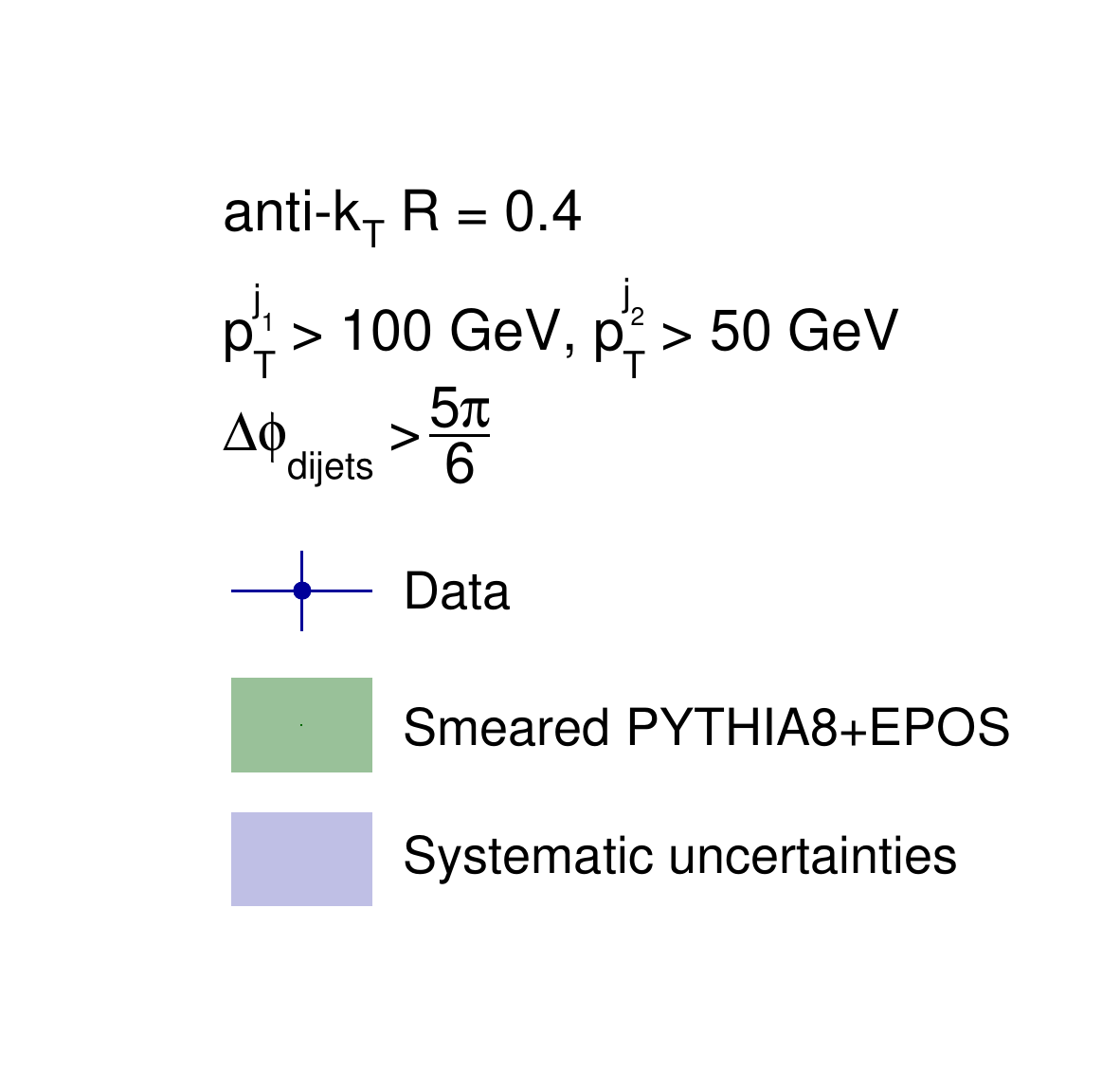} 
    \includegraphics[width=0.44\textwidth]{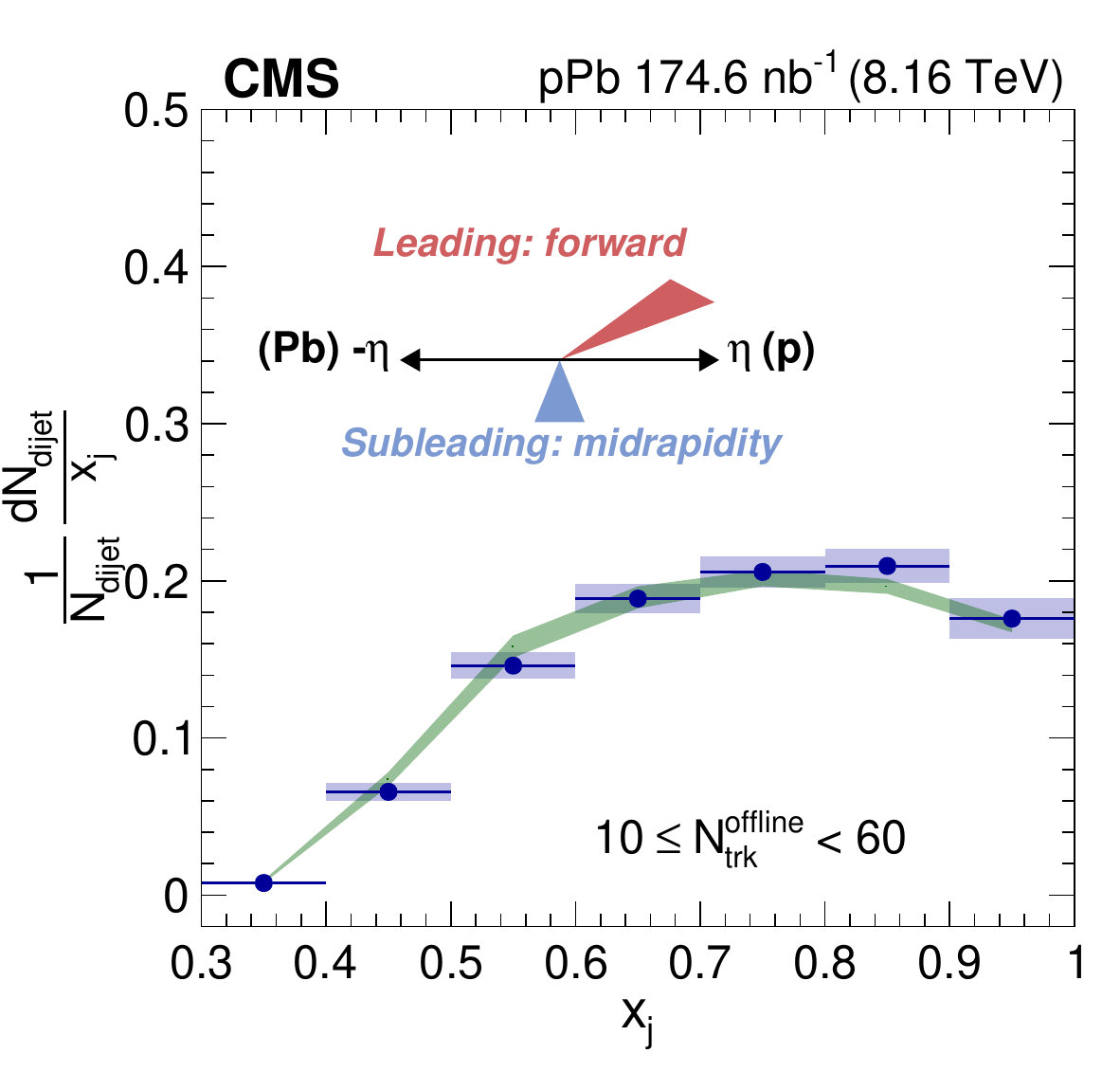} \\
    \includegraphics[width=0.44\textwidth]{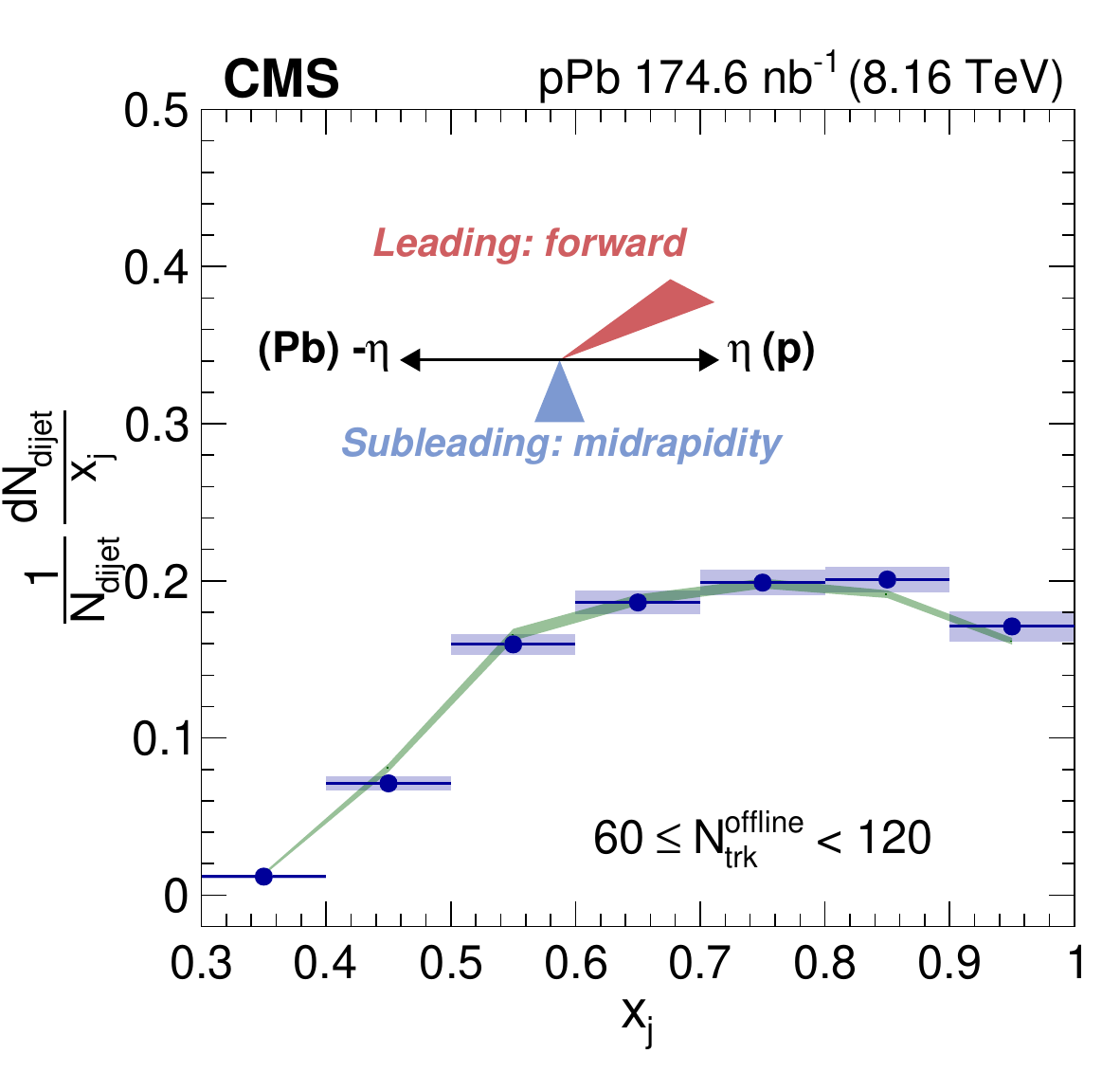} 
    \includegraphics[width=0.44\textwidth]{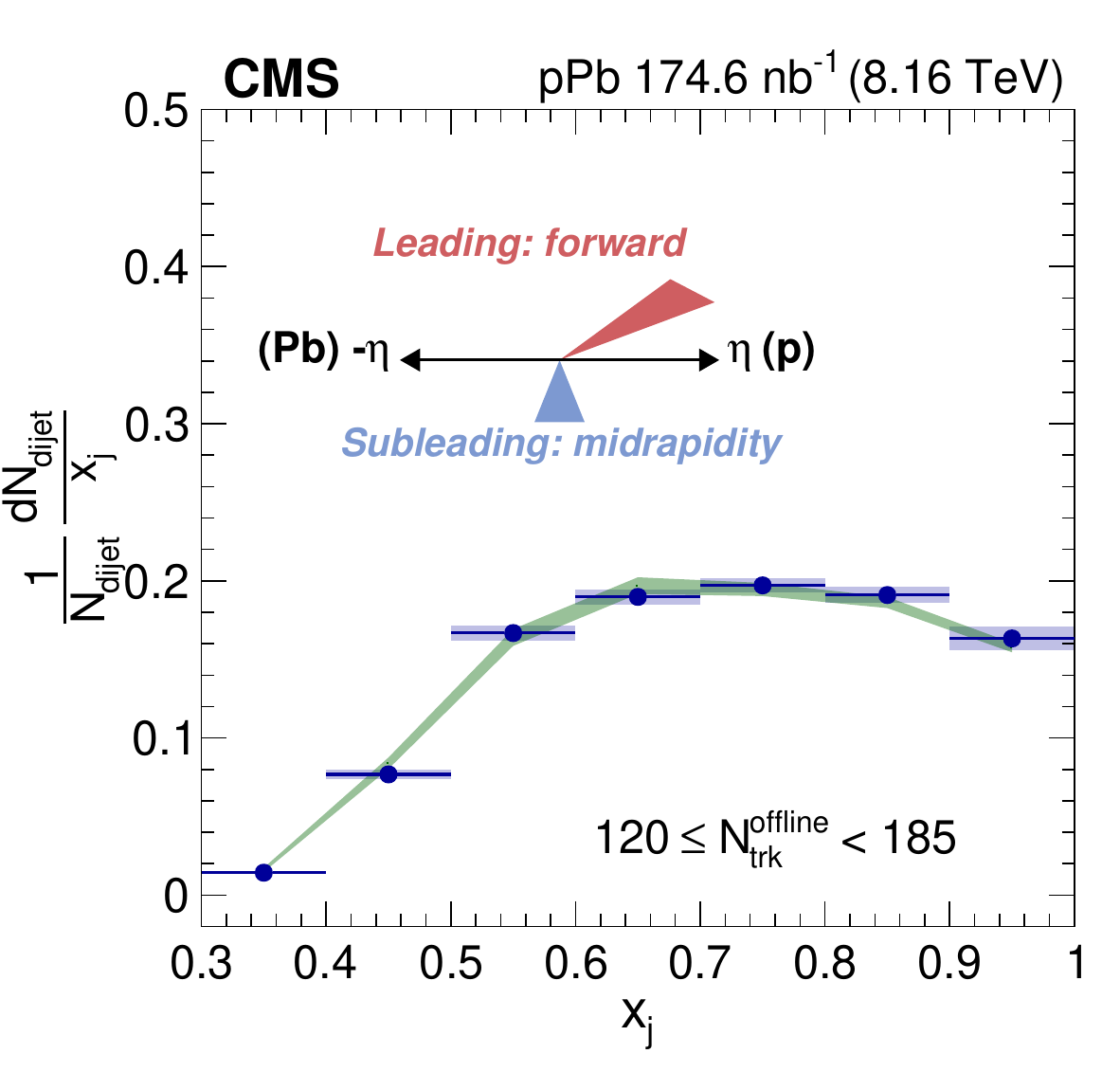} \\
    \includegraphics[width=0.44\textwidth]{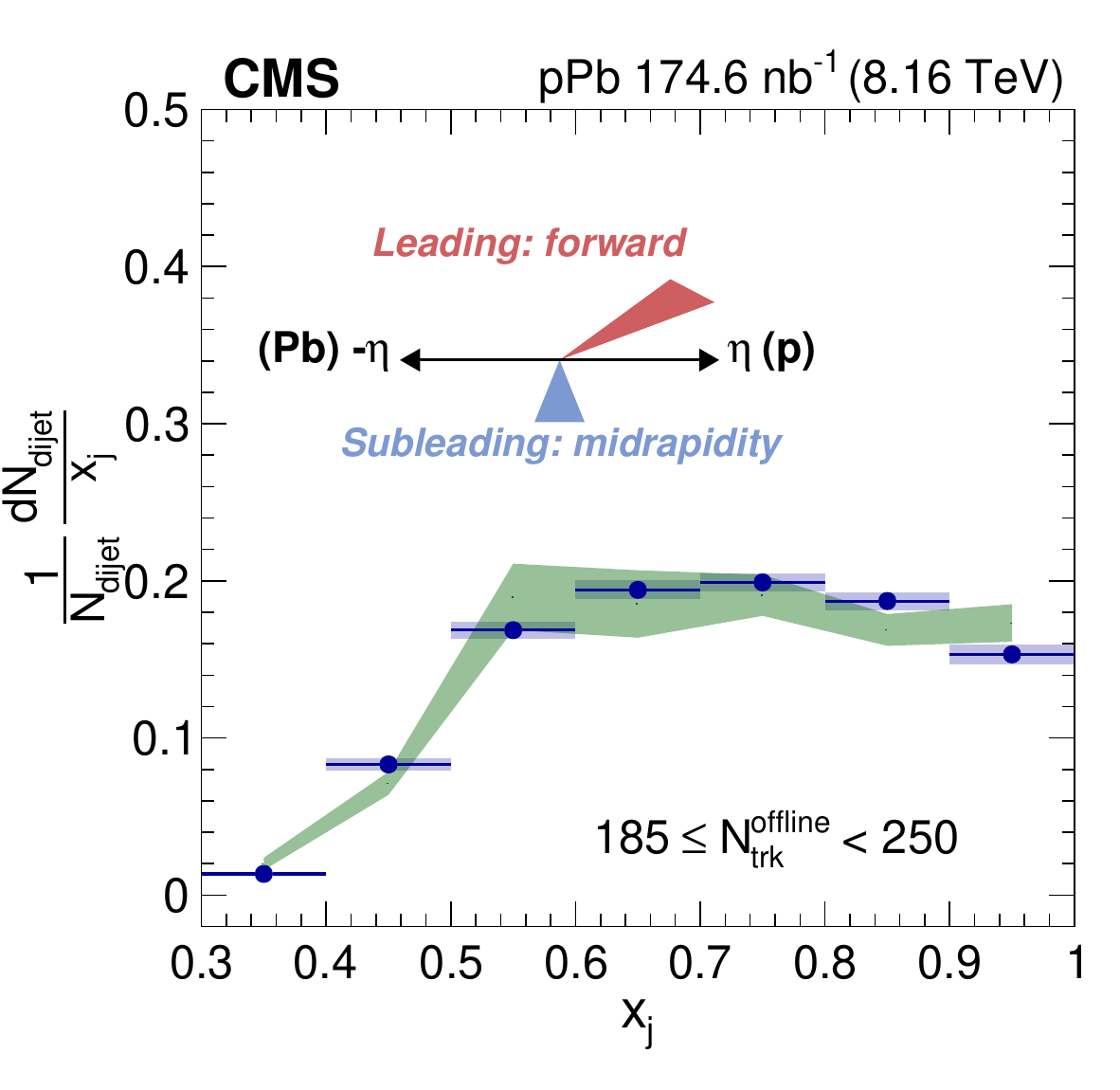} 
    \includegraphics[width=0.44\textwidth]{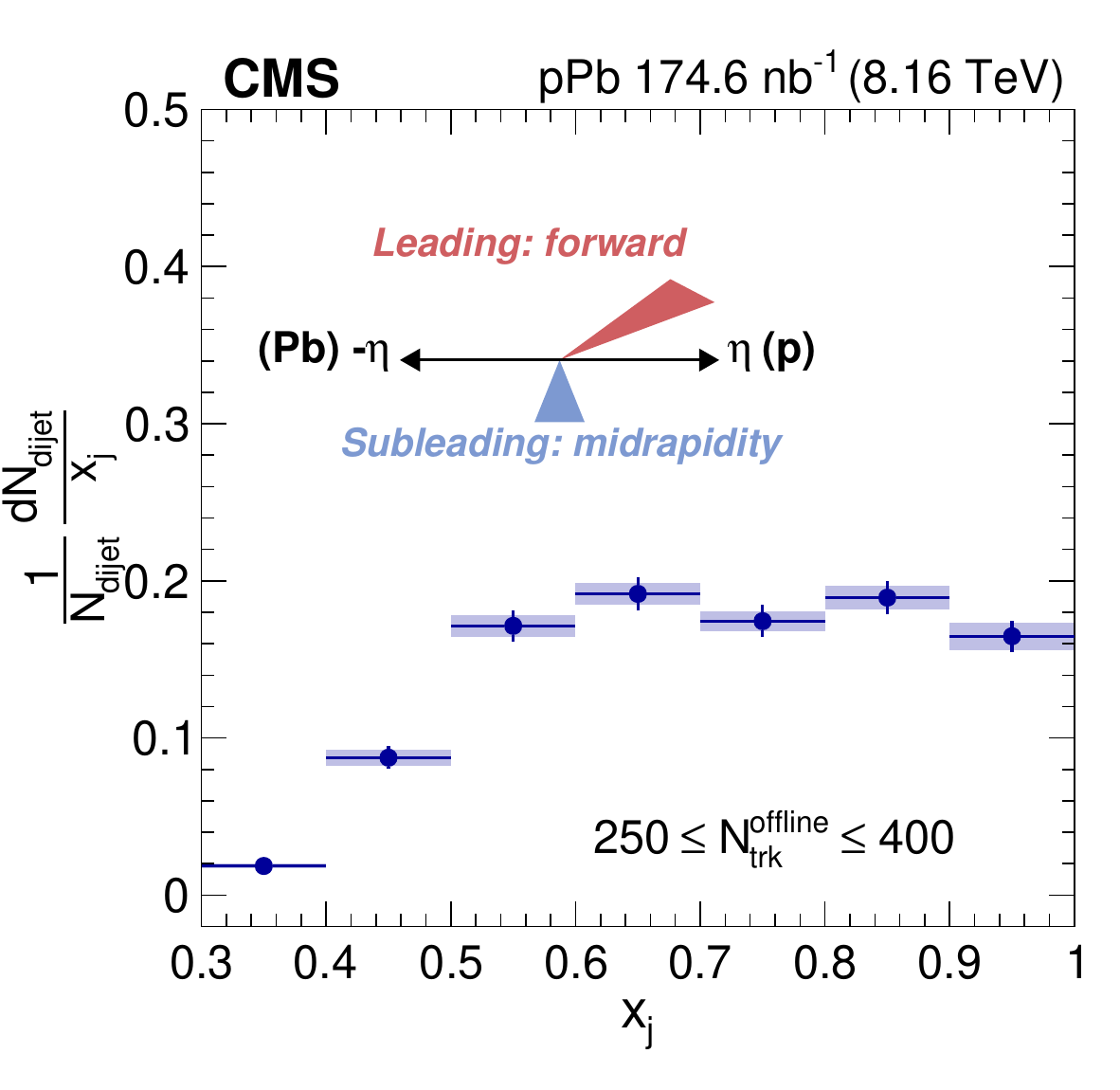} 
    \caption{The measured (raw) \xj distributions, from lower to higher multiplicities and for leading and subleading jets in the forward and midrapidity regions, respectively, are shown. Comparisons with the reconstruction-level {{\PYTHIA}8}+\EPOS simulation (green band) are also displayed.}
    \label{fig:xjmultrec_fwd_mid}
\end{figure}

\begin{figure}[!hbpt]
  \centering
    \includegraphics[width=0.44\textwidth]{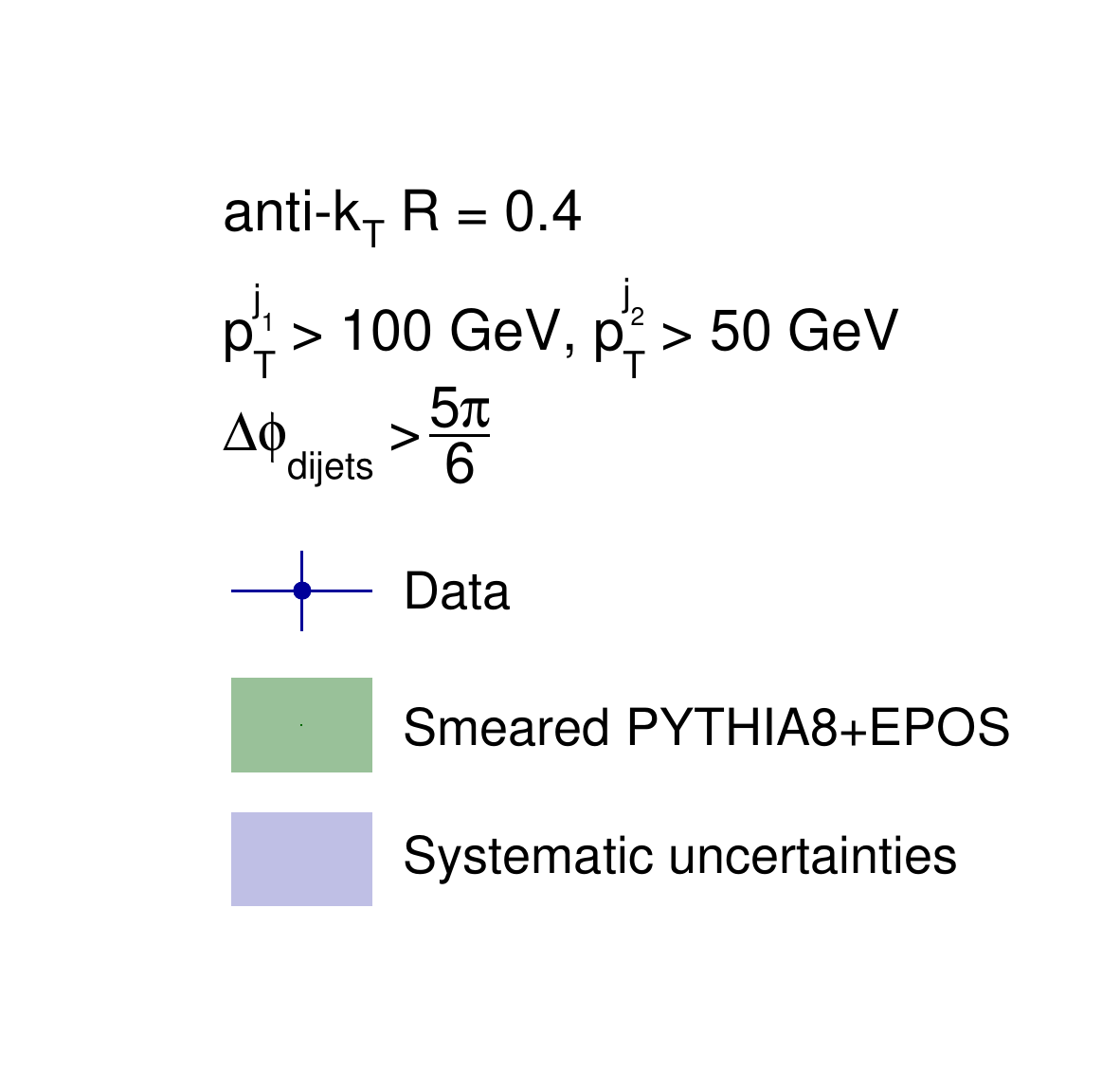} 
    \includegraphics[width=0.44\textwidth]{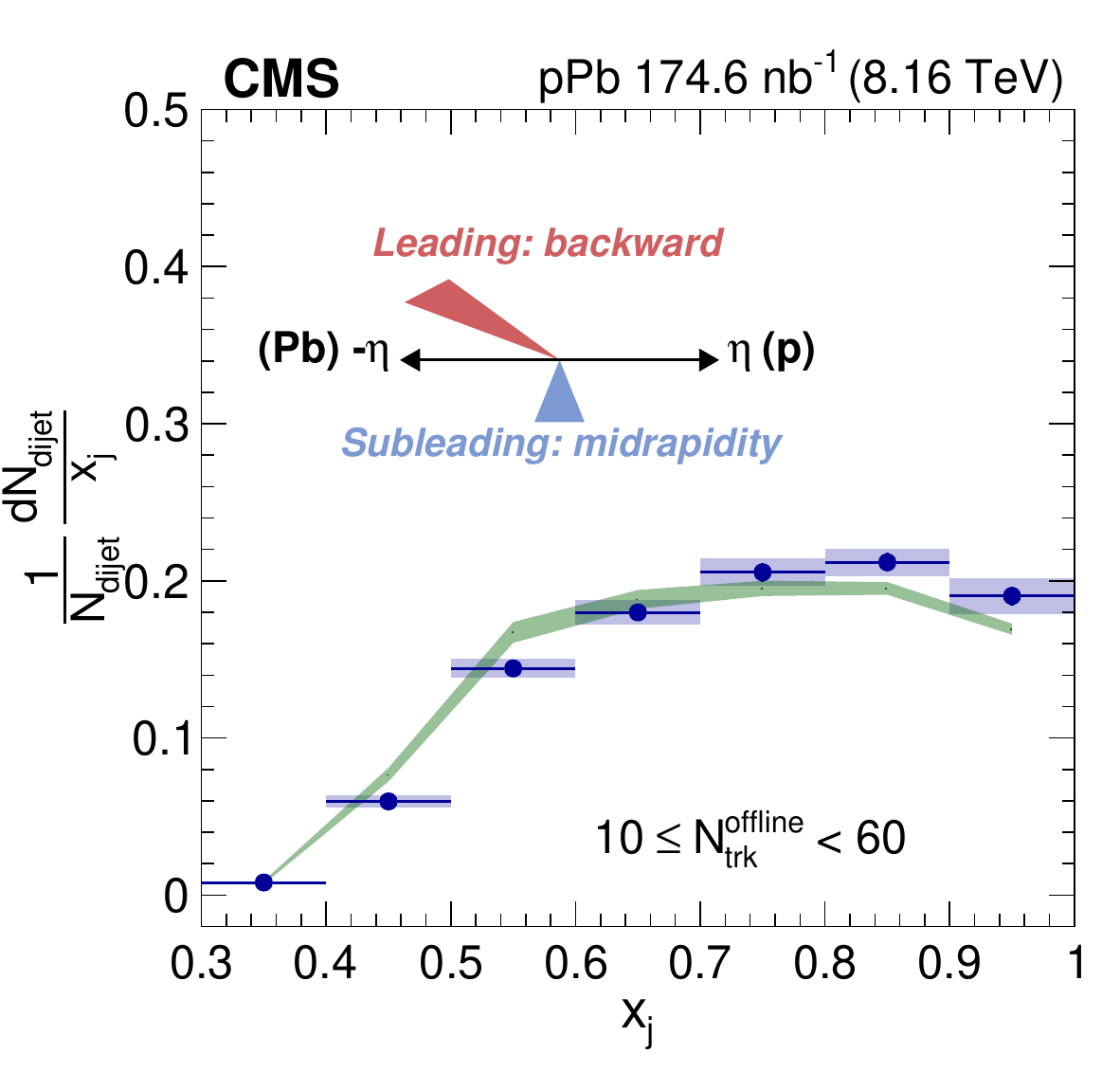} \\
    \includegraphics[width=0.44\textwidth]{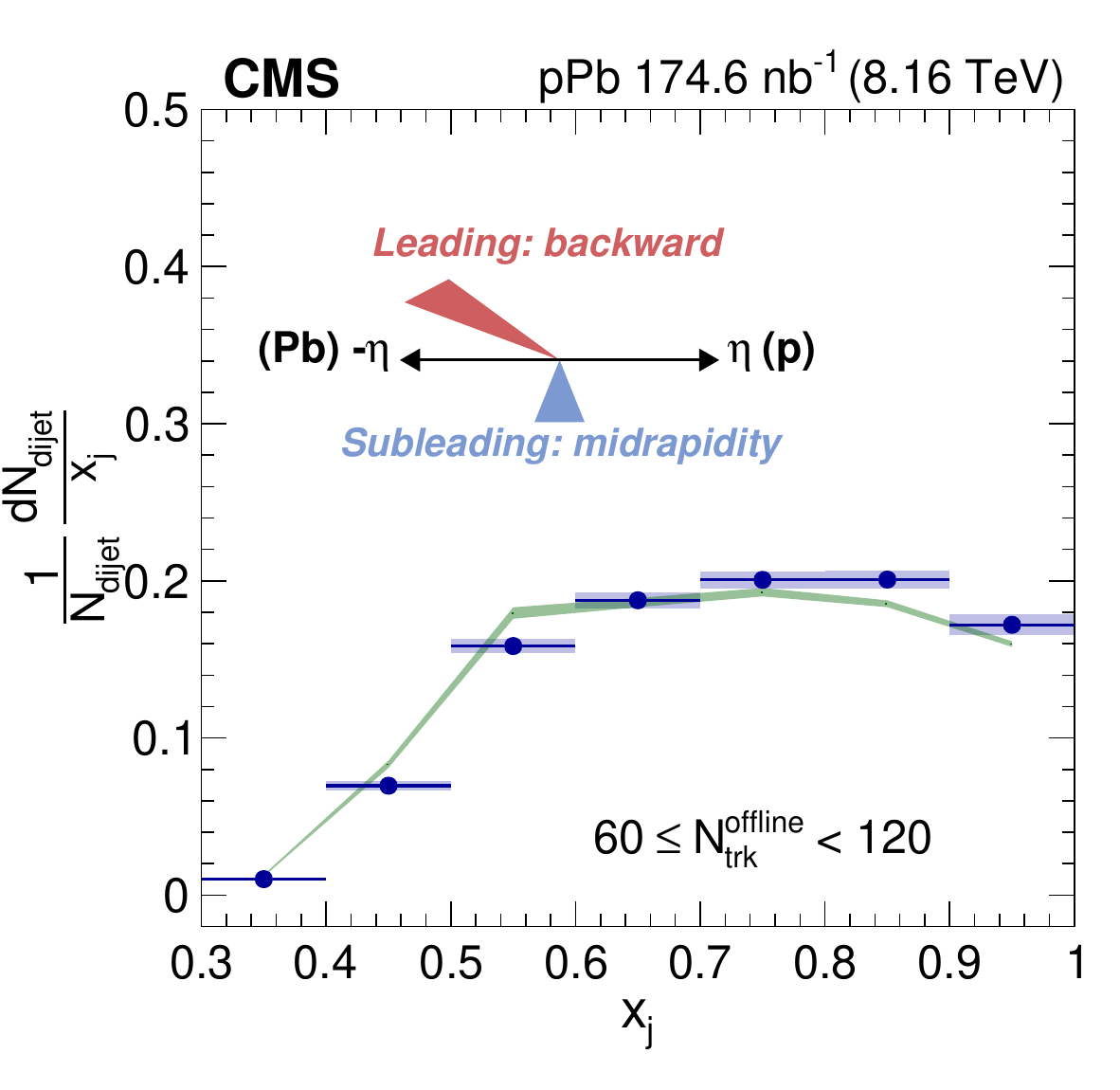} 
    \includegraphics[width=0.44\textwidth]{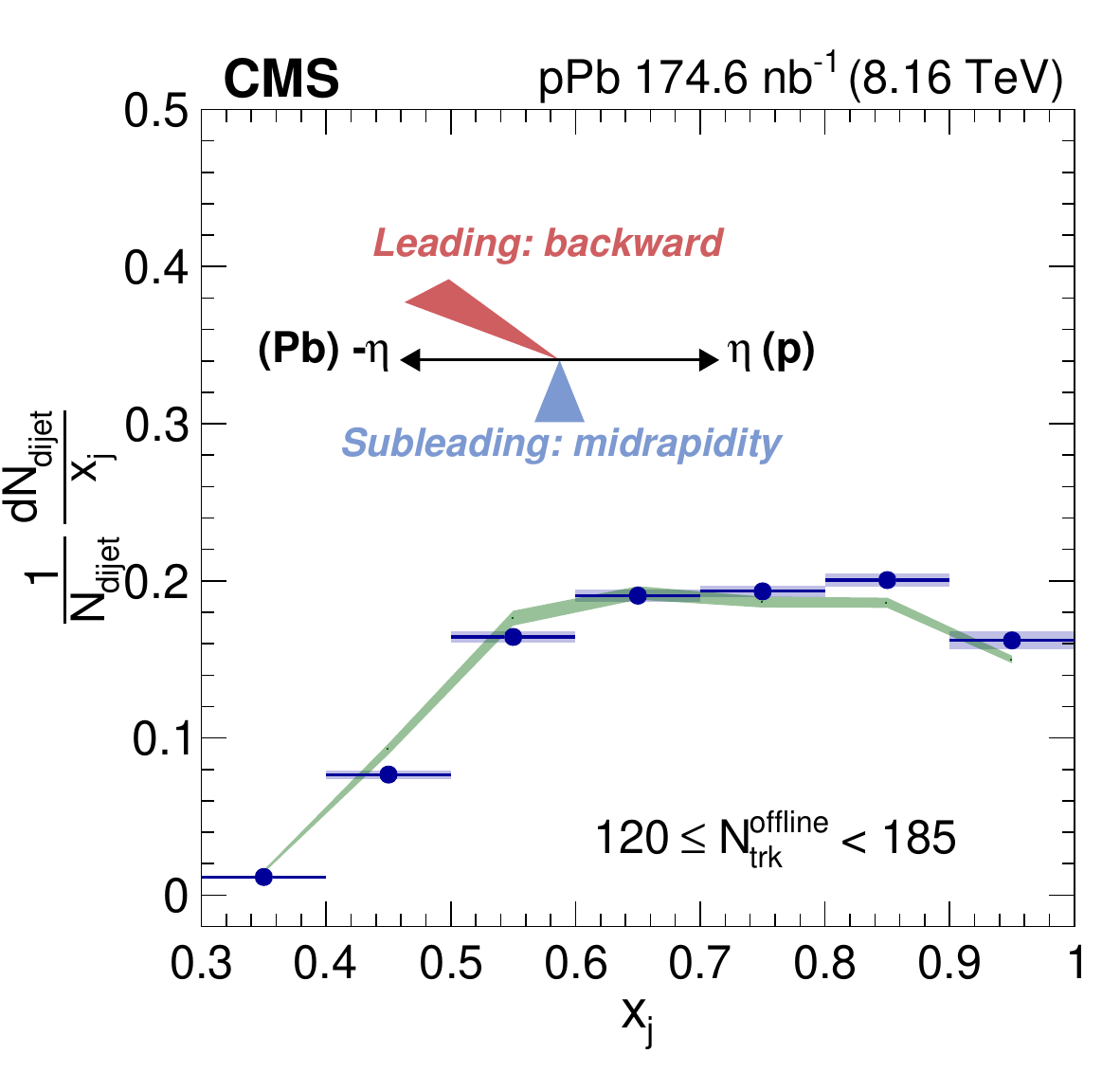} \\
    \includegraphics[width=0.44\textwidth]{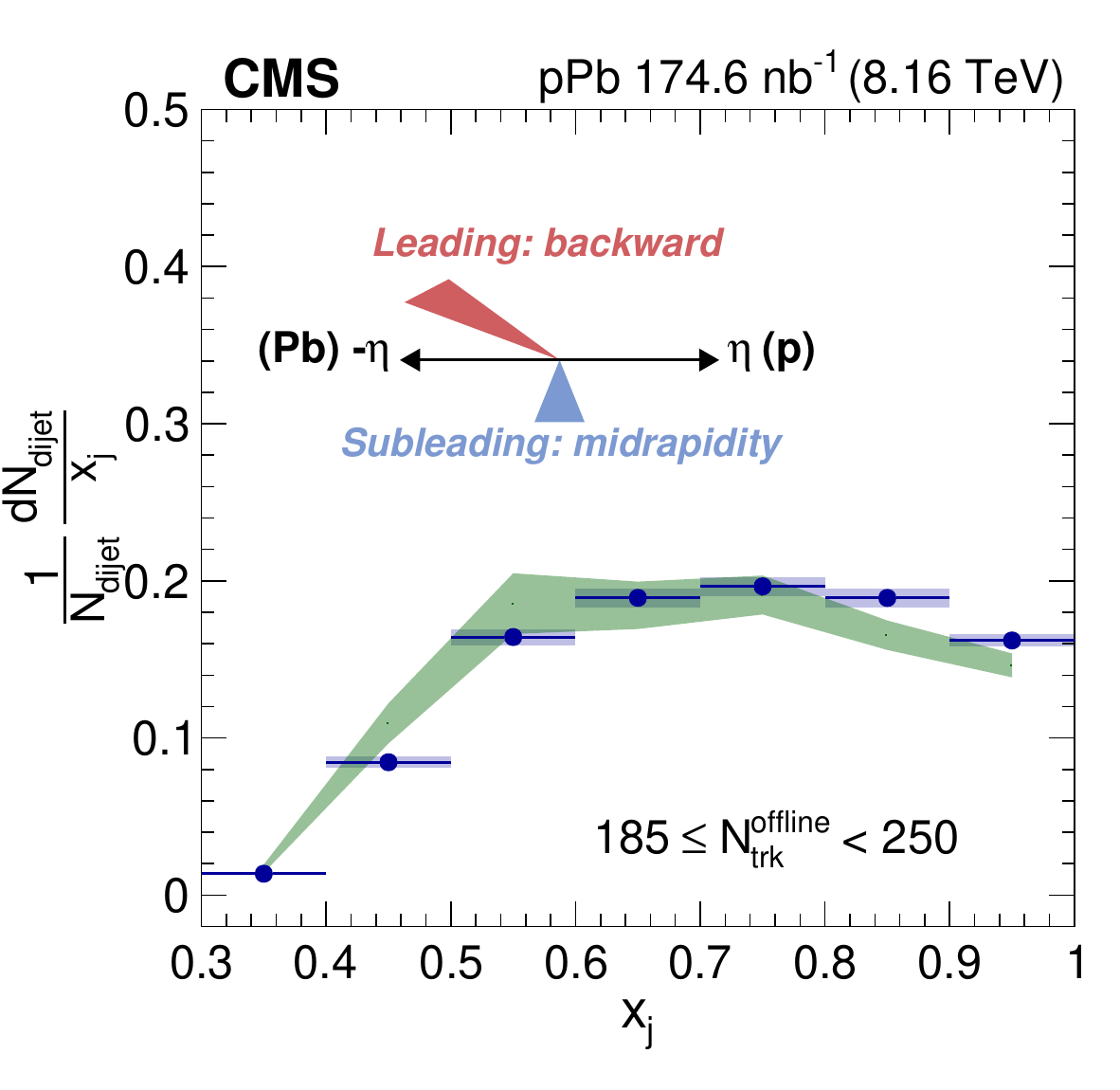} 
    \includegraphics[width=0.44\textwidth]{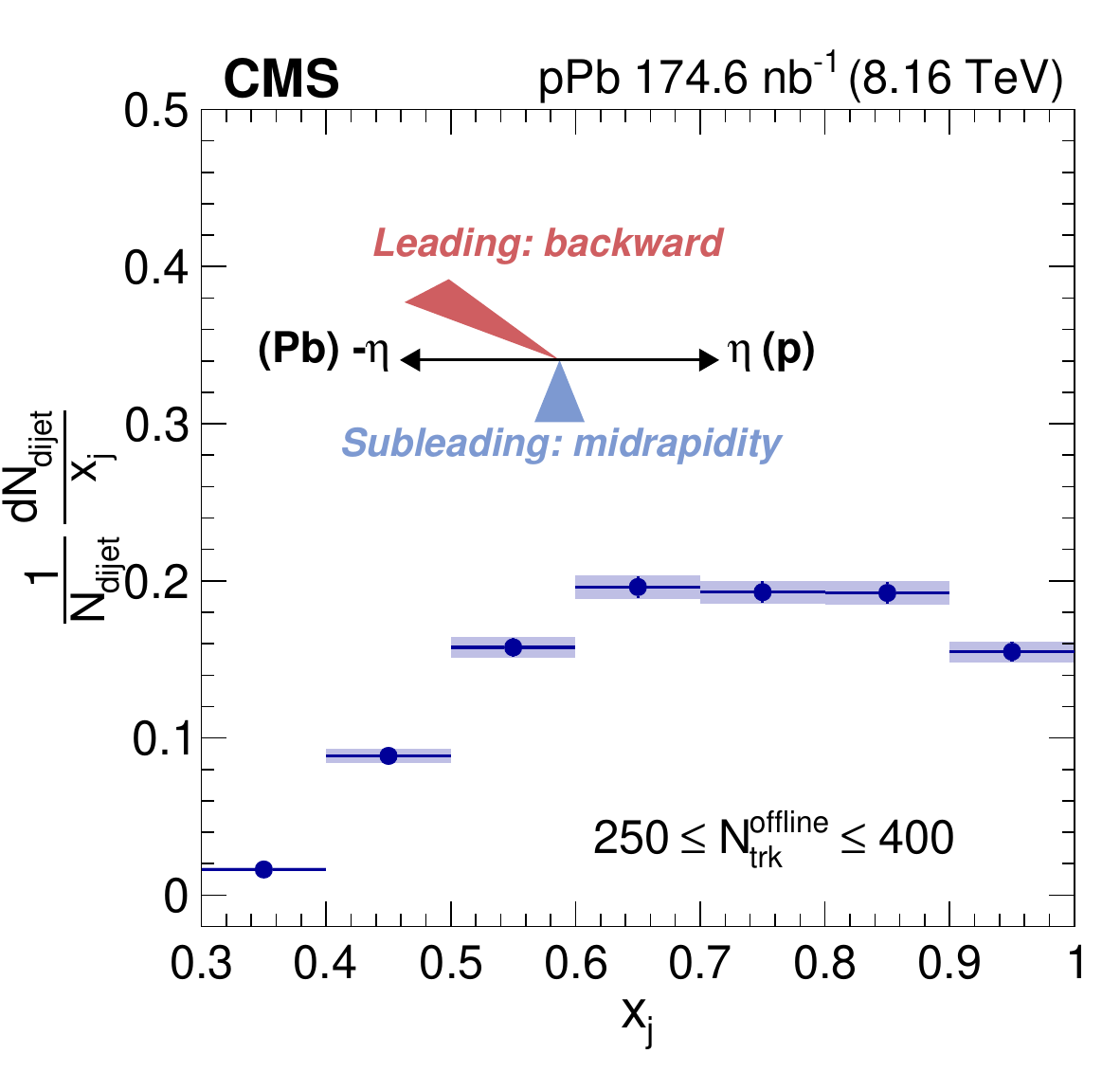} 
    \caption{The measured (raw) \xj distributions, from lower to higher multiplicities and for leading and subleading jets in the backward and midrapidity regions, respectively, are shown. Comparisons with the reconstruction-level {{\PYTHIA}8}+\EPOS simulation (green band) are also displayed. }
    \label{fig:xjmultrec_bkw_mid}
\end{figure}

\begin{figure}[!hbpt]
  \centering
    \includegraphics[width=0.44\textwidth]{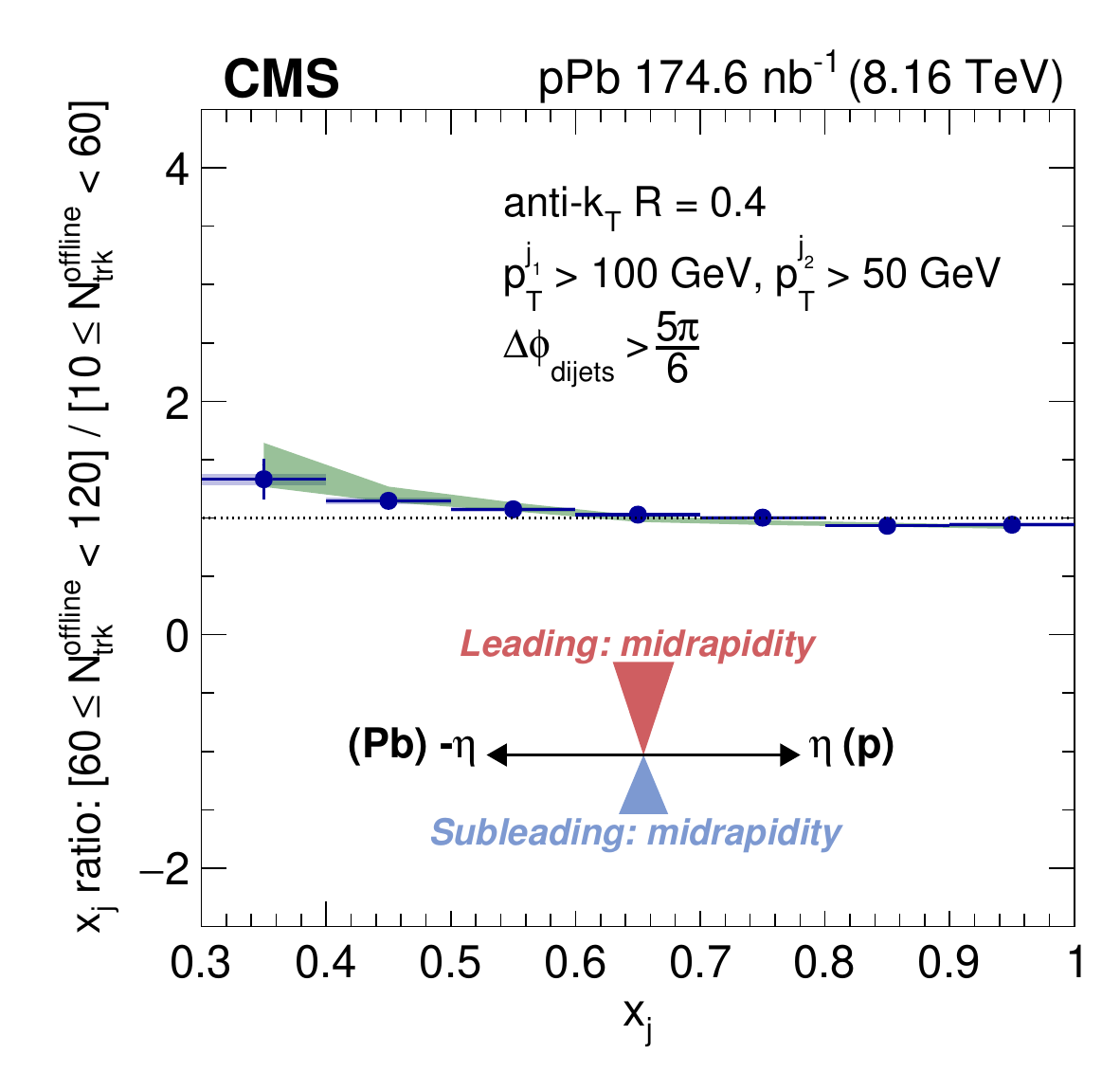} 
    \includegraphics[width=0.44\textwidth]{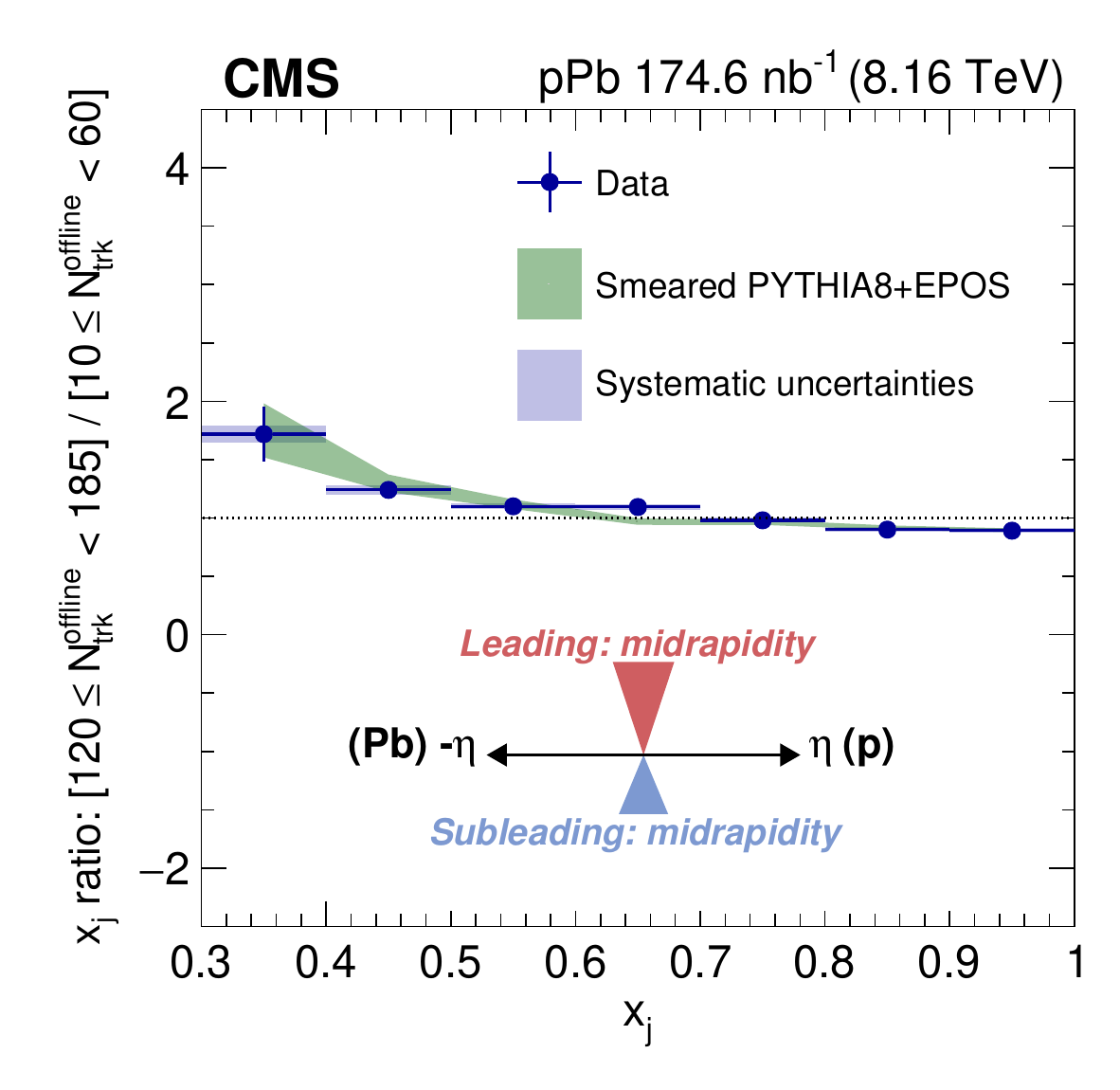} \\
    \includegraphics[width=0.44\textwidth]{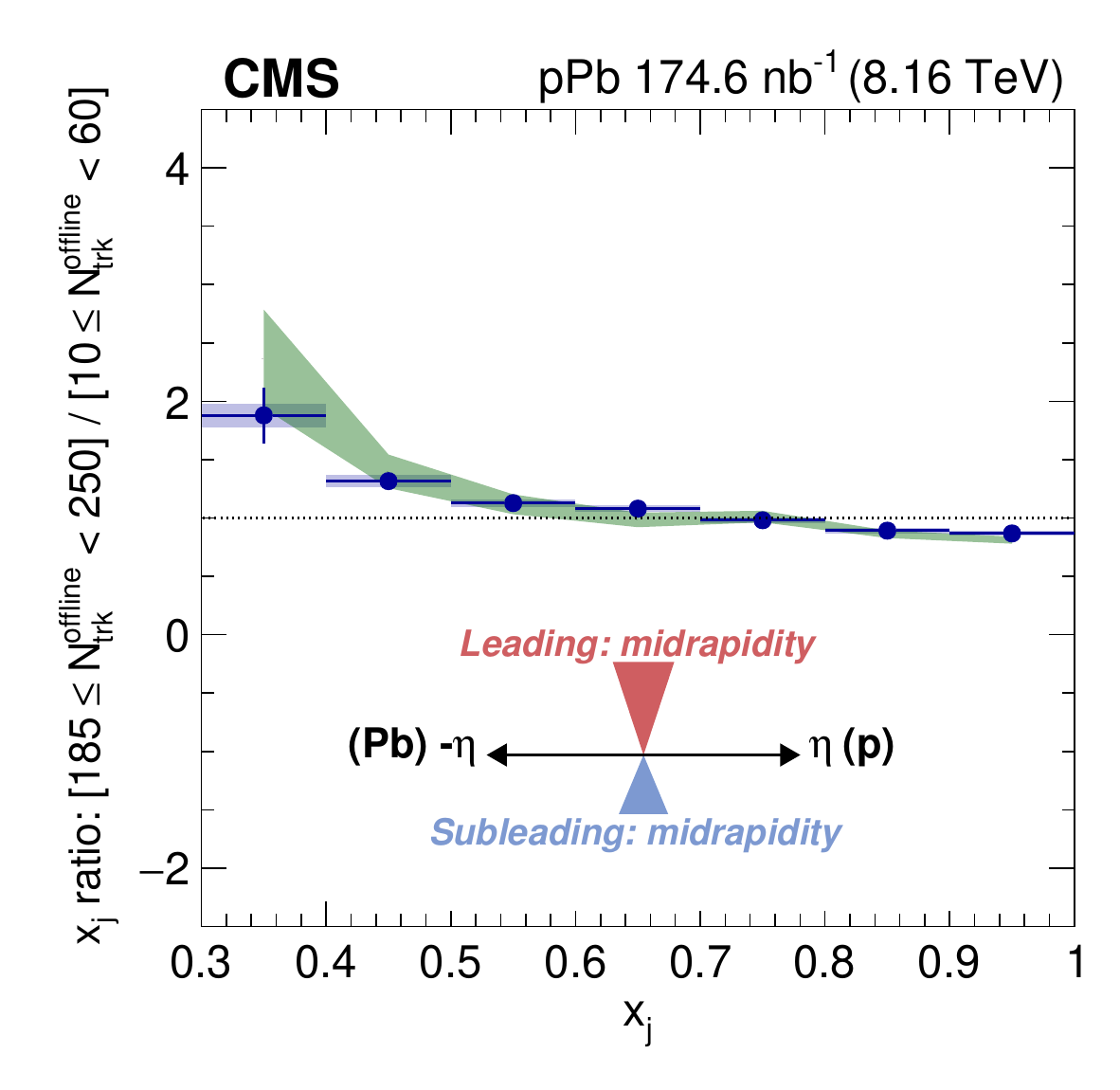} 
    \includegraphics[width=0.44\textwidth]{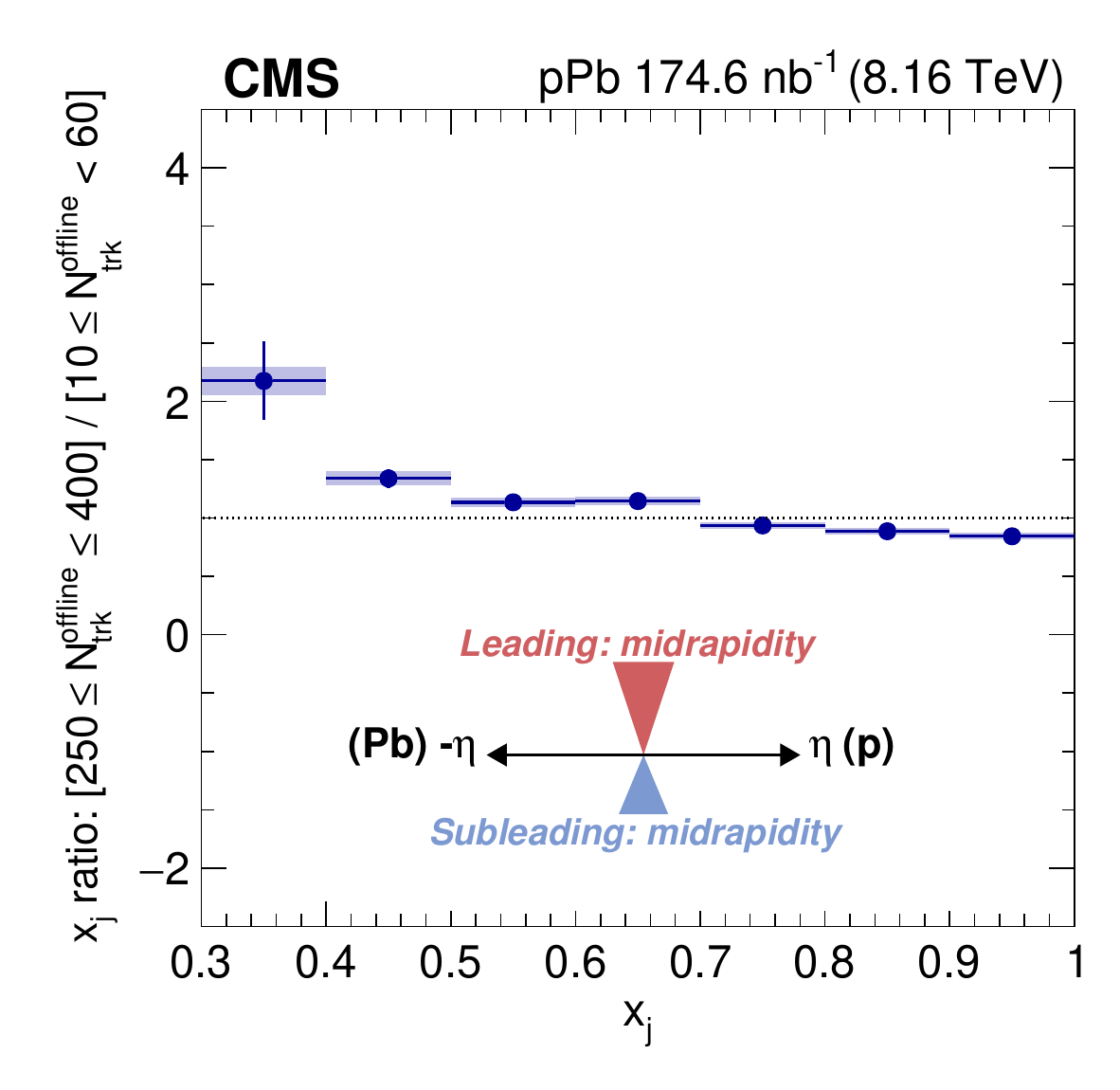} 
    \caption{The measured (raw) \xj ratio in the range $10 \leq \Ntroff < 60$, from lower to higher multiplicities for both leading and subleading jets in the midrapidity region, are shown. Comparison with the reconstruction-level {{\PYTHIA}8}+\EPOS simulation (green band) are also displayed. }
    \label{fig:xjratiomultrec_mid_mid}
\end{figure}

\begin{figure}[!hbpt]
  \centering
    \includegraphics[width=0.44\textwidth]{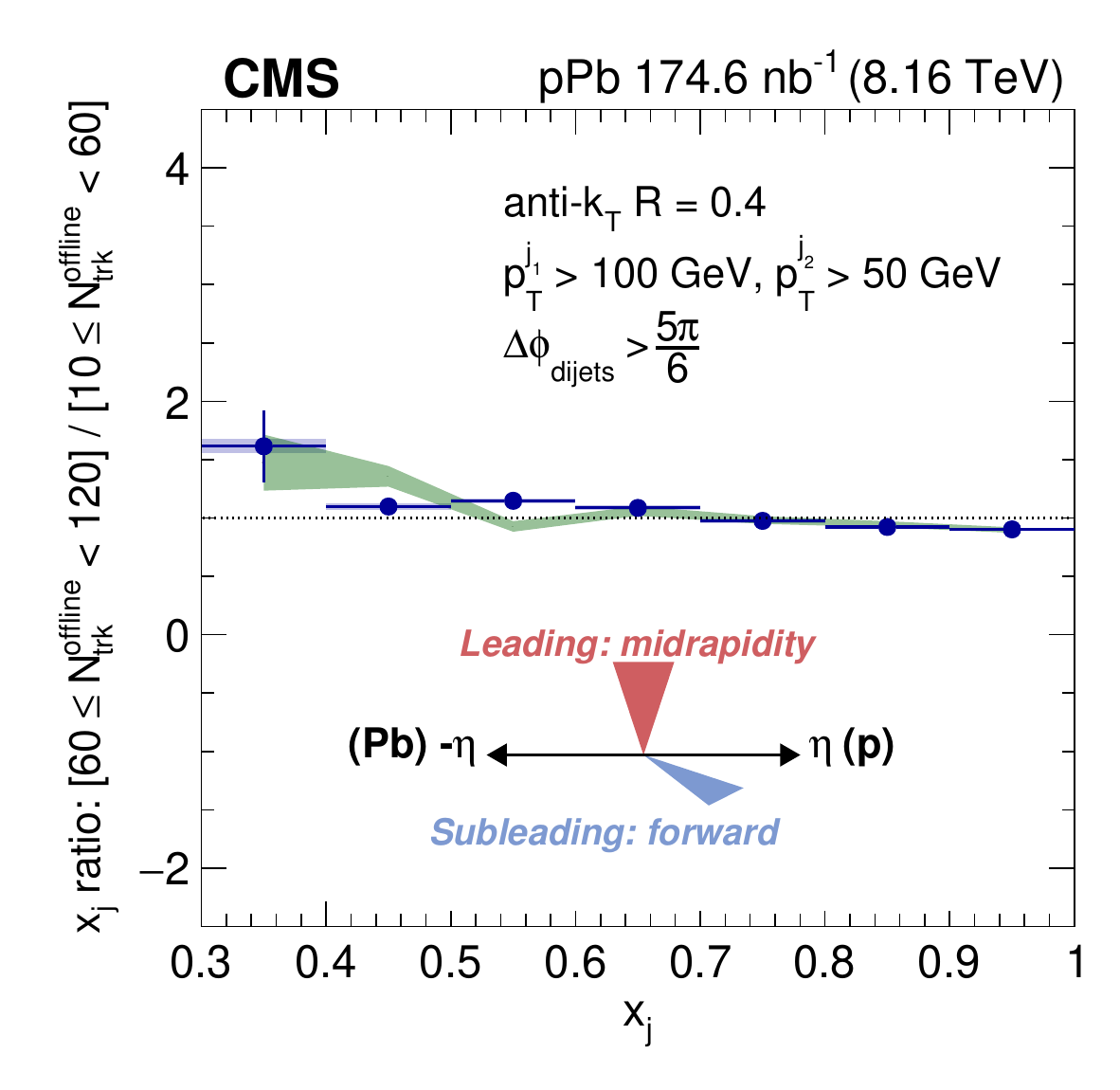} 
    \includegraphics[width=0.44\textwidth]{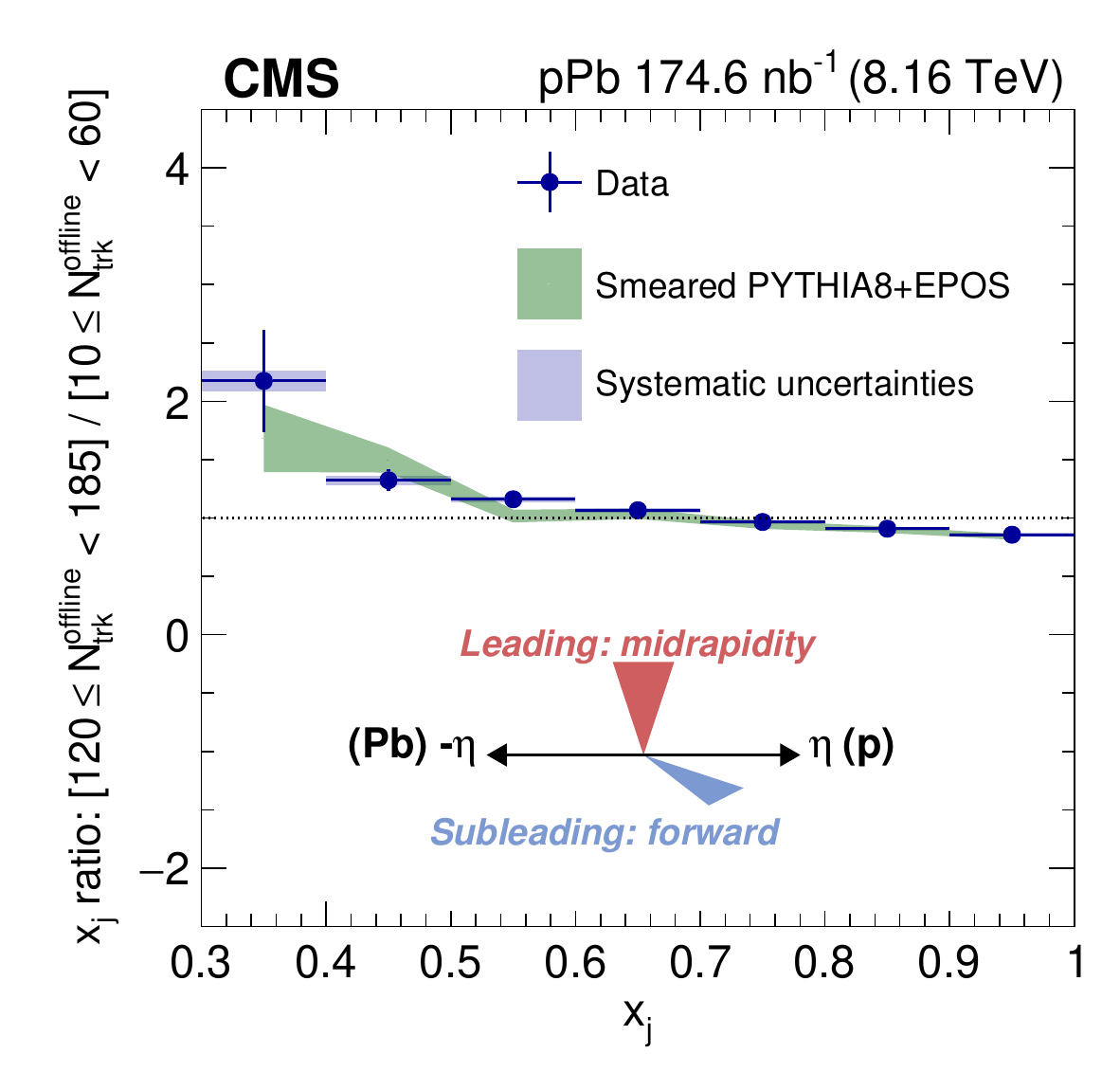} \\
    \includegraphics[width=0.44\textwidth]{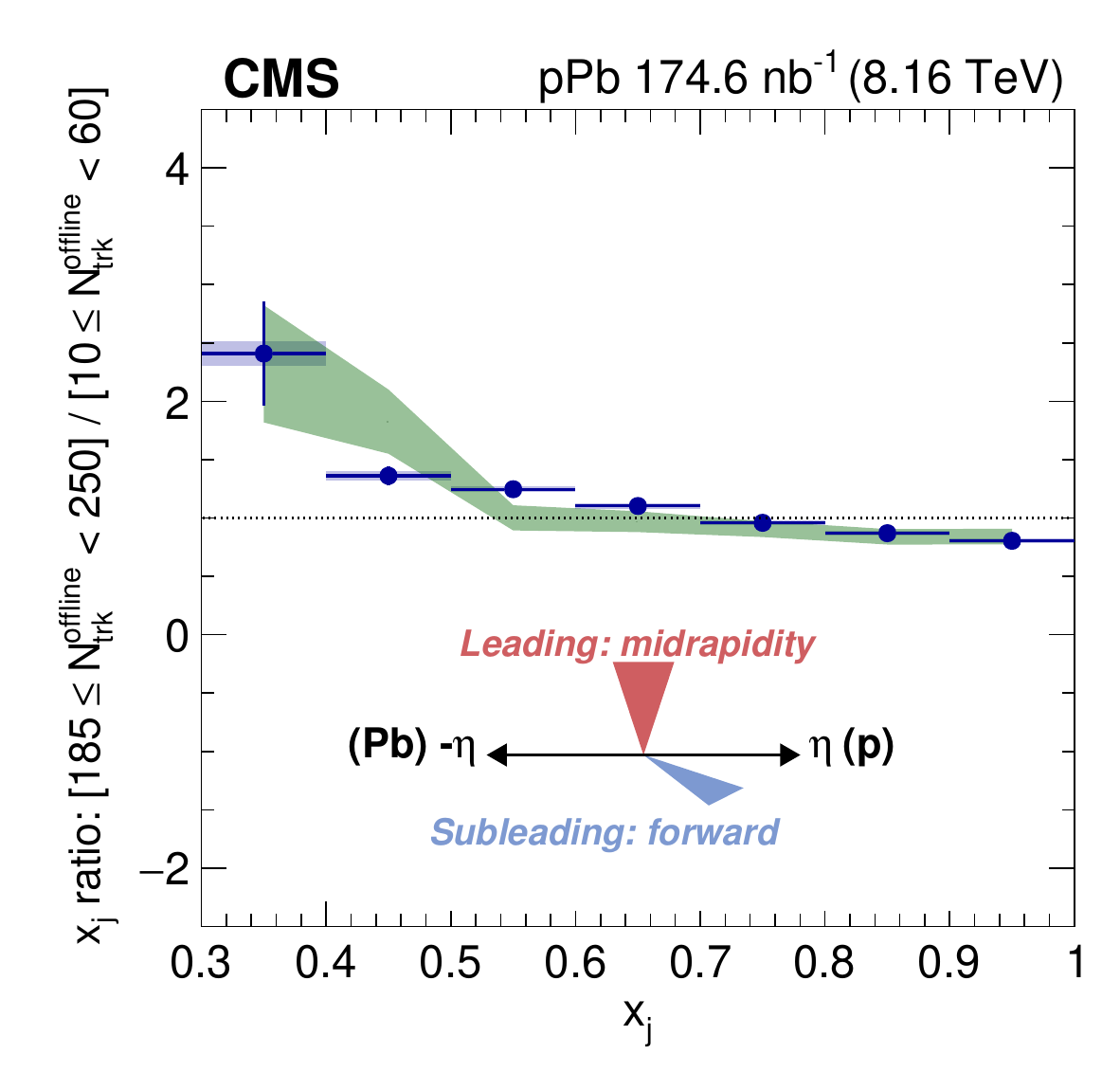} 
    \includegraphics[width=0.44\textwidth]{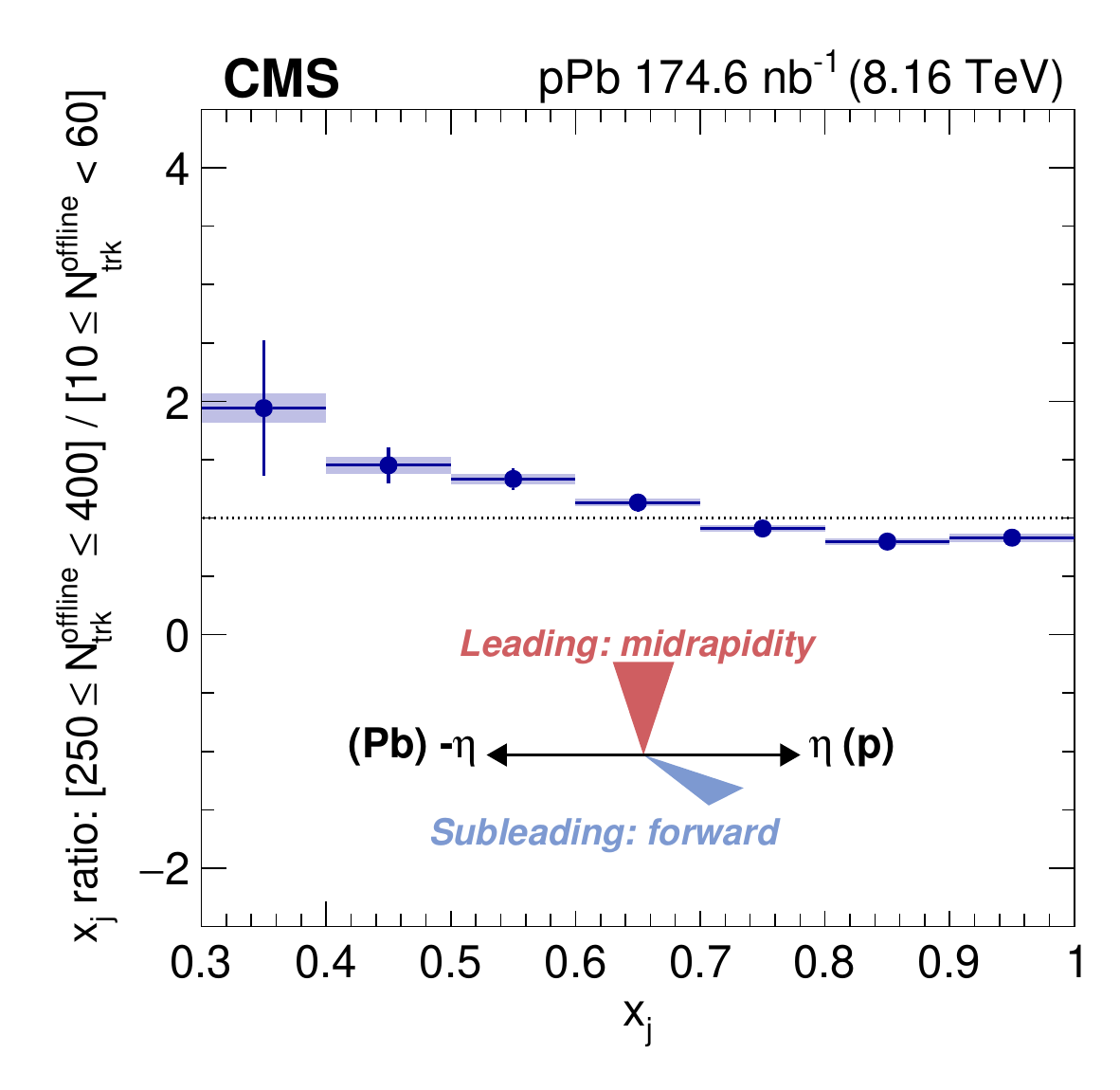} 
    \caption{The measured (raw) \xj ratio in the range $10 \leq \Ntroff < 60$, from lower to higher multiplicities for leading and subleading jets in the midrapidity and forward regions, respectively, are shown. Comparison with the reconstruction-level {{\PYTHIA}8}+\EPOS simulation (green band) are also displayed. }
    \label{fig:xjratiomultrec_mid_fwd}
\end{figure}

\begin{figure}[!hbpt]
  \centering
    \includegraphics[width=0.44\textwidth]{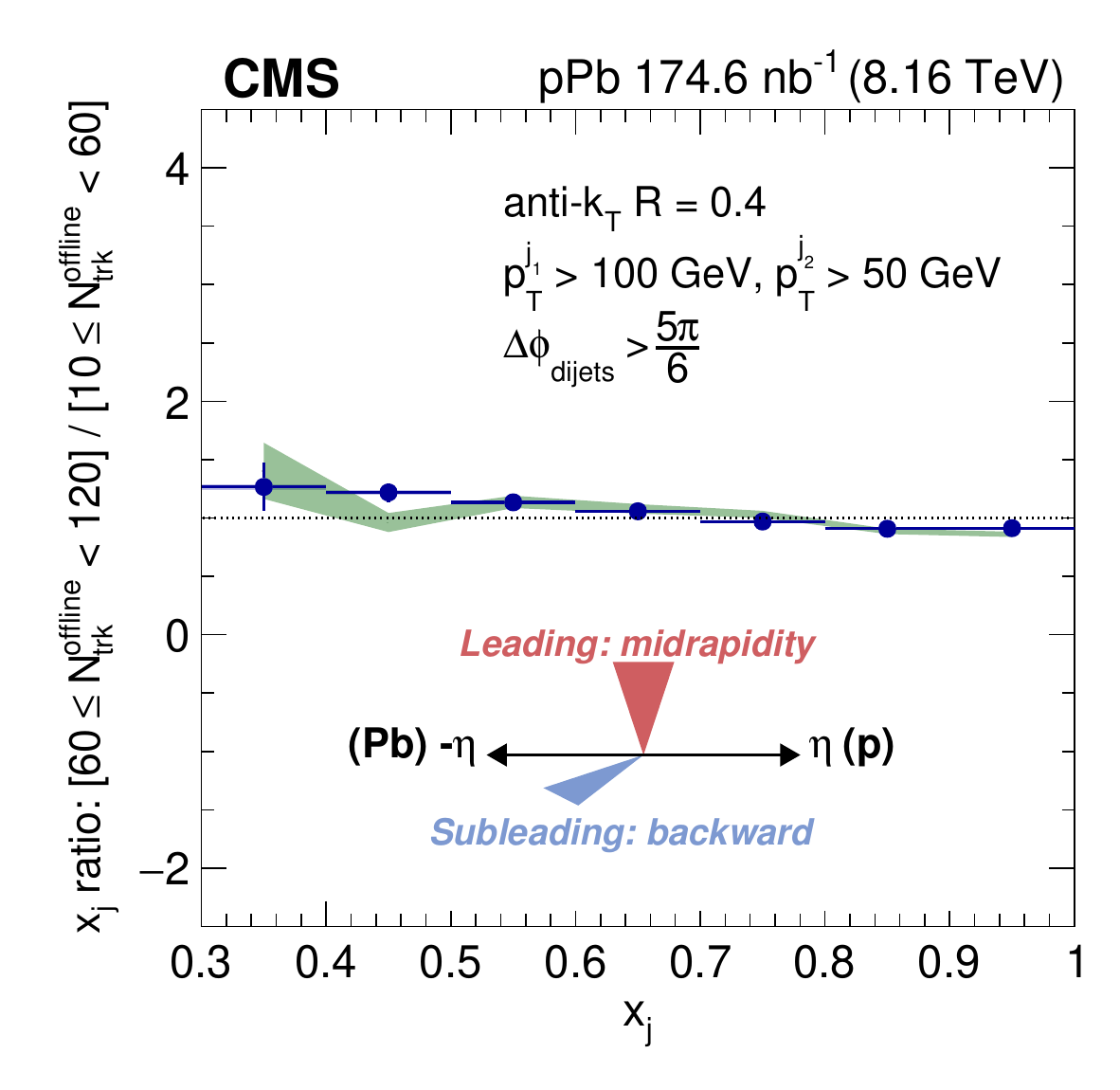} 
    \includegraphics[width=0.44\textwidth]{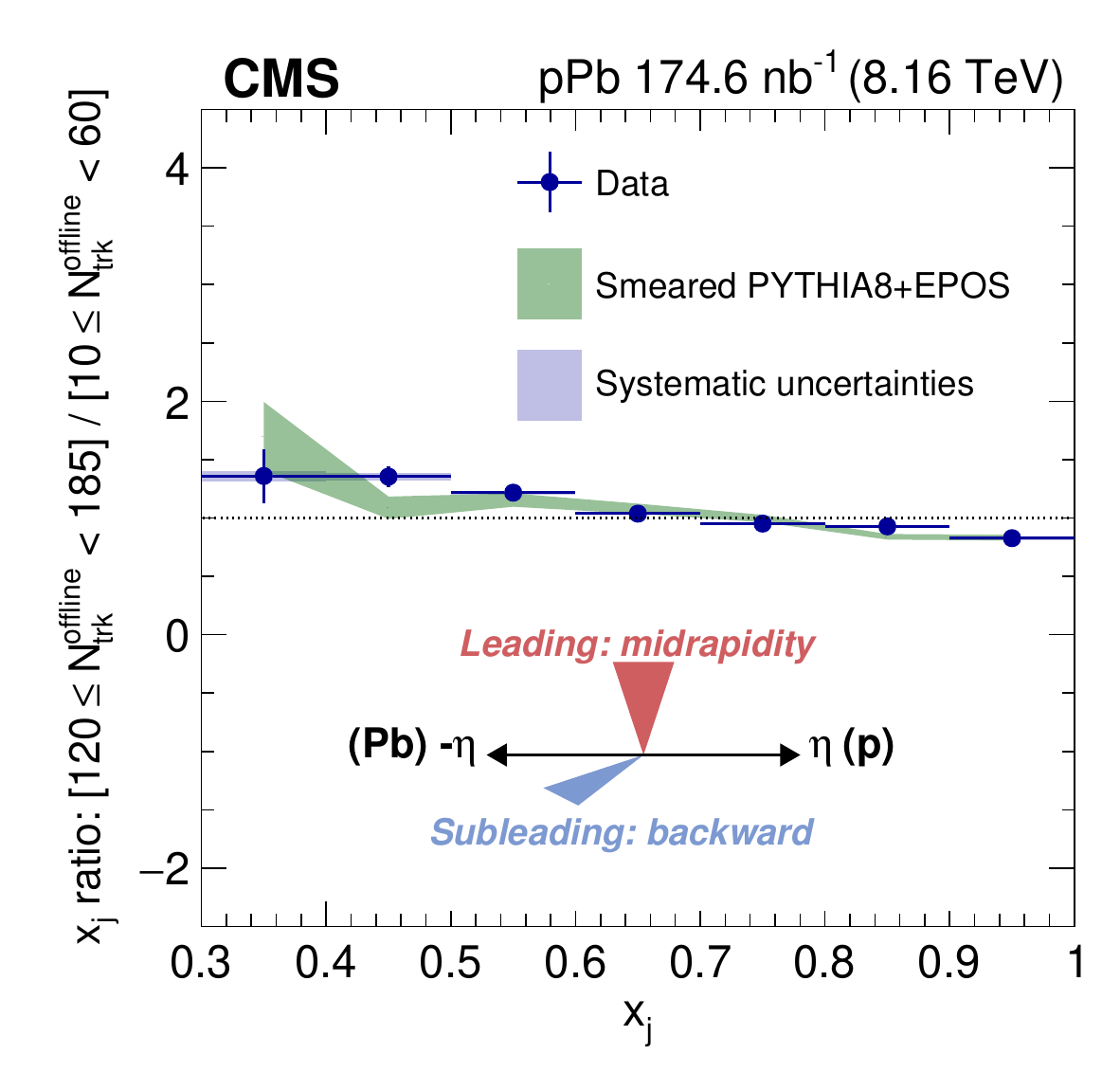} \\
    \includegraphics[width=0.44\textwidth]{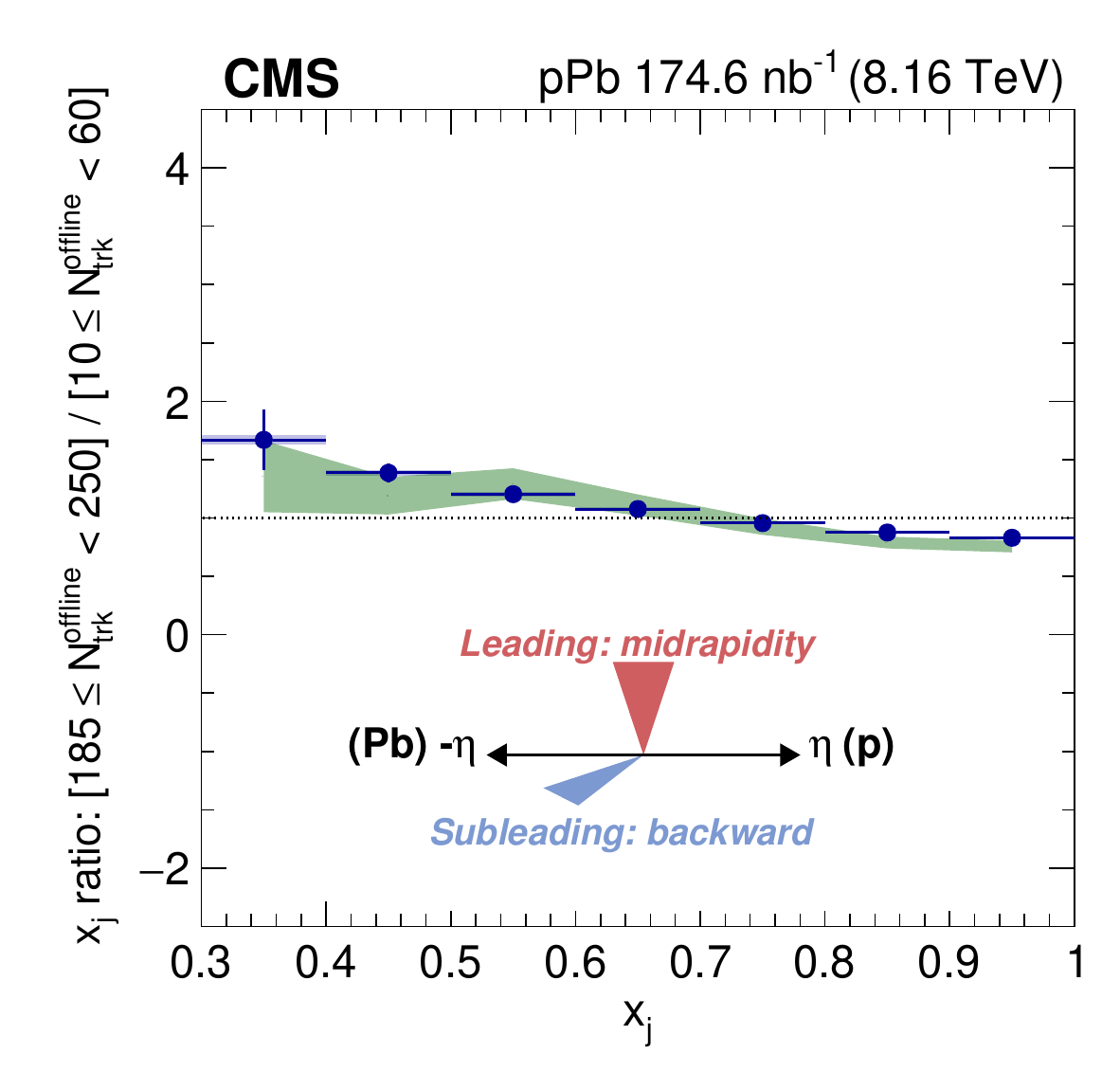} 
    \includegraphics[width=0.44\textwidth]{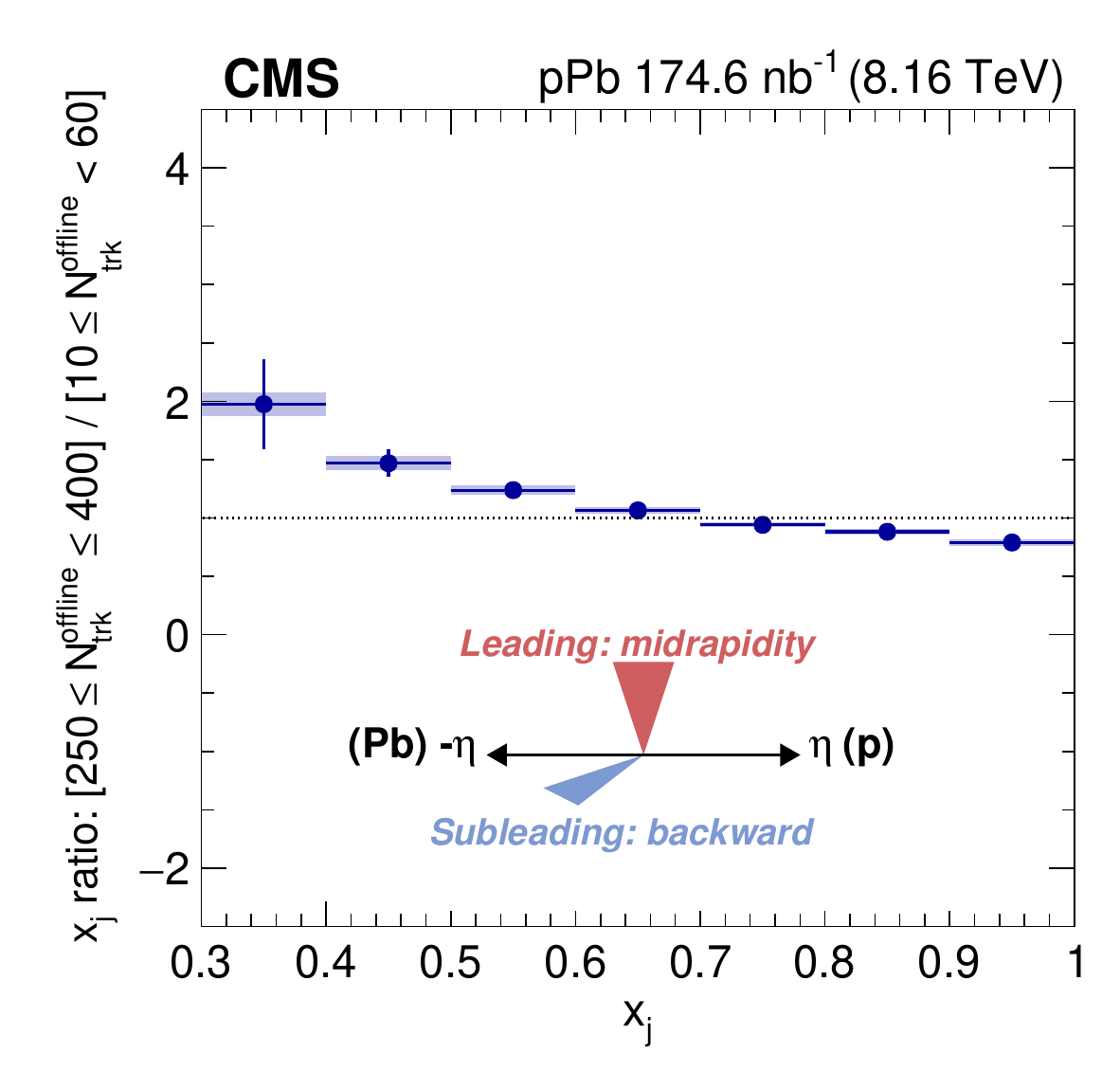} 
    \caption{The measured (raw) \xj ratio in the range $10 \leq \Ntroff < 60$, from lower to higher multiplicities for leading and subleading jets in the midrapidity and backward regions, respectively, are shown. Comparison with the reconstruction-level {{\PYTHIA}8}+\EPOS simulation (green band) are also displayed. }
    \label{fig:xjratiomultrec_mid_bkw}
\end{figure}

\begin{figure}[!hbpt]
  \centering
    \includegraphics[width=0.44\textwidth]{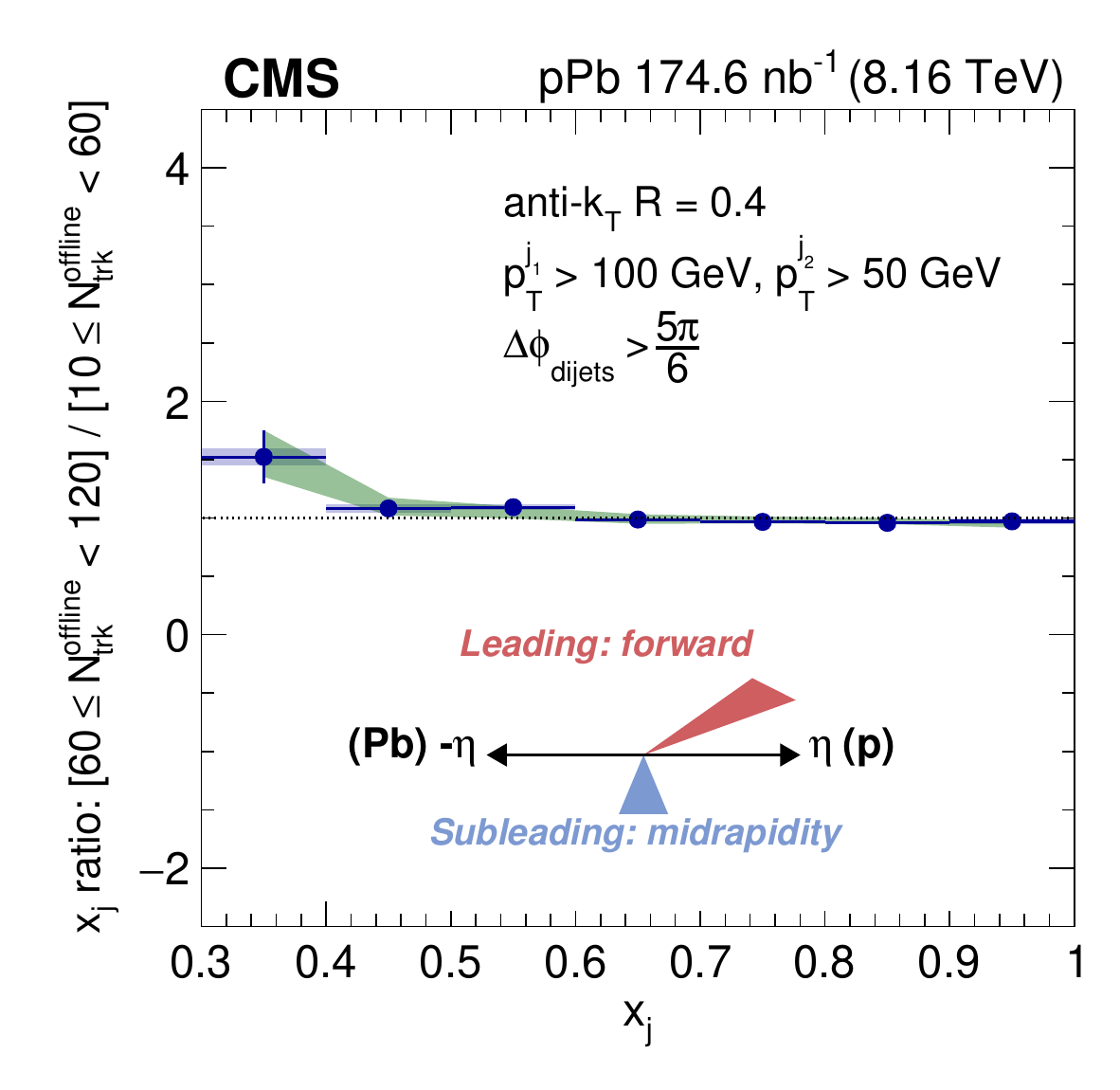} 
    \includegraphics[width=0.44\textwidth]{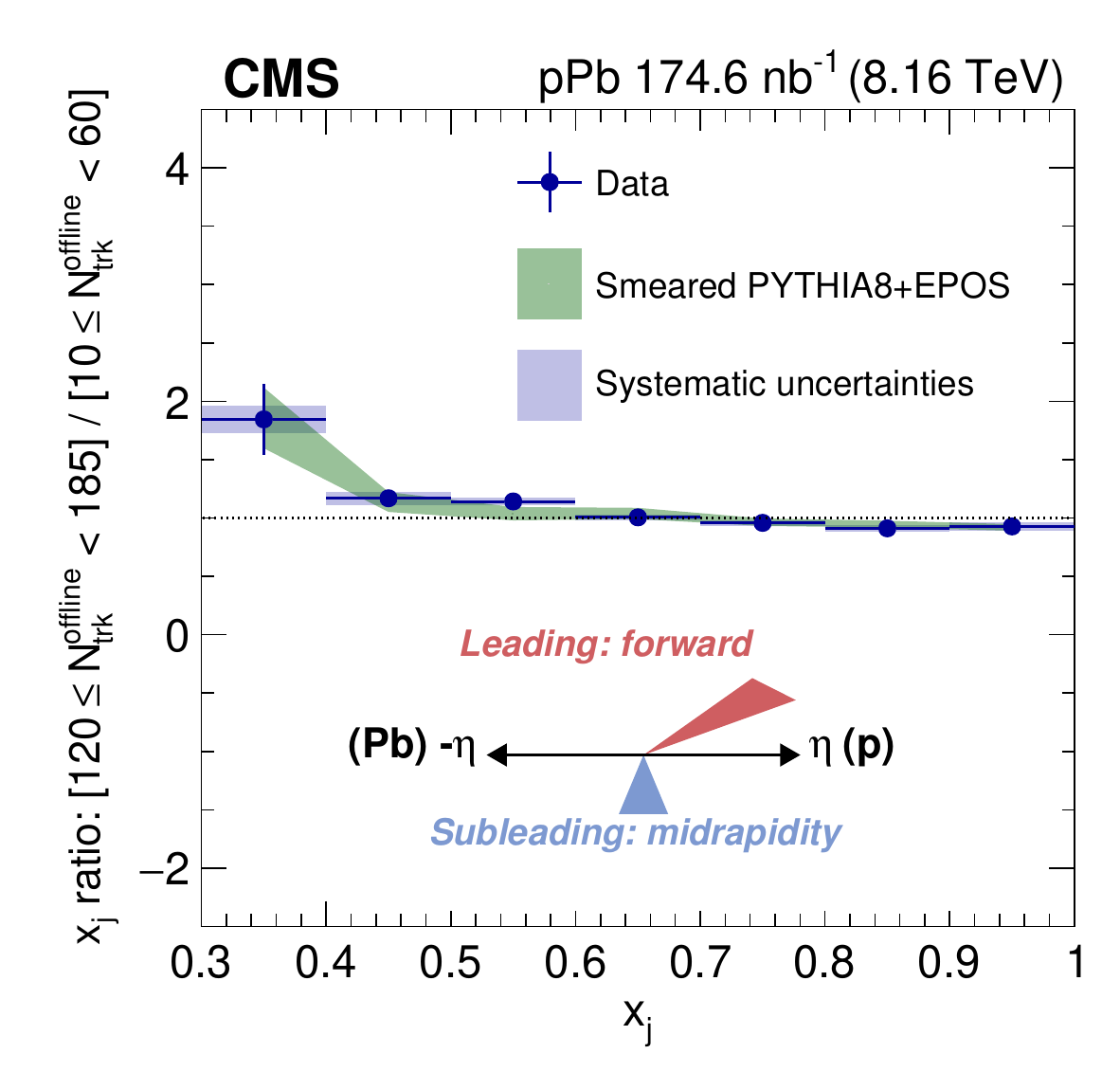} \\
    \includegraphics[width=0.44\textwidth]{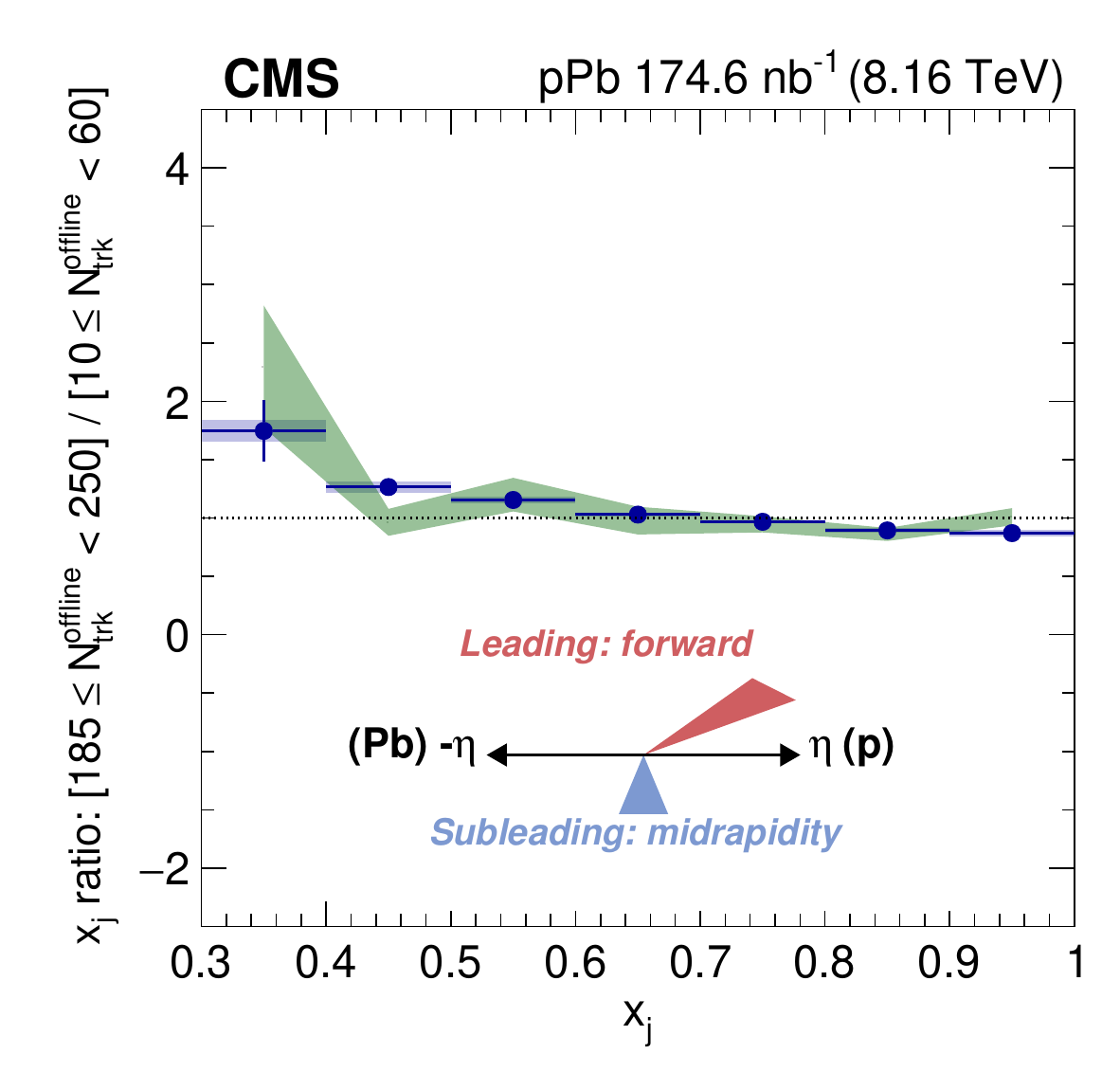} 
    \includegraphics[width=0.44\textwidth]{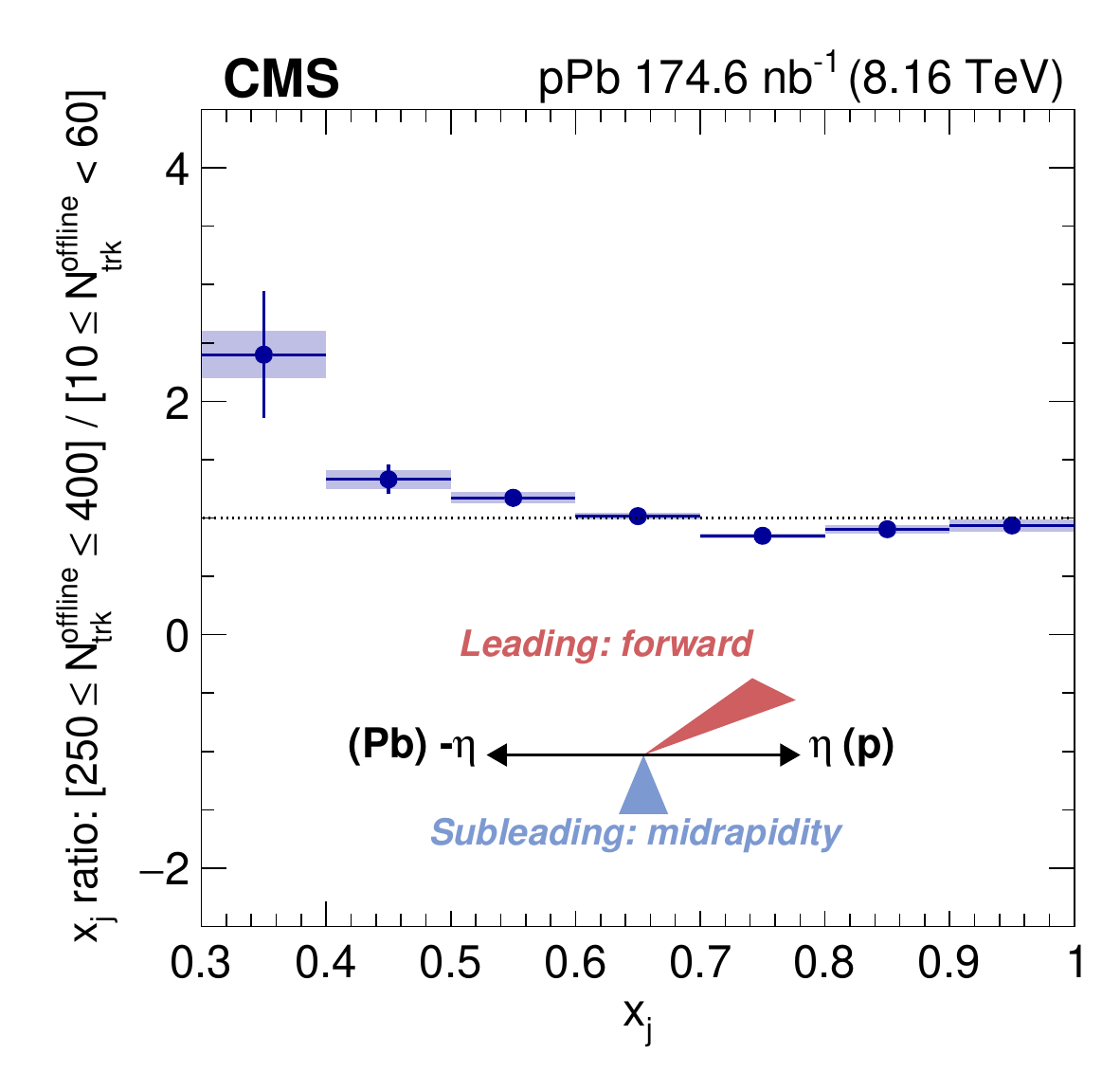} 
    \caption{The measured (raw) \xj ratio in the range $10 \leq \Ntroff < 60$, from lower to higher multiplicities for leading and subleading jets in the forward and midrapidity regions, respectively, are shown. Comparison with the reconstruction-level {{\PYTHIA}8}+\EPOS simulation (green band) are also displayed.}
    \label{fig:xjratiomultrec_fwd_mid}
\end{figure}

\begin{figure}[!hbpt]
  \centering
    \includegraphics[width=0.44\textwidth]{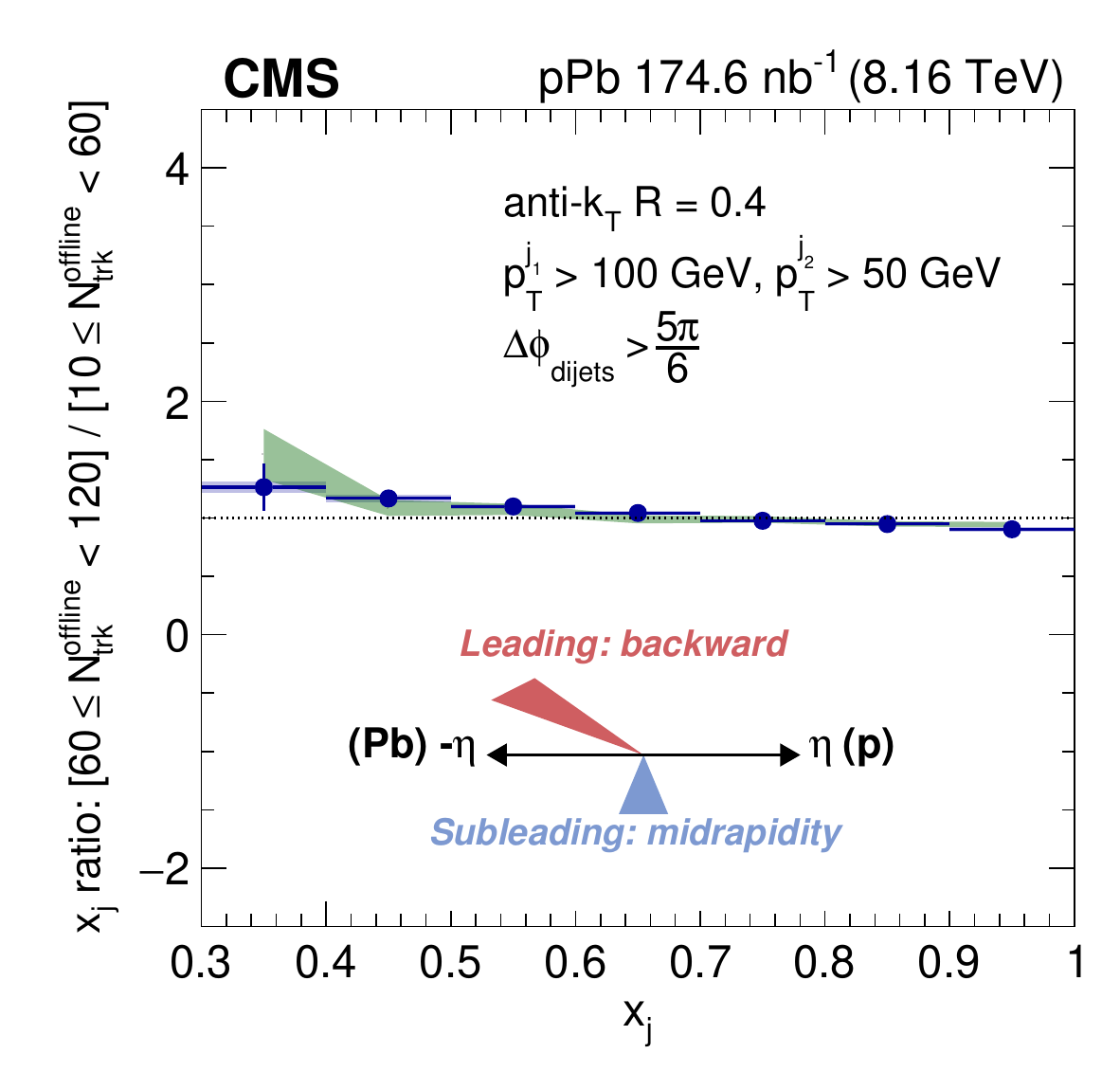} 
    \includegraphics[width=0.44\textwidth]{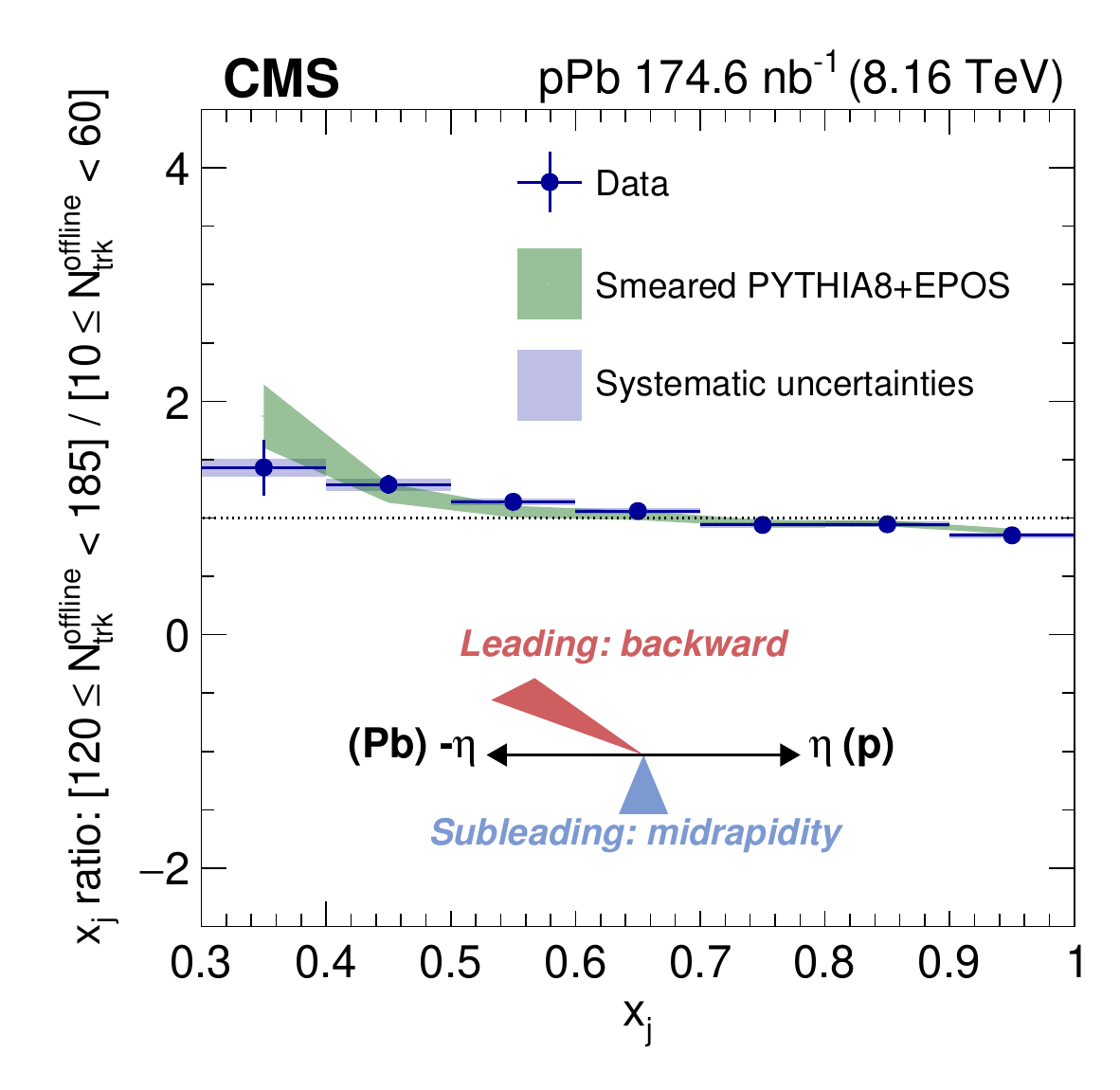} \\
    \includegraphics[width=0.44\textwidth]{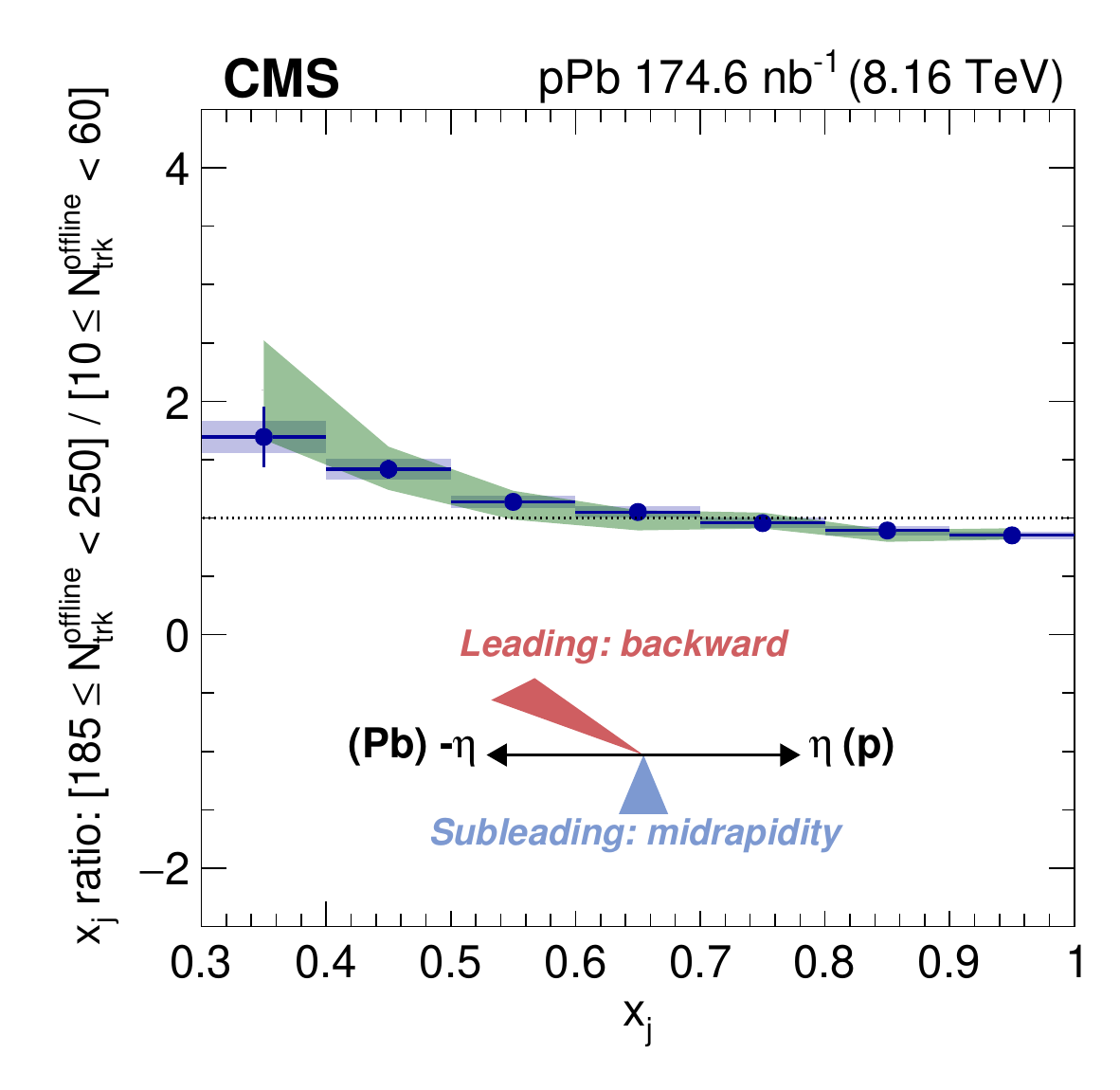} 
    \includegraphics[width=0.44\textwidth]{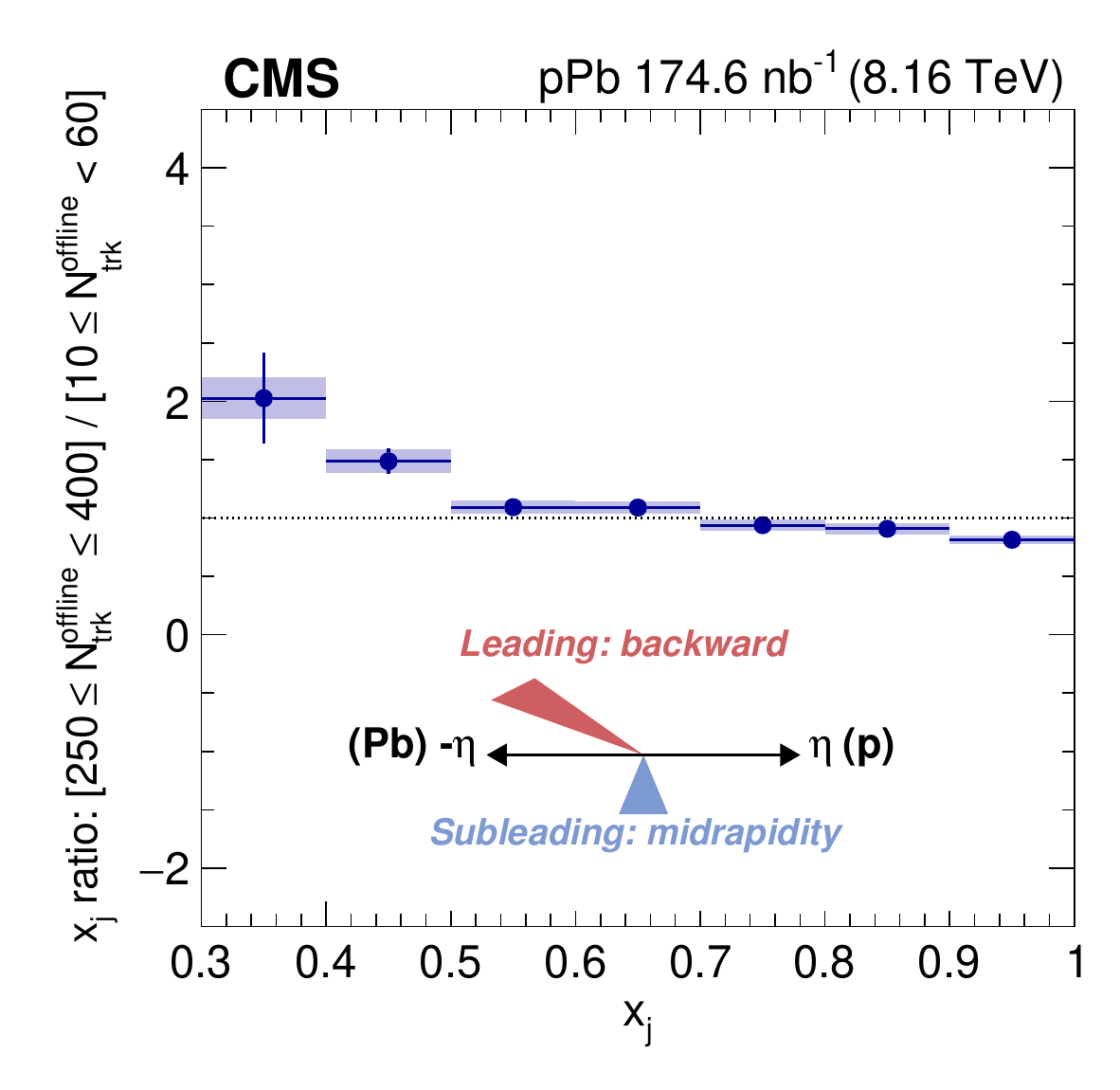} 
    \caption{The measured (raw) \xj ratio in the range $10 \leq \Ntroff < 60$, from lower to higher multiplicities for leading and subleading jets in the backward and midrapidity regions, respectively, are shown. Comparison with the reconstruction-level {{\PYTHIA}8}+\EPOS simulation (green band) are also displayed. }
    \label{fig:xjratiomultrec_bkw_mid}
\end{figure}

\begin{figure}[!hbpt]
  \centering
    \includegraphics[width=0.44\textwidth]{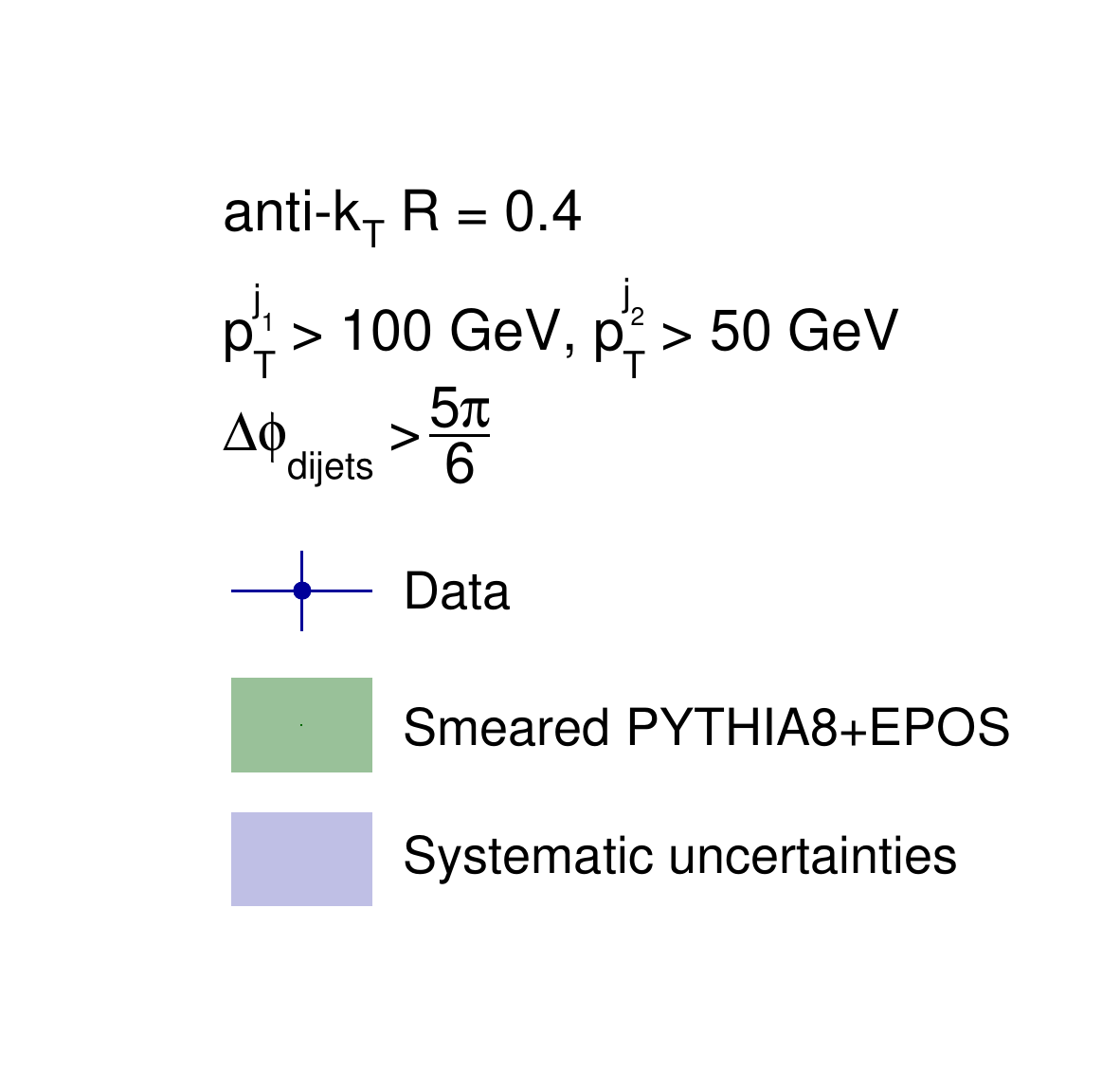} 
    \includegraphics[width=0.44\textwidth]{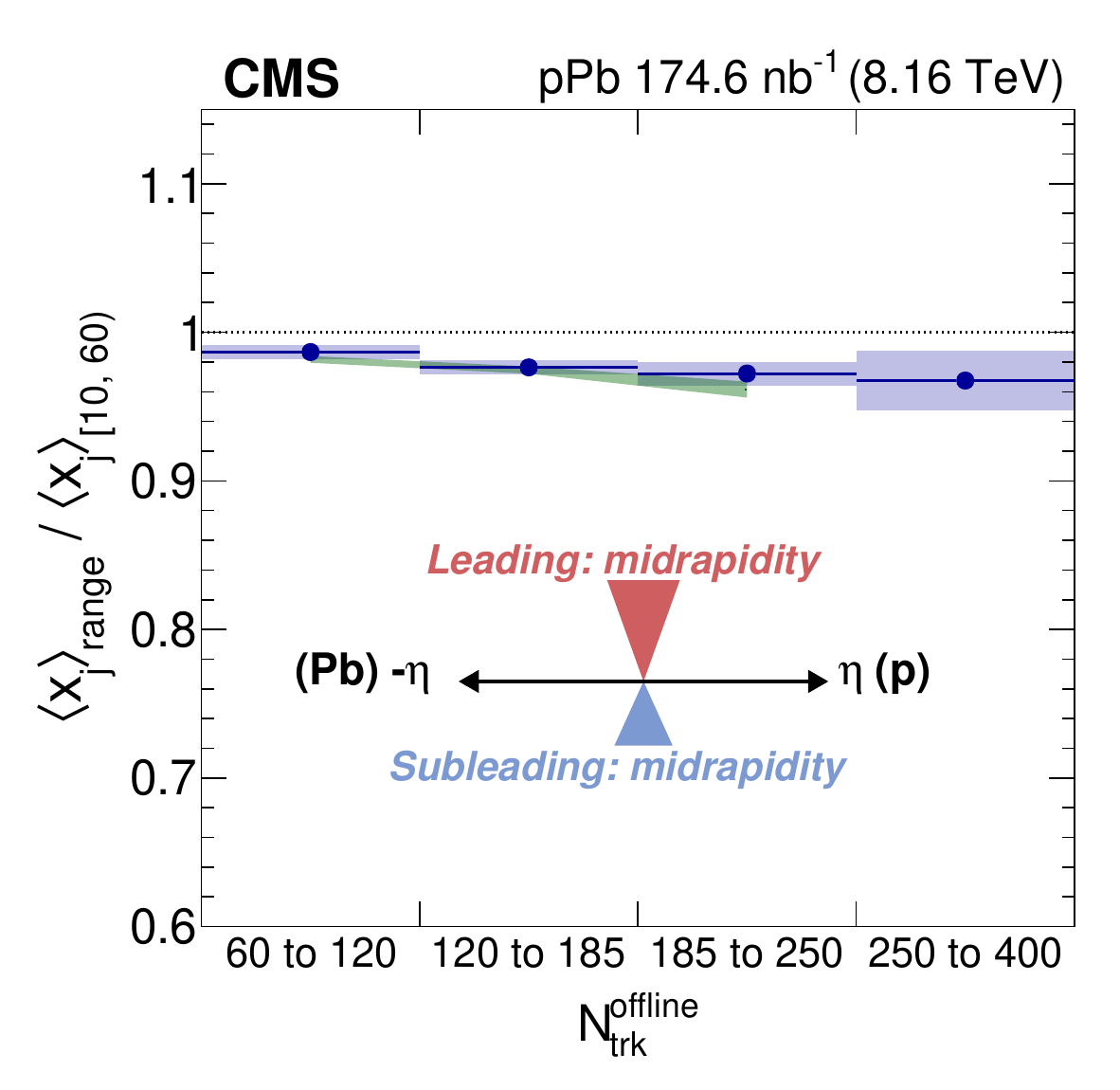} \\
    \includegraphics[width=0.44\textwidth]{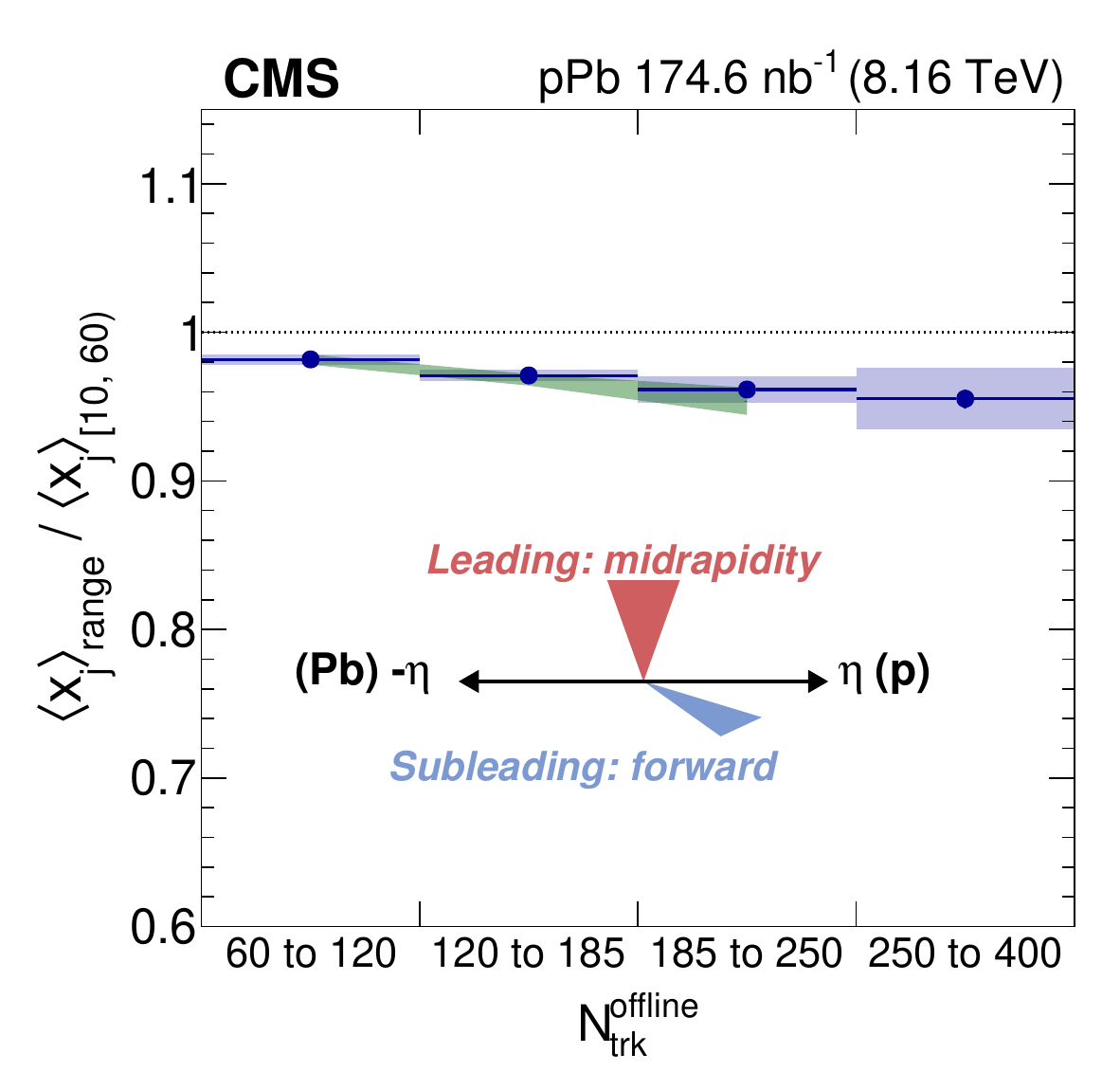} 
    \includegraphics[width=0.44\textwidth]{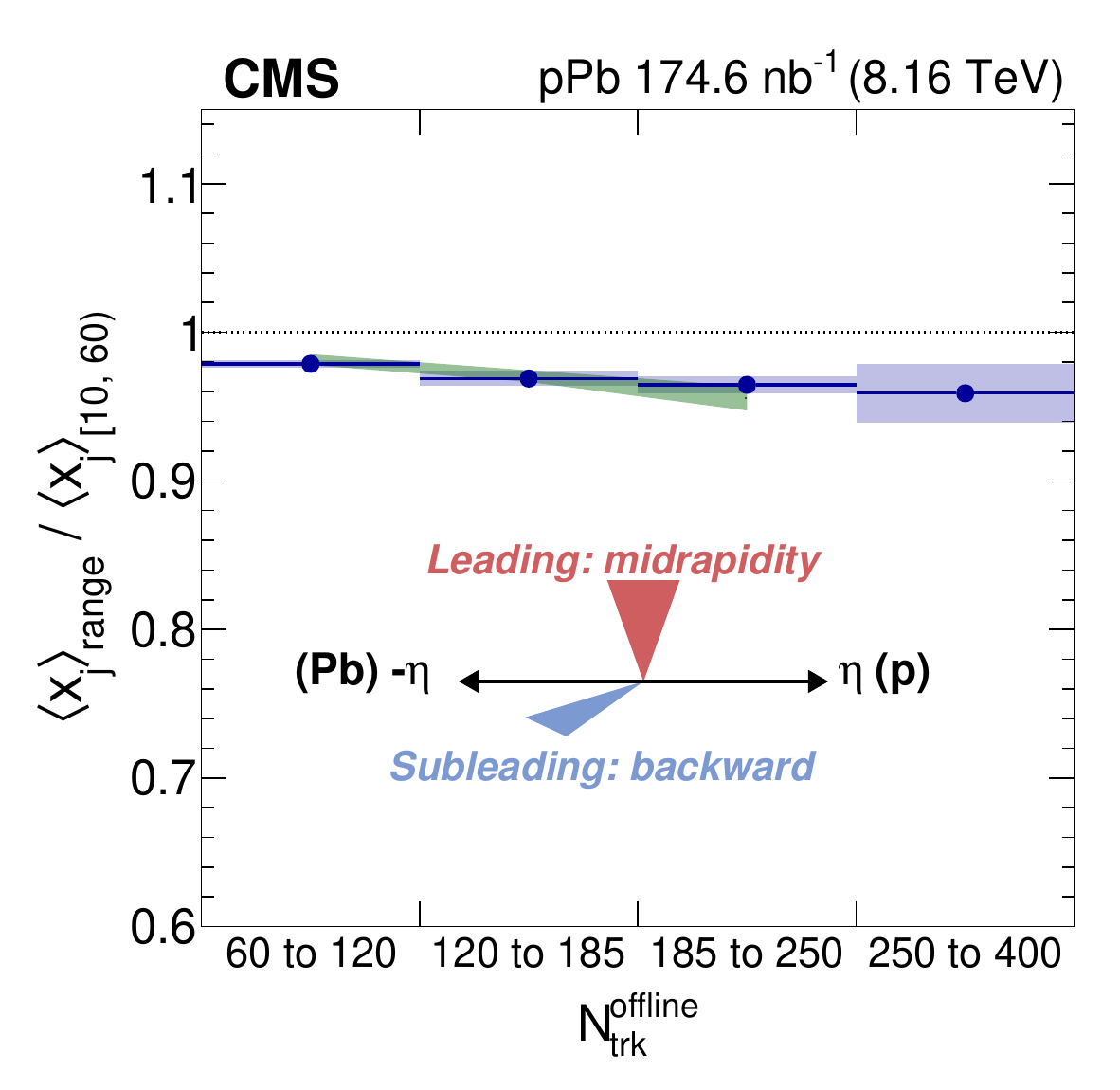} \\
    \includegraphics[width=0.44\textwidth]{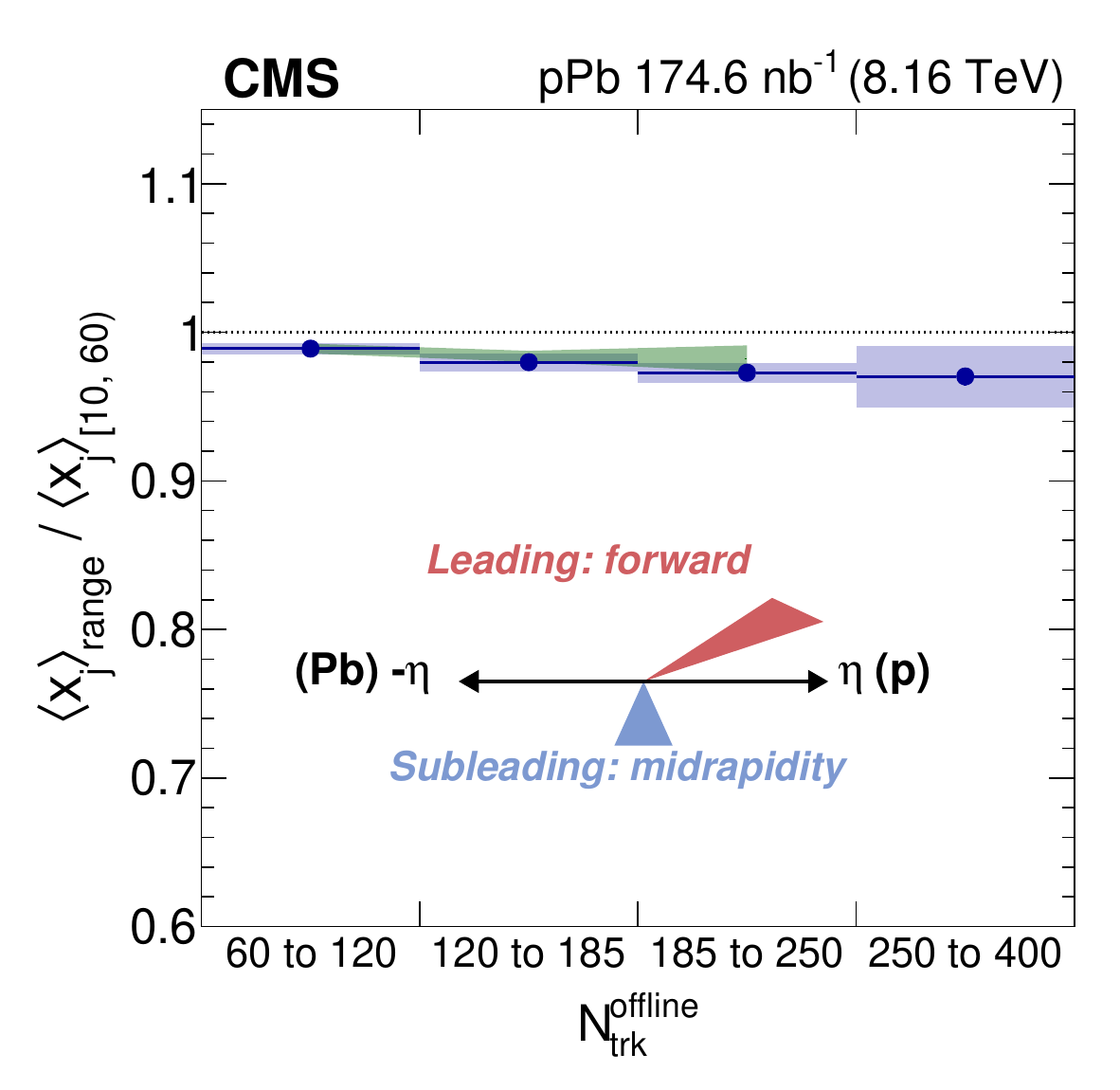} 
    \includegraphics[width=0.44\textwidth]{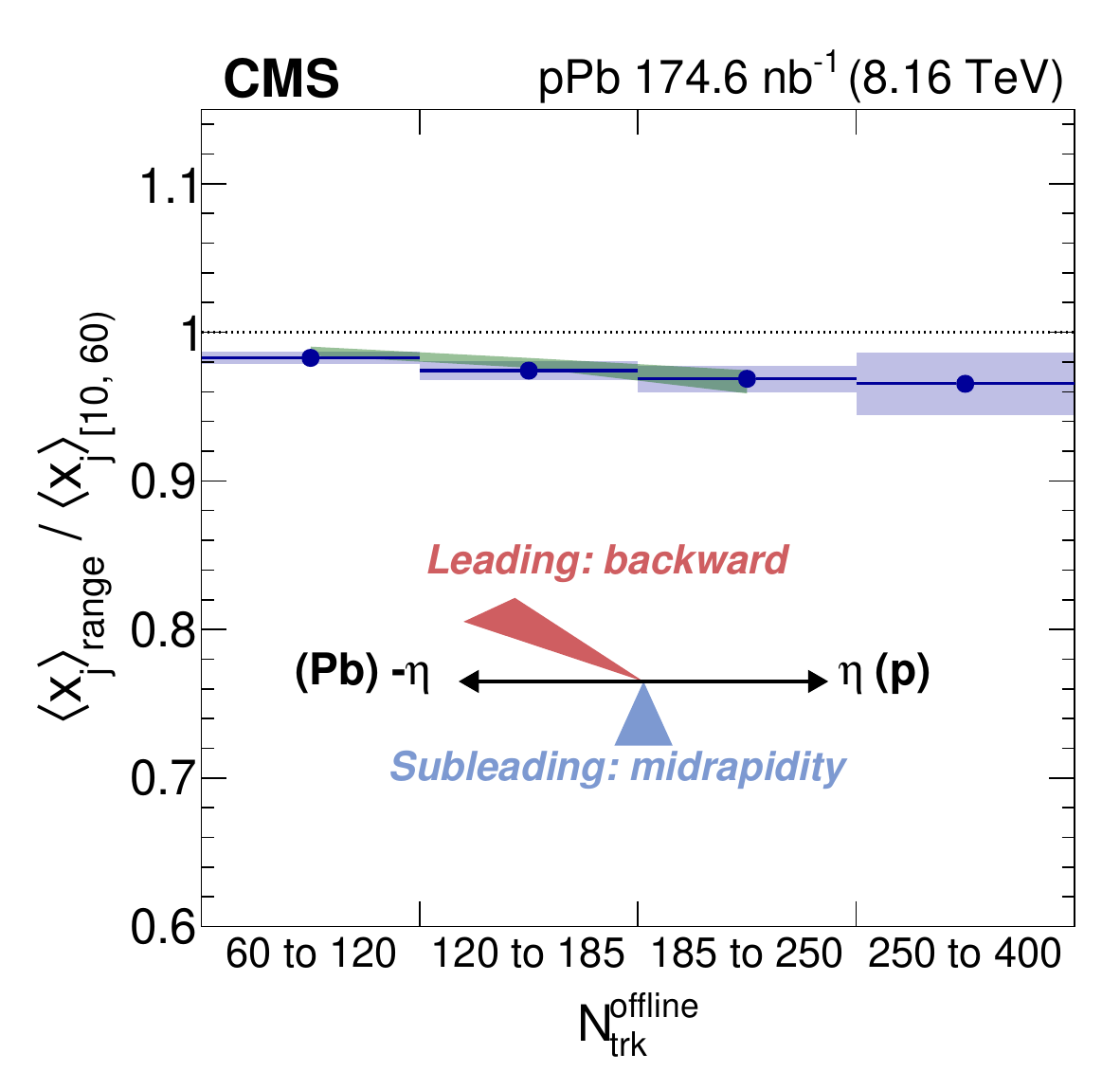} 
    \caption{Mean \xj ratio, \xja, in the range $10 \leq \Ntroff < 60$ as measured (raw) for different multiplicities, for different leading and subleading jet combinations for the midrapidity, forward, and backward regions, are shown. Comparison with the reconstruction-level (green band) {{\PYTHIA}8}+\EPOS are also displayed. The $\xja_{\text{range}}$ corresponds to the higher multiplicity ranges shown in the x-axis.}
    \label{fig:xjmeanratioreco}
\end{figure}

\cleardoublepage \section{The CMS Collaboration \label{app:collab}}\begin{sloppypar}\hyphenpenalty=5000\widowpenalty=500\clubpenalty=5000
\cmsinstitute{Yerevan Physics Institute, Yerevan, Armenia}
{\tolerance=6000
V.~Chekhovsky, A.~Hayrapetyan, V.~Makarenko\cmsorcid{0000-0002-8406-8605}, A.~Tumasyan\cmsAuthorMark{1}\cmsorcid{0009-0000-0684-6742}
\par}
\cmsinstitute{Institut f\"{u}r Hochenergiephysik, Vienna, Austria}
{\tolerance=6000
W.~Adam\cmsorcid{0000-0001-9099-4341}, J.W.~Andrejkovic, L.~Benato\cmsorcid{0000-0001-5135-7489}, T.~Bergauer\cmsorcid{0000-0002-5786-0293}, K.~Damanakis\cmsorcid{0000-0001-5389-2872}, M.~Dragicevic\cmsorcid{0000-0003-1967-6783}, C.~Giordano, P.S.~Hussain\cmsorcid{0000-0002-4825-5278}, M.~Jeitler\cmsAuthorMark{2}\cmsorcid{0000-0002-5141-9560}, N.~Krammer\cmsorcid{0000-0002-0548-0985}, A.~Li\cmsorcid{0000-0002-4547-116X}, D.~Liko\cmsorcid{0000-0002-3380-473X}, I.~Mikulec\cmsorcid{0000-0003-0385-2746}, J.~Schieck\cmsAuthorMark{2}\cmsorcid{0000-0002-1058-8093}, R.~Sch\"{o}fbeck\cmsAuthorMark{2}\cmsorcid{0000-0002-2332-8784}, D.~Schwarz\cmsorcid{0000-0002-3821-7331}, M.~Sonawane\cmsorcid{0000-0003-0510-7010}, W.~Waltenberger\cmsorcid{0000-0002-6215-7228}, C.-E.~Wulz\cmsAuthorMark{2}\cmsorcid{0000-0001-9226-5812}
% ...
\par}
\vskip\cmsinstskip
\dag:~Deceased\\
$^{1}$Also at Yerevan State University, Yerevan, Armenia\\
$^{2}$Also at TU Wien, Vienna, Austria\\
% ...
\end{sloppypar}
\end{document}